\documentclass [11pt,twoside,] {uwthesis}

\usepackage{alltt,multirow,booktabs,natbib,float,setspace}  %
\usepackage{amssymb,amsfonts,amsmath,graphicx,algorithm,algpseudocode}

\renewenvironment{glossary}
  {\begin{list}{}{\setlength\itemindent{\parindent}
   \def\makelabel##1{\large{##1:}\hfill}}}
  {\end{list}}

\newcommand{\qed}{\nobreak \ifvmode \relax \else
      \ifdim\lastskip<1.5em \hskip-\lastskip
      \hskip1.5em plus0em minus0.5em \fi \nobreak
      $\Box$\fi}
\newenvironment{proof}[1][Proof]{\begin{trivlist}
\item[\hskip \labelsep {\bfseries #1}]}{\end{trivlist}}
\newtheorem{thm}{Theorem}
\newtheorem{lem}{Lemma}
\newtheorem{prop}{Proposition}
\newtheorem{corollary}{Corollary}

\newtheorem{defn}{Definition}
\newtheorem{conj}{Conjecture}

\begin{document}

%

\prelimpages

%
%
\Title{Up-and-Down and the Percentile-Finding Problem}
\Author{Assaf Peretz Oron}
\Year{2007}

\Program{Statistics}

{\Degreetext{A dissertation
  submitted in partial fulfillment of\\
  the requirements for the degree of}
\date{\today}
 \titlepage
 }
\setcounter{footnote}{0}

%
%

\Chair{Peter D. Hoff}{Associate Professor}{Statistics}

\Signature{Peter D. Hoff}
\Signature{Paul D. Sampson}
\Signature{Barry E. Storer}
\signaturepage

%

\setcounter{page}{-1}





\doctoralabstractquoteslip


\setcounter{page}{-1}
{\small
\abstract{
\onehalfspacing
Up-and-Down (U\&D), a popular sequential design for estimating threshold percentiles in binary experiments, has not been a major subject of statistical research since the 1960's. U\&D application practices have stagnated, and significant gaps in understanding its properties persist. The first part of my work aims to fill critical gaps in U\&D theory. All U\&D variants studied generate Markov chains of treatments. New results concerning stationary distribution properties are proven. A commonly used U\&D variant known as ``fixed forced-choice staircase'' or ``k-in-a-row'' (KR), is proven to have a single-mode stationary distribution -- contradicting recent remarks about it in literature. Moreover, KR is shown to be the best-performing design in terms of convergence rate and estimation precision, compared with other designs targeting the same percentiles. A second focus of this study is nonparametric U\&D estimation. An improvement to isotonic regression called ``centered isotonic regression'' (CIR), and a new averaging estimator called ``auto-detect'' are introduced and their properties studied. Interval estimation solutions are also developed for both estimators.

Bayesian percentile-finding designs, most notably the continual reassessment method (CRM) developed for Phase I clinical trials, are also studied. CRM is believed to converge to exclusive allocation to a single optimal level, but to date, this belief had not been substantiated by extensive proofs. Here I add several proofs, expanding the understanding of when CRM might converge. In general, CRM convergence depends upon random run-time conditions -- meaning that convergence is not always assured. Small-sample behavior is studied as well.  It is shown that CRM is quite sensitive to outlier sub-sequences of thresholds, resulting in highly variable small-sample behavior between runs under identical conditions. Nonparametric CRM variants exhibit a similar sensitivity.

Ideas to combine the advantages of U\&D and Bayesian designs are examined. A new approach is developed, using a hybrid framework, that evaluates the evidence for overriding the U\&D allocation with a Bayesian one. This new design, called ``Bayesian Up-and-Down'' (BUD), is shown numerically to be more robust than CRM, and moderately outperforms either CRM or U\&D in terms of estimation.
}
}
%
%
\tableofcontents
\listoffigures
\listoftables

\acknowledgments{This work started out in fall 2003 as a routine consulting task at the University of Washington Departments of Statistics' consulting service, headed by Paul D. Sampson. I thank Prof. Sampson, the consulting client Dr. M.J. Souter (who approached us with the idea of using Up-and-Down for his experiment), consulting team leader Nancy Temkin of the Biostatistics department, and colleague David Dailey for arousing my initial interest and curiosity regarding the Up-and-Down design.

During 2004 I was able to continue working on Up-and-Down and support Dr. Souter's anesthesiology experiment, thanks to the work's designation as a dependent data (``stochastic'') preliminary exam project, under the guidance of Prof. Sampson. Profs. Galen Shorack and Matthew Stephens were members of the exam committee.

In summer 2005 I decided to turn the project into my dissertation, even though I was working with my advisor, Prof. Peter Hoff, on a rather different topic at the time. I thank Prof. Hoff and graduate coordinator Prof. Jon Wellner for the openness and flexibility they showed in supporting the transition.

Thanks to my full doctoral committee - Profs. Hoff, Sampson, Wellner, GSR Nancy K. Rivenburgh of Communications and Margaret O'Sullivan Pepe and Barry E. Storer of Biostatistics and the Fred Hutchinson Cancer Research Center. Thanks to Marloes Maathuis who provided valuable and patient help with isotonic regression theory.

Finally, thanks to my fellow students -- especially to my study pal Veronica Berrocal -- and to department staff, who were there to lend a helping hand along the way. Former graduate student advisor, the late Kristin Sprague (1947-2006), is fondly remembered.
}

\dedication{To the memory of my grandparents

\begin{center}
\textsc{Lotte Lieberman-Oppenheim (1908-1989)}

\textsc{Julius Oppenheim (1904-1993)}

\textsc{Esther Berger-Hochman (1904-1980)}

\textsc{Yehoshua-Shiyye Hochman (1906-2005)}
\end{center}

and my step-grandmother
\begin{center}

\textsc{Mania Kalmanovich-Hochman (1917-2007)}
\end{center}

To my parents Rachel and Mosh\'{e} Oron and parents-in-law Bat Sheva and Shimon Eliyahu,

and most of all to my wife Orna Eliyahu-Oron and sons Daniel, Guy and Ben,

who have tolerated and supported me through this journey.
}

\textpages
\onehalfspacing
\chapter{Preface and Introduction}\label{ch:pref}

\section{Preface}
In many binary-response experiments, researchers are not interested in finding the entire treatment-response curve, but only a single representative value, which can be described via a percentile of the underlying threshold distribution $F$. This percentile can represent (for example) the sensory threshold, the $LD_{50}$ of a toxic substance, or the maximum-tolerated-dose (MTD) of a medication. Such experiments are typically limited to a sample of several dozen trials or even less, and with little prior knowledge about the form of $F$. An additional, rather common constraint is that treatment levels are limited to a fixed, discrete, finitely-spaced set.

Statisticians have been developing designs to answer this type of challenge for at least two generations. Yet, interestingly, if an applied researcher were to seek statistical consulting on this problem, he or she would receive wildly differing recommendations, depending upon which field they work in or which statistician happened to be the consultant. Moreover, an applied statistician wishing to consult a client on this issue would discover, upon a cursory literature search, that any clear, well-tested and practical guidelines (if they exist) are drowning in a sea of contradictory ``best practices'' recommendations, antiquated references, unverified claims and controversy. {\bf In short: until quite recently, there was no up-to-date, standard reference for this problem.}\footnote{Apparently, I am not the only person to have noticed this. \citet{Chevret06} is a recent attempt to fill the gap. \citet{PaceStylianou07}'s review is an attempt to update the anesthesiology community with recent statistical developments relevant to their needs.}

These are not hypothetical statements; this is exactly what happened to me in late 2003, when I was enlisted to help Dr. M. J. Souter design an anesthesiology percentile-finding experiment \citep{Oron04,BhanankerEtAl07}. My search for solid and reliable solutions has evolved into the dissertation you are reading. After this tortuous journey, I find it important to repeat in this introduction a well-known truism, which is somehow forgotten in this field's prevalent patterns of debate. Here it is, in the form of a disclaimer:

\begin{quote}
Constrained by the binary response, a small sample size, an unknown response threshold distribution and a discrete treatment set, {\bf the quality of information the experiment can provide is limited}. Failed experiments (however one chooses to define "failure") are far from unlikely. The degree of confidence expressed by many statisticians promoting this or that method is exaggerated.\footnote{\citet{Fisher07}, in a recent \emph{Anesthesiology} editorial, voiced similar concerns regarding the use of `Up-and-Down' methods.}
\end{quote}
Keeping this perspective in mind, there is quite enough room for theoretical insight and methodological developments. The next section will expand and clarify these statements.

\section{Introduction}
\subsection{Conceptual Framework}\label{sec:prefconcept}
As the opening paragraph explains, we are looking for a treatment value $x$ that will generate positive response with probability $p$. We assume that this probability is monotone increasing in $x$. Furthermore, we assume that the response is deterministically triggered via an underlying population (sub)distribution of response thresholds $F(t)$, with a density $f(t)$.\footnote{This last assumption is often wrong, especially in psychophysical experiments to find the sensory threshold of a single subject. In that field, it is common to apply fixes such as added parameters for the false positive and false negative rates. However, the simplifying assumption allows us to treat $x$ and $t$ as one and the same, and focus on the core features of the problem.} We aim to estimate $Q_p\equiv F^{-1}(p)$, known as the target percentile, or simply {\bf the target}.

An excellent, perhaps ideal experiment, would be to \emph{directly sample the thresholds themselves}, $\{T_n\}\sim f$. If this is possible, then without knowledge of $F$ a good target estimate would be $T_{(np)}$, the sample $np$-th order statistic (or $100p$-th percentile). For large $n$, $T_{(np)}$ is approximately normal with asymptotic variance
\begin{equation}\label{eq:varTq}
Var\left[T_{(np)}\right]\approx\frac{p(1-p)}{nf^2(Q_p)}.
\end{equation}
This can be perceived as a rough precision benchmark for the percentile-finding problem.

Our conditions, however, are far from ideal. Rather than the thresholds themselves, we observe binary responses that can be interpreted as {\bf censored current-status information:} at a given trial indexed $i$ with treatment $x_i$, we know that $t_i\leq x_i$ if the response was {\em `yes'} and $t_i>x_i$ if it was {\em `no'}. If we pool all our treatments to the same value $x_0$, we can estimate $F(x_0)$ with variance $F(x_0)\left[1-F(x_0)\right]/n$. This is not a very impressive feat; for example, unless $F(x_0)$ is close to $0$ or $1$, we need $n\approx 100$ to get a $95\%$ CI of $\pm 0.1$ for $\hat{F}(x_0)$. And the best we can achieve this way, is knowing whether $Q_p>x_0$, $Q_p<x_0$, or (if we are lucky) $Q_p\approx x_0$. It is no wonder that in the related problem of binary-response opinion polls (which are analogous to estimating $F$ at a single specified $x_0$), the minimal acceptable sample size is several hundred.

If we try to cover a larger range of treatments by splitting the sample equally between $m$ levels $(l_1\ldots l_m) $, we will obtain $m$ even less precise point estimates, with many observations taken quite far from $Q_p$ -- indicating a waste of resources. The MLE with no knowledge of $F$'s shape is not much more than an interpolation between these rather blunt point estimates (this is examined more closely in Chapter \ref{ch:est}). Taking the range-covering approach further, if we could let $x$ vary randomly and continuously, the nonparametric MLE of $F$ for current-status data converges -- both globally and locally -- only at an $n^{1/3}$ rate \citep{KimPollard90,GroenenboomWellner92}.

In short, the problem's properties imply that our prospects for solidly successful target estimation are not very promising.

A way out of this quagmire would be some ``intelligent'' sequential scheme, that would lead us to collect most of our information in the vicinity of $Q_p$ -- even though before our first trial we have little knowledge where that target might be -- and then allow for reasonably useful estimation.\footnote{Observing (\ref{eq:varTq}), the concentration of observations around target is tantamount to artificially increasing $f(Q_p)$ -- with a linear effect on estimation error.} Again, as hinted previously, statisticians and others have suggested many such schemes, but the most widely used methods fall almost exclusively into one of two groups. These are Markovian or {\bf Up-and-Down (U\&D)} designs, and Bayesian schemes.

I shall revisit this generic sampling perspective occasionally throughout the dissertation.\footnote{See Sections~ \ref{sec:undconcept},\ref{sec:estconcept},\ref{sec:crminterp}, \ref{sec:crmconcept} and 5.3.1. To complete this bird's-eye view of the dissertation, I also suggest Section \ref{sec:estopt} -- which summarizes my U\&D-related findings in the form of practical recommendations -- and the conclusions.} Typically, these will also be the only places where the response threshold distribution might be referred to as $F(t)$. Otherwise, because of the threshold-triggering assumption spelled out above, I will refer to the threshold CDF as $F(x)$: an (indirect) function of treatments.

\subsection{Some Definitions and Constraints}

Before we proceed further, it may be a good idea to define the two design approaches. The following definition is broad enough to include all methods discussed in the thesis, and is in the spirit of \citet{Storer01,Potter02} and \citet{IvanovaFlournoy06}.

\begin{defn}\label{def:zero} (i) An {\bf `Up-and-Down' (U\&D) or algorithm-based design} is any percentile-finding, binary-response sequential experimental design with treatments and responses $\{x_n\},\{y_n\}$, respectively; with fixed, discrete treatment levels $\{l_m\}$ and sequential treatment allocation rules, which are functions of $\{x_n\},\{y_n\}$ and possibly also $\phi$, a set of fixed parameters.

(ii) A {\bf Bayesian or model-based scheme} is as in (i) above, but the allocation rules are functions of $\{x_n\},\{y_n\}, \phi$, and a statistical model $G\left(\{x_n\},\{y_n\}\mid\phi,\theta\right)$, where $\theta$ is a set of data-estimable parameters.
\end{defn}

As Definition~\ref{def:zero} indicates, the dissertation will only deal with designs constrained to a fixed discrete set $\{l_m\}$. This excludes methods such as stochastic approximation and its descendants \citep{RobbinsMonro51,Lai03}. Additionally, we will usually assume that $\{l_m\}$ is evenly spaced (whether on the original or on a transformed scale), with level spacing $s$.\footnote{In some cases we will assume $\{l_m\}$ is finite and in others not, depending upon context. Usually the difference between the two cases is more notational than substantial.} Other simplifying assumptions are that the underlying threshold CDF $F$ is continuous strictly increasing, that it has a density $f$, and that thresholds in the experiment's trials are i.i.d. draws from $f$.

These assumptions will save a great deal of burdensome notation, and should help us focus on essential design and estimation properties rather than on terminology.\footnote{Most results discussed here can be easily extended to cases when one suspects $F$ to behave otherwise, or wishes to run the experiment on an unevenly spaced $\{l_m\}$.}

\subsection{Audience and Scope}
Percentile-finding is an application encountered in many research fields. However, two fields have dominated the design debate. In one of them (psychophysics or sensory studies), it is because the field's most routine experiment - determination of sensory thresholds - is modeled as a percentile-finding problem. Proper treatment of the sheer wealth and breadth of percentile-finding designs used in psychophysics may take several books. Fortunately, the constraints outlined above already filter out a large part of these designs.

The second field (Phase I clinical trials) is dominant not so much because of its prevalence, or because of the wealth of designs. In fact, the vast majority of Phase I trials still appear to use conservative protocols, which most statisticians (myself included) view as outdated. The `Phase I' application has dominated debate mostly because of the large amount of resources and statistical attention commanded by clinical trials in general.

The typical practical constraints in these two fields are very different. In psychophysics, most experimental runs are carried out on a single subject. Sample size and trial cost are usually not a problem, and in principle the treatment can vary continuously (though most designs prefer to stick to a discrete set). The sensory threshold itself is also defined quite flexibly: anything from the median to the $80$th percentile may be commonly encountered. By contrast, Phase I trials are limited to small sample size ($<40$), and are usually carried out in cohorts of $3$ patients. The target is typically between the $20$th and $33$rd percentile, and due to toxicity concerns it is preferably approached from below.

In between, the ``vanilla'' engineering and scientific applications are usually median-targeting, size-constrained, and for lack of statistical attention have continued to plough along with 1940's to 1960's methodology. The three application types are summarized in Table~\ref{tbl:applications}.

\begin{table}
\begin{center}
\caption[Most Common Percentile-Finding Applications]{Side-by-side comparison of the properties of three most prominent types of percentile-finding applications mentioned in the text.\label{tbl:applications}}
{\small
\begin{tabular}{p{3.5cm}p{3.cm}p{3.cm}p{3.cm}}
\toprule
{\bf Application} & {\bf Psychophysics} & {\bf Phase I} & {\bf General Engineering} \\
\toprule
{\bf Target} & Median or above & $20$th to $33$rd Percentile & Median \\
\midrule
{\bf Cohort Size} & 1 & 2 -- 6 & 1 \\
\midrule
{\bf Major Constraints} & Simplicity, response fatigue/reliability  & Toxicity, medical ethics, volunteers, duration & Simplicity, cost, duration, toxicity (in some cases) \\
\midrule
{\bf Currently Popular Designs} & A large variety, including U\&D and Bayesian designs & `3+3' protocol, with Bayesian designs increasingly popular & U\&D \\
\bottomrule
\end{tabular}
}\end{center}
\end{table}

My approach has been to steer clear of complete commitment to a single field's specific needs. The rationale is that this is precisely a statistical researcher's role: to look at the generic problem, examine and formulate solutions to it - solutions useful for all applications. From these generic solutions, it should not be difficult to tailor specific applications. All that being said, given the medical motivation that introduced me to the problem, the terms and concerns encountered in medical literature are more prevalent in this thesis than those of other fields.

\subsection{Thesis Layout}
The thesis focuses on the most veteran family of U\&D designs, which includes the original 1940's method \citep{DixonMood48}. Chapter~\ref{ch:und} tackles the designs themselves, while Chapter~\ref{ch:est} is devoted to estimation. In recent decades, Bayesian model-based designs known under acronyms such as QUEST and CRM \citep{WatsonPelli83,OQuigleyEtAl90} have quickly eclipsed U\&D in the number of statisticians studying them (though not in their prevalence of use). These designs are discussed in Chapter~\ref{ch:crm}.
There have been some attempts to combine the two approaches, and I have devoted quite a bit of energy to this as well. Combined approaches are discussed in Chapter~\ref{ch:bud}, which is followed by the conclusions. {\bf To minimize confusion between the thesis terminology and that of other publications, I will refer to the algorithm-based design family simply as U\&D, and to the model-based family as CRM or ``Bayesian''.}

Rather than precede the thesis body with a long and tedious introduction to all these subjects, I have opted to start each chapter with an introductory section, which reviews the current state of knowledge on that chapter's topic. Such sections are followed by sections with theoretical results, and then sections with new methodological developments. In Chapter \ref{ch:est}, the dissertation's largest chapter, there is also a separate section presenting numerical results \footnote{In other chapters, numerical results are embedded within the sections.}, a section reporting and analyzing the results of the motivating anesthesiology experiment, which has been recently completed \citep{BhanankerEtAl07} -- and finally, a section summarizing Chapters~2-3 with a list of practical recommendations. At the end of Chapters 2, 3 and 4 the reader can find glossaries for these chapters.

\subsection{The Role of Simulation}
In fall 2003, while still on the anesthesiology consulting job, team leader Nancy Temkin suggested that I use computer simulation to see how different U\&D designs pan out. I proceeded to do so, unwittingly following in the footsteps of Wetherill, who in the 1960's traveled across the ocean to make use of computing resources not available at the time in Great Britain \citep{Wetherill63,WetherillEtAl66}. This was such a novelty back in England, that the mere use of simulations became a major gossip topic during the customary Royal Society discussion appearing in \cite{Wetherill63}. Ever since then, for better or worse, simulation has played a central role in evaluating percentile-finding designs and estimators. Sometimes this approach is taken to the extreme: recently, \citet{OQuigleyZohar06} in their mini-review of CRM claimed that numerical simulation should be the main approach in studying and evaluating designs for the small-sample percentile-finding problem.

{\bf I could not disagree more.}

At their best, simulations can perform a role analogous to that of lab experiments -- proving or disproving a theory, demonstrating a theoretical result, and revealing novel patterns that inspire further theoretical research. Much of the painstaking work carried out by Garc\'{i}a-Perez and colleagues over a decade, examining percentile-finding designs in psychophysics, belongs to this category \citep{GarciaPerez98,AlcalaQuintana04,GarciaPerez05}.

Not at their best, simulations can be a lazy man's escape from doing theory, producing individual trees that are then mistaken for a forest. A classic example is Wetherill's ``reversal'' estimator (see Chapter~\ref{ch:est}), chosen in a large part because of favorable simulation outcomes \citep{WetherillEtAl66}. As Section~\ref{sec:est2} will show, in general this estimator is in fact inferior.

At their worst, simulations can be abused to further irrelevant agendas. Here the classic example is the rise of Bayesian designs for Phase I trials. Researchers have used simulations to ``prove'' this approach's advantage over others, most notably U\&D \citep{OQuigleyChevret91,BabbEtAl98}. Immediately upon presenting the first such simulation study, the authors openly called upon regulatory authorities to discontinue use of U\&D \citep{OQuigleyChevret91}. While this call was not heeded, such simulations and the confidence with which they were presented have doubtlessly played a role in shutting most of the statistical world off to U\&D. But in both cases the comparisons have been ``rigged'' in several ways against U\&D (see section \ref{sec:crm1} for details). More neutral comparisons \citep{GarciaPerez05} have yielded an almost diametrically opposite conclusion. I hope this thesis would help explain why.

During my work on this problem, I have spent thousands of human and CPU hours on simulation. I tried to keep simulations ``at their best'' as defined above. Many times I have found myself to be lazy, discovering that $2$ months of simulations have saved me... $10$ minutes of theory. I have been quite wary, and hopefully have succeeded, in keeping myself from abusing simulations, as happened in the cases mentioned above.

As the thesis materialized, I have tended to base more and more of my results on theory and not on simulation. This has been a mostly successful endeavor. Simulation in this thesis is confined mainly to illustrating the pros and cons of different designs and point estimators under different scenarios, or for such essential technical tasks as evaluating the coverage of interval estimators. There are one or two exceptions, where the numerical pattern strongly suggests a theoretical result, but my quest for that result has been incomplete (U\&D convergence comparison in Section \ref{sec:und2}, smoothed isotonic regression in Chapter \ref{ch:est}). Here I have to cling to the well-worn, job-security clich\'{e}: hopefully future work will find that proof.

\chapter[Up-and-Down Designs]{The Up-and-Down Design and Some Variants}\label{ch:und}

\section{Description and Current Knowledge}\label{sec:und1}
\subsection{Basic Terminology}
Let $X$ be some experimental treatment with a binary response $Y(X)\sim \mathsf{ Bernoulli}\left(F(X)\right)$. We assume $F$ is continuous and strictly increasing. Researchers look for $Q_p\equiv F^{-1}(p)$, the $100p$-th percentile (or $p$-th quantile) of $F$.  Treatments are administered sequentially: both treatments and responses are indexed $x_i, y_i,\ i=1\ldots n$. Following trial $i$, the next treatment $x_{i+1}$ is determined by some subset of all previous treatments and responses $\left\{x_1\ldots x_i, y_1\ldots y_i\right\}$, and by the specific method's transition rules. In U\&D methods, experimental treatments are restricted to a fixed set of levels $l_u, u=1,\ldots m$, with level spacing $s$.\footnote{However, since $F$ is continuous, we shall sometimes refer to $x$ as a continuous variable when studying properties of $F$.} The subsequent discussion focuses on targets at or below the median, but all results can be trivially translated to targets above the median.

\subsection{History}
The up-and-down (U\&D) sequential method was developed in the 1940's \citep{DixonMood48,vonBekesy47,AndersonEtAl46}. It is designed to estimate the median threshold, under constraints of moderate sample size and a discrete set of treatment levels. It is still in use today in its original applications - finding the median sensitivity of explosives to shock loading \citep{ChaoFuh03} and estimating sensory thresholds \citep{Treutwein95,GarciaPerez98}. It can also be found in a wide array of applied research fields - including fatigue testing in metallurgy and material science \citep{LagodaSonsino04}, testing of dental restorative materials \citep{ScherrerEtAl03}, breakdown voltage estimation in electrical engineering \citep{KomoriHirose00}, animal toxicity response \citep{SunderamEtAl04}, and anesthesiology \citep{CapognaEtAl01,DroverEtAl04}. In numerous applications, U\&D is considered a standard method \citep{JSME81,ASTM91,OECD98, NIH01}. Note on terminology: the term ``up-and-down'' used to refer only to the median-finding design, but hereafter I shall use it in its more recent interpretation - any percentile-finding design that can be seen as an extension of the original method. That original method will be called `simple up-and-down' (SU\&D).

Methodological research during U\&D's early years \citep{BrownleeEtAl53,Derman57,Wetherill63,Dixon65,WetherillEtAl66,Tsutakawa67} generated accepted practices for design and estimation. The first proper identification of the U\&D treatment sequence as a Markov chain is credited to \citet{Tsutakawa67block,Tsutakawa67}. Paradoxically, just as this discovery was made, and as U\&D was proliferating into more and more fields, statistical interest in the method had begun to dwindle. From the mid-1970's until the late 1980's, very few U\&D theoretical or methodological studies were published. This gap has probably contributed to the rather uneven pattern of usage observed today. Different fields adhere to different practices, whose rationale has often been lost in the mists of time.

Sample size is a case in point. \citet{DixonMood48} originally recommended a sample size of $n\thickapprox 40$ for median estimation. Shortly afterwards, there was a trend to tailor the method for smaller and smaller sample sizes, $n<10$ \citep{BrownleeEtAl53,Dixon65}. Nowadays, some military industries still use $n\thickapprox 40$ for median-sensitivity estimation (H. Dror, personal communication, 2006). Some groups in anesthesiology use a fixed sample size of $30$ for ED$_50$ estimation \citep{CapognaEtAl01,CamorciaEtAl04}. In the same field, \citet{PaulFisher01} report that many anesthesiologists use the fourth reversal as a stopping rule -- roughly equivalent to a dozen treatments (reversals will be defined later in this section). Meanwhile, many researchers in anesthesiology and elsewhere use $n<10$ for the same goal \citep{Lichtman98,SunderamEtAl04,DroverEtAl04}.

Since around 1990, methodological interest in U\&D has been gradually picking up again. Practitioners in fields using U\&D have expressed increased interest and skepticism about the method, leading to extensive studies -- typically numerical -- and to ad-hoc improvement suggestions \citep{GarciaPerez98,ZhangKececioglu98,BraamZwaag98,LinLeeLu01,PaulFisher01,PollakEtAl06}. Several statistical teams have entered the fray as well, mostly in the context of Phase I clinical trials, which target the 20th to 33rd percentiles \citep{Storer89}. This recent research has yielded results that incorporate advances in the study of Markov chains \citep{DurhamFlournoy94,DurhamFlournoy95,DurhamEtAl95,Gezmu96,GiovagnoliPintacuda98,IvanovaEtAl03,BortotGiovagnoli05,GezmuFlournoy06} and in estimation \citep{KomoriHirose00,ChaoFuh01,StylianouFlournoy02,StylianouEtAl03}.  Much of this progress is summarized by \citet{IvanovaFlournoy06} and \citet{PaceStylianou07}. All the same, some of the U\&D's basic properties are still seldom discussed.

But perhaps before everything else, the simple question of which U\&D variant to choose for a given application, especially for non-median targets, has not been properly addressed. Since the 1950's researchers have creatively proposed many dozens of non-median U\&D variants (e.g., \citet{WetherillEtAl66,Storer89,DurhamEtAl95,GarciaPerez98,IvanovaEtAl03,BortotGiovagnoli05}). Yet, only a handful of articles actually compare different U\&D methods' performance, and usually only via limited-scope simulations.\footnote{The only exception that comes to mind is \citep{BortotGiovagnoli05}, who prove that BCD is optimal among first-order Markovian methods. But applied researchers do not care about a method's Markov-chain order; they would like to know which method is best for estimating a given percentile, under certain practical constraints.} There are also many innovative non-U\&D sequential designs to achieve the same goal \citep{OQuigleyEtAl90,RosenbergerGrill97}, and one may legitimately question whether U\&D is the best platform. But a comparison of U\&D with other approaches means little, if the U\&D design and estimation methods used are clearly inferior to other available U\&D methods, of which the researchers were not aware.

Another interesting aspect of the historical gap in U\&D research is the method's perceived nature. U\&D was originally tied to a parametric assumption: the thresholds were assumed to be normally distributed \citep{DixonMood48,Dixon65}. Nowadays U\&D is almost universally seen as a nonparametric design. There has been little theoretical exploration into the ramifications of this change in perception, regarding design and estimation as actually practiced in the field.

\subsection{Transition Probabilities, Balance Equations and Stationary Formulae}

The original, median-targeting simple up-and-down (SU\&D) starts at an arbitrary level, and then moves up or down $1$ level, following {\it `no'} or {\it `yes'} responses, respectively. This makes SU\&D a lattice random walk, with `up' and `down' transition probabilities $p_u=1-F(l_u)$ and $q_u=F(l_u)$, respectively.\footnote{On the boundaries, trivial ``reflecting'' conditions are imposed; this type of boundary conditions is assumed throughout the text, but explicit boundary details are omitted for brevity.} Assuming $F(l_u)\in(0,1)\ \forall u$, there is a stationary distribution \citep{Tsutakawa67,DurhamFlournoy95}, obeying:
\begin{equation}\label{eqn:sudgamma}
\begin{array}{cccr}
   \pi_u p_u & = & \pi_{u+1}q_{u+1}, & u=1\dots m-1 \\
\gamma_u & \equiv & \frac{\pi_{u+1}}{\pi_u}= & \frac{p_u}{q_{u+1}}=\frac{1-F(l_u)}{F(l_{u+1})}=\frac{1-F(x)}{F(x+s)  }
\end{array},
\end{equation}
with straightforward normalization to determine $\pi_{1}$. The symbol $\gamma_u$, hereafter termed ``the stationary distribution profile'', is monotone decreasing in $u$. Therefore, the stationary distribution has a single mode, including the possibility that the mode is on a boundary \citep{DurhamFlournoy95}. The U\&D treatment sequence is known as a random walk with a central tendency \citep[see e.g.,][Ch. 3.4]{Hughes95}. \citet{DurhamFlournoy94} prove that the stationary mode is at most one spacing interval away from the median.

This is the place to note that U\&D's basic properties, such as a single mode near target, are retained even under an infinite (unbounded) set of treatment levels. Since the profiles $\{\gamma_m\}$ are monotone decreasing, with reference to the frequency at the mode the stationary frequencies form a sequence that decreases in a faster-than-geometric rate to either direction. Therefore, the overall sum of frequencies is a convergent series, even if carried to infinity, and $\mathbf{\pi}$ always exists with a mode near target. Hence, distinguishing between a bounded set of treatments indexed $1\ldots m$, and an unbounded one indexed $u,-\infty<u<\infty$ presents only a minor technical challenge and not a conceptual difference. Furthermore, as will be seen in Chapter \ref{ch:est}, in some sense the unbounded-design version of U\&D is the ``pure'' one.

\citet{Derman57} suggested modifying U\&D via randomization in order to target non-median percentiles. However, the method now known as {\bf `biased-coin' design (BCD)} was developed independently by \citet{DurhamFlournoy94,DurhamFlournoy95}. It is similar to SU\&D, except that after a {\it `no'} response the next treatment is determined using a random draw: with probability $\Gamma/(1-\Gamma)$ we go up one level, otherwise the level is unchanged ($\Gamma \in (0,0.5]$). Now $r_u$, the probability of remaining at the same level for another trial, can be nonzero.  The resulting transition probability rule as a function of $F$ is
\begin{equation}\label{eqn:bcdp}
\begin{array}{ll}
p_u=\left[1-F(l_u)\right]\frac{\Gamma}{1-\Gamma},& u=1\ldots m-1;\\
r_u=\left[1-F(l_u)\right]\frac{1-2\Gamma}{1-\Gamma},& u=2\ldots m-1;\\
q_u=F(l_u),& u=2\ldots m
 \end{array}.\end{equation}
The stationary distribution obeys \citep{DurhamFlournoy95}
  \begin{equation}\label{eqn:bcdgamma}
 \gamma_u=\frac{1-F(l_u)}{F(l_{u+1})}\frac{\Gamma}{1-\Gamma}.
\end{equation}
Again, the walk has a central tendency and a single mode; it was shown \citep{DurhamFlournoy94} that the stationary mode is at most one spacing interval away from $Q_{\Gamma}$, which is the method's designated target.

Another non-median U\&D method is known as ``forced-choice fixed staircase'' or {\bf `k-in-a-row' (KR)} \citep{WetherillEtAl66,Gezmu96}.\footnote{In fairness, I should have used the term `staircase' to refer to this method, since this is how most of its users (in psychophysics) know it. However, `k-in-a-row' is shorter and more descriptive. Moreover, rigorously speaking, all $3$ U\&D `flavors' discussed here answer the definition `forced-choice fixed staircase' (as contrasted with `adaptive staircase', in which the `up' and `down' steps have different level spacing; see e.g. \citet{GarciaPerez98}).} Here there is no random draw; instead, we must observe exactly $k$ consecutive {\it `no'}s at a given level before moving up (the `down' rule remains as in SU\&D). For $k=1$, KR reduces to SU\&D. Even though KR is by far the most widely used non-median U\&D method and is arguably easier to administer than BCD (requiring no random draw), its Markov chain properties are more complex and have rarely been studied. KR's transition probability matrix and stationary distribution were described by Gezmu in her unpublished dissertation \citep{Gezmu96}. In numerical runs, KR has repeatedly exhibited advantages over similar-targeted BCD designs \citep[see also][]{Storer89,IvanovaEtAl03}. However, Gezmu and her collaborators have come to doubt the method's theoretical properties, most notably the existence of a single stationary mode \citep{IvanovaFlournoy06}. This and other properties will be explored in the next section.

One aspect of KR that has been known since its inception \citep{WetherillEtAl66} is its target: it is the solution to the equation \footnote{Paradoxically, \citet{WetherillEtAl66} derived KR's target in an erroneous way: they calculated the target of a GU\&D method to be presented below, and attributed it to KR.}
 \begin{equation}\label{eqn:krtarget}
 \begin{array}{c}
   [1-F(Q_p)]^{k}=1-[1-F(Q_p)]^{k} \\
 p\equiv F(Q_p)=1-\left(\frac {1}{2}\right)^{1/k}
\end{array}.
\end{equation}
Unlike BCD, KR can target only a discrete set of percentiles; for $k=2, 3, 4,$ these are approximately $Q_{0.293}, Q_{0.206}$ and $Q_{0.159}$, respectively.\footnote{In sensory studies the method is inverted: one {\it `no'} response triggers an `up' move, while $k$ consecutive {\it `yes'} responses are required for a `down' move. Hence the targets are $Q_{0.707}, Q_{0.794}, Q_{0.841}$, etc.}

\begin{figure}
\begin{center}
\includegraphics[scale=.8]{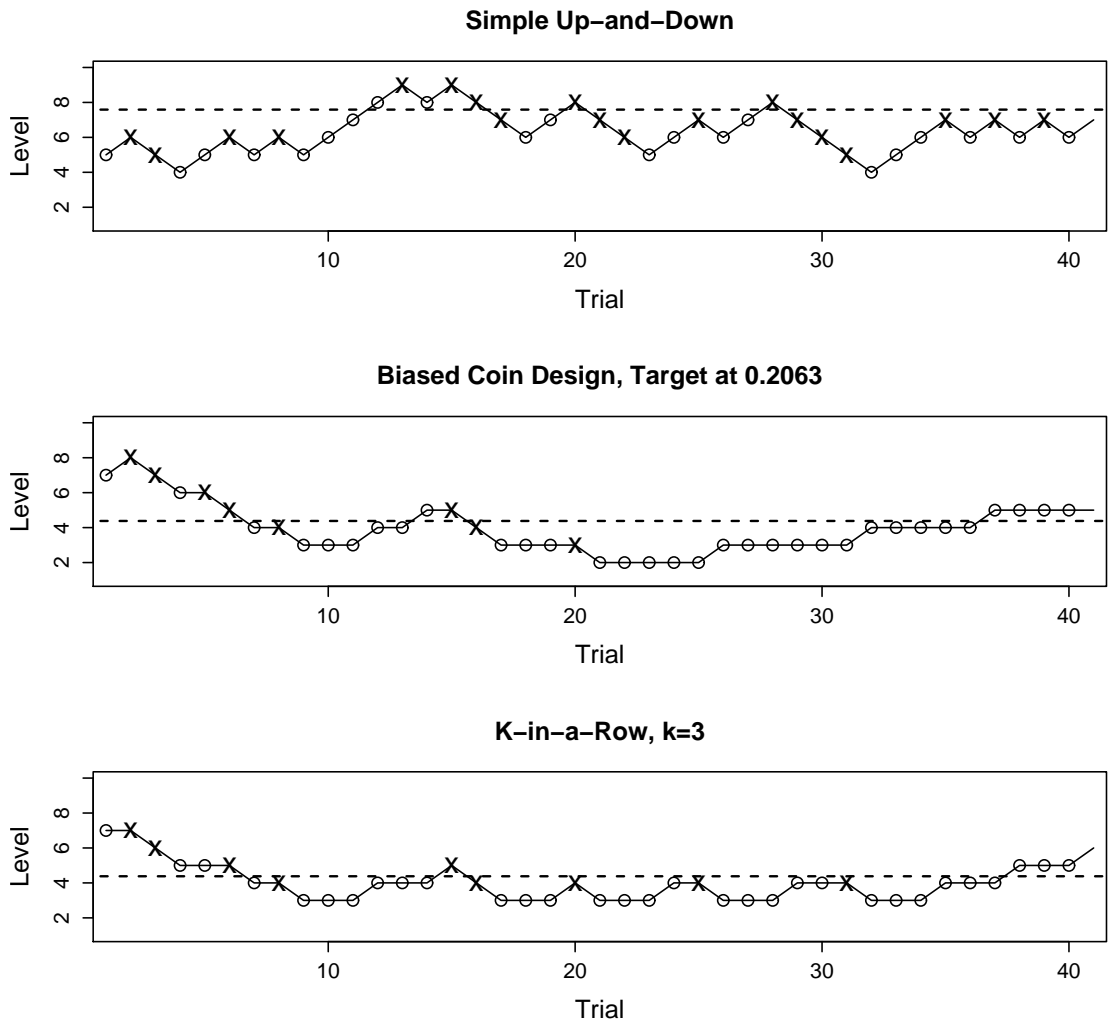}
\caption[Illustration of Up-and-Down Runs]{\small{Sample numerical runs of SU\&D (top), BCD (center) and KR (bottom) experiments. The latter two were designed to have the same target ($\Gamma=0.2063, k=3$). Targets are indicated by horizontal dashed lines. {\em `Yes'} and {\em `no'} responses are indicated by `x' and `o' marks, respectively. The design had $10$ levels, and thresholds were drawn from a Weibull distribution with shape parameter at $2$, $F=0.7$ at the top level and $0$ one spacing interval below the bottom level.}\label{fig:basic}}
\end{center}
\end{figure}

SU\&D, BCD ($\Gamma=0.206$) and KR ($k=3$) are illustrated in Fig. \ref{fig:basic} - where all three were run numerically on the same set of $40$ thresholds. As the figure shows, how closely an individual experiment's chain tracks around $Q_p$ depends upon $F$'s properties (the distribution used for Fig. \ref{fig:basic} is relatively shallow), and also upon the {\em ``luck of the draw''}: for example, between trials $15$ and $31$, low thresholds were frequently encountered, and therefore the SU\&D and KR chains remained mostly below target. For the BCD chain this was compounded by `unlucky' randomization draws: between trials $21$ and $37$, treatment escalations took $5$, $6$ and $5$ successive {\em `no'}s, respectively, while the conditional escalation probability per trial (given a  {\em `no'} response) was in fact a little over $1/4$. Note also that the $n+1$-th treatment is determined by $n$-th outcome, a property used for averaging estimators (see Chapter \ref{ch:est}).

In {\bf group `up-and-down' (GU\&D)} \citep{Tsutakawa67} instead of a single trial, at each stage a cohort of $k$ simultaneous independent trials with the same treatment is performed. Probabilistically, GU\&D does not use a binary-outcome trial, but rather the binomial outcome $Y_{G}(x)\sim \mathsf{ Binomial}\left(k,F(x)\right)$, the number of {\it `yes'}\  responses observed in the cohort. GU\&D transition rules stipulate a move up if $Y_{G}\leq a$, a move down if $Y_{G}\geq b$, and staying at the same level otherwise (obviously, $0\leq a<b\leq k$).  As $k$ increases, the increasing number of possible combinations of $a$ and $b$ provides a large variety of targets (there are $k(k+1)/2$ possible designs with cohort size $k$). Generic stationary formulae and other key properties of GU\&D are detailed by  \citet{GezmuFlournoy06}. More recently, \citet{IvanovaEtAl07} prove that GU\&D's mode lies at most $1$ spacing unit away from target.

Here we focus mostly on the GU\&D subset with $a=0, b=1$ (hereafter: GU\&D$_{(k,0,1)}$), whose transition probabilities are equal to those of SU\&D performed on a transformed CDF
\begin{equation}\label{eq:gudtrans}
H(x)\equiv 1-[1-F(x)]^{k}
\end{equation}
(and with each single `transformed trial' representing a cohort of $k$ actual trials). GU\&D$_{(k,0,1)}$ targets are identical to those of KR with the same $k$, thus enabling direct comparison between the three U\&D variants described here.

Another interesting GU\&D design family is GU\&D$_{(k,a,k-a)}$, which targets the median (SU\&D itself can be re-defined as GU\&D$_{(1,0,1)}$ and therefore belongs to this family too). \citet{GezmuFlournoy06} found numerically that for a fixed cohort size $k$, convergence accelerates with increasing $a$, but the stationary distribution's mode becomes more shallow \citep[see also comment in][]{IvanovaEtAl07}. Finally, as \citet{Storer89}, \citet{IvanovaFlournoy06} and \citet{Ivanova06} note, the commonly used phase I cancer trial `3+3' method can be modeled as an initial GU\&D$_{(3,0,2)}$ stage, changing into a size $6$ cohort design before stopping. The convergence of GU\&D$_{(3,0,2)}$ will be examined in the next section. That design's target is $Q_{0.347}$.

\subsection{The Convergence-Stationarity Tradeoff, Reversals and `Down-shift' Schemes}

In his seminal work, \citet{Wetherill63} noticed that U\&D designs face an inherent tradeoff. Choosing a large spacing reduces the expected number of trials until the chain first reaches target, and hence speeds convergence. On the other hand, a smaller spacing generates a tighter distribution around target, once it is reached. The most obvious solution is to begin with a large spacing, then reduce it at some point (most conveniently by a factor of $2$).\footnote{This can be seen as a crude attempt to utilize some of the advantages of stochastic approximation \citep{RobbinsMonro51}, under the constraint of discrete levels. It can also be viewed as a simplified version of simulated annealing (V. Minin, personal communication, 2007).} I shall refer to these spacing-reduction designs as {\bf ``down-shift'' schemes}. The question is, what would be a good point to make the shift, if such a point exists.

This is where the notion of ``reversal'' comes into play. A reversal is a point in the experiment, where the response is different from the previous, i.e. a reversal at $i$ means that $y_i\neq y_{i-1}$. For SU\&D, reversals are where the treatment chain reverses direction, hence the name. For the U\&D variants described above, that is not necessarily the case; however, the term ``reversal'' has been used for them under the same definition. \citet{BrownleeEtAl53} were the first to suggest discarding all data up to the first reversal, using `common sense' arguments. \citet{Wetherill63} suggested ``down-shifting'' at the first reversal, an idea that has recurred in different forms since then. Interestingly, ``down-shift'' schemes triggered by the first reversal have been prominent in recent attempts to integrate U\&D into model-based Phase I designs \citep{Storer89,Storer01,Potter02}. In some fields, reversals are also used as a stopping rule, with the experiment length specified in reversals instead of trials \citep{GarciaPerez98,PaulFisher01}.

The benefit of using the first reversal as a down-shift point has not been verified theoretically. Numerically, both \citet{GarciaPerez98} and this author \citep{Oron05} have concluded that down-shift schemes are not a guarantee for performance improvement. More generally, the theoretical properties of reversals have not been studied since \citet{Wetherill63}.

\begin{figure}
\begin{center}
\includegraphics[scale=.8]{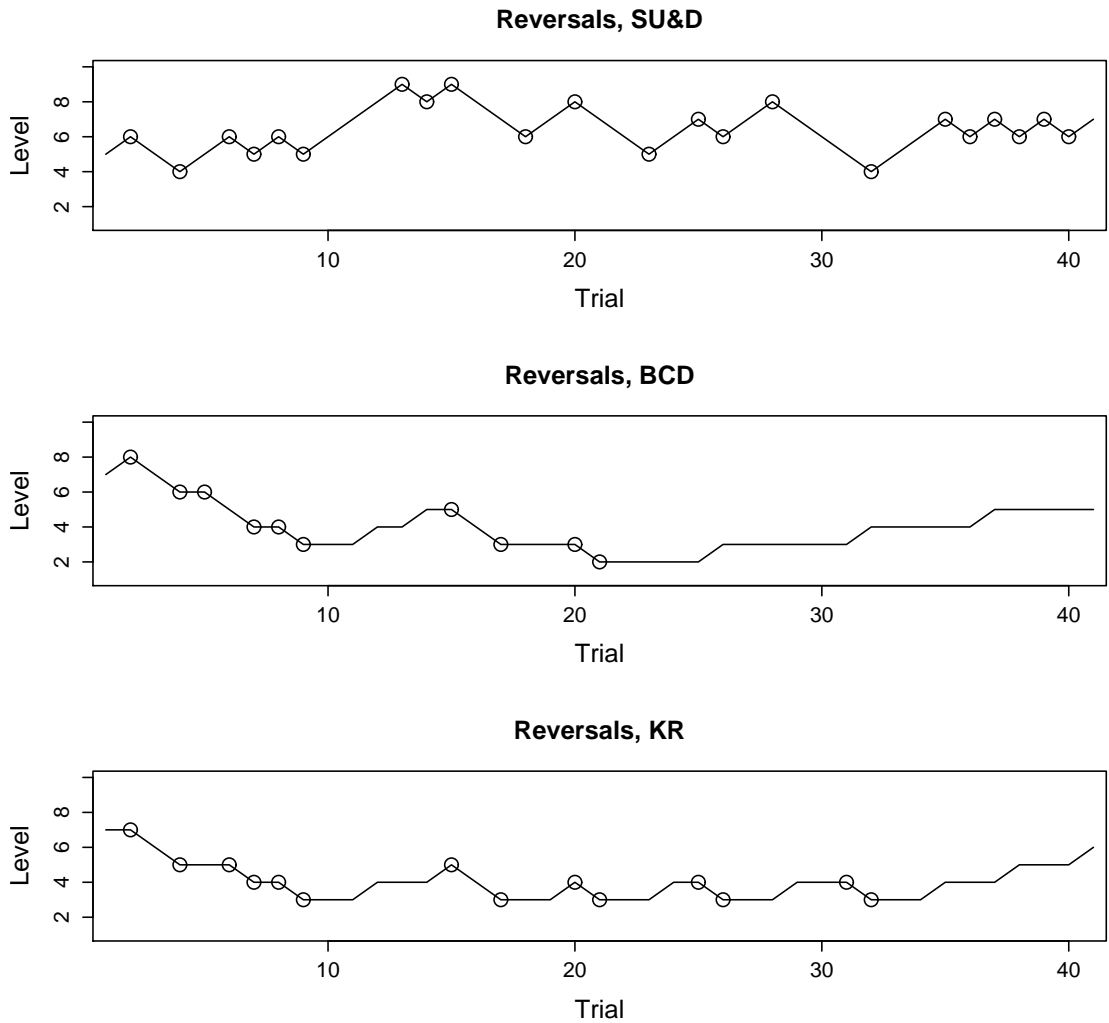}
\caption[Up-and-Down Reversals]{\small{Illustration of U\&D reversals (marked in circles), using the same numerical runs from Fig. \ref{fig:basic}. SU\&D reversals (top) can be identified visually as direction changes. For the other two designs, reversals occur at the beginning and end of each uninterrupted descent (of course, for the mirror-image versions of these designs targeting above-median percentiles, reversals would occur around ascents rather than descents).}\label{fig:revbasic}}
\end{center}
\end{figure}

\subsection{Other Up-and-Down Descendants}

Beyond the straightforward modifications described above, U\&D has spawned a wide variety of related designs over the years. \citet{Wetherill63} was the first to lay out a catalogue of such designs. In psychophysics, a popular family of related designs mandates a different step size for `up' and `down' transitions \citep{GarciaPerez98,GarciaPerez05}. If the step size ratio is irrational, the range of levels becomes continuous. Another related family is that of Poly\`{a} urn schemes \citep{RosenbergerGrill97}. Here the transition probabilities vary as a function of prior responses, according to a predefined algorithm. That family has been less extensively studied.

Given the significant gaps in knowledge about the simpler Markov-chain U\&D designs, these non-Markovian U\&D descendants are outside the scope of my dissertation.

\section{Theoretical Study}\label{sec:und2}
\subsection{Conceptual Prelude}\label{sec:undconcept}
The U\&D design family achieves a remarkable feat: it converts censored, binary-response (and therefore indirect) threshold sampling -- a rather unfriendly turf for percentile estimation -- back into direct sampling from a quantitative distribution on the threshold/treatment scale. This distribution is $\pi$, the stationary distribution of treatments. Even better, $\pi$ appears to be centered near $Q_p$. These properties are even more remarkable, considering that the method was developed not from a deep understanding of Markov chains, but rather using ad-hoc intuitive innovation.

The properties of $\pi$ suggest a very natural estimator for $Q_p$: the empirical mean. How well such an estimator performs, depends on how quickly we can assume to be sampling from $\pi$, and on how well $\pi$ is centered around target. These will be some of the topics discussed below. But first, for lack of a standard reference, some foundational definitions of basic concepts are needed -- as well as a careful study of KR's properties.

Notational comment: from here on, I will use $F_u$ as a shorthand for $F(l_u)$.

\subsection{Definition of Target and Mode}

The concept of U\&D target has been somewhat ill-defined in research. Most \citep [e.g.][]{GezmuFlournoy06} are content with a general intuitive notion, similar to that of the method's originators. \citet{GiovagnoliPintacuda98} attempted a more rigorous approach; however, they end up defining target only for a narrow (and somewhat contrived) subset of U\&D designs, which does not cover KR or GU\&D. I have found the following definition useful, broad and rigorous, and in the spirit of the original idea:

\begin{defn}\label{def:targ} Consider an `Up-and-Down' design with a stationary distribution $\pi$, and with (marginal) stationary `up' transition probabilities $\left.p(x)\right| _{\pi}$ continuous monotone decreasing in $x$, and `down' probabilities $\left.q(x)\right| _{\pi}$ continuous monotone increasing in $x$, respectively. The design's {\bf target} $Q_p$ is defined as the $x$ value such that
\begin{equation}\label{eqn:deftarget}
\left.p(x)\right| _{\pi}=\left.q(x)\right| _{\pi}.
\end{equation}
\end{defn}

\citet{IvanovaEtAl07}, in a concurrent paper, have suggested a similar definition for GU\&D's target. In words, the target is the $x$ value from which marginal stationary `up' and `down' probabilities would be equal, had a design level been placed exactly there. It is straightforward to see that this definition does recover SU\&D's and BCD's targets as the median and $Q_{\Gamma}$, respectively. It is also equivalent to the definition appearing in \citet{GiovagnoliPintacuda98} on the subset discussed there.
The reason for including the terms ``stationary'' and ``marginal'' in the definition will become clear below.

This is also a good occasion for defining the notion of a single mode. The definition is followed by a useful result about the mode's location w.r.t $Q_p$.
\begin{defn}\label{def:mode} Let $\pi$ be an U\&D (marginal) stationary distribution. Then $\pi$ will be said to have {\bf a single mode}, if the set of indices $M=\left\{u:\pi_u>\pi_v\right.$  $\forall v<u$ and $\pi_u>\pi_{v^{'}}$ $\left.\forall v^{'}>u\right\}$ has either a single member or two members with adjacent indices $u,u+1$.
\end{defn}
\begin{lem}\label{lem:mode} If a U\&D (marginal) stationary distribution's profile $\gamma_u(x)$ is continuous and strictly monotone decreasing in $x$, then

(i)  The distribution has a single mode.

(ii) If there exist two adjacent levels $l_{u^*},l_{u^*+1}$, such that $Q_p\in\left[l_{u^*},l_{u^*+1}\right]$, then the mode is either at $l_{u^*}$ or at $l_{u^*+1}$.
\end{lem}
\begin{proof} (i) If $\gamma_1\leq 1$ or $\gamma_{m-1}\geq 1$, then the mode is obviously on the lower or upper boundary, respectively. In the remaining (and more interesting) case, there has to be some $u^*$ such that $\gamma_{u^*-1}\geq 1\geq\gamma_{u^*}$, with equality holding for at most one of the relationships. By the definitions of $\gamma$ and of the mode, the mode is at $l_{u^*}$, possibly with either $l_{u^*-1}$ or $l_{u^*+1}$ (but not both). Due to monotonicity, this mode is unique.

(ii) Note that the following inequality always holds for any $u$ (except on the lower boundary -- a case that can be omitted here w.l.o.g.), under the specified conditions:
\begin{equation}\label{eqn:gammas1}
\gamma_{u-1}>\frac{p_u}{q_u}>\gamma_u.
\end{equation}
Therefore, for all $x\leq Q_p-s$, $\gamma(x)>1$ and for all $x\geq Q_p$, $\gamma(x)<1$. Recalling (i), this means that the mode $l_{u^*}$ has to maintain $l_{u^*}\in\left[Q_p-s,Q_p+s\right]$. Together with $\gamma$'s monotonicity, this forces the mode to be within $s$ of $Q_p$.\qed\end{proof}

\subsection{Basic Properties of `K-in-a-row'}

We now return to the `k-in-a-row' design (KR). Unlike SU\&D, BCD and GU\&D, KR does not generate a first-order random walk. A KR run can be described as a $k$-th order random walk with two sets of transition rules, only one of which applies at each trial depending on the string of $k$ most recent treatments \citep{Gezmu96}:
\begin{small}
$$
\begin{array}{ll}
 x_{i}=l_u \textrm{, but }\exists j, i-k<j<i\textrm{ s.t. }x_{j}\neq l_u: & \left\{
  \begin{array}{ll}
    p_u=0\\
    r_u=1-F_u,&u=2\ldots m \\
    q_u=F_u,&u=2\ldots m
  \end{array}\right.\\
  x_{i}=x_{i-1} \ldots =x_{i-k+1}=l_u: & \left\{
  \begin{array}{ll}
    p_u=1-F_u,&u=1\ldots m-1 \\
    r_u=0 ,&u=1\ldots m-1 \\
    q_u=F_u,&u=2\ldots m
  \end{array}\right.  \\
\end{array},
$$
\end{small}

However, this description does not explicitly acknowledge the counting of $k$ `no' responses before an up transition. For this and other purposes, it is more useful to view KR as generating a Markov chain whose treatments $\{x_n\}$ are paired with a sequence of {\bf internal states} $\{w_n\}$. Each internal state can take one of $k$ possible values $\tau=0\ldots k-1$, for a total of $mk$ states \citep[{\em cf.} e.g.][Ch.6]{Weiss94}. Under this internal-state formulation the transition rules become
\begin{small}
\begin{equation}\label{eqn:krpinternal}
\begin{array}{l}
 \left\{\begin{array}{lll}
    p_{u,\tau}=0,&u=1\ldots m,&\tau<k-1\\
    r_{u,\tau}=1-F_u,&u=2\ldots m,& \tau<k-1 \\
    q_{u,\tau}=F_u,&u=2\ldots m,& \tau<k-1
  \end{array}\right.\\
    \left\{\begin{array}{ll}
    p_{u,\tau}=1-F_u,&u=1\ldots m-1,\tau=k-1 \\
    r_{u,\tau}=0 ,&u=1\ldots m-1,\tau=k-1 \\
    q_{u,\tau}=F_u,&u=2\ldots,\tau=k-1 m
  \end{array}\right.\\
  \left\{\begin{array}{ll}
x_{i+1}=x_{i} \Rightarrow w_{i+1}=w_{i}+1 \\
x_{i+1}\neq x_{i} \Rightarrow w_{i+1}=0
\end{array}\right.
\end{array}.\end{equation}\end{small}

Note that the terms $p_u,q_u,r_u$ still pertain to transition between physical treatment levels, and not between internal states. A few more notes on the internal-state formulation:
\begin{itemize}
\item The $w_i$ are uniquely determined by the $x_i$; hence, after the experiment is over it suffices to record $\{x_n\}$, and then $\{w_n\}$ can be reconstructed.
\item In the same vein, a subject receiving a given treatment $x_i$ is oblivious to $w_i$, which has no practical effect upon trial $i$ itself.
\item For this reason and also for comparison purposes, we ultimately focus upon {\bf marginal probabilities and distributions}, i.e. values summed over all internal states. The marginal stationary frequency of level $u$ will be denoted $\pi_u\equiv \sum_{\tau}{\pi_{u,\tau}}$.
\end{itemize}

\begin{thm}\label{thm:kr1} (i) KR's marginal stationary distribution profile $\{\gamma_u\}$ is given by
\begin{equation}\label{eqn:krgamma}
\gamma_u=\frac{F_u\left[1-F_u\right]^{k}}{F_{u+1}\left\{1-\left[1-F_u\right]^{k}\right\}}
\end{equation}

(ii) Let the stationary marginal `up' probability of a KR design be the weighted average $\left.p_u\right| _{\pi}=\left(\sum_{\tau}{\pi_{u,\tau}p_{u,\tau}}\right)\pi_u$. Then
\begin{equation}\label{eqn:krpm}
  \left.p_u\right| _{\pi} = \frac{F_u[1-F_u]^{k}}{1-[1-F_u]^{k}}
\end{equation}.

(iii) KR's marginal stationary distribution has a single mode, which is at most 1 spacing unit away from $Q_p$ under the conditions specified in Lemma \ref{lem:mode} (ii).
\end{thm}
\begin{proof}
(i) Taking the internal-state approach, the balance equations between adjacent treatment levels may be written as
   \begin{equation}\label{eqn:kradjacent}
   \pi_{u,k-1}[1-F_u]=\pi_{u+1}F_{u+1}.
\end{equation}

 Now, transitions between internal states of the same treatment level are possible only for single upward increments, each with probability $1-F_u$. Therefore, to maintain balance at stationarity, the internal state frequencies must obey $\pi_{u,\tau+1}=[1-F_u]\pi_{u,\tau}\textrm{, }\tau=0 \ldots k-2$ - a diminishing geometric sequence. This enables us to calculate relative internal-state frequencies, and specifically the upper state:
\begin{equation}\label{eqn:kgeometric}
  \frac{\pi_{u,k-1}}{\pi_u} = \pi_{u,0}[1-F_u]^{k-1} \left/\left\{\frac{\pi_{u,0}\left\{1-[1-F_u]^{k}\right\}}{F_u}\right\}\right. = \frac{[1-F_u]^{k-1}F_u}{\left\{1-[1-F_u]^{k}\right\}}.
\end{equation}
Plugging back into (\ref{eqn:kradjacent}), we obtain (\ref{eqn:krgamma}).

(ii) This result is immediate from (\ref{eqn:krgamma}).

(iii) Differentiating $\left.p_u\right| _{\pi}$ w.r.t to $F$ shows that it is monotone decreasing as $F$ increases (a more detailed derivation can be found in Appendix A). Since $F$ itself is monotone increasing in $x$, and since $\left.q_u\right| _{\pi}=F_u$ is monotone increasing in $x$, $\gamma(x)$ is monotone decreasing. Therefore, by Lemma \ref{lem:mode}, there is a single stationary mode at most $1$ spacing unit away from target.\qed
\end{proof}

A somewhat different proof for result (i) appears in \citet{Gezmu96}.

Under stationarity, the frequencies of internal states within each treatment level form a decreasing geometric sequence. This means that the base ($w=0$) internal state is the most common. Intuitively, this is due to the fact that each level transition (whether up or down) resets the internal state to zero. It may be of interest to look at the distribution of these first visits - i.e., the stationary distribution of a KR chain for which each sojourn at a treatment level is collapsed to a single trial corresponding to the base internal state.
\begin{corollary}\label{cor:zerostate}
Consider a subset of any KR experimental chain, composed only of base-state trials, i.e. $\left\{x_i:\ \ w_i=0\right\}$. The stationary profile of this subset is
\begin{equation}\label{eqn:zerostate}
\gamma_{u,0}=\frac{[1-F_u]^{k}}{1-[1-F_{u+1}]^{k}},
\end{equation}
That is, a profile identical to that of a GU\&D$_{(k,0,1)}$ design with the same $k$.
\end{corollary}
\begin{proof}
Using the geometric-series formula as in (\ref{eqn:kgeometric}),
$$
\gamma_{u,0}=\frac{\pi_{u+1,0}}{\pi_{u,0}}=
\frac{\pi_{u+1}}{\pi_{u}}\frac{F_{u+1}\left\{1-[1-F_u]^{k}\right\}}{F_u\left\{1-[1-F_{u+1}]^{k}\right\}}.
$$
Plugging in $\gamma_u$ leads to most terms canceling out, yielding (\ref{eqn:zerostate}).\qed\end{proof}

Conceptually, we can explain this surprising identity: after deleting all trials with $w_i>0$, one is left with an SU\&D-like (or GU\&D$_{(k,0,1)}$-like) chain, in the sense that it never remains at the same level. Now, from a given level's base state, the probability of the next transition being an \emph{'up'} one is $[1-F_u]^{k}$, and the move must be down otherwise - yielding exactly the transition probabilities of GU\&D$_{(k,0,1)}$ (see (\ref{eq:gudtrans})). This result will prove rather useful later on.

\subsection{Location of the Stationary Mode}
We have established that for all U\&D variants discussed here, the stationary mode is one of the two levels closest to target. It turns out we can do even better.

\begin{thm}\label{thm:modesud}
(i) SU\&D stationary mode is located at the level whose $F$ value is closest to $0.5$.

(ii) GU\&D$_{(k,0,1)}$ stationary mode is located at the level whose $H$ value (with $H(x)$ defined in (\ref{eq:gudtrans})) is closest to $0.5$.

(iii) GU\&D$_{(k,a,k-a)}$ stationary mode is located at the level whose $F$ value is closest to $0.5$.
\end{thm}
\begin{proof} (i) Let $Q_p\in [l_{u^*},l_{u^*+1}]$ w.l.o.g. Then we can write $F_{u^*}=0.5-\Delta p_1,F_{u^*+1}=0.5+\Delta p_2$, with $\Delta p_1,\Delta p_2\geq 0$. Now
\begin{equation}\label{eqn:sudmode}
\gamma_{u^*}=\frac{p_{u^*}}{q_{u^*+1}}=\frac{0.5+\Delta p_1}{0.5+\Delta p_2}.
\end{equation}
Therefore, $\gamma_{u^*}>1$ if and only if $\Delta p_1>\Delta p_2$, with equality only if the two are equal. Since the mode is at one of these two levels, our proof is complete.\qed

(ii) This part follows immediately from (i) and from the definition of $H(x)$.

(iii) First, note that for the binomial cohort-outcome r.v. $Y_G$ defined in the previous section, $\Pr\left(Y_G\leq a | F=0.5-\Delta\right)=\Pr\left(Y_G\geq k-a | F=0.5+\Delta\right)$. Moreover, both sides of the equation are monotone increasing in $\Delta$. Now using the same $\Delta p_1,\Delta p_2$ notation of part (i),
\begin{equation}\label{eqn:gudmode}
\gamma_{u^*}=\frac{\Pr\left(Y_G\leq a | F=0.5-\Delta_1\right)}{\Pr\left(Y_G\geq k-a | F=0.5+\Delta_2\right)}.
\end{equation}
Again, $\gamma_{u^*}>1$ if and only if $\Delta p_1>\Delta p_2$, with equality only if the two are equal.\qed
\end{proof}

For median-finding U\&D, the stationary mode is the closest to target on the response scale. For GU\&D$_{(k,0,1)}$, this is true on the transformed scale of $H(x)$. Since $F=1-(1-H)^{1/k}$ is convex in $H$, on the original response scale the transformed midpoint lies closer to $F_{u^*}$, the lower of the two levels. This means that the ``attraction basin'' of the upper level is larger, and therefore if the target lies midway between levels on the $F$ scale, the mode would revert upward to $l_{u^*+1}$. However, this asymmetry is relatively small: to first order in $\Delta p_1,\Delta p_2$, the two ``attraction basins'' are equal in size.

If one wishes to measure closest treatment on the $x$ scale, exact results depend upon the form of $F(x)$. However, to first order in $s$, it is straightforward to show that the treatment midpoint is the boundary between the two ``attraction basins'' for all designs covered by Theorem \ref{thm:modesud}.

The mode location of the two remaining non-median designs is less symmetric.

\begin{thm}\label{thm:modebcdkr}
(i) For BCD designs, and using the same terminology as in Theorem \ref{thm:modesud}, the stationary mode would revert to $l_{u^*}$ unless $F_{u^*+1}$ is closer to $p$ by a factor greater than $(1-p)/p$.

(ii) For KR designs, the stationary mode would revert to $l_{u^*}$ if  .
\begin{equation}\label{eqn:krmode1}
\left(p-\Delta p_1\right)\left(1-p+\Delta p_1\right)^k<\left(p+\Delta p_2\right)\left[1-\left(1-p+\Delta p_1\right)^k\right],
\end{equation}
and vice versa.
\end{thm}
\begin{proof} (i) This follows directly from substituting $p_{u^*},q_{u^*+1}$ from (\ref{eqn:bcdp}) in the formula for $\gamma_{u^*}$.

(ii) This follows directly from substituting the analogous KR formulae found in the previous subsection.\qed
\end{proof}

The KR formula is again more complicated; analysis to first order in $\Delta p_1,\Delta p_2$ yields the approximation
\begin{equation}\label{eqn:krmode2}
\frac{\Delta p_1}{\Delta p_2}<\frac{1-p}{(2k+1)p-1}
\end{equation}
for the mode to revert to $l_{u^*}$. For $k=2,3$ this yields approximately $1.52$, $1.79$, respectively, for the $\Delta p_1/\Delta p_2$ ratio at the ``basin boundary''. The analogous factors for BCD are $2.41$, $3.85$, respectively. So both designs tend to create a mode below target more often than above it -- a tendency opposite that of GU\&D$_{(k,0,1)}$ -- but KR's tendency is quite a bit milder than BCD's. This explains the numerical observations of \citet{Gezmu96}, who noted that KR is `better centered' on target than same-target BCD (see e.g. Fig.~\ref{fig:peaked}). As above, on the treatment scale these same ratios can be used as approximations to first order in $s$.

\subsection{More Stationary Properties}\label{sec:stationary}

\subsubsection{Peakedness of non-median-target designs}\label{sec:peaked1}
Due to limitations related to the unknown location of target relative to the fixed design levels, the notion of `peakedness' is not easily converted to more familiar terminology used to describe dispersion (e.g. precision or variance), but it carries a similar meaning within the U\&D context. The more `peaked' a stationary U\&D distribution is around its mode, the closer the method is to the ideal design, and percentile estimation precision will generally improve.  Here is a rigorous definition of `peakedness', following \citet{GiovagnoliPintacuda98}.

\begin{defn}\label{def:peaked} (i) For two `Up-and-Down' designs, {\bf "all other things being equal"} will mean that both target the same percentile, the threshold distribution $F$ is the same, the treatment levels $\left\{l_u\right\}$ are the same and the initial conditions are the same.

(ii) \citep{GiovagnoliPintacuda98} Let designs $1$ and $2$ be two U\&D designs. Then, all other things being equal,  if $\gamma^{(1)}_u\geq\gamma^{(2)}_u$ for $l_u\leq Q_p$ while $\gamma^{(1)}_u\leq\gamma^{(2)}_u$ for $l_u\geq Q_p$ (i.e., design 1's stationary distribution profile is steeper), then design $1$ and its stationary distribution are called {\bf more `peaked'}.
\end{defn}
Thus, for example, to compare KR and BCD we need to examine
\begin{equation} \label{eqn:krbcdratio}
\frac{\gamma^{KR}_u}{\gamma^{BCD}_u}=\frac{F_u\left(1-F_u\right)^{k-1}}{1-\left(1-F_u\right)^{k}}\frac{1-p}{p}
.\end{equation}
\begin{thm} For any $k>1$, all other things being equal, KR designs are more `peaked' than BCD designs.
\end{thm}
\begin{proof} On target, the ratio in (\ref{eqn:krbcdratio}) is exactly $1$. Therefore, it suffices to show that (\ref{eqn:krbcdratio}) is monotone decreasing in $F$. Careful differentiation yields this result.\qed
\end{proof}
\begin{prop} For KR and GU\&D$_{(k,0,1)}$ with the same $k$, neither design can be said to be more `peaked' than the other.
\end{prop}
\begin{proof} The ratio analogous to (\ref{eqn:krbcdratio}) is
$$
\frac{\gamma^{KR}_u}{\gamma^{GUD_{(k,0,1)}}_u}=\frac{F_u\left[1-\left(1-F_{u+1}\right)^{k}\right]}{F_{u+1}\left[1-\left(1-F_u\right)^{k}\right]}\leq 1
.$$ The inequality results from the concavity of $1-\left(1-F)\right)^{k}$ in $F$. GU\&D$_{(k,0,1)}$'s stationary distribution profile is steeper to the left of the peak, while KR's is steeper to the right of the peak.\qed
\end{proof}

\begin{figure}
\begin{center}
\includegraphics[scale=.8]{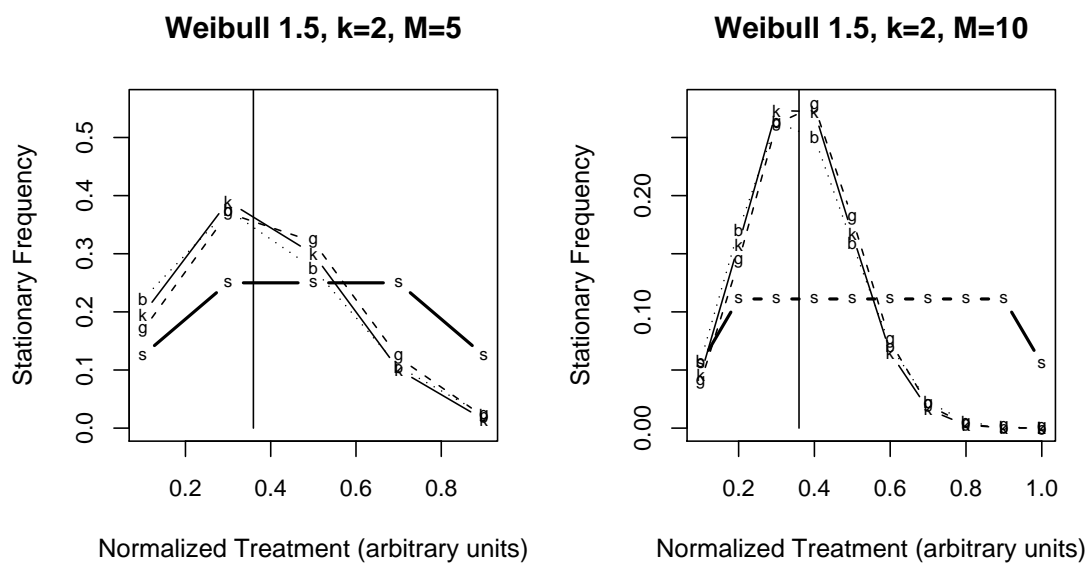}

\caption[Stationary Distribution Comparison]{Stationary distributions for KR ('k' marks and solid lines), BCD ('b' marks and dotted lines) and GU\&D ('g' marks and dashed lines) - all targeting $Q_{0.293}$, marked with a vertical line. $F$ is Weibull with shape parameter $1.5$, and scale normalized so that $F(1)=0.8$. Shown are a coarser design with $m=5$ (left) and a finer design with $m=10$ (right). Also plotted is a standard non-sequential treatment allocation of $1/(m-1)$ to interior levels and $1/2(m-1)$ to boundary levels ('s' marks and thick lines). The vertical axes are normalized to account for the factor $2$ difference in $m$, so that the area under the curves is similar in both plots. }\label{fig:peaked}
\end{center}
\end{figure}

Fig.~\ref{fig:peaked} provides a numerical example of `peakedness'. The differences between methods do not appear to be dramatic. On the other hand, if the spacing is too coarse (e.g. $m=5$ on the left) the stationary peak is very broad and the method's stationary properties offer only a modest improvement over standard, non-sequential treatment-response designs. As the spacing becomes finer (e.g. $m=10$ on the right), a sharp peak forms: under stationarity, all $3$ methods allocate $50-55\%$ of treatments to the two levels closest to target. Thus, finer spacing dramatically reduces the chance for treatments allocated too far from target. In the scenario illustrated in Fig. $1$, with $m=10$, $\Pr_\pi (x\geq 0.7)$ ($x=0.7$ roughly corresponds to the $60$th percentile here) is $0.02$ to $0.03$ depending upon method, while with $m=5$ the analogous probabilities are $0.12$ to $0.14$. The figure also shows (for $m=10$) BCD and KR's tendency to have the mode below target, and GU\&D$_{(k,0,1)}$'s opposite tendency. Here $\Delta p_1=0.061$ and $\Delta p_2=0.042$, so KR's ``basin boundary'' falls just above target on the $F$ scale, while BCD's and GU\&D$_{(k,0,1)}$'s boundaries are at a larger distance above and below target, respectively.

\subsubsection{Peakedness of median-target GU\&D designs}\label{sec:peaked2}
Another result related to GU\&D `peakedness' was numerically observed by \citet{GezmuFlournoy06}: for median-targeting GU\&D$_{(k,a,k-a})$ $(k>2)$, the stationary distribution appeared more peaked with decreasing $a$. Here is a general proof.
\begin{thm} \label{thm:gudpeaked} For GU\&D$_{(k,a,k-a})$ designs $(k>2)$, and all other things being equal, `peakedness' increases with decreasing $a$.\end{thm}

\begin{proof}Consider design $1$ with $a$ and design $2$ with $a^{'}$, such that $a<a^{'}$. Since the transition probabilities are `drawn' from the same binomial model, they can be represented thus: $p^{'}_u=p_u+\delta p_u,q^{'}_u=q_u+\delta q_u$, where $\delta p_u\equiv Pr(a<Y_{G}\leq a^{'}\mid F_u), \delta q_u\equiv Pr(k-a^{'}\leq Y_{G}<k-a\mid F_u) $. The `peakedness' ratio is then
\begin{large}
\begin{equation} \label{eqn:gudbtpeak1}
\frac{p_u}{q_{u+1}}{\frac{q^{'}_{u+1}}{p^{'}_u}} = \frac{(q_{u+1}+\delta q_{u+1})p_u}{(p_u+\delta p_u)q_{u+1}}=\frac{1+\frac{Pr(k-a^{'}\leq Y_{G}< k-a|F_{u+1})}{Pr(Y_{G}\geq k-a|F_{u+1})}}{1+\frac{Pr(a<Y_{G}\leq a^{'}|F_u)}{Pr(Y_{G}\leq a|F_u)}}.
\end{equation}\end{large}
If $F_u+F_{u+1}=1$ (meaning that the target is exactly midway between the two levels if distance is measured on the response scale), the ratio is exactly $1$. Focusing on the fraction on the right-hand-side, the ratio in the numerator is decreasing with $F$ while the one in the denominator is increasing. Therefore, the ratio would be $<1$ with both levels below target and vice versa, and thus by definition design $1$ is more `peaked'. \qed \end{proof}

\subsection{Convergence Rates}\label{sec:conv}

For U\&D methods, the term ``convergence rate'' can be interpreted in a number of ways. In view of theoretical results about the stationary mode, one may be interested in the convergence of the probability that the empirical mode is the stationary mode, or more generally in the convergence of empirical frequencies to their stationary values. On the other hand, in view of the prevalence of averaging-based estimators, there is practical interest in the convergence of the empirical mean to the stationary mean. Fortunately, the three methods discussed here generate simple Markov chains with finite state spaces, and therefore all convergence rate comparisons should yield equivalent results. Here we focus on the empirical mean, which is the simplest, most intuitive, and carries a direct practical significance for most applications.

As \citet{GezmuFlournoy06} note, for finite-state Markov chains the exact treatment distribution after $i$ trials can be calculated, given knowledge of initial conditions and of $P$, the design's transition probability matrix (tpm):
\begin{equation} \label{eqn:exactprogression}
\rho^{(i)\dag}=\rho^{(1)\dag}P^{i-1},
\end{equation}
where $\rho^{(1)}$ is an initial probability vector over the levels, $\dag$ is the transpose operator and $P^{i-1}$ is $P$ raised to the $i-1$-th power. As $n\to\infty$, $\rho^{(n)}\to\pi$, regardless of $\rho^{(1)}$. Convergence rates can be estimated using the tpm. The tpm's of the three Markovian methods discussed here are stochastic and irreducible. Hence, the real parts of their eigenvalues are bounded between $-1$ and $1$, and are usually labeled in decreasing order: $1=\lambda_{0}\geq Re(\lambda_{1})\geq \ldots \geq Re(\lambda_{m-1})\geq -1$. If the tpm is also reversible, that is $\pi_{i}P_{ij}=\pi_{j}P_{ji}\textrm{ }\forall i,j$, then the eigenvalues are all real. In that case, it can be shown that the total variation distance between $\rho^{(n)}$ and $\pi$ converges to zero with a rate \citep{DiaconisStroock91}
\begin{equation}\label{eqn:diaconis}
\|P^{n}\rho^{(1)}-\pi\|^{2}_{var}\propto \left\{\sum_u\left|\rho^{(n+1)}_u-\pi_u\right|\right\}^2 \leq c\lambda_{\#}^{2n},
\end{equation}
where $c$ is a constant and $\lambda_{\#}\equiv \max\left(\lambda_{1},\mid\lambda_{m-1}\mid\right)$. Thus, the rate of convergence is governed by the second-largest eigenvalue. For finite-state Markov chains, it is straightforward to show that the empirical mean would converge at half the rate of the total variation distance.

Unfortunately, comparing the $3$ methods cannot be trivially reduced to an eigenvalue problem: KR tpm's (listing all internal sub-states as separate states) are $mk\times mk$ irreversible matrices \citep{Gezmu96}. On the other hand, all sub-states of the same external KR level actually represent the same experimental treatment. Therefore, a possible fix is to compare convergence by observing KR's $k\times k$ reversible matrix of marginal transition probabilities, using (\ref{eqn:krpm}). This is analogous to assuming that in KR experiments internal-state balance is achieved much faster than external-level balance.

One relationship can be directly proven without resorting to eigenvalues.
\begin{prop}\label{prop:krgudconv} Consider a KR and a GU\&D$_{(k,0,1)}$ design with the same $k$. Then, all other things being equal, the KR design converges faster to its stationary distribution.\end{prop}
\begin{proof}
Recall Corollary \ref{cor:zerostate} and the discussion following it, and observe the zero-state-only KR subchain. Its transition probabilities are identical to those of GU\&D$_{(k,0,1)}$. However, while the latter must carry out a cohort of $k$ trials for each transition, the KR subchain may take anywhere from $1$ to $k$ trials -- on the average, less than $k$. Therefore, it would require less trials to achieve the same probabilistic convergence.\qed\end{proof}

In order to investigate convergence rates more closely, an ``all other things being equal'' numerical time-progression was performed, as in (\ref{eqn:exactprogression}).\footnote{Two sets of initial conditions were examined: beginning at $l_1$ (similar to typical Phase I conditions) or at $l_m$, over a variety of right-skewed, left-skewed and symmetric threshold distributions, $3$ values of $s$ and $k=1,2,3$ for BCD, KR and GU\&D $_{(k,0,1)}$; a second comparison was performed between BCD and GU\&D $_{(3,0,2)}$. Table~\ref{tbl:conv1} shows a subset of the data. The adjectives `tight' and `disperse' for the logistic and log-normal scenarios indicate smaller or larger scale parameters, respectively. Convergence rates were estimated via an exponential fit on the difference between the mean at time $i$ and the stationary mean. All calculations -- here and in the rest of the dissertation -- were performed in R \citep{RLang}.} A summary of results for convergence `upwards' from $l_1$ and some targets appears in Table~\ref{tbl:conv1} ('downward' convergence was faster in most scenarios, but the overall pattern was not substantially different; data omitted here). Shown are the number of trials needed for the ensemble mean of the $x_i$'s to converge $99\%$ of the way from $l_1$ to the stationary mean. These numbers can be seen as a practical cutoff, beyond which the treatment chain is `as good as' a sample from $\pi$, at least for averaging purposes (one reason for choosing $99\%$ is because halving the numbers on Table~\ref{tbl:conv1} conveniently yields the number of trials needed to converge a more lenient $90\%$ of the way).

\begin{table}
\begin{center}
\caption[Up-and-Down Convergence Summary]{Comparative summary of convergence calculations for several design scenarios and targets. Shown are the number of trials needed to achieve 99\% convergence of the ensemble mean of $x_i$ to the stationary mean, beginning with the entire probability mass on $l_1$ at trial $1$. Numbers were rounded up to the nearest integer, except for GU\&D, for which they were rounded up to the nearest multiple of $k$.}\label{tbl:conv1}
{\small
\begin{tabular}{clcccccccc}
&\multicolumn{9}{c}{\large{\bf ``Trials to 99\% Convergence''}}\\
  \toprule
  &\multirow{2}{*}{\bf Distribution}&\multicolumn{4}{c}{Weibull \tiny{(Shape Parameter)}}&\multicolumn{2}{c}{Logistic}&\multicolumn{2}{c}{Log-Normal}  \\
   && 1 \tiny{(Exp.)} & 1.5 & 3.5 & 5 & `Tight' & `Disperse' & `Tight' & `Disperse' \\
  \toprule
\multirow{6}{*}{\bf$m=5$}
 & SU\&D  &  12 & 9  & 6  & 10 & 7 &  9 &  6  & 9\\
\cmidrule{2-10}
&BCD, $\Gamma=0.293$ & 19 &17 &15 &17 &14 &17  &13 & 16 \\
 & KR, $k=2$ &{\bf 17}&{\bf 14} &{\bf 10}&{\bf 10}&{\bf 10}&{\bf 14}&{\bf 8}&{\bf 13} \\
&GU\&D $_{(2,0,1)}$ & 20 &16 &12 &{\bf 10} &12 &16  &12 & 16 \\
\cmidrule{2-10}
&BCD, $\Gamma=0.347$&{\bf 18}&{\bf 15}&{\bf 12}&{\bf 12}&{\bf 12}&{\bf 15}&{\bf 10}&{\bf 14}\\
&GU\&D $_{(3,0,2)}$&36&27&15&15&15&24&15&24\\
\midrule
\multirow{6}{*}{$m=10$}
&SU\&D&34&26&14&12&13&22&14&26\\
\cmidrule{2-10}
&BCD, $\Gamma=0.293$ &44&38&30&31&28&39&23&33\\
&KR, $k=2$&{\bf 39}&{\bf 32}&{\bf 21}&{\bf 21}&{\bf 20}&{\bf 32}&{\bf 17}&{\bf 27}\\
&GU\&D $_{(2,0,1)}$ &46&36&24&22&22&36&20&32\\
\cmidrule{2-10}
&BCD, $\Gamma=0.347$&{\bf 44}&{\bf 35}&{\bf 24}&{\bf 23}&{\bf 22}&{\bf 34}&{\bf 19}&{\bf 31}\\
&GU\&D $_{(3,0,2)}$&81&57&33&30&33&54&27&51\\
\bottomrule
\end{tabular}
}\end{center}
\end{table}
Some observations from Table~\ref{tbl:conv1}:
\begin{enumerate}
\item SU\&D converges faster than any non-median method examined. This is intuitively expected, since SU\&D requires a only single trial before transition in either direction.
\item For a given target response rate $p$, convergence rates are most strongly affected by a combination of distribution properties and spacing - specifically, by the difference in $F$ values between adjacent levels around target. For example, the ``shallowest'' CDF (exponential thresholds and $m=10$) causes the slowest convergence, and vice versa.
\item Among the three non-median methods, KR is solidly ahead, while BCD is slowest. Based on these and other numerical runs not shown here, one could expect BCD to take around $20-40\%$ more trials than same-target KR to reach the same degree of convergence. The performance gap increases with $k$.
\item The number of trials to 99\% convergence can be quite substantial when compared with the design limitations of applications such as Phase I clinical trials. Moreover, the Markovian method most closely resembling the `3+3' design (GU\&D$_{(3,0,2)}$) converges very slowly - even slower than BCD. It takes GU\&D$_{(3,0,2)}$ $50-90\%$ more trials than KR ($k=2$) to achieve the same degree of convergence, even though the former's target is closer to the median. Furthermore, recall that these are ensemble averages and not single-run statistics.
\end{enumerate}
Table~\ref{tbl:conv1}'s convergence rate estimates are in rough agreement with eigenvalue predictions for BCD and GU\&D (data not shown). KR rates show an agreement with the `marginalized' tpm eigenvalue, indicating that indeed it is the between-level convergence that is rate-limiting.

Note that in agreement with Proposition \ref{prop:krgudconv}, KR converges faster than GU\&D$_{(k,0,1)}$. Furthermore, same-targeted BCD is the slowest-converging. This has been observed universally, over a much wider range of scenarios than is reported here. I have yet to find a rigorous proof why. Can a heuristic explanation be found?

The KR-BCD convergence gap recalls the observation by \citet{GezmuFlournoy06} that for median-targeting GU\&D with fixed $k$, convergence speeds up with increasing $a$. In both cases, the (marginal) $p_u,q_u$ of the faster method are always as large or larger than those of the slower method -- and in such a way that the $\left|p_u-q_u\right|$ difference is also larger. This means that both the ``probability flux'' in each direction, and the net flux in direction of target, are larger for the faster method. This is formalized in the conjecture below.

\begin{conj}\label{thm:fasterconverge} Consider two U\&D designs: design 1 with $p,q,r$ and design 2 with $p',q',r'$, and all other things being equal. If $p_u\geq p^{'}_u,q_u\geq q^{'}_u$ for all levels, and $\mid p_u-q_{u+1}\mid\geq\mid p^{'}_u-q^{'}_{u+1}\mid$ when $l_u,l_{u+1}$ are both below or above target, then design 1 converges faster.
\end{conj}

Note the similarity to and difference from the `peakedness' definition, Definition \ref{def:peaked}. Here and there we compare up and down probabilities between adjacent levels. However, `peakedness' is related to the probability {\bf ratio}, while the convergence conjecture is related to the probability {\bf difference} (or `flux'). In the median-targeting GU\&D case, increasing $a$ increases the difference while making the ratio smaller; therefore for this design family one faces a tradeoff between fast convergence and sharp stationary distribution. On the other hand, KR maintains both a greater ratio and a larger difference between up and down probabilities, compared with BCD -- so it has a performance edge on both aspects.

\subsection{Bias of Early Reversals}
\subsubsection{First Reversal}
From this chapter's theoretical analysis, it is now clear that the number of trials to the first reversal is a random variable. Of practical interest is the location of first reversal, denoted $x_{R_1}$, conditional upon starting point $x_1$. Those who use the first reversal as a cutoff or transition point, implicitly assume that this location is close to target. Let us examine this for GU\&D$_{(k,0,1)}$ and KR. For both of them, assuming a start from the lowest level $l_1$, the distribution will be
\begin{equation}\label{eqn:Pr1strevup}
Pr\left(x_{R_1}=l_u\mid x_1=l_1\right)=\left[1-(1-F_u)^k\right]\prod_{v=1}^{u-1}(1-F_v)^k.
\end{equation}
This distribution is reminiscent of a generalized geometric r.v., with the reversal defined as `success', and with monotonically decreasing `failure' (i.e., passage) probabilities at each trial. If $s$ is sufficiently small w.r.t. the slope of $F$, then in the vicinity of $Q_p$ the `up' passage probability is approximately $0.5$. This means that conditional upon reaching $l_{u^*}$, the level immediately below $Q_p$, before the first reversal, the expected additional number of levels traversed is less than $1$ - landing us very close to target. However, the expected location of $x_{R_1}$ is a weighted average of this and of the expectation conditional upon {\bf not} reaching $l_{u^*}$:
\begin{equation}\label{eqn:E1strevup}
\begin{array}{l} E\left[x_{R_1}=l_u\mid x_1=l_1\right]=E\left[x_{R_1}=l_u\mid x_1=l_1,x_{R_1}\geq l_{u^*}\right]Pr\left(x_{R_1}\geq l_{u^*}\mid x_1=l_1\right)\\
+E\left[x_{R_1}=l_u\mid x_1=l_1,x_{R_1}<l_{u^*}\right]Pr\left(x_{R_1}<l_{u^*}\mid x_1=l_1\right).\end{array}
\end{equation}
Therefore, under typical design scenarios, if $Q_p>l_3$, $x_{R_1}$ should be expected to be well below target. Now, suppose $x_1=l_u$, with $l_1<l_u<l_{u^*}$. Then (\ref{eqn:E1strevup}) still holds, but now the second term includes the possibility that the chain's first move would be {\it downward}. So again, if $x_1$ is $2$ or more levels below target, $E\left[x_{R_1}\right]$ should also be below target. The same conclusions hold for BCD since eventual transition probability without a reversal is also $0.5$ on target, and is monotone decreasing with increasing $x$. On the other hand, if $x_1$ is close to target, the first reversal will be approximately `right on'.

If the experiment begins at the top level $l_m$, for SU\&D the outcome will be similar to that when starting from bottom. For the non-median methods, downward transition probabilities will become $0.5$ at the median - long before reaching target. Therefore, if $x_1$ is at or above the median, the upward bias of $x_{R_1}$ should be even greater than the downward bias when starting below target.

This asymmetry suggests that perhaps the interpretation of reversal for non-median-target designs may be wrong. Perhaps, instead of ``a point where a change in response is observed'' we should define reversal as a point where a change in chain direction is observed, i.e. a move up when the last move was down (ignoring trials that mandate no change). For SU\&D the two definitions are equivalent. It is easy to see that for non-median-target designs the second definition causes a symmetry in behavior, as far as first reversal bias is concerned. However, a closer look reveals that direction changes are more likely than reversals to occur far from target (to either side; more about this on Section 3.2). All in all, the potential benefit from this changed definition is not clear, and is outweighed by the added complication. Hence it is more useful to understand the properties of reversals as they are currently defined, and plan our design accordingly.

To sum it up, the location of $x_{R_1}$ is in general biased towards $x_1$, and is not guaranteed to be close to target unless $x_1$ itself is close.

\subsubsection{Subsequent Reversals}
So we have just learned that unless $x_1$ is very close to target, the point of first reversal $x_{R_1}$ is likely to be on the same side of target as $x_1$ (though closer on the average). Since $x_{R_2}$ is between $x_1$ and $x_{R_1}$, the second reversal is an even worse choice for cutoff or transition. However, the expected net movement away from target between first and second reversals, is obviously smaller than the expected gain between starting point and first reversal (since probabilities for moving away from target are smaller than for moving towards it). This makes {\bf the third reversal} an interesting cutoff-point candidate. The next potential candidate is naturally the fifth reversal. Choosing a cutoff point beyond the fifth reversal may be too late for many sample-size restricted applications.
\newpage
\section*{Glossary for Chapters 1-2}
\addcontentsline{toc}{section}{\em{Glossary}}

Glossary is alphabetically ordered, starting with acronyms, then mathematical symbols (Roman letters first, Greek following). There are additional glossaries for Ch. 3 and Ch.4-5, but this one is the most extensive.

Note: in order to keep the glossary manageable and informative, I chose to omit some symbols which make a one-time appearance somewhere in the text, if they have no general significance.

\begin{glossary}
\item[BCD] The Biased-Coin Up-and-Down design developed by \citet{DurhamEtAl95}.
\item[GU\&D] The Group Up-and-Down design first developed by \citet{Tsutakawa67}.
\item[KR] The ``k-in-a-row'' Up-and-Down design, first developed by \citet{WetherillEtAl66}; Markov-chain properties first described properly by \citet{Gezmu96}. The name KR is taken from \citet{IvanovaEtAl03}; I use it because it better distinguishes KR from other U\&D variants.
\item[SU\&D] The original median-targeting Up-and-Down design described by \citet{DixonMood48}.
\item[U\&D] Up-and-Down. In this thesis, I use the name U\&D to refer to the family of designs that generate Markov chains whose stationary distribution is centered on the target percentile. Includes (though not limited to) BCD, GU\&D, KR and SU\&D.
\\[1cm]
\item[$a,b$] The lower and upper transition criteria parameters in group U\&D designs. If $\leq a$ `yes' responses are observed in the $k$-sized cohort, a move `up' is mandated, and if $\geq b$ are observed a move `down' is mandated.
\item[$F$] The CDF of thresholds; throughout the thesis, assumed to be strictly increasing and to have a density.
\item[$f$] The threshold density; first derivative of $F$.
\item[$F_u$] The value of $F$ at treatment level $l_u$.
\item[$H$] A transformed CDF of thresholds for a certain family of GU\&D designs (see text for details). The concept of ``transformed CDF'' was first offered in this context by \citet{WetherillEtAl66}, where it was used erroneously for KR-type designs.
\item[$i$] Used for indexing trials and responses.
\item[$k$] The number of consecutive `yes' ('no') responses needed for a move `up' ('down') in below-(above-)median KR designs; the cohort size in group U\&D designs.
\item[$l_u$] Treatment level, indexed $u$.
\item[$m$] The number of treatment levels (if finite)
\item[$N$] Ensemble size (i.e., number of individual runs) used in simulations.
\item[$n$] Sample size of a single experiment.
\item[$p$] The target response rate.
\item[$P$] The transition probability matrix (tpm) of an U\&D design.
\item [$p_u,q_u,r_u$ \ {\small(indexed)}] The U\&D `up', `down' and `stay the same' probabilities, respectively, at level $l_u$.
\item[$Q_p$] The $p$-th quantile (or $100p$-th percentile) of $F$. $Q_p$, with $p$ specifically the target response rate, is the experiment's target.
\item[$R_j$] Index denoting the location (between $1$ and $n$) of the $j$-th reversal point.
\item[$s$] The spacing between adjacent treatment levels (assuming levels are evenly spaced).
\item[$T$] The response-threshold r.v. Viewing a binary-response experiment as thresholds being triggered by the treatments, underlies the entire statistical approach to the percentile-finding problem. Individual thresholds are then assumed to come from some (generally unknown) distribution with a density $f$.
\item[$t$] Actual values that $T$ takes.
\item[$u,v,u^{'},v^{'}$] Used for indexing treatment levels.
\item [$w_i$] The internal state at trial $i$ in a KR experiment, so that the current state of the experiment is fully described by the pair $\left(x_i,w_i\right)$. $w$ can take integer values from $0$ to $k-1$.
\item [$x$  \ {\small(generic)}] The treatment as an independent variable (in the algebraic sense).
\item [$x_i$ \ {\small(indexed)}]  The treatment value at trial $i$ of an U\&D design.
\item [$y_i$ \ {\small(indexed)}]  The binary response at trial $i$.
\item [$Y_{G}$] The cohort response variable in group U\&D experiments. It is binomial rather than binary.
\\[1cm]
\item[$\Gamma$] The BCD ``biased-coin'' parameter. It is equal to $p$, but is retained here for terminological compatibility with other publications.
\item[$\gamma_u$] The ``stationary profile'': the ratio between adjacent stationary frequencies, $\pi_{u+1}/\pi_u$.
\item[$\gamma(x)$] The ``stationary profile'' as an algebraic function of the independent variable $x$.
\item[$\lambda_u$] The $u$-th eigenvalue of the tpm $P$, arranged in decreasing order.
\item[$\pi$] The stationary and/or limiting distribution of treatment allocations.
\item[$\rho^{(i)}$] The probability distribution of treatments (across levels) at trial $i$.
\item[$\tau$] Used for indexing internal states in KR designs.
\end{glossary}

\chapter{Nonparametric Up-and-Down Estimation}\label{ch:est}

\section{Existing Estimators and Current Knowledge}\label{sec:est1}
Nearly all U\&D target estimators can be divided into two groups: \textbf{averaging estimators}, making direct use of the treatment chain $\{x_n\}$ only; and \textbf{response-based estimators}, making use of both treatments and responses. This chapter will discuss each type separately, in alternating order. Only the numerical study (Section~\ref{sec:estsim}) will compare them to each other. As the chapter's title indicates, the focus is solely on nonparametric estimators.

\subsection{Averaging Estimators}
\subsubsection{History and Basic Description}
Early U\&D estimators were modified averages of the treatment chain, with the modification coefficients based on normal theory \citep{DixonMood48,BrownleeEtAl53,Dixon65}. \cite{Tsutakawa67} studied the properties of an average of all trials beginning with $x_2$, and found a nonparametric expression for its variance (see further below). However, the U\&D estimator of choice in most fields is {\bf reversal averaging}, introduced by Wetherill and aptly named $\bar{w}$ \citep{WetherillEtAl66}.\footnote{Interestingly, Wetherill's reasoning for favoring reversals over all treatments was less theoretical than numerical: this estimator performed somewhat better in simulation.} The estimator $\bar{w}$ is a somewhat modified arithmetic average of $x$ at reversal points $\{x_{R_j}\}$ (reversals were defined in Section~\ref{sec:und1}):
\begin{equation}\label{eq:wbar}
\bar{w}=\frac{\sum_j x_{R_j}+x_{R_j-1}}{2n_R},
\end{equation}
where $n_R$ is the number of reversals. A straightforward reversal average is known as $\hat{w}$, introduced by \citet{Choi71}. Estimators belonging to the $\bar{w}$ or $\hat{w}$ type (i.e., some average of treatments at reversals only, beginning from a certain cutoff point) appear to be the most commonly used U\&D estimators in median-finding applications across most relevant fields, and in psychophysics experiments using KR. From here on I will refer to this estimator family as $\hat{w}$.\footnote{It should be noted that according to \citet{PaceStylianou07}'s recent meta-analysis of 16 U\&D anesthesiology experiments published in 2000-2006, half the studies still used \citet{DixonMood48}'s original estimator, with $\hat{w}$ only second in popularity.} As \citet{Choi71} pointed out, for SU\&D $\bar{w}$ is identical to the simpler $\hat{w}$ when $n_R$ is even, and differs from it by $\pm s/2n_R$ when $n_R$ is odd. For KR and other non-median methods, this is not the case, but the two are still very closely related.

Considering its prevalent use, $\hat{w}$ has received very scant theoretical attention in its $40$ years of existence. In the 1980's, \citet{Kershaw85,Kershaw87} looked at stationary properties, especially variance, and concluded that $\hat{w}$ is in fact inferior to Dixon and Mood's original estimator. Kershaw's conclusions were largely ignored, and $\hat{w}$ continues to be the most popular U\&D estimator. More recently, \citet{PaulFisher01} conducted a numerical study of $\hat{w}$ in the context of anesthesiology experiments. They concluded that $6$ or more reversals are needed for reliable estimation, contrary to a prevalent practice by some groups in that field that use only $4$.

\citet{Dixon65} claimed without proof, that SU\&D's empirical mean (apparently under stationarity) is unbiased if $F$ has a symmetric density. \citet{GarciaPerez98} simulated stationary SU\&D and KR runs (using the last $9000$ reversals from runs having $10000$ reversals each), finding that for KR $\hat{w}$ is biased towards the median (a bias that is of course zero for SU\&D).

Finally, the empirical mode can be seen as related to averaging estimators. Markov-chain studies \citep{DurhamEtAl95,GiovagnoliPintacuda98} prove that the empirical mode converges to the stationary mode, which is guaranteed to be one of the $2$ levels closest to target.  However, we saw in Section~\ref{sec:und2} that the stationary mode may end up being only the second-closest level to target, especially for BCD and KR designs. Additionally, the mode is restricted to be at a treatment level, while averages can vary on a finer scale and hence are more precise. It is therefore not surprising that numerical studies find this estimator performs poorly \citep{DurhamEtAl97,StylianouFlournoy02,Oron05}. The empirical mode will not be discussed further.

\subsubsection{Mitigating The Starting-Point Effect}
Perhaps the most obvious problem with averaging estimators is starting-point bias. The first treatment $x_1$ is completely arbitrary (even if determined with good judgment). Then, somewhere during the experiment, the chain begins meandering around target. But at what point does the chain become ``good enough'' to include in an averaging estimator? Since during U\&D's formative years it was not understood to generate a Markov chain, solution approaches had developed gradually and intuitively. \citet{BrownleeEtAl53} recommended discarding $x_1$ (they were also the first to note that $x_{n+1}$ is uniquely determined by the experiment, and incorporated it into the estimate instead). \citet{Dixon65} and then \citet{WetherillEtAl66} recommended starting from the first reversal. Much more recently, \citet{GarciaPerez98}, based on extensive KR simulations, recommended starting from the 2nd or 3rd reversal. In practice, most researchers use all reversals for $\hat{w}$, but some discard the first~$1-4$.\footnote{\citet{GarciaPerez98} tabulates a meta-analysis of KR vision research experiments published in the mid-1990's. The vast majority of the 82 studies used some form of reversal averaging. Out of 60 reversal-averaging studies for which information was available about initial-sequence removal, 32 removed no reversals, 14 removed the first 1-2 reversals, 12 removed 3 or more reversals, and 2 used some other criterion.}

In the previous chapter, we have seen that in general the first reversal point is quite likely to occur while the starting-point bias is still substantial. We will revisit this issue, and novel ways to tackle it, later on.

\subsubsection{Mitigating The Boundary Effect}
When the treatment set is finite and $Q_p$ lies too close to a design boundary, averaging estimators will display a strong bias away from that boundary (this point will be explained in more detail in the next section). \citet{GarciaPerez98} offers an interesting fix for this problem, dubbed ``layover boundary conditions'': one continues the experimental bookkeeping (for averaging purposes) as if there was no boundary. However, in case levels beyond the boundary are indicated, the boundary treatment is administered. For example, if $l_m=1$, $s=0.1$ and the transition rules indicate an upward move, the next treatment will be recorded as $1.1$ but administered at $1$. Afterwards, the chain will have to move back down ``through'' $1$ before reaching $0.9$.

Layover boundary conditions are equivalent to assuming that $F$ remains constant outside the design boundaries. The treatment chain becomes infinite-state, but since transition probabilities beyond the true boundaries are equal to those at the boundaries, the stationary frequencies of the `phantom' levels constitute a diminishing geometric sequence, and so $\pi$ still exists. This issue, too, will be inspected more closely later on.

\subsection{Response Based Estimation: Isotonic Regression}\label{sec:ir1}
Response-based averages use $\{\hat{F}_m\}$, the empirical binomial point estimates of $F$ at design points, created by tabulating the responses:
\begin{equation}\label{eq:Fhat}
\hat{F}_u\equiv\frac{\sum_{i=1}^n y_i\textbf{1}\left[x_i=l_u\right]}{\sum_{i=1}^n\textbf{1}\left[x_i=l_u\right]}.
\end{equation}
Probit and logit regression are occasionally enocountered in U\&D applications \citep{Storer93,CamorciaEtAl04}. Being parametric, they exhibit the usual pros and cons, with their quality depending mostly on whether the model family provides a fair approximation of $F$. Parametric U\&D estimators will not be discussed here.

A nonparametric response-based estimator - isotonic regression, which is in fact the NPMLE - was recently introduced to U\&D in the Phase I context, and has been recommended by the NIH \citep{StylianouFlournoy02,StylianouEtAl03,NIH01}. This innovation has yet to reach all other fields using U\&D; and perhaps for the better, since it was not directly compared with $\hat{w}$ which it aims to replace. This comparison appears further below in the numerical study.

Isotonic regression (IR) is a well-known nonparametric solution for a variety of statistical problems \citep{BarlowEtAl72,RobertsonEtAl88}. IR can be produced via a couple of lengthy-named algorithms (pool-adjacent-violators-algorithm, or PAVA; least-convex-minorant), but in fixed treatment-response design context it boils down to a set of binomial point estimates:  the original point estimate $\hat{F}_u$ wherever there is no monotonicity violation, and an estimate pooled from the responses at adjacent levels in case of a violation. The procedure is described in Algorithm 1 below.

\begin{algorithm}[h]
\caption{Pool-Adjacent-Violators Algorithm (PAVA), Adapted for Treatment-Response Designs}
\begin{algorithmic}\label{alg:PAVA}

\Procedure{PAVA}{$\{z_m\}, \{n_m\}$}
\Statex
\While {$\textbf{H}\equiv\left\{u: 1\leq u<m,z_u>z_{u+1}\right\}\neq\emptyset$}
    \State $i\gets\min (\textbf{H})$
    \State  $M\gets 1$
    \While {$z_i>z_{i+M}$}
        \State $z_i=\ldots =z_{i+M}\gets
\overline{z}_{i:i+M}\equiv\sum_{u=i}^{i+M}\tilde{n}_uz_u$, where $\tilde{n}_u\equiv n_u/\left(\sum_{j=i}^{i+M}n_j\right)$
        \State $M\gets M+1$
    \EndWhile
\EndWhile
\Statex
\State \textbf{Return} $\{z_m\}$.
\EndProcedure
\end{algorithmic}
\end{algorithm}

PAVA replaces any subsequence of $\{\hat{F}_m\}$ it deems ``violating'' by a sequence that is a repetition of a single value (which in the case of treatment-responses designs is simply the overall proportion of `yes' responses over the subsequence). This means that the IR estimate of $F$ is constant over any interval of $x$ values covered by a violating subsequence. In intervals between non-violating design points, IR provides no unique estimate. Since in order to find the percentile one has to perform an inverse estimation, we need a (forward) estimate of $F$ over the entire range. In treatment-response applications this problem is commonly solved by linear interpolation \citep[e.g.][]{DilleenEtAl03,StylianouFlournoy02}.\footnote{apparently, in some fields people still believe that a straight line is the simplest way to connect two points. So statisticians have their work cut out for them :)}.

\begin{table}[h]
\begin{center}
\caption[Isotonic Regression Example]{Summary tables illustrating how IR works in practice on treatment-response data. Data are taken from two simulated GU\&D$_{(2,0,1)}$ experiments with the same threshold distribution and $n=32$. Yes/no summaries from sequences of violating points are pooled together to a single point estimate, which is then used for all original points in the sequence. \label{tbl:ir_demo}}
\small{
\begin{tabular}{l@{\extracolsep{.1cm}}rrrrrr@{\extracolsep{1cm}}r@{\extracolsep{.2cm}}rrrrr}
\toprule
{\bf Treatment} & \multicolumn{3}{c}{\bf Raw Input} &\multicolumn{3}{c}{\bf IR Output}
&  \multicolumn{3}{c}{\bf Raw Input} &\multicolumn{3}{c}{\bf IR Output} \\
& Yes & No & $\hat{F}$ & Yes & No & $\hat{F}$ & Yes & No & $\hat{F}$ & Yes & No & $\hat{F}$ \\
\midrule
$0.17$ & $0$ &  $4$ &  $0.00$ & $0$ &  $4$ &  $0.00$ & $1$ &  $7$ &  $0.13$ & $1$ &  $7$ &  $0.13$ \\
$0.33$ & $3$ &  $9$ &  $0.25$ & $3$ &  $9$ &  $0.25$ & $\mathbf{4}$ &  $\mathbf{8}$ &  $\mathbf{0.33}$ &\multirow{2}{*} {$\mathbf{6}$} & \multirow{2}{*} {$\mathbf{14}$} & $\mathbf{0.30}$\\
$0.50$ & $\mathbf{3}$ &  $\mathbf{7}$ &  $\mathbf{0.30}$ & \multirow{2}{*} {$\mathbf{4}$} & \multirow{2}{*} {$\mathbf{10}$} &  $\mathbf{0.28}$ & $\mathbf{2}$ &  $\mathbf{6}$ &  $\mathbf{0.25}$  & & & $\mathbf{0.30}$\\
$0.67$ & $\mathbf{1}$ &  $\mathbf{3}$ &  $\mathbf{0.25}$ &    &   & $\mathbf{0.28}$ & $4$ &$0$ &$1.00$ & $4$ &$0$ &$1.00$ \\
$0.83$ & $1$ &  $1$ &  $0.50$ & $1$ &  $1$ &  $0.50$ & & & & &  \\
\bottomrule
& \multicolumn{6}{c}{\bf (Simulation Run 14)}  &   \multicolumn{6}{c}{\bf (Simulation Run 9)} \\
  \end{tabular}
}\end{center}
\end{table}

\begin{figure}[h]
\begin{center}
\includegraphics[scale=.8]{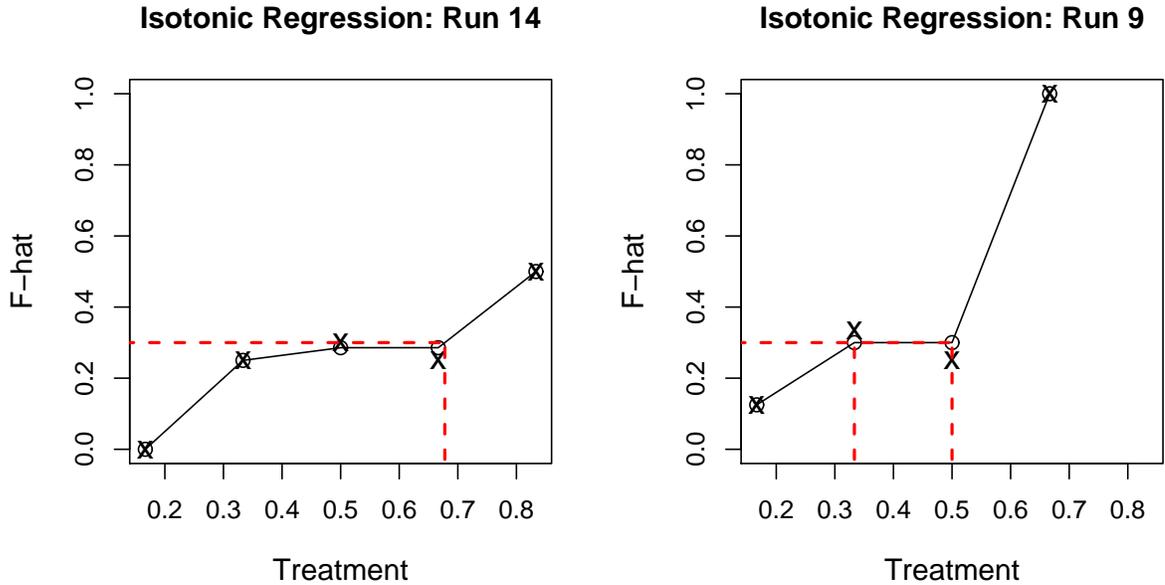}
\caption[Isotonic Regression Example Illustrated]{Graphs illustrating how IR works in practice on treatment-response data. Data are the same as shown on Table~\ref{tbl:ir_demo}. Raw $\hat{F}$ values are in `X' marks, and the IR output values are in circles connected by solid lines. The dashed red lines indicate the inverse-interpolation process. Note that in Run 9 (right), the IR estimate for $Q_{0.3}$ is not unique, and could be any value in $[1/3,1/2]$.\label{fig:ir_demo}}
\end{center}
\end{figure}

The operation of IR on treatment-response data is illustrated in Table~\ref{tbl:ir_demo} and Fig.~\ref{fig:ir_demo}. Shown are two simulated runs, in each of which there was a single violating pair. Assuming the target is $Q_{0.3}$, in Run 14 (LHS of both table and figure) $\hat{F}_3=p$ exactly; however, $\hat{F}_3>\hat{F}_4$. Therefore, technically we cannot determine a unique estimate via linear interpolation between $\hat{F}$ values; and conceptually, neither $\hat{F}_3$ nor $\hat{F}_4$ can be fully ``trusted'' without more information. IR resolves this by pooling the data from points $3,4$ together; the final estimate for $Q_{0.3}$ is to the right of point $4$. In Run 9 (RHS) we see an (extreme, but possible whenever $p$ is rational) example for IR's limitations: the flat stretch produced by the process lies exactly at $F=0.3$, and therefore IR does not provide a unique inverse estimate without some additional convention (taking the highest/lowest/midpoint of possible estimates, etc.). These two numerical examples will be revisited in Section~\ref{sec:est3}.

Why is IR the NPMLE for our application? Let us examine U\&D's nonparametric likelihood; it can be written as
\begin{equation}\label{eq:likelihood0}
\begin{array}{l}
\mathcal{L}=\Pr\left(\{x_n,y_n\}\right)=\prod_{i=2}^n \Pr\left(x_i|x_1\ldots x_{i-1},y_1\ldots y_{i-1}\right)\Pr\left(y_i|x_1\ldots x_i,y_1\ldots y_{i-1}\right)\\
=const.\times \prod_{i=2}^n \Pr(y_i|x_i)\\
=const.\times \prod_{u=1}^m F_u^{n_u\hat{F}_u}(1-F_u)^{n_u\left(1-\hat{F}_u\right)},
\end{array}
\end{equation}
where the $\hat{F}_u$ were defined above in (\ref{eq:Fhat}). The derivation uses the following properties:
\begin{itemize}
\item Given each treatment $x_i$, the corresponding response $y_i$ depends on no other treatment or response, since the underlying threshold sampling is i.i.d.;
\item Given all previous treatments and responses, $x_i$ can be found either deterministically, or (for BCD) using a separate, independent random draw which is unrelated to $F$.
\end{itemize}

Hence, the ``const'' appearing in the equation is either a product of functions of the biased-coin parameter $\Gamma$ (for BCD), or simply $1$ (for all other designs).

The form of (\ref{eq:likelihood0}) is identical to a product of binomial likelihoods, except for having different constant coefficients (the binomial likelihood has combinatorial coefficients). A product of binomials is the nonparametric likelihood of a standard non-sequential treatment-response experiment. Note that the nonparametric approach yields a likelihood that can be interpreted parametrically, as having $m$ parameters representing the true values $\{F_m\}$.

Now, since the log-likelihood takes the form

\begin{equation}\label{eq:likelihood1}
\log{\mathcal{L}}=const.+\sum_{u=1}^{m}n_u\left\{\hat{F}_u\log{F_u}+(1-\hat{F}_u)\log{[1-F_u]}\right\},
\end{equation}
the unconstrained NPMLE is of course just $\{\hat{F}_m\}$. Under monotonicity constraints, proceeding recursively from the first monotonicity-violating pair of $\hat{F}$ values and treating one pair at a time, it can be easily shown that the constrained NPMLE is identical to PAVA's output. More detailed derivations appear in standard references \citep{BarlowEtAl72,RobertsonEtAl88}.

The ``flat'' intervals produced by PAVA in case of a monotonicity violation in $\{\hat{F}_m\}$ appear to hurt estimation performance, both forward and inverse. I will inspect this suspicion more closely in the next section.

\subsection{Interval Estimation}
Nonparametric U\&D interval estimation has been even more seldom studied than point estimation. However, interest in this topic as well has been picking up recently.
\subsubsection{SD-Based Approaches}
\citet{DixonMood48}'s original averaging estimator came with a variance estimator based on normal theory. \citet{PollakEtAl06}, using numerical simulations of material-fatigue SU\&D testing, show that this variance estimator performs very poorly, and is strongly dependent upon the spacing $s$. They suggest a 3-parameter empirical correction function fitted to their data.

In anesthesiology, some researchers using SU\&D with $\hat{w}$ report confidence intervals. In calculating standard errors, they seem to count each pair of reversals as a single independent observation, and assume asymptotic normality of the average, hence the method's name: ``independent paired reversals'' \citep{CapognaEtAl01,CamorciaEtAl04}.\footnote{To be honest, up till now (summer 2007) I have yet to find a full documentation, nor a reference for ``independent paired reversals''. The observation reported here is based on ``reverse engineering'' of reported estimates in anesthesiology articles using the raw data, and I am not $100\%$ sure that it is correct.}

\citet{Tsutakawa67} pursued a Markov-chain based approach. He noted that a stationary Markov chain can be split into subchains, identified by successive {\bf hitting times} at a given state - i.e., the points at which the treatment chain visits the state. The lengths of these subchains, $\{\nu_j\}$, are i.i.d. random variables, as is any algebraic functional on the treatments, calculated separately per subchain. \citet{Tsutakawa67} then goes on, basing his calculations on \citet[Ch. 1]{Chung60}, and claims that
\begin{equation}\label{eq:tsukavar1}
\frac{\sum_j \nu_j^2\left(\bar{x}_j-\bar{x}\right)^2}{\sum_j \nu_j^2}\to Var(\bar{x})
\end{equation}
as $n\to\infty$. Therefore, he proposes the LHS of (\ref{eq:tsukavar1}) as an estimator of the variance for the empirical-mean estimator. However, from data tables in that article, as well as my own numerical trials, it seems that the LHS approaches the limit from below, and therefore the resulting CI's are too optimistic.

Another approach to averaging-estimator confidence intervals was presented by \citet{Choi90}. Starting from first principles, the variance of the average of $n$ r.v.'s $\{V_1,\ldots V_n\}$ is
\begin{equation}\label{eq:varsumplain}
Var\left[\bar{V}_n\right]=\frac{\sum_i Var(V_i)+2\sum_{i<j}Cov(V_i,V_j)}{n^2}.
\end{equation}
Now, being part of a Markov chain, $x$ values at reversals are clearly dependent. Assuming that they are identically distributed and that dependence follows an AR(1) autoregressive form, \citet{Choi90} obtains formulae for the s.e.\  of $\hat{w}$ as a function of the threshold s.d. $\sigma$ and the autocorrelation coefficient $\rho$. Additionally assuming normally-distributed thresholds, CI's can be calculated. \citet{Choi90} found them to be overly optimistic, but better than Dixon and Mood's original CI's.

\subsubsection{Bootstrap Approaches}
Recently, use of the bootstrap for U\&D CI's was explored -- both for parametric estimation  \citep{ChaoFuh01} and for for isotonic regression \citep{StylianouEtAl03}. The bootstrap approach for U\&D is to simulate chains of length $n$, with transition probabilities determined via some estimate of $F$ (IR or a parametric regression). In both cases, however, the bootstrap was used not for CI's around $Q_p$, but for CI's around the forward estimate of $F$ at the closest design level. \citet{ChaoFuh01} used the ordinary bootstrap and bootstrap-$t$, while \citet{StylianouEtAl03} used the bias-corrected bootstrap \citep[see][Ch. 5]{DavisonHinkley95}. \citet{PollakEtAl06} explored using the bootstrap combined with their empirical fix to \citet{DixonMood48}'s formula, and reported good results.

\subsubsection{Use of Current-Status Theory}
Recalling that U\&D is a special case of current-status data, there is a wealth of theoretically-derived interval estimation knowledge  in that field \citep{BanerjeeWellner05}. However, U\&D appears to be too special a case: under the typical current-status assumptions, treatments are random with a continuous CDF. This leads to an $n^{1/3}$ rate of convergence everywhere for the NPMLE. U\&D, though its treatments are random, is more akin to standard treatment-response designs in that treatments are limited to a discrete finite grid -- while the continuity of treatments is a key assumption in most current-status CI discussions. As a result, the point estimates $\{\hat{F}_m\}$ converge with an $n^{1/2}$ rate, and the linearly-interpolated IR used in U\&D converges at the same rate - but not to $F$, but rather to $\bar{F}$ - a linear interpolation of $F$ between the design points. Therefore, it appears that current-status CI theory is not applicable to U\&D interval estimation.

\section{Theoretical Study}\label{sec:est2}
\subsection{Conceptual Prelude}\label{sec:estconcept}
We have at our disposal two nonparametric approaches to U\&D estimation. The response-based one is generic to any binary-outcome design. It does not make direct use of the properties of $\pi$, but benefits indirectly from having more precise $\hat{F}$ estimates in the vicinity of $Q_p$. Averaging estimators, on the other hand, are intimately related to the peculiar properties of U\&D sampling.

Considering the stationary properties already discussed here, one can say that some sort of empirical treatment mean is {\bf the ``natural estimator'' for U\&D designs}. I put forward this rather strong statement, in order to re-emphasize the conversion trick that underpins averaging estimation. U\&D converts censored sampling back into direct sampling -- which, as suggested in Section \ref{sec:prefconcept}, would be far more desirable than indirect, binary censored sampling.

However, to utilize the potential of averaging estimators, we need to better understand the properties of $\pi$ and of random-walk sampling from it. Unfortunately, since the averaging estimators currently in use were developed without the Markovian perspective, they do not utilize the full estimation potential. In particular, choosing to average reversal points means that we are not sampling from $\pi$ itself, but from its ``sibling'': the stationary distribution of reversals. Much of this section will be devoted to untangling the peculiar properties of such sampling, and to suggesting other approaches.

Meanwhile, even though isotonic regression is not ``natural'' in the sense discussed above, this also means that it suffers less from U\&D's peculiarities and limitations, such as starting-point bias or boundary effects. Therefore we expect it to be more robust, and will look at its properties more carefully later on.

\subsection{Estimation Properties of Treatment Means}\label{sec:bias}
\subsubsection{Bias of the Stationary Mean}\label{sec:bias}
We have already seen that for median-targeting designs and for GU\&D$_{(k,0,1)}$, the stationary mode is on the closest treatment to target on the response scale - which is also the closest on the threshold or treatment scale (to first order in $s$). BCD and KR tend to shift the mode somewhat to the left of target. One could expect, therefore, that with the former variants the stationary mean $\mu_\pi$ would show little or no bias w.r.t. $Q_p$, while with the latter there would be a downward bias, or more generally, a bias away from the median.

More insight into stationary biases can be gained via a telescopic-series analysis. First, note that in the limit of infinitely fine design ($s\rightarrow 0$), the stationary biases of BCD and KR, too, should vanish. Therefore, we may expect the bias magnitude to be related to $s$. In the subsequent discussion, we assume that there are no design boundaries (i.e., the treatment set is potentially infinite) that $F$ is twice continuously differentiable in $x$, and also -- for convenience -- that $Q_p$ falls exactly on $(l_{u^*}$ (letting $Q_p$ land midway between design levels is nearly as convenient).\footnote{Numerical runs (data not shown) seem to indicate that biases related to design offset w.r.t to target are much smaller than the spacing-related bias discussed here.} One can then express the stationary mean as a telescopic series around target:
\begin{eqnarray} \label{eqn:taylortelescope}
\mu_\pi\equiv\sum_{u}\pi_ul_u & = & Q_p+s\sum_{j}j(\pi_{u^*+j}-\pi_{u^*-j}) \\
\nonumber & = & Q_p+s\pi_{u^*}\sum_{j}j\left[\prod_{v=u^*}^{u^*+j-1}\gamma_{v}-\prod_{v=u^*-j}^{u^*-1}\gamma_{v}^{-1}\right].
\end{eqnarray}
Since the stationary frequencies diminish in a faster-than-geometric rate away from $Q_p$, the sum is typically dominated by the first $1-3$ summand pairs. The pairs are composed of opposite-sign terms, which may nearly cancel out. We now carry out a detailed calculation of the first pair for SU\&D.

First, note that if $F$ has a symmetric density, for SU\&D the terms in each pair will cancel each other \emph{exactly}. So apart from design-offset biases, SU\&D averaging estimation is indeed unbiased as claimed by \citet{Dixon65}. For asymmetric distributions, we would need standard Taylor expansions around target:

\begin{eqnarray} \label{eqn:taylorbasic}
\frac{1}{F_{u^*+1}} & = & \frac{1}{F_{u^*}}-\frac{sf_{u^*}}{F_{u^*}^2}+ \frac{s^2\left(2f_{u^*}^2-f^{'}_{u^*}F_{u^*}\right)}{2F_{u^*}^3}+\ldots\\
\nonumber \frac{1}{1-F_{u^*-1}} & = & \frac{1}{1-F_{u^*}}-\frac{sf_{u^*}}{(1-F_{u^*})^2}+ \frac{s^2\left(2f_{u^*}^2+f^{'}_{u^*}(1-F_{u^*})\right)}{2(1-F_{u^*})^3}+\ldots,
\end{eqnarray}
where $f$ is the threshold density and $f^{'}$ its derivative. Substituting this and setting $F_{u^*}=0.5$, the first summand pair then becomes
\begin{eqnarray} \label{eqn:sudtaylorbias}
\gamma_{u^*}-\gamma_{u^*-1}^{-1} & = & 0.5\left[2-4sf_{u^*}+2s^2\left(4f_{u^*}^2-f^{'}_{u^*}\right)\right]+\ldots\\
\nonumber & & - 0.5\left[2-4sf_{u^*}+2s^2\left(f^{'}_{u^*}+4f_{u^*}^2\right)\right]+\ldots\\
\nonumber & \approx & -2s^2f^{'}_{u^*}.
\end{eqnarray}
The first-order terms cancel out, and we are left with a bias that is second-order in $s$ with a sign opposite that of $f'(Q_{0.5})$ (for unimodal or monotone-decreasing threshold densities, this means a bias in the direction of $F$'s skew). It is quite straightforward to show that all summand pairs in (\ref{eqn:taylortelescope}) for SU\&D give similar negative second-order terms when thus expanded. Therefore, it appears that SU\&D's stationary mean is a biased estimator of $Q_{0.5}$ for asymmetric densities. However, this bias can be kept very small under reasonably fine spacing.

The same analysis for BCD yields
\begin{eqnarray} \label{eqn:bcdtaylorbias}
\gamma_{u^*}-\gamma_{u^*-1}^{-1} & = & \frac{\Gamma}{F_{u^*+1}}-\frac{1-\Gamma}{1-F_{u^*-1}}\\
\nonumber & \approx & \frac{2p-1}{p(1-p)}sf_{u^*}<0,
\end{eqnarray}
and for KR
\begin{eqnarray} \label{eqn:krtaylorbias}
\gamma_{u^*}-\gamma_{u^*-1}^{-1} & = & p\left[\frac{1}{F_{u^*+1}}-\frac{1-\left(1-F_{u^*-1}\right)^k}{F_{u^*-1}\left(1-F_{u^*-1}\right)^k}\right]\\
\nonumber & \approx & 2\frac{(k+1)p-1}{p(1-p)}sf_{u^*}<0.
\end{eqnarray}
In both cases there is a downward (i.e., away from median) first-order bias. BCD's bias term is larger than same-targeted KR, by a factor of about $5/3$ (the exact ratio as a function of $k$ is found by substituting KR's target (\ref{eqn:krtarget}) for $p$). GU\&D$_{(k,0,1)}$ ``inherits'' the properties of SU\&D, meaning that its bias is only second-order and therefore usually smaller than that of the other two methods. For symmetric or upward-skewed threshold distributions (a realistic assumption when dealing with positive thresholds), GU\&D$_{(k,0,1)}$'s bias is positive, i.e., towards the median.

So we have seen that stationary biases are approximately first-order or second-order in the spacing, depending upon the design. All these finds are in line with numerical observations such as Fig.~\ref{fig:peaked} and with the theory in Section~\ref{sec:und2}. Numerical results indicate that for all four variants, stationary biases are quite moderate except when using very coarse spacing.

\subsection{Averaging All Treatments or Only Reversals?}
The simplest candidate to replace the reversal average $\hat{w}$ is some average of {\bf all} treatments (possibly starting from a certain cutoff point); let us call this latter estimator $\hat{v}$. \citet{WetherillEtAl66}'s simulations showed $\hat{w}$ as having smaller MSE than $\hat{v}$, and this is why it was chosen. Since then, other simulations \citep[e.g., ][]{Kershaw85} have shown the opposite. Can we make a more theoretical comparison between the two options?

\subsubsection{Reversals as a Filter}
Assuming stationarity, the frequency of reversals at each level should also become stationary. However, the stationary distribution of reversals is not necessarily identical to $\pi$. We can calculate it using $\pi$ and the balance equations:
\begin{eqnarray}\label{eqn:reverspi_generic}
\pi^{(rev.)}_u&=&\frac{\left[\pi_{u-1}p_{u-1}+\pi_{u}r_{u}\right]F_u+\pi_{u+1}q_{u+1}\left(1-F_u\right)}{\sum_v \pi^{(rev.)}_v}\\
\nonumber & = & \frac{\pi_{u}\left[(q_u+r_u)F_u+p_u\left(1-F_u\right)\right]}{\sum_v \pi^{(rev.)}_v}\\
\nonumber & = & \frac{\pi_{u}\left[p_u+F_u-2p_uF_u\right]}{\sum_v \pi^{(rev.)}_v},
\end{eqnarray}
where we implicitly assumed that `up' and `down' transitions occur only after a `no' or a `yes', respectively, and that staying at the same level (if possible) also follows a `no'. Conveniently enough, the (relative) stationary reversal frequency at a given level is a linear function of $\pi$ at the same level. This means, that picking only reversals can be viewed as a secondary, linearly filtered sampling from $\pi$.

For SU\&D, (\ref{eqn:reverspi_generic}) translates to
\begin{equation}\label{eqn:reverspi_sud}
\pi^{(rev.)}_u\propto\pi_u\left[F_u^2+\left(1-F_u\right)^2\right],
\end{equation}
where the normalizing proportionality factor has been omitted. So the filter is distorting: levels far from the median are over-sampled compared with those near the median, by a factor of up to $2$ - which means that the stationary variance of reversals is larger than that of the entire chain. If the threshold density is exactly symmetric, the sampling does not induce a bias to the averaging estimator. But if it is asymmetric, the sampling induces a slight bias in direction of the distribution's skew -- i.e., in the same direction as the stationary-mean bias.

For BCD, $\pi^{(rev.)}$ is proportional to
\begin{equation}\label{eqn:reverspi_bcd}
\pi^{(rev.)}_u\propto\pi_u\frac{\Gamma}{1-\Gamma}(1-F_u)(1-2F_u)+F_u,
\end{equation}
which reduces to (\ref{eqn:reverspi_sud}) under $\Gamma=0.5$. The sampling coefficient's minimum point, as a function of $F$, is at
\begin{equation}\label{eqn:reverspi_bcdmin}
F^{(min.)}=\frac{4\Gamma-1}{4\Gamma}.
\end{equation}
The minimum moves to the left of $\Gamma$  as $\Gamma$ decreases; it is positioned at zero for $\Gamma=0.25$. Thus, for most or all of $F$'s range, sampling probability increases with $F$ - meaning that BCD reversal averaging induces a bias towards the median. Here, too, we expect $\pi^{(rev.)}$ to have larger variance, since the filtering is more aggressive near target.

For KR we get
\begin{equation}\label{eqn:reverspi_kr}
\pi^{(rev.)}_u\propto\pi_u\frac{F_u\left[ 1-2F_u(1-F_u)^k\right]}{1-(1-F_u)^k}.
\end{equation}
Again, the formula reduces to (\ref{eqn:reverspi_sud}) for $k=1$. KR sampling has the same qualitative properties as BCD (i.e., bias towards the median and a larger stationary variance), albeit milder. The filtered sampling coefficient has a minimum slightly to the left of target.

Note that for BCD and KR, $\hat{w}$'s additional bias is in the opposite direction of $\mu_\pi$'s stationary-mean bias. This leads to the interesting question whether the two cancel out. First-order analysis indicates that the reversal bias may be of comparable magnitude, but is generally smaller than the stationary-mean bias. The exact interplay between the two would depend upon specific distribution and spacing conditions.

\subsubsection{Variance Comparison}
One way to generically describe the variance of a dependent-sample average is
\begin{equation}\label{eq:segeneric}
\sigma_{\bar{x}}=\frac{\sigma_{\pi}}{\sqrt{n_{eff}}},
\end{equation}

where $n_{eff}$ is an effective sample size, usually smaller than $n$. We have seen that under stationarity, $\sigma_{\pi}$ for reversals is larger than for all trials. What about $n_{eff}$? This is a measure of the information content in the sample about the distribution of trials (or reversals). But reversals are a subset of the sample used for $\hat{v}$; their information content cannot be larger. Therefore, $Var\left(\hat{w}\right)>Var\left(\hat{v}\right)$.

Instead of this intuitive and not quite rigorous approach, it is also possible to obtain the same result via a direct and more tedious approach for some of the designs.
\begin{prop}\label{prop:varvw} For SU\&D and GU\&D$_{(k,0,1)}$ designs, under stationarity $Var\left(\hat{w}\right)>Var\left(\hat{v}\right)$.
\end{prop}
\begin{proof} We slice the chain into subchains, each beginning with a reversal and ending right before the next reversal. For each subchain indexed $j$, let $a_j$ be the subchain's treatment mean. A simple average of $a_j$ is equal to $\bar{w}$ which is known to be equivalent to $\hat{w}$. Meanwhile, $\hat{v}$ is an average of the $a_j$'s, weighted by the subchain length $n_j$. Which one has smaller variance?

If shorter subchains have a higher probability of being farther away from $\mu_a$, the population mean of the $a_j$'s (which is equal to $E[\hat{w}]$), then definitely $Var\left(\hat{w}\right)>Var\left(\hat{v}\right)$, since the latter down-weights them. Now, at a given level $l_u$, the length of a subchain ending with the next reversal is a generalized geometric r.v. with increasing ``success'' probabilities (''success'' being a reversal). However, the subchain-terminating ``success'' probabilities are always larger in the direction away from $Q_p$, which (ignoring boundary conditions) is very close to $\mu_a$. Therefore, subchains are expected to be shorter, the farther their center is from $\mu_a$, and hence $Var\left(\hat{w}\right)>Var\left(\hat{v}\right)$.\qed
\end{proof}

For BCD and KR, this approach encounters some difficulties. Now $\hat{v}$ is not a weighted average of $a_j$'s, but somewhat smaller than that average because ascent subchains are ``bottom-heavy'' (recall that $\hat{w}$ is biased upwards w.r.t $\mu_\pi$). Subchains still tend to be shorter when they are oriented away from $Q_p$, but because of the upward bias we are not quite sure how close $Q_p$ is to $\mu_a$. However, since the biases (of $\mu_\pi$ w.r.t to $Q_p$ and of $\mu_a$ w.r.t $\mu_\pi$) are opposite in sign, there is little reason to expect a different relationship between the two variances for BCD and KR.

\subsection{Isotonic Regression and Bias}
As detailed in Section~\ref{sec:est1}, whenever isotonic regression (IR) departs from simple linear interpolation, it produces constant intervals. Besides being problematic for inverse estimation of quantiles, since we assumed $F$ to be continuous strictly increasing this may generate biases. Specifically, if $F$ is as assumed, then any estimator $z(t)$ producing a constant interval $z=c\ \ \forall t\in[t_1,t_2]$ is either unbiased only at a single point on $[t_1,t_2]$, or is biased across the entire interval. Common sense suggests that if IR is based on unbiased estimates $\{\hat{F}_m\}$, then the former case is far more likely. This can in fact be proven in some specific scenarios, such as normal errors with variance inversely proportional to the IR weights.

The proof requires two preparatory lemmas.

\begin{lem}\label{lem:PAVAindlast} Assume that at each treatment level $l_u$ we have a raw point estimate $\hat{F}_u$, which is normal with mean $F_u$ and variance $V_0/n_u$, and that all point estimates are mutually independent. Then for each sequence of points indexed $u\ldots u+M$ which are replaced by a single value in the PAV algorithm (hereafter: a violating interval), the pooled IR estimate is independent of the degree of the last violation in the sequence, i.e.
\begin{equation}\label{eqn:homoindep}
\tilde{F}_{u:u+M}\perp\left(\tilde{F}_{u:u+M-1}-\hat{F}_{u+M}\right),
\end{equation}
where $\tilde{F}_{u:u+M}$ denotes a weighted average of $\hat{F}$ values from $l_u$ through $l_{u+M}$.\end{lem}
\begin{proof}
Let $Y\equiv\left(Y_1,Y_2\right)^{'}$ be independent normal r.v.'s with variance
$$
\textbf{V}\equiv \left(
                    \begin{array}{cc}
                      V_1 & 0 \\
                      0 & V_2 \\
                    \end{array}
                  \right).
$$ Then $Z$, a linear transformation of $Y$, $Z=\textbf{A}Y$, will be normal with variance $Var[Z]=\textbf{AVA}^{'}$. Specifically, $Z=\left(Y_2-Y_1,\frac{KY_1}{V_1}+\frac{KY_2}{V_2}\right)^{'}$, with $K\neq 0$ some constant, will have a variance
$$
\left(              \begin{array}{
                    cc}
                      V_1+V_2 & 0 \\
                      0 & \frac{K^2}{V_1}+\frac{K^2}{V_2}\\
                    \end{array}
                  \right).
$$
Therefore, $Z_1\perp Z_2$.

Now take a violating interval extending from $l_u$ to $l_{u+M}$, and set $Y_1=\tilde{F}_{u:u+M-1}, Y_2=\hat{F}_{u+M}$. Then

$$Var\left(\hat{F}_{u+M}\right)=\frac{V_0}{n_{u+M}},$$ while
$$
Var\left(\tilde{F}_{u:u+M-1}\right)=\frac{V_0}{\left(\sum_{v=u}^{u+M-1} n_v\right)^2}\sum_{v=u}^{u+M-1} n_v=\frac{V_0}{\sum_{v=u}^{u+M-1} n_v}.
$$ The two are independent, and
$$
\tilde{F}_{u:u+M}=K\left[\left(\sum_{v=u}^{u+M-1} n_v\right)\tilde{F}_{u:u+M-1}+n_{u+M}\hat{F}_{u+M}\right],
$$ i.e., an average of the two with weights inversely proportional to variances. By the proof above, this average is independent of the difference $Y_2-Y_1$, which in our case denotes the degree of last violation in the sequence.\qed
\end{proof}

\begin{lem}\label{lem:pavaindependence} Under the same assumptions, the weighted average $\tilde{F}_{u:u+M}$ is independent of the event that the points indexed $u\ldots u+M$ are part of a violating sequence starting at $u$.
\end{lem}
\begin{proof} By induction. Let the event in question be denoted $\widetilde{viol}_{u:u+M}$. For $M=1$, the identity follows directly from Lemma \ref{lem:PAVAindlast}.

For $M>1$, assume $\tilde{F}_{u:u+M-1}\perp \widetilde{viol}_{u:u+M-1}$, and prove for $M$. Due to the independence of the $\hat{F}$-s, trivially $\tilde{F}_{u:u+M}\perp \widetilde{viol}_{u:u+M-1}$. Now the event $\widetilde{viol}_{u:u+M}=\widetilde{viol}_{u:u+M-1}\cap\{\hat{F}_{u+M}<\tilde{F}_{u:u+M-1}\}$. But by Lemma \ref{lem:PAVAindlast}, $\tilde{F}_{u:u+M}\perp\{\hat{F}_{u+M}<\tilde{F}_{u:u+M-1}\}$. Therefore it is also independent of the intersection of these two events and the proof is complete.\qed
\end{proof}
These two lemmas show that weighted averages such as those produced by the PAV algorithm, are invariant to monotonicity violations (under the specified distributional assumptions). This clears the way to a simple and useful fix to IR's flat-interval problem.

\begin{thm}\label{thm:updown} Under the same assumptions as above, for each violating interval from $l_u$ to $l_{u+M}$, PAVA produces an estimate which is biased upward at $l_u$ and downward at $l_{u+M}$ from its unconditional expectations $F_u,F_{u+M}$, respectively.\end{thm}
\begin{proof} From Lemma \ref{lem:pavaindependence},
$$
E\left[\tilde{F}_{u:u+M}\mid \widetilde{viol}_{u:u+M}\right]=E\left[\tilde{F}_{u:u+M}\right]=\overline{F}_{u:u+M},
$$
the weighted average of $F$ over the violating design points. By the strict monotonicity of $F$, $F_u<\overline{F}_{u:u+M}<F_{u+M}$.\qed
\end{proof}

So, for these particular distribution assumptions on point estimates, we have ``nailed'' the IR estimate to be unbiased somewhere within the flat interval. It follows naturally, that percentile (inverse) estimates around the interval must be biased away from the interval to either direction.

Of course, in a treatment-response case the binomial estimates $\{\hat{F}_m\}$ are not normal. However, the scenario covered by Theorem~\ref{thm:updown} approximates their asymptotic behavior as $n\rightarrow\infty$, especially if the binomial variance factor $F(1-F)$ is changing slowly over the violating interval.

Therefore, it may be useful to see where in the interval, according to that scenario, the IR estimator is unbiased. The answer is quite simple:

\begin{thm}\label{thm:unbiasedpoint} Under the same assumptions as above, and if $F$ is twice continuously differentiable in $x$, then $\tilde{F}_{u:u+M}$ -- the PAVA estimate for a violating interval from $l_u$ to $l_{u+M}$ -- is unbiased to (Taylor) first order in segment length $l_{u+M}-l_u$, only at the weighted-average point
\begin{equation}\label{eqn:weightedx}
\overline{l}_{u:u+M}\equiv\sum_{v=u}^{u+M}\omega_vl_v,\textrm{ where  }\omega_v\equiv\frac{n_v}{\sum_{j=u}^{u+M}n_j}.
\end{equation}
\end{thm}
\begin{proof} First, from $F$'s continuity, strict monotonicity and Theorem \ref{thm:updown} we know that there is exactly one point at which $\tilde{F}_{u:u+M}$ is the true value of $F$, and that this point is in $(l_u,l_{u+M})$.

Now $\tilde{F}_{u:u+M}$ can be written as a first-order Taylor expansion from a single reference point $x*\in (l_u,l_{u+M})$:
\begin{equation}\label{eqn:x*}
\begin{array}{l}
\tilde{F}_{u:u+M}= F(x*)+\sum_{v=u}^{u+M}\omega_vF'(x*)(l_v-x*)+\frac{1}{2}\sum_{v=u}^{u+M}\omega_vF''(\xi_v)(l_v-x*)^2\\
=F(x*)+F'(x*)\left(\overline{l}_{u:u+M}-x*\right)+\frac{1}{2}\sum_{v=u}^{u+M}\omega_vF''(\xi_v)(l_v-x*)^2,
\end{array}
\end{equation}
where $\{\xi_v\}$ are points between $x*$ and $l_v, v=u\ldots u+M$. At $x*=\overline{l}_{u:u+M}$, the first-order term vanishes, and therefore to first order,
$$
E[\tilde{F}_{u:u+M}]\approx F(\overline{x}_{i:i+M}).
$$
Clearly, this is also the only location in the interval that removes the first-order term.\qed
\end{proof}

This result will be used in the next section to improve upon IR.

\section{New Methodologies}\label{sec:est3}
\subsection{Averaging Estimators}
\subsubsection{Overview}
Averaging estimators are not just commonly used in U\&D; surprisingly or not, their performance is also pretty hard to beat under the popular SU\&D and KR designs. The convergence of Markov chains implies $\bar{x}_n\rightarrow\mu_\pi$, the stationary mean, with asymptotically normal behavior and without any further parametric assumptions on $F$. Averaging estimators typically have the smallest variance, and more often than not it is the variance term that dominates estimation performance. Averaging estimators' major drawback, as the previous section indicated, has to do with biases. Here is a summary of potential averaging estimators biases:
\begin{enumerate}
\item Bias towards the starting point $x_1$.
\item Bias away from the closest boundary.
\item Theoretical bias of $\mu_\pi$ itself, as exposed in Section~2.2.
\item If a reversal-only average is used, an additional reversal-sampling bias.
\end{enumerate}

Since there are so many biases in potentially different directions, sometimes they neutralize each other to a certain extent. This is why the reversal-only average $\hat{w}$ occasionally outperforms the all-treatment average $\hat{v}$ in numerical simulations. This phenomenon will be observed numerically later on, in Section~\ref{sec:estsim}.

Items 2 and 3 on this list can be managed by choosing a sufficiently fine spacing (which unfortunately increases the magnitude of item 1) -- and by not imposing unnecessary boundaries. Additionally, for KR designs there is a potentially interesting fix to item 3 (stationary bias). Recall from Section~\ref{sec:und2} that KR's sub-chain of base states (i.e., picking only the first trial of each visit to a treatment level) has an identical stationary distribution to that of GU\&D$_{(k,0,1)}$. This means that averaging only over the zero-state sub-chain may help us get rid of KR's first-order stationary bias.

for general designs, the first item on the list -- the starting-point effect -- has received the most attention in devising new estimators.

Let $\hat{w}_r$ be the average of all reversal points from reversal $r$ onward; according to \citet{GarciaPerez98}, this is the most popular estimator in vision research for KR designs. Our analysis up to now indicates that it should be outperformed by $\hat{v}_r$, the corresponding all-treatment average starting from the same reversal point. How should $r$ be chosen? This is a classic bias-variance dilemma: the farther out we push the cutoff point, the starting-point bias w.r.t $\mu_\pi$ decreases, but the variance increases.

Or perhaps there is another way out? Are reversals a good cutoff choice at all? The previous section showed that under stationarity reversals have a larger variance than a randomly-chosen treatment. This means their expected distance from $\mu_\pi$ is larger. Since stationarity is in fact a best-case assumption, it seems that reversals are not a very good choice for cutoff.

\subsubsection{A New Adaptive Estimator}

\begin{figure}
\begin{center}
\includegraphics[scale=.8]{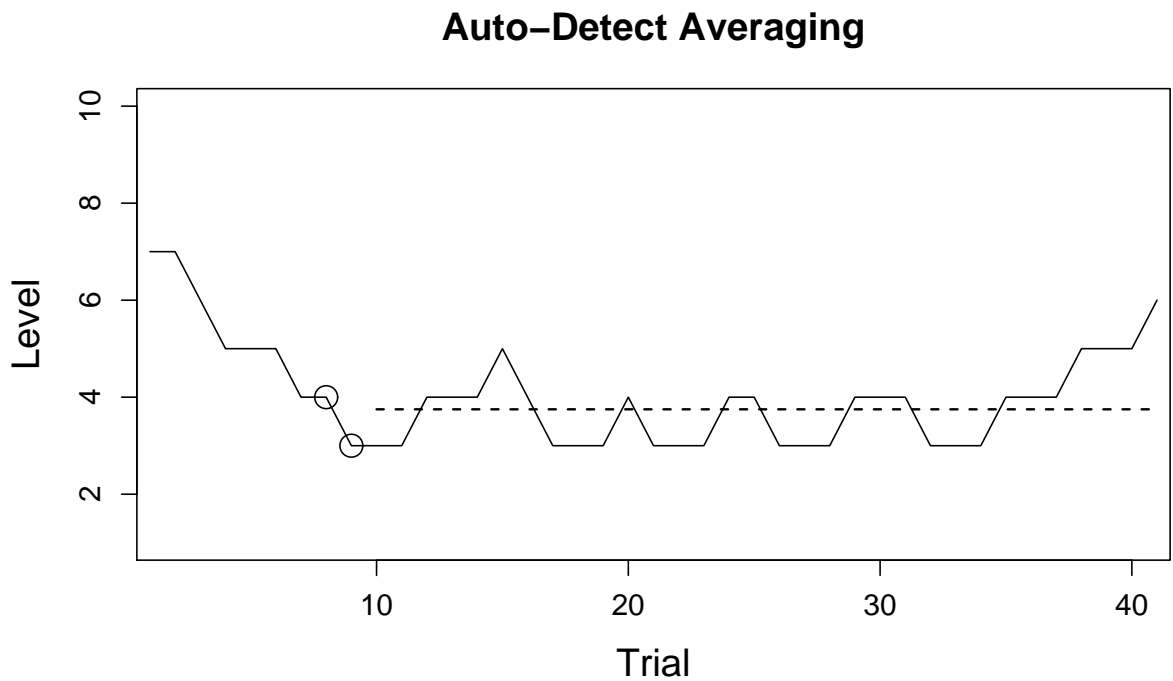}
\caption[Auto-Detect Estimator]{Illustration of the auto-detect estimator. The chain first crosses the ``tail-average'' line on trial $9$, meaning that trial $x_9$ is the first one smaller than the mean of remaining trials (marked with a horizontal dashed line), while $x_1\ldots x_8$ were all larger than the analogous tail means. The auto-detect estimate $\hat{v}_{AD}$ is then the average of $x_8\ldots x_{41}$. Trials $8$ and $9$ are circled. The chain used for illustration is the KR run from Fig.~\ref{fig:basic}, but in principle this method could be applied to any sequential experiment where a starting-point effect is suspected, including non-U\&D ones.}\label{fig:auto}
\end{center}
\end{figure}

Another approach seeks to adaptively identify the length of the initial transition phase. The underlying assumption is that the latter phase of the chain resembles a sample from $\pi$. If so, the average of the chain's ``tail end'' is close to $\mu_\pi$. Meanwhile, the ``head'' or starting point is arbitrarily fixed to $x_1$.\footnote{A similar situation is encountered in MCMC simulations, where a variety of diagnostics have been developed to determine stationarity. However, MCMC usually involves a very large sample and the ability to run several parallel instances of the same run -- privileges not available in the small-sample percentile-finding application.} The {\bf auto-detect estimator $\hat{v}_{AD}$} chops off this biased ``head'', exactly at the first point in which the chain traverses to the other side of the ``tail end'' average, compared with the starting point - and then proceeds to average the rest of the chain. Symbolically this estimator can be described as
\begin{equation}\label{eq:auto}
\hat{v}_{AD}\equiv\bar{x}_{c(n):n}\equiv\frac{\sum_{i=c(n)}^{n}x_i}{n-c(n)+1},\textrm{ \\where }
c(n)=\min_{1<i<n}\left\{i:\textrm{sgn}\left(x_i-\bar{x}_{i+1:n}\right)\neq\textrm{sgn}\left(x_1-\bar{x}_{2:n}\right)\right\}-1.
\end{equation}
The location of the cutoff point is illustrated in Fig.~\ref{fig:auto}

Sometimes, the ``tail end'' of the chain is an ``unlucky draw'' containing an excursion far from target; sometimes the entire chain converges more slowly than expected by the experiment's designers. The auto-detect method would usually identify such cases by placing the transition point relatively late. From an overall performance perspective, using a late auto-detect point is not advisable; the probability that the auto-detect scheme has been fooled by an excursion -- conditional upon the value of $c(n)$ -- increases sharply as $c(n)$ approaches $n$.  one should therefore set a maximum cutoff $c(n)_{crit.}$, so that if $c(n)>c(n)_{crit.}$ we use $c(n)_{crit.}$ as our cutoff. This critical value should definitely be less than $n/2$. Since it can be argued that the ``tail end'' needs to be sufficiently longer than the transition phase in order to serve as a reference, $c(n)_{crit.}$ values around $n/4$ or $n/3$ seem reasonable. Thus, the auto-detect estimator would be guaranteed to include at least the final two-thirds or three-quarters of the treatment chain.

If one wishes to be even more efficient, this critical value can be compared with the number of trials required to reach the boundaries from $x_1$ (if there are no physical boundaries, visualize ``virtual boundaries'' around the range in which you realistically expect $Q_p$ to fall). Allow for one ``switchback'' of falling back one level towards $x_1$ before resuming progression. If such boundary-related predictions can be made, then resulting cutoff should be compared with $n/3$ or $n/4$, and the smaller of the two chosen for $c(n)_{crit.}$ For example, if the design is KR with $k=2$ and the boundaries extend $5$ levels in each direction, then one needs $5+(k+1)=8$ trials to reach the bottom end including one ``switchback'', but $5k+(k+1)=13$ trials to reach the top end. Then, if $n>52$, one can use $c(n)_{crit.}=13$ instead of $n/3$ or $n/4$, and improve precision.\footnote{Clearly, this indicates that for below-median targets $x_1$ should be positioned above the middle of the ``effective range''; see Section~\ref{sec:estopt}.}

Conversely, if sample size is flexible, the auto-detect properties can be used to formulate a stopping rule. The experiment should continue until $c(n)\leq n/3$, or until $n-c(n)$ passes a certain value - in both cases promising a sufficient sample size of approximately stationary sampling.

For KR designs, we may also wish to look at $\hat{v}_{AD,0}$ -- the auto-detect estimator taken only on the zero-state sub-chain. It promises to mitigate both the starting-point and stationary-mean biases simultaneously. This idea has not been explored in detail in this thesis, and may merit further examination.

\subsubsection{A Geometrically Weighted Estimator}
Instead of trying to locate a cutoff point, we can weight the chain according to convergence theory. In Section~2.2, we saw that
\begin{equation}\label{eq:diaconmu}
\left|\mu_{n+1}-\mu_\pi\right|\propto \lambda_{\#}^n,
\end{equation}

where $\mu_{n+1}\equiv E\left[x_{n+1}\right]$, and $\lambda_{\#}$ is the second-largest eigenvalue of $P$. This justifies the following weighted-average estimate:
\begin{equation}\label{eq:GW}
\hat{Q}_{GW}\equiv\frac{\sum_{i=k+1}^{n}w_ix_i}{\sum_{i=k+1}^{n}w_i}\textrm{, \\where }
w_i=g\left(\lambda_{\#}^{-i}\right).
\end{equation}
The weights should be a function that starts near zero, then increases until some transition point where we suspect the bias term becomes small compared to the variance term. The location of this point can be estimated by comparing the standard error of some raw form of $\hat{v}$ with some order-of-magnitude estimate for starting-point bias.

Of course, we do not know $P$ and its eigenvalues; however, $P$ can be nonparametrically estimated from $\hat{F}$. $\hat{Q}_{GW}$ provides a nice complement to $\hat{v}_{AD}$: the former uses responses to evaluate convergence, while the latter tracks the treatment chain.

Two implementation notes:
\begin{itemize}
\item The $\hat{F}$-derived $\hat{P}$ tends to produce optimistic rates, or even no rates (in case there is only one point estimate of $F$ falling in $(0,1)$). It is therefore recommended to weight $F$ with a prior estimate, say uniform. A Bayesian saturated model, or any other parametric model, are also an option (see Section~4.1). However, there is no reason to `overkill' this estimate of $P$ with sophisticated procedures, as it is only used as an aid to determine weights.
\item Another implementation option is to multiply the decay factor by an `acceleration' factor somewhat greater than $1$. This ensures that the initial run's weight is indeed smaller than its impact on bias.
\end{itemize}

At bottom line, $\hat{Q}_{GW}$ has many more ``moving parts'' than other estimators. Since numerical trials did not indicate an overall performance improvement over the simpler options described above, it will not be discussed again below.

\subsubsection{Mitigating the Boundary Effect?}
If $Q_p$ is less than $2s$ away from a hard boundary, bias source 2 on the list at the beginning of the section can overwhelm all the rest. The best fix is prevention: not to place a boundary, unless it is physically impossible to continue moving up or down. If treatments are positive, consider placing $\{l_m\}$ on a log-transformed scale (as is the default in sensory studies). If they are proportions, consider a log-ratio transform. This is not as cumbersome as it sounds: phrase the treatment levels as $u$ parts component B to one part component A, etc. Yet another alternative is to halve the spacing if a boundary seems to play a dominant role; a trigger for halving could be the $2$nd visit to the same boundary, or the first transition decision violates the boundaries. Of course, this would have an impact only if the halving occurs relatively early in the experiment.

Sometimes none of these fixes are available, and we are stuck with a real physical boundary. One existing fix was described in Section~\ref{sec:est1} \citep{GarciaPerez98} - continue administering treatments on the boundary, as a proxy for levels beyond it. This fix has the property of over-assigning treatments on the boundary. An alternative, milder fix would be to impute $\{x_n\}$ following the experiment by adding virtual treatments at $l_{m+1}$ wherever a move `up' was mandated from $l_m$, and at $l_0$ wherever a move `down' from $l_1$ was mandated (these virtual `treatments' would of course never be administered, only added to the average). the number of treatments added would follow the transition rule (e.g., for KR, $k$ virtual treatments would be added each visit to $l_0$, but only $1$ at $l_m$). This imputation is equivalent to assuming $F=0,1$ at the two boundaries. The imputation fix requires no change in experiment planning.

It turns out, again, that neither of these fixes works as well as one might expect. This is because both increase the averaging-estimator variance -- an increase which is not necessarily offset by a substantial bias reduction. It seems that, if the chain runs dangerously close to a boundary, one should turn to response-based estimation -- which, coincidentally, is our next topic.

\subsection{Centered Isotonic Regression}
Due to the theoretical results on IR bias, culminating in Theorem~\ref{thm:unbiasedpoint}, we may attempt to modify isotonic regression by replacing each constant interval with a single point at the location prescribed by (\ref{eqn:x*}). To be coherent, the new estimator would perform the same action at intervals that were originally constant as well. Algorithm~2 describes this estimator, dubbed ``Centered Isotonic Regression'' (CIR).

\begin{algorithm}[!h]
\caption{Centered Isotonic Regression (CIR) for Treatment-Response Designs}
\begin{algorithmic}\label{alg:CIR}

\Procedure{CIR}{$\{l_m\},\{z_m\},\{n_m\}$}
\Statex
\While {$\textbf{H}\equiv\left\{u: 1\leq u<m,z_u>z_{u+1}\right\}\neq\emptyset$}
    \State $v\gets\min (\textbf{H})$
    \State  $M\gets 1$
    \While {$z_v>z_{v+M}$}
        \State $z_v\gets\overline{z}_{v:v+M-1}\equiv\sum_{u=v}^{v+M-1}n_uz_u/\sum_{u=v}^{v+M-1}n_u$
        \State $M\gets M+1$
    \EndWhile
    \State $l_v\gets\overline{l}_{v:v+M-1}\equiv\sum_{u=v}^{v+M-1}n_ul_u/\sum_{u=v}^{v+M-1}n_u$
    \State $n_v\gets\sum_{u=v}^{v+M-1}n_u$
    \State Remove points $\{(l_{v+1},z_{v+1},n_{v+1})\ldots(l_{v+M-1},z_{v+M-1},n_{v+M-1})$
\EndWhile

\Statex
\State \textbf{Return} $\{(l_m,z_m,n_m)\}$.
\EndProcedure
\end{algorithmic}
\end{algorithm}

CIR's point-estimate output is used to estimate $F$ and its quantiles via linear interpolation, including at original design points which have been removed. The following observations may help clarify some more major similarities and differences between PAVA and CIR.
\begin{enumerate}

\item CIR produces the same set of unique $z$ values (point estimates) as PAVA, but under CIR each value appears only once.
\item Unlike PAVA, CIR requires $x$ values as inputs, and modifies them in the same manner as the $z$ values. As a result, each unique weighted-average $z$ value is assigned to the corresponding weighted-average $x$ value.\footnote{The point placement prescribed by (\ref{eqn:x*}) is optimal for the normal-error scenario and seems to work well for the treatment-response case as well. For other scenarios (e.g. a current-status dataset copmosed of 1's and 0's only), the formula for the optimal point may be different.}
\item CIR identifies ties, too, as violations since they violate strict monotonicity.
\item In the absence of strict monotonicity violations, the two algorithms return the original data.
\item In the presence of strict monotonicity violations, CIR returns a smaller number of points than the input.
\end{enumerate}

Due to the last property, sometimes a boundary point may be removed by CIR. Suppose w.l.o.g. that the original $l_1$ has been removed, and now some $x^\#>l_1$ is the smallest $x$ value for which a CIR estimate exists. Then complementing the CIR output with a constant interval on $[l_1,x^\#]$ is equivalent to the IR output on the same interval, and should be regarded as the default. However, in case prior knowledge exists regarding $F$ on or outside the boundaries (e.g., $F(0)=0$), then it should be incorporated either into CIR's input (when an appropriate weight can be assigned) or into its output.

Conceptually, the CIR process is analogous to saying: ``We have this sequence of point estimates, each perfectly valid on its own, but as a sequence they are suspect due to monotonicity violations. We can pool them together to get a single, more believable estimate; however, this is still a point estimate only, and deserves to be placed at a single point rather than occupy a complete interval.''

Under the normal-error scenario described in Section~\ref{sec:est2} and some additional assumptions, it can be proven that CIR yields a smaller overall estimation MSE than IR, but the calculation is rather cumbersome and uninteresting. In nearly all numerical U\&D simulations I have carried out, CIR clearly outperforms IR for quantile estimation (some data will be presented below in Section~\ref{sec:estsim}). The only case when CIR is not recommended, is when the researchers suspect $F$ may resemble a staircase function. In that case, IR's piecewise-constant output is obviously more appropriate then CIR's linear smoothing.

\begin{table}[h]
\begin{center}
\caption[Centered Isotonic Regression Example]{Summary tables illustrating how CIR works in practice on treatment-response data. Data are identical to that shown on Table~\ref{tbl:ir_demo}.\label{tbl:cir_demo}}
{\small

\begin{tabular}{p{1.6cm}@{\extracolsep{\fill}}rrrrrrr@{\extracolsep{0.5cm}}r@{\extracolsep{\fill}}rrrrrr}
\toprule
{\bf Treatment} & \multicolumn{3}{c}{\bf Raw Input} &\multicolumn{4}{c}{\bf CIR Output}
&  \multicolumn{3}{c}{\bf Raw Input} &\multicolumn{4}{c}{\bf CIR Output} \\
& Yes & No & $\hat{F}$ & Yes & No & $x$ & $\hat{F}$ & Yes & No & $\hat{F}$ & Yes & No & $x$ & $\hat{F}$ \\
\midrule
$0.17$ & $0$ &  $4$ &  $0.00$ & $0$ &  $4$ & $0.17$ & $0.00$ & $1$ &  $7$ &  $0.13$ & $1$ &  $7$ & $0.17$ &  $0.13$ \\
$0.33$ & $3$ &  $9$ &  $0.25$ & $3$ &  $9$ & $0.33$ & $0.25$ & $\mathbf{4}$ &  $\mathbf{8}$ &  $\mathbf{0.33}$ &\multirow{2}{*} {$\mathbf{6}$} & \multirow{2}{*} {$\mathbf{14}$} & \multirow{2}{*} {$\mathbf{0.40}$}& \multirow{2}{*} {$\mathbf{0.30}$}\\
$0.50$ & $\mathbf{3}$ &  $\mathbf{7}$ &  $\mathbf{0.30}$ & \multirow{2}{*} {$\mathbf{4}$} & \multirow{2}{*} {$\mathbf{10}$} & \multirow{2}{*} {$\mathbf{0.55}$}&  \multirow{2}{*} {$\mathbf{0.28}$} & $\mathbf{2}$ &  $\mathbf{6}$ &  $\mathbf{0.25}$  & & & &\\
$0.67$ & $\mathbf{1}$ &  $\mathbf{3}$ &  $\mathbf{0.25}$ &       & & & & $4$ &$0$ &$1.00$ & $4$ &$0$ & $0.67$ &$1.00$ \\
$0.83$ & $1$ &  $1$ &  $0.50$ & $1$ &  $1$ & $0.83$ & $0.50$ & & & & & &  \\
\bottomrule
& \multicolumn{7}{c}{\bf (Simulation Run 14)}  &   \multicolumn{7}{c}{\bf (Simulation Run 9)} \\
  \end{tabular}
}
\end{center}
\end{table}

\begin{figure}[h]
\begin{center}
\includegraphics[scale=.8]{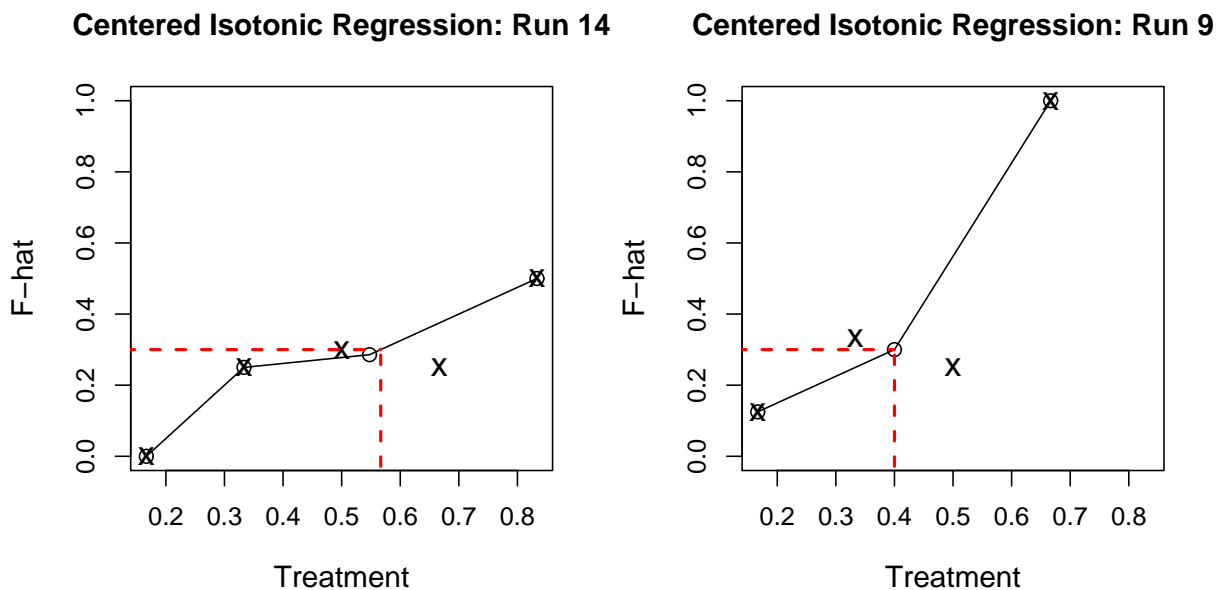}
\caption[Centered Isotonic Regression Example Illustrated]{Graphs illustrating how CIR works in practice on treatment-response data. Data are the same shown on Table~\ref{tbl:cir_demo}. Raw $\hat{F}$ values are in `X' marks, and the CIR output values are in circles connected by solid lines. The dashed red lines indicate the inverse-interpolation process. Compare with Fig.~\ref{fig:ir_demo}.\label{fig:cir_demo}}
\end{center}
\end{figure}

We now revisit the two data examples from Table~\ref{fig:ir_demo} and Fig.~\ref{fig:ir_demo}. Table~\ref{tbl:cir_demo} and Fig.~\ref{fig:cir_demo} show the same data, estimated via CIR. Since CIR's output is strictly monotone (at least between its output point estimates), the quantile estimate for Run 9 (RHS) is now unique. Note that CIR is not a risk-free fix. For example, in Run 9 the CIR estimate of $F$ at $l_3$ is $0.56$, while $\hat{F}_3$ was only $0.25$. This is because the new estimate is a weighted average of $\hat{F}_2,\hat{F}_3$ and $\hat{F}_4$ -- the latter being equal to $1$. From a least-squares perspective (i.e., least-square distance from the original point estimates), CIR is clearly not better than IR. However, with respect to the true $F$ the pulling-in of neighboring $\hat{F}$ values to smooth out the estimate in violating intervals (where point estimates are more suspect) seems to pay off.

\subsubsection{A Combined Estimator?}
Averaging estimators and CIR appear to complement each other: the former typically have smaller variance, and the latter smaller bias. This suggests the simple fix of using their average as a combined estimator. Unfortunately this fix is not successful, because the two estimator types are highly correlated (numerical ensemble correlations are typically around $0.8$ or more). This means that a simple average of CIR and an averaging estimator is likely to have higher variance than the averaging estimator alone. Since estimation error is dominated by the variance term (often over $90\%$ of the MSE is due to variance, see below in Section~\ref{sec:estsim}), the improvement in bias may be insufficient to offset the variance hit.

Why are averaging estimators and CIR so correlated? Suppose w.l.o.g. that we observe a sequence of above-average thresholds. The response rate at most treatments will be below the population expected values, pushing $\hat{F}$ downward and CIR inverse estimates upward. At the same time, the negative responses will also push the treatment chain itself upward, and with it the averaging estimators.

\subsection{Interval Estimation}
\subsubsection{Confidence Intervals for Averaging Estimators}

For averaging estimators, our rationale is that we sample from $\pi$ and try to estimate the standard error of the running sample mean. Trials before the cutoff point (be it a reversal or the auto-detect point) are discarded. A simple estimate for the standard error could be of the form
\begin{equation}\label{eq:se}
S.E._{\hat{v}}=\frac{S}{\sqrt{\hat{n}_{eff}}},
\end{equation}
where $S$ is the sample standard deviation of the sub-chain used for averaging, and $\hat{n}_{eff}$ is an estimate of the effective sample size -- which should be smaller than the sub-chain length, since the sampling is dependent. We can estimate $n_{eff}$ using the approach offered by \citet{Choi90}, again with some convenient simplifications. We assume an AR(1) structure. The variance of an averaging estimator $\hat{v}$ is then
\begin{equation}\label{eq:varv1}
\frac{\tilde{n}^2Var(\hat{v})}{\sigma_{\pi}^2}=\left[\tilde{n}+2\sum_{i<j}\rho^{j-i}\right]=\left[\tilde{n}+\frac{2\rho}{1-\rho}\sum_{i=1}^{n-1}1-\rho^i\right]\textrm{\\}
=\tilde{n}\frac{1+\rho}{1-\rho}-\frac{2\rho}{1-\rho}\left[1-\rho^{\tilde{n}}\right]
\end{equation}
\begin{equation}\label{eq:varv2}
\frac{\tilde{n}Var(\hat{v})}{\sigma_{\pi}^2}\approx\frac{1+\rho}{1-\rho},
\end{equation}
where $\rho$ is the first-order autocorrelation and $\tilde{n}$ is the chain length used for averaging. The approximation in (\ref{eq:varv2}) is conservative. So our autocorrelation-based sample size estimate is $\hat{n}_{eff}=\tilde{n}(1-\hat{\rho})/(1+\hat{\rho})$. Since the variance is estimated from the data, we should use a $t$-multiplier with $\hat{n}_{eff}-1$ d.f. (or $1$ d.f. if $\hat{n}_{eff}<2$).

A rather different approach follows \citet{Tsutakawa67}'s observation that intervals between hitting times are i.i.d. \citet{Tsutakawa67} used the complicated formula (\ref{eq:tsukavar1}) to directly estimate the variance; however, as mentioned earlier this appears to be overly optimistic. Instead, we estimate $\hat{v}$'s variance from first principles. Let $N_j$ be the length of hitting-time interval $j$, and let $A_j$ be the treatment mean over the interval (Ideally, we use hitting times at the level closest to target, $l_{u^*}$). Then each r.v. is i.i.d. across intervals.\footnote{We employ a convention that avoids gaps or overlaps, e.g. that each interval begins with the treatment following the hitting time. Additionally, for KR each sub-state should be viewed as a separate state. Hence, we look only at the first visit to each level (i.e., at visits to $l_{u,0}$ only), since those are the most numerous.}  Now, neglecting possible ``leftovers'' before the first interval and after the last, averaging estimators can be expressed as
$$
\hat{v}=\frac{\sum_j a_jn_j}{\tilde{n}},
$$
where $\tilde{n}$, as before, is the sample size used for averaging (i.e., from the cutoff point onward). Neglecting the randomness of $\tilde{n}$,\footnote{For fixed-sample applications this is not a bad assumption. Of course, an approach using hitting-times or reversals (which are related to them) as stopping rules will have to account for $\tilde{n}$'s variability.} we focus on the numerator. For a single interval,
\begin{eqnarray}\label{eq:varajnj}
Var(A_jN_j) & \approx & E\left[Var(A_jN_j|N_j)\right]+Var\left(E[A_jN_j|N_j]\right)\\
\nonumber  & < & E\left[N_j^2\left(\sigma_{\pi}^2+(\mu_\pi-l_{u^*})^2\right)\right]+Var(N_j)\max(\mu_\pi,l_{u^*})^2,
\end{eqnarray}
where $\sigma_{\pi}^2$ is the stationary variance and $l_{u^*}$ is the level used for determining the intervals. Summing over all intervals, normalizing and replacing population parameters by estimates, we obtain the estimate
\begin{equation}\label{eq:varoron}
S.E._{\hat{v}} = \frac{h\left\{\overline{n_h^2}\left(\tilde{S}^2+(\hat{v}-l_{\widehat{u^*}})^2\right)+\max(\hat{v},l_{\widehat{u^*}})^2S_h^2\right\}}{\tilde{n}^2},
\end{equation}
where $l_{\widehat{u^*}}$ is the level closest to the point estimate, at which the hitting-times are counted; $h$ is the number of hitting-time intervals, $\tilde{S}^2$ is the sample variance of the chain used for averaging, and $\overline{n_h^2}$ and $S_h^2$ are the sample mean-square and variance of interval lengths, respectively. As above, we use a $t$-multiplier with $h-1$ d.f. (or $1$ d.f. if $h<2$).

Based on bot
h theoretical appeal and numerical trials, I chose the second, hitting-time-based S.E. estimate (\ref{eq:varoron}). In general it is quite conservative. One reason for choosing the more conservative option, is that $\hat{v}$ estimates the stationary mean and not $Q_p$, and therefore the interval will inevitably be somewhat shifted. Sometimes, when the shift is too large (i.e., one of the bias sources mentioned at the top of this section is too dominant), no interval would do the job. This again points to the importance of averaging-estimator bias reduction.

\subsubsection{Confidence Intervals for CIR}

CIR's intervals (and also IR's) can also be approximated using a simpler and more direct approach than the bootstrap. Consider the CIR interpolated quantile estimate $\tilde{Q}_p$. Suppose w.l.o.g. that the it falls between the CIR-output treatment values $\tilde{l}_u$ and $\tilde{l}_{u+1}$. The estimation variance can be approximated by
\begin{equation}\label{eq:varcir1}
Var\left(\tilde{Q}_p^{(CIR)}\right)\approx\left[\frac{\tilde{l}_{u+1}-\tilde{l}_u}{\tilde{F}_{u+1}-\tilde{F}_u}\right]^2Var\left(\tilde{F}(\tilde{Q}_p)\right),
\end{equation}
Now, the forward estimate of $F$ is simply a weighted average of binomial point estimates. Obviously
$$
Var\left(\tilde{F}(\tilde{Q}_p)\right)<\max\left(Var(\tilde{F}_u),Var(\tilde{F}_{u+1})\right).
$$
Nearly always, the level with higher variance is the one with a smaller sample size, assume w.l.o.g. it is $\tilde{l}_u$. Recalling that the $\{n_m\}$ are random,
\begin{eqnarray}\label{eq:varfu}
Var(\tilde{F}_u)  & \approx &   E\left[Var(\tilde{F}_u|n_u)\right]+Var\left(E[\tilde{F}_u|n_u]\right)\\
\nonumber & = & F_u(1-F_u)E\left[\frac{1}{n_u}\right].
\end{eqnarray}
Note that the second term on the RHS is zero because $E[\tilde{F}_u]$ remains (at least to first order) $F_u$ regardless of sample size. Estimating the expectation of $1/n_u$ is generally infeasible without knowledge of $F$, but due to Jensen's inequality we can see that the variance is always larger than in the fixed-assignment case, as it should be. Note that this expression still does not account for the total variance in the CIR inverse estimate, since different runs may lead the estimate to fall in different intervals, i.e., different values of $u$.

In any case, a simple way to continue from here is by using binomial quantiles for $\tilde{F}_u$. We convert the standard piecewise-constant binomial quantile function, whose properties are strongly dependent upon the specific (and random) value of $n_u$, into a strictly increasing $\bar{q}(n_u,p)$, which is the linear interpolation between the exact-probability points $\{(p_z,z)\},z=0,\ldots n_u+1$, where $p_z$ are defined via $Pr(Z\leq z)=p_z$. This piecewise-linear quantile curve is then re-centered around the expectation $\bar{n}p$ for symmetry. One way to then account for the inflated variance of (\ref{eq:varfu}), is to offset the curve by $1$ to each direction away from $n_up$. Finally, divide by $n_u$ to normalize.

\begin{figure}
\begin{center}
\includegraphics[scale=.8]{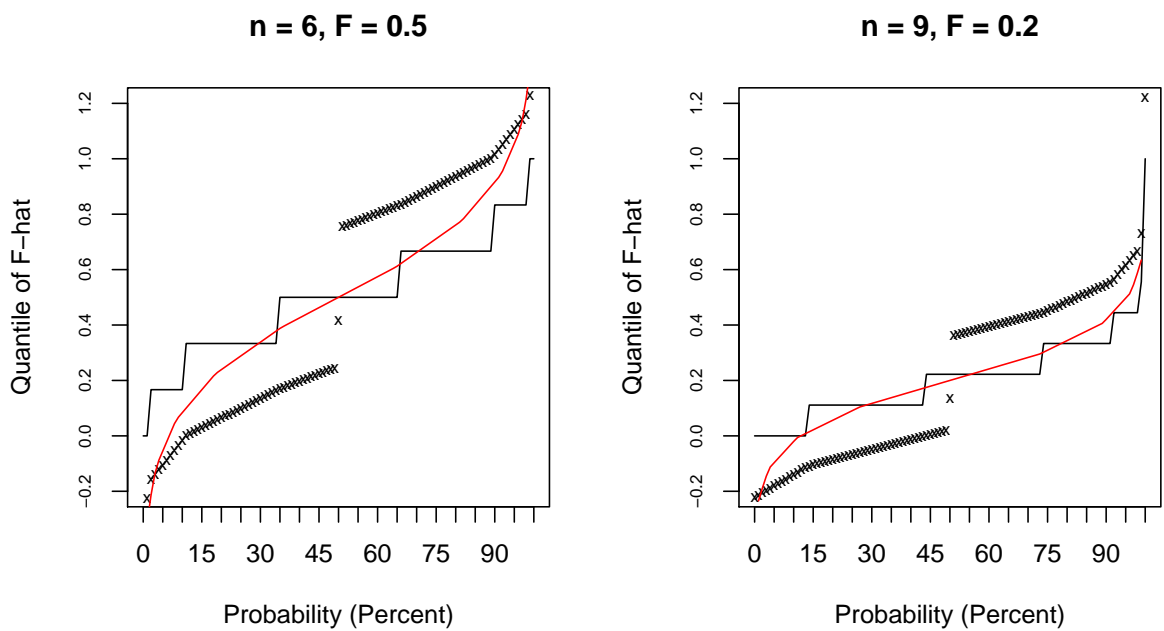}
\caption[Quantile Functions for CIR Interval Estimation]{Illustration of possible quantile functions used for CIR interval estimation. Shown are scenarios with $n_u=6,p=0.5$ (left) and $n_u=9,p=0.2$ (right). Depicted are standard binomial quantiles (black line), the linearized binomial (black `x' marks) and linearized symmetrized Poisson (red line). The latter two are described in the text.}\label{fig:smoothq}
\end{center}
\end{figure}

The resulting curve is always more conservative than standard binomial quantiles (see Fig.~\ref{fig:smoothq}); however, at extreme percentiles it still suffers from the limitations of the binomial quantile function, and does not increase fast enough, considering that the experiment's sequential nature may generate heavy-tailed sampling distributions. Numerical trials show that the linearized function $\bar{q}$ does not provide enough differentiation between the $95$th percentile and $97.5$th percentiles, and therefore $90\%$ CI's are overly conservative or $95\%$ CI's lack in coverage (or both).

Recall that rare-event binomial probabilities can be approximated by Poisson probabilities with rate $n_up$, and note that the Poisson variance is always greater than the approximated binomial's variance. Therefore, using Poisson for the binomial tail is generally conservative for our purposes. We employ the same linearization method as above in order to avoid step-function artifacts, and additionally symmetrize the quantile curve by using a mirror image of the higher-probability tail in lieu of the less conservative lower-probability tail.

The binomial-derived and Poisson-derived functions are compared on Fig.~\ref{fig:smoothq}, for two different small-sample scenarios. Even though the Poisson curve may be less conservative than the standard binomial for some middle percentiles, and is generally less conservative than the linearized binomial for all but the most extreme percentiles, it does provide a better (and more realistic) distinction between percentiles on the tail. If negative values for binomial or Poisson quantiles (as happens on the lower tail) seem counter-intuitive, recall that we are not really sure between which two treatment levels the true quantile falls.

\subsection{Down-Shift Schemes Revisited}
In the previous chapter I mentioned the popularity of designs for which transition rules change at some point, most commonly the first reversal. In particular, down-shift schemes cut the spacing by half at the first reversal. The analysis presented since then indicate that a reversal, and surely the first reversal, does not possess good properties for a transition point. They are quite likely to be far removed from target. Numerical experimentation confirms this: while under some conditions a reversal-based down-shift scheme provides for estimation performance improvement, overall it is less robust. Using the third reversal improves robustness, compared with the first reversal, though it does not solve the issue of reversal variance under stationarity.

The hitting-time analysis performed earlier suggests another direction. Clearly, levels near target are hit more frequently (in probability). Therefore, one can use trial $x_{h_j}$ as a transition point, where $h_j$ is the index where the most frequently-hit level is visited for the $j$-th time. The natural choice for $j$ appears to be $j=3$: a smaller value does not preclude a possible early excursion, and a larger value pushes the transition too far into the experiment. Numerical runs indicate that this scheme is roughly equivalent to using the third reversal in a reversal-based scheme; both provide overall performance (as measured via empirical MSE) somewhere between that of the original and the halved spacing. Due to space limitations, down-shift schemes are not part of the extensive numerical study presented in the next section.
\section{Numerical Study}\label{sec:estsim}

\subsection{Simple Comparisons}
\subsubsection{IR and CIR}

The theoretical results about IR and CIR in Section~\ref{sec:est2} were for a related limit case of normal errors; however, the logic underlying them should hold for a wider array of scenarios and hence we may expect CIR to perform better. Even though CIR is a bias-reducing correction on an individual run basis, in ensemble-level simulation analysis the benefit should manifest itself in estimation \emph{variance} reduction. For different runs, IR's flat intervals may occur to the right or left of target, throwing the inverse estimate off either upward or downward of target, respectively. Over a large ensemble, the individual-run offsets may balance each other out -- but would inflate ensemble variance instead.

\begin{figure}[!h]
\begin{center}
\includegraphics[scale=.75]{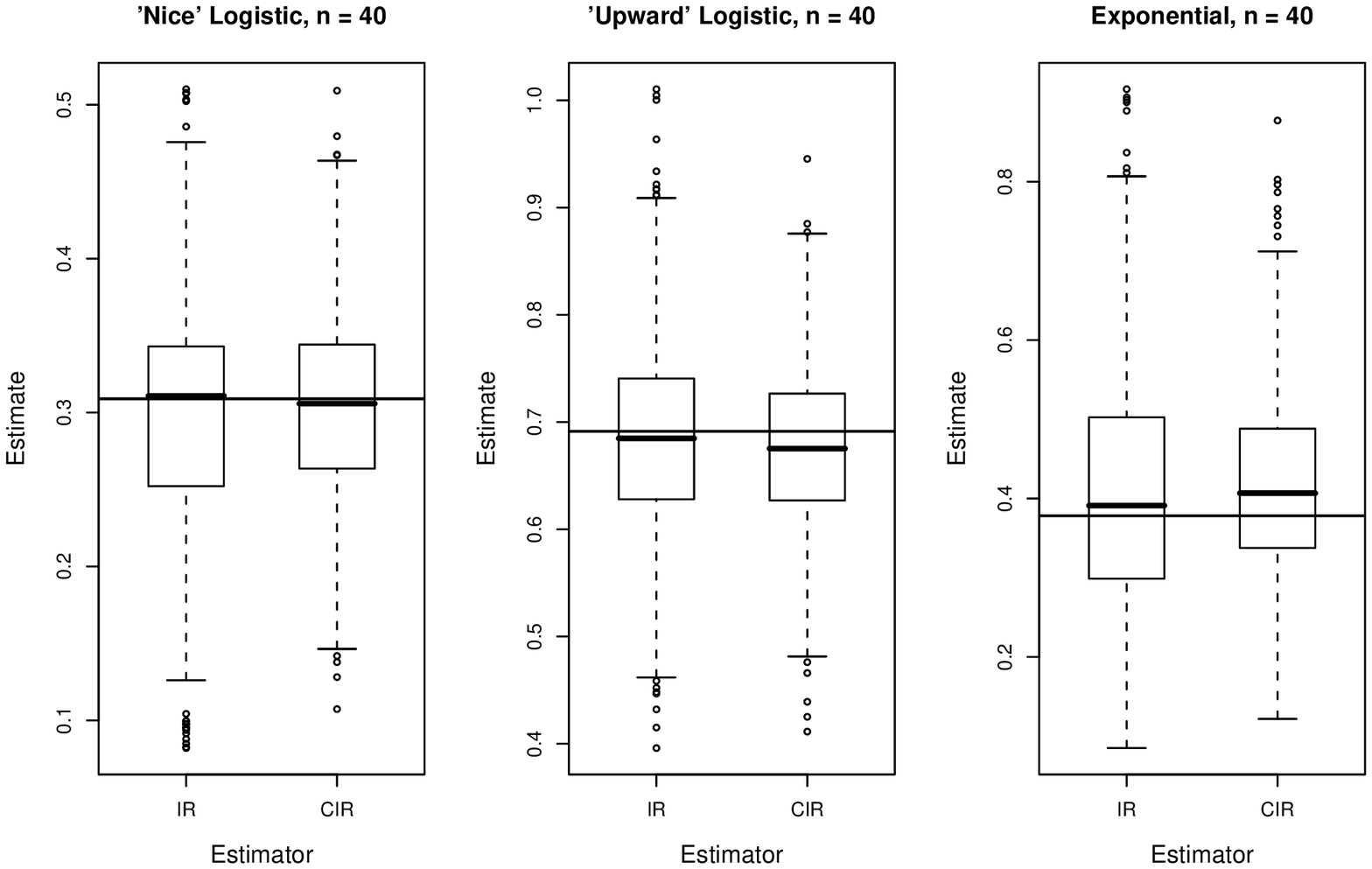}
\caption[IR-CIR Comparison]{Comparison of isotonic regression (IR) and centered isotonic regression (CIR) under the $3$ scenarios summarized in Table~\ref{tbl:sim_cir} and a sample size of $40$. Boxplots reflect only those runs in which the two methods yielded different estimates. In all plots, IR is on the left. Targets are marked by solid horizontal lines.\label{fig:simcir}}
\end{center}
\end{figure}

This is indeed what Table~\ref{tbl:sim_cir} and Fig.~\ref{fig:simcir} show. The empirical efficiency gains of CIR over IR are quite substantial.\footnote{Keep in mind that the two estimate differ only over about half the runs; if one compares MSE only over runs producing monotonicity-violations, the efficienty gain is about double that shown on the table.} In general, they increase as $s$ increases (data not shown). For exponential thresholds one can discern an average upward shift associated with CIR (right-most frame in Fig.~\ref{fig:simcir}), and for logistic thresholds a downward shift (middle frame). This is because, $F$ being concave around target in the exponential case and convex in the logistic case, the forward IR/CIR would under/over-estimate $F$, respectively -- resulting in an inverse (quantile-estimation) bias in the opposite direction. IR's flat intervals mitigate this bias, while CIR's interpolation increases it. However, this bias increase is more than compensated by the variance reduction.

\begin{table}[!h]
\begin{center}
\caption[Isotonic Regression vs. Centered Isotonic Regression Performance]{Comparison of isotonic regression (IR) and centered isotonic regression (CIR), from KR ($k=2$) simulations over $3$ distributional conditions and $2$ sample sizes. The `Violation Frequency' column indicates what proportion of runs (out of $2000$) had monotonicity violations in a region that affects estimation -- meaning that for the remaining runs the IR and CIR estimates were identical. E.E.R. is the `Empirical Efficiency Ratio': the ensemble MSE of IR estimates w.r.t the true $Q_p$, divided by the MSE of CIR estimates. All statistics were taken over the entire ensemble of $2000$ runs. Bias and standard deviation are in the same arbitrary normalized units of treatments. Spacing was $0.1$, meaning that $m=10$.\label{tbl:sim_cir}}
\begin{tabular}{p{2.5cm}ccrrrrc}
  \toprule
  \multirow{2}{2cm}{\bf Distribution}&\multirow{2}{1.5cm}{\bf Sample Size}&\multirow{2}{2cm}{\bf Violation Frequency}&\multicolumn{2}{c}{{\bf Bias}}&\multicolumn{2}{c}{{\bf Std. Dev.}}&\multirow{2}{1.5cm}{\bf E.E.R.}\\
\cmidrule{4-7} &&& IR & CIR & IR & CIR &  \\
\midrule
\multirow{2}{2cm}{{\bf `Nice' Logistic}}  & $20$ &$50.8\%$ &$-0.003$&$0.008$&$0.091$ &$0.081$ & $1.24$\\
                                      & $40$ &$50.6\%$ &$-0.004$&$-0.001$&$0.071$ &$0.062$ & $1.30$\\
\midrule
\multirow{2}{2cm}{\bf `Upward' Logistic}& $20$ &$59.0\%$ &$-0.015$&$-0.035$&$0.110$ &$0.097$ & $1.15$\\
                                    & $40$ &$59.8\%$ &$-0.005$&$-0.011$&$0.086$ &$0.076$ & $1.25$\\
\midrule
\multirow{2}{*}{\bf Exponential}& $20$ &$42.5\%$ &$0.039$&$0.045$&$0.152$ &$0.132$ & $1.28$\\
                                    & $40$ &$46.3\%$ &$0.022$&$0.030$&$0.132$ &$0.113$ & $1.32$\\
\bottomrule
  \end{tabular}
\end{center}
\end{table}

\subsubsection{Reversal-only and All-trial Averaging}

\begin{figure}[!h]
\begin{center}
\includegraphics[scale=.75]{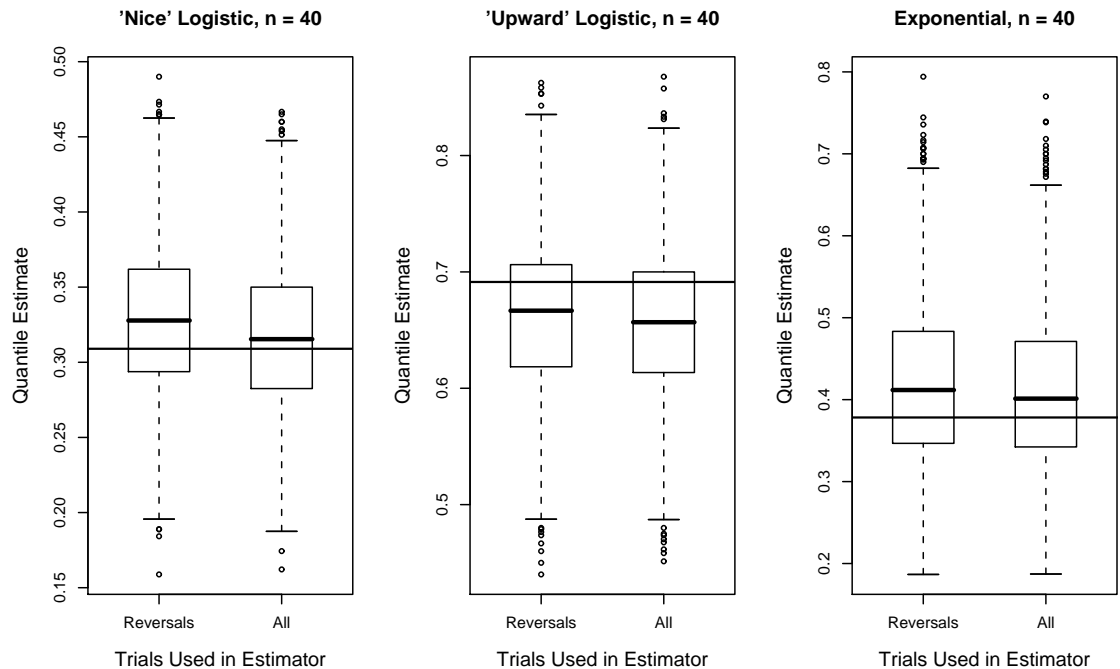}
\caption[Reversal Averaging vs. All Trials Comparison]{Comparison of reversal-averaging from reversal $1$ ($\hat{w}_1$) and averages of all treatment from reversal $1$ ($\hat{v}_1$), from KR ($k=2$) simulations over the $3$ scenarios summarized in Table~\ref{tbl:sim_rever1} and a sample size of $40$. In all plots, reversal-only averaging is on the left.\label{fig:simrev}}
\end{center}
\end{figure}

For averaging estimators, we expect reversal-averages $\hat{w}$ to be shifted upward and have larger variance than otherwise-identical all-trial averages $\hat{v}$. This is indeed demonstrated in Table~\ref{tbl:sim_rever1} and Fig.~\ref{fig:simrev}. In one scenario, $\hat{w}$'s upward bias offsets some of the downward starting-point bias, yielding roughly equal performance for the two estimators. An important detail to note is that, in spite of averaging-estimators' plethora of bias sources, it is still the variance term that dominates performance.

\begin{table}[!h]
\begin{center}
\caption[Reversal Averaging vs. All-Trial Averaging Performance]{Comparison of reversal-averaging from reversal $1$ ($\hat{w}_1$) and averages of all treatment from reversal $1$ ($\hat{v}_1$), from KR ($k=2$) simulations over $3$ distributional conditions and $2$ sample sizes. Details are as in Table~\ref{tbl:sim_rever1}.\label{tbl:sim_rever1}}

\begin{tabular}{p{2.5cm}crrrrc}
  \toprule
  \multirow{2}{2cm}{\bf Distribution}&\multirow{2}{1.5cm}{\bf Sample Size}&\multicolumn{2}{c}{{\bf Bias}}&\multicolumn{2}{c}{{\bf Std. Dev.}}&\multirow{2}{1.5cm}{\bf E.E.R.}\\
\cmidrule{3-6} && $\hat{w}_1$ & $\hat{v}_1$ & $\hat{w}_1$ & $\hat{v}_1$ &  \\
\midrule
\multirow{2}{2cm}{{\bf `Nice' Logistic}}  & $20$ &$0.036$&$0.022$&$0.068$ &$0.066$ & $1.24$\\
                                      & $40$  &$0.019$&$0.009$&$0.052$ &$0.049$ & $1.20$\\
\midrule
\multirow{2}{2cm}{\bf `Upward' Logistic}& $20$  &$-0.069$&$-0.071$&$0.099$ &$0.093$ & $1.06$\\
                                    & $40$  &$-0.037$&$-0.044$&$0.070$ &$0.068$ & $0.96$\\
\midrule
\multirow{2}{*}{\bf Exponential}& $20$ &$0.057$&$0.047$&$0.112$ &$0.111$ & $1.09$\\
                                    & $40$ &$0.041$&$0.033$&$0.097$ &$0.095$ & $1.10$\\
\bottomrule
\end{tabular}
\end{center}
\end{table}

\subsubsection{BCD , KR, GU\&D$_{(k,0,1)}$}
I end this simulation \emph{hors d'{\oe}uvre} with some cursory method comparisons. From my theoretical work, one would expect KR to show a clear advantage over BCD for averaging estimators, but also for CIR due to its sharper stationary distribution. As Table~\ref{tbl:sim_bcdkr} shows, the averaging-estimator gap is more pronounced ($25\%$ to $40\%$ difference in MSE); the CIR performance gap between designs is much less dramatic, but is still observed across the board (including in many numerical runs not shown here).

\begin{table}[!h]
\begin{center}
\caption[BCD vs. KR Estimator Performance]{Comparison of estimator performance from BCD ($\Gamma=0.293$) and KR ($k=2$) simulations. The two estimators compared are $\hat{v}_1$ (top), and CIR (bottom). The E.E.R. column quantifies KR's performance edge over BCD. Details are as in Table~\ref{tbl:sim_cir}.\label{tbl:sim_bcdkr}}

\begin{tabular}{lp{2.5cm}crrrrc}
  \toprule
  \multirow{2}{*}{\bf Estimator}&\multirow{2}{2cm}{\bf Distribution}&\multirow{2}{1.5cm}{\bf Sample Size}&\multicolumn{2}{c}{{\bf Bias}}&\multicolumn{2}{c}{{\bf Std. Dev.}}&\multirow{2}{1.5cm}{\bf E.E.R.}\\
\cmidrule{4-7} &&& BCD & KR & BCD & KR &  \\
\midrule
\multirow{6}{*}{\bf $\hat{v}_1$}&\multirow{2}{2cm}{{'Nice' Logistic}}  & $20$ &$0.031$&$0.022$&$0.074$ &$0.066$ & $1.34$\\
                                 &     & $40$  &$0.011$&$0.009$&$0.056$ &$0.049$ & $1.30$\\
\cmidrule{2-8}
&\multirow{2}{2cm}{'Upward' Logistic}& $20$  &$-0.084$&$-0.071$&$0.107$ &$0.093$ & $1.36$\\
 &                                   & $40$  &$-0.055$&$-0.044$&$0.079$ &$0.068$ & $1.41$\\
\cmidrule{2-8}
&\multirow{2}{*}{Exponential}& $20$ &$0.058$&$0.047$&$0.123$ &$0.111$ & $1.27$\\
 &                                   & $40$ &$0.041$&$0.033$&$0.107$ &$0.095$ & $1.31$\\
\midrule[1pt]
\multirow{6}{*}{\bf CIR}&\multirow{2}{2cm}{{'Nice' Logistic}}  & $20$ &$0.009$&$0.008$&$0.081$ &$0.080$  & $1.01$\\
                                 &     & $40$  &$-0.003$&$-0.001$&$0.063$ &$0.062$  & $1.03$\\
\cmidrule{2-8}
&\multirow{2}{2cm}{'Upward' Logistic}& $20$  &$-0.047$&$-0.035$&$0.098$ &$0.097$ & $1.11$\\
 &                                   & $40$  &$-0.017$&$-0.011$&$0.077$ &$0.076$ & $1.05$\\
\cmidrule{2-8}
&\multirow{2}{*}{Exponential}& $20$ &$0.050$&$0.045$&$0.134$ &$0.132$ & $1.05$\\
 &                                   & $40$ &$0.032$&$0.030$&$0.115$ &$0.113$ & $1.04$\\
 \bottomrule
\end{tabular}
\end{center}
\end{table}

The GU\&D$_{(k,0,1)}$ vs. KR comparison is less clear-cut (Table~\ref{tbl:sim_gudkr}). KR does have an edge with averaging estimators: in spite of GU\&D's smaller stationary bias, the GU\&D experiment is composed of a small number of same-treatment cohorts -- causing a ``grainy'' averaging estimator with increased variance. And again, it is variance that dominates performance. Using CIR, the two designs run neck-and-neck with GU\&D having a slight edge in these particular scenarios.

The bottom line from these comparisons (and others, not shown here but reported in earlier work: \citet{Oron05}), is that KR appears preferable over BCD -- especially if we can make averaging estimators work. Between GU\&D$_{(k,0,1)}$ and KR, the decision would probably depend upon whether a cohort design is required, desired or feasible -- since the performance differences are small. Also, clearly for BCD and GU\&D, CIR appears to be the preferred estimator.

\begin{table}[!h]
\begin{center}
\caption[GU\&D vs. KR Estimator Performance]{Comparison of estimator performance from GU\&D$_{(2,0,1)}$) and KR ($k=2$) simulations. The two estimators compared are $\hat{v}_{AD}$ (top), and CIR (bottom). The E.E.R. column quantifies KR's performance edge over GU\&D. Details are as in Table~\ref{tbl:sim_cir}.\label{tbl:sim_gudkr}}
\begin{tabular}{lp{2.5cm}crrrrc}
  \toprule
  \multirow{2}{*}{\bf Estimator}&\multirow{2}{2cm}{\bf Distribution}&\multirow{2}{1.5cm}{\bf Sample Size}&\multicolumn{2}{c}{{\bf Bias}}&\multicolumn{2}{c}{{\bf Std. Dev.}}&\multirow{2}{1.5cm}{\bf E.E.R.}\\
\cmidrule{4-7} &&& GU\&D & KR & GU\&D & KR &  \\
\midrule
\multirow{6}{*}{\bf $\hat{v}_{AD}$}&\multirow{2}{2cm}{{Lognormal}}  & $18$ &$-0.028$&$-0.045$&$0.132$ &$0.124$ & $1.05$ \\
                                 &     & $32$  &$0.007$&$-0.007$&$0.115$ &$0.106$ & $1.16$ \\
\cmidrule{2-8}
&\multirow{2}{2cm}{Normal}& $18$  &$-0.006$&$-0.027$&$0.112$ &$0.102$ & $1.12$ \\
 &                                   & $32$  &$0.008$&$-0.013$&$0.094$ &$0.085$ & $1.20$\\
\cmidrule{2-8}
&\multirow{2}{*}{Gamma}& $18$ &$0.001$&$-0.023$&$0.103$ &$0.099$ & $1.03$\\
 &                                   & $32$ &$0.013$&$-0.010$&$0.082$ &$0.079$ & $1.08$\\
\midrule[1pt]
\multirow{6}{*}{\bf CIR}&\multirow{2}{2cm}{{Lognormal}}  & $18$ &$-0.011$&$-0.018$&$0.138$ &$0.140$  & $0.96$\\
                                 &     & $32$  &$0.015$&$0.012$&$0.122$ &$0.121$  & $1.02$\\
\cmidrule{2-8}
&\multirow{2}{2cm}{Normal}& $18$  &$-0.006$&$-0.015$&$0.111$ &$0.111$ & $0.99$\\
 &                                   & $32$  &$-0.001$&$-0.007$&$0.091$ &$0.092$ & $0.98$\\
\cmidrule{2-8}
&\multirow{2}{*}{Gamma}& $18$ &$0.001$&$-0.005$&$0.104$ &$0.106$ & $0.96$\\
 &                                   & $32$ &$0.007$&$0.002$&$0.081$ &$0.084$ & $0.93$\\
 \bottomrule
  \end{tabular}
\end{center}
\end{table}

\subsection{Detailed Estimator Study}
We can be reasonably confident that CIR performs better than IR, that the KR design has an edge over similarly-targeted BCD, and that reversal-averaging tends to shift the estimate upward and has a larger variance than all-trial-averaging. This, however, leaves many design decision questions unanswered.

Many factors affect estimation performance -- spacing, the starting-point offset $x_1-Q_p$, the distribution's shape and scale, the location of boundaries, the sample size and so forth. I have chosen to demonstrate the sensitivity of various estimators to the first two factors (spacing and the starting-point effect), by neutralizing all others or holding them constant.

In order to make the results intuitively meaningful and more general, both $s$ and $x_1-Q_p$, as well as estimation bias and standard deviation, were normalized by $F$'s standard deviation. The following series of figures shows 2-D contour maps of estimator performance. The comparison is between $\hat{v}_1$ (top left), $\hat{w}_1$ (top right), $\hat{v}_{AD}$ (bottom left) and CIR (bottom right).

\subsubsection{SU\&D, Normal Thresholds}

\begin{figure}[!h]
\begin{center}
\Large{\textsf{    SU\&D, Normal Thresholds, Empirical Bias, $n=30$}}
\\[0.75cm]
\includegraphics[scale=.9]{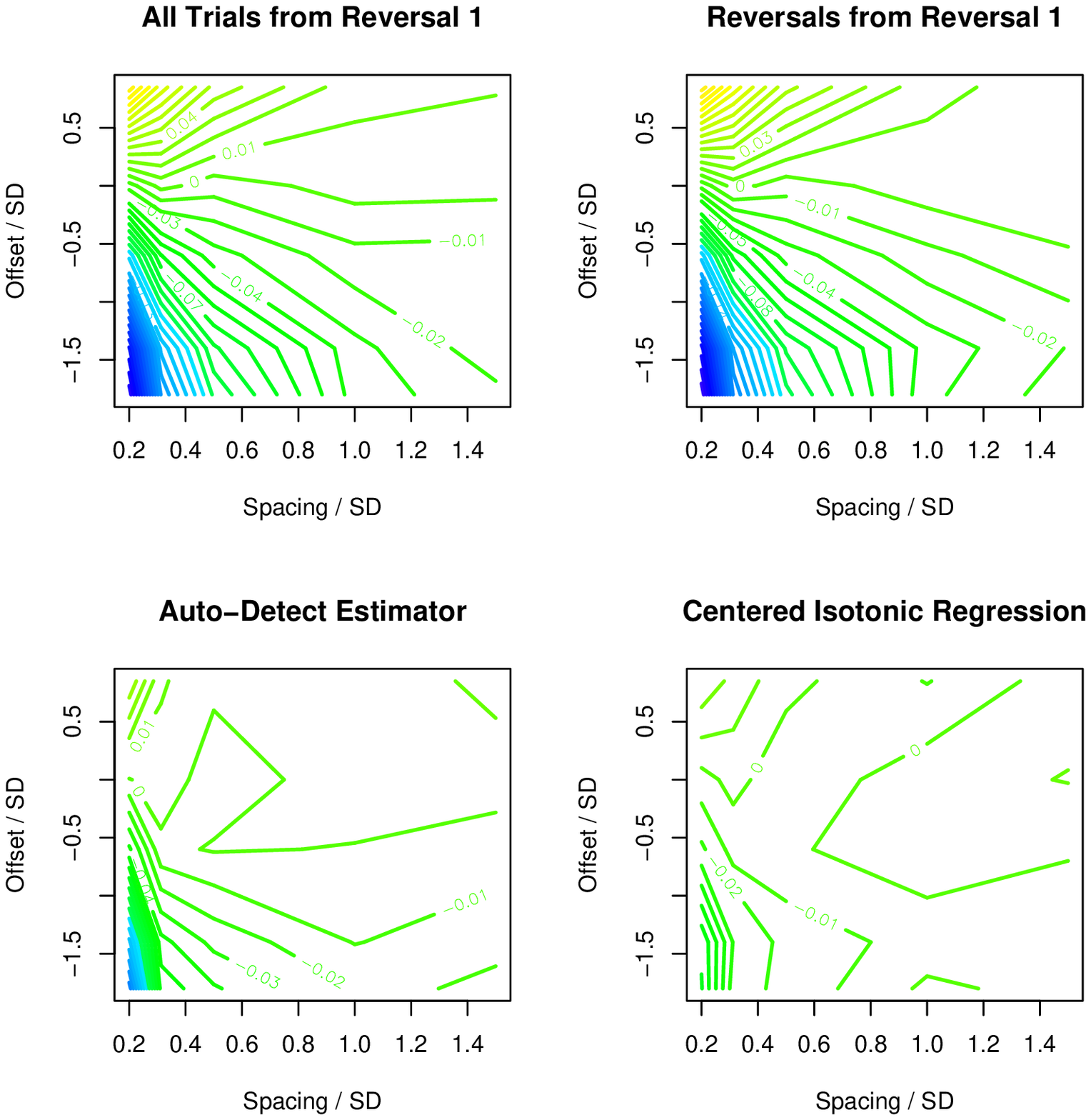}
{\small
\caption[Estimator Sensitivity: SU\&D, Normal Thresholds, Bias]{Sensitivity of selected estimators to normalized spacing $s/\sigma$ (horizontal axis) and normalized starting-point offset $(x_1-Q_p)/\sigma$ (vertical axis). Contours are of SU\&D ensemble {\bf bias} normalized by $\sigma$, for $n=30$. Contours are color-coded: blue indicates negative bias, yellow positive bias, and green indicates near-zero bias. In each page, four estimators are shown: $\hat{v}_1$ (top left), $\hat{w}_1$ (top right), $\hat{v}_{AD}$ (bottom left) and CIR (bottom right). The threshold distribution is normal. Each contour plot uses $5$ by $5$ empirical bias statistics, each of which was calculated from $2000$ runs.\label{fig:sim_biasnorm}}
}
\end{center}
\end{figure}

We begin from the most favorable U\&D scenario, for which the method was originally designed. The normal scenario is favorable for several reasons:
\begin{itemize}
\item $f$ is symmetric, and hence for SU\&D the stationary bias of $\mu_\pi$ is negligible;
\item The curvature of $F$ around target is very mild, and therefore CIR is also expected to have very little bias;
\item $f$ is thin-tailed, and therefore within-run excursions are less common.
\end{itemize}

\begin{figure}[!here]
\begin{center}
\Large{\textsf{    SU\&D, Normal Thresholds, Empirical SD, $n=30$}}
\\[0.75cm]
\includegraphics[scale=.9]{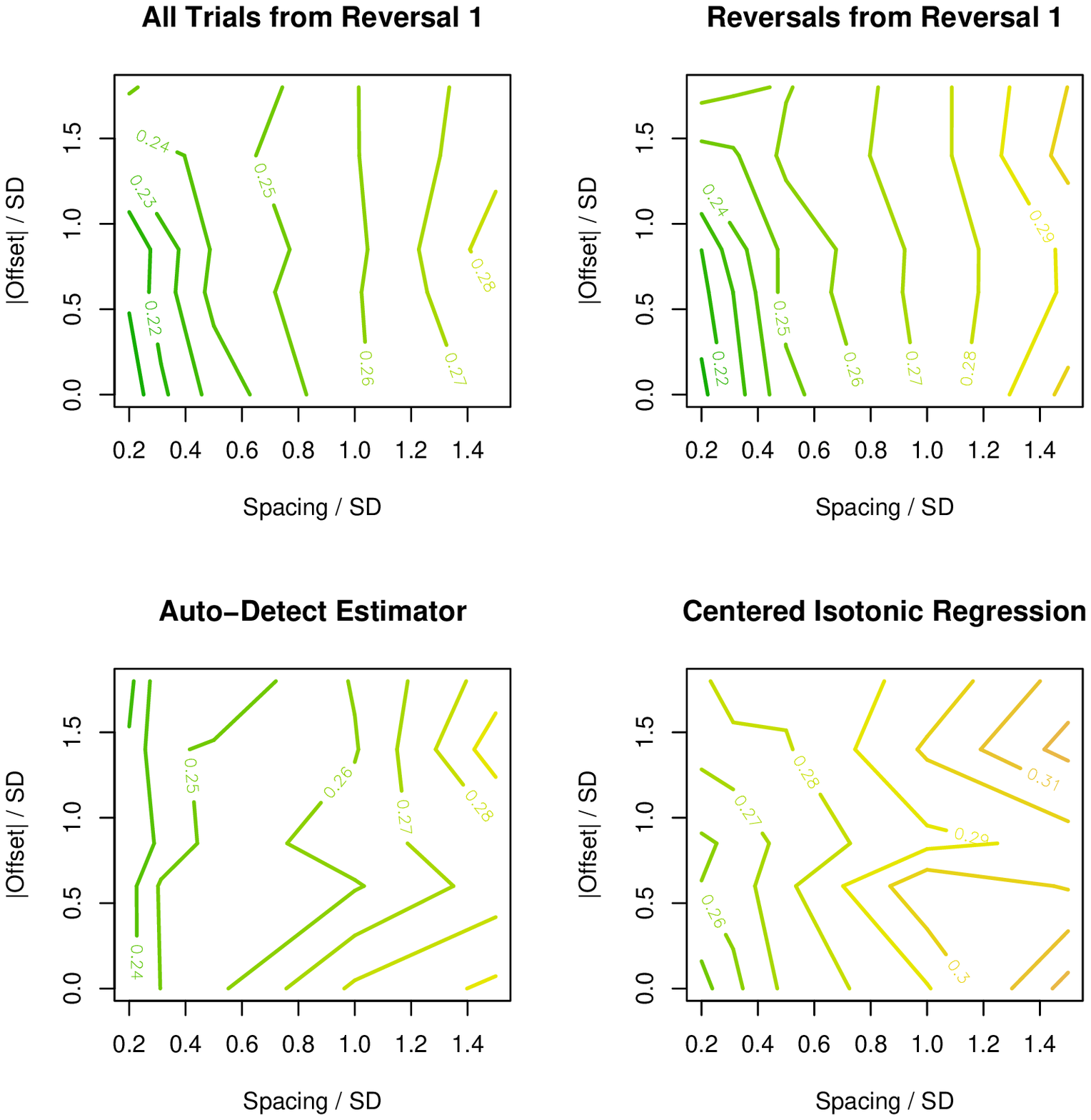}
\caption[Estimator Sensitivity: SU\&D, Normal Thresholds, SD]{Sensitivity of selected estimators to normalized spacing $s/\sigma$ (horizontal axis) and to \emph{absolute} offset $|Q_p-x_1|/\sigma$ (vertical axis). Contours are of SU\&D ensemble {\bf standard deviation} normalized by $\sigma$, for $n=30$. Contours are color-coded: as the SD increases, color changes from green through yellow to brown. All other details are as in Fig.~ \ref{fig:sim_biasnorm}.\label{fig:sim_sdnorm}}
\end{center}
\end{figure}

\begin{figure}[!here]
\begin{center}
\Large{\textsf{    SU\&D, Normal Thresholds, Empirical MSE, $n=30$}}
\\[0.75cm]
\includegraphics[scale=.9]{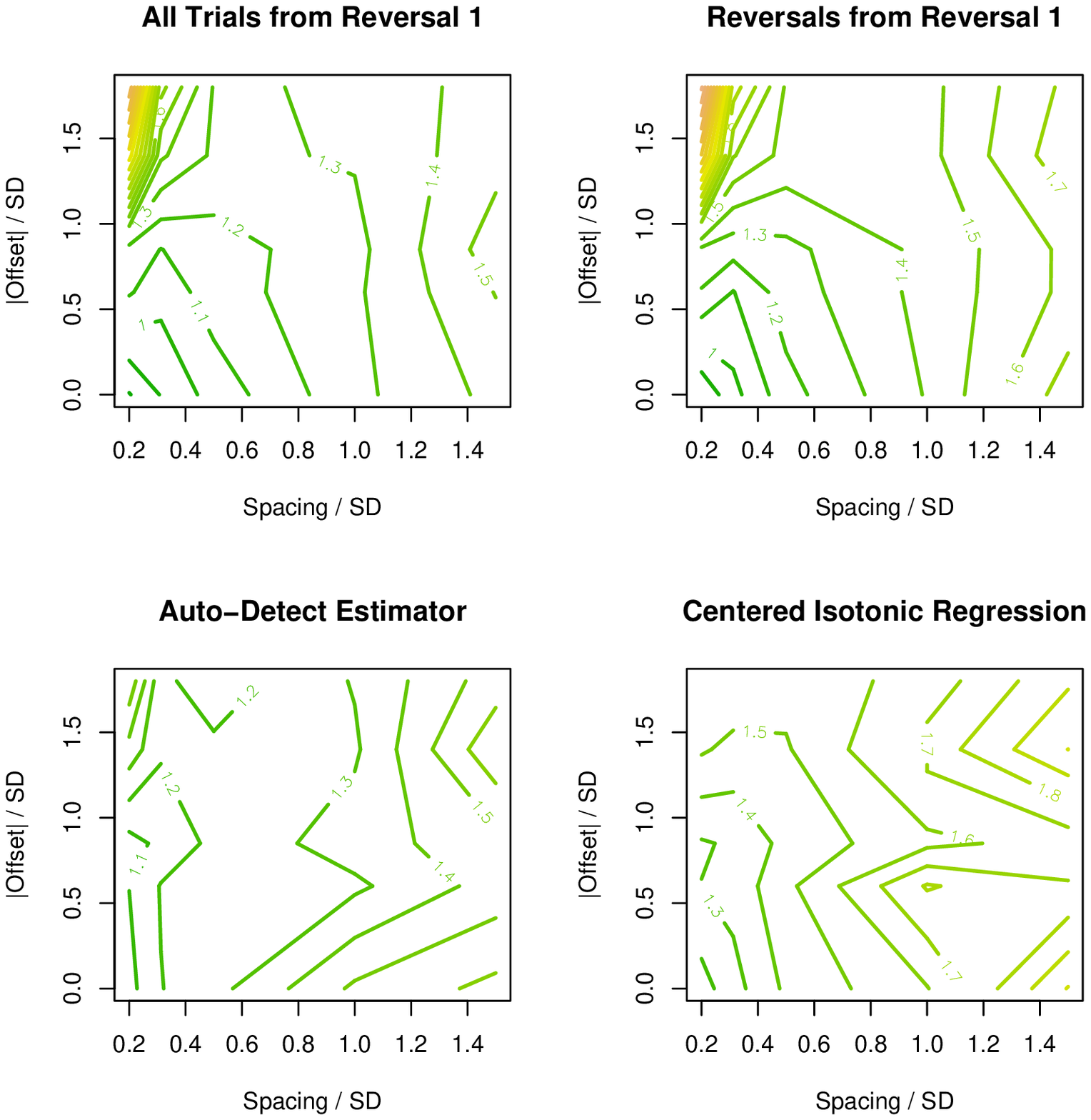}
\caption[Estimator Sensitivity: SU\&D, Normal Thresholds, MSE]{Sensitivity of selected estimators to normalized spacing $s/\sigma$ (horizontal axis) and to \emph{absolute} offset $|Q_p-x_1|/\sigma$ (vertical axis). Contours are of SU\&D ensemble {\bf MSE}, normalized by the asymptotic percentile-estimation error (\ref{eq:varTq}), for $n=30$. Color-coding is as in Fig.~\ref{fig:sim_sdnorm}. All other details are as in Fig.~\ref{fig:sim_biasnorm}.\label{fig:sim_msenorm}}
\end{center}
\end{figure}

\begin{figure}[!here]
\begin{center}
\Large{\textsf{   KR ($k=3$), Exponential Thresholds, Empirical MSE, $n=40$}}
\\[0.75cm]
\includegraphics[scale=.9]{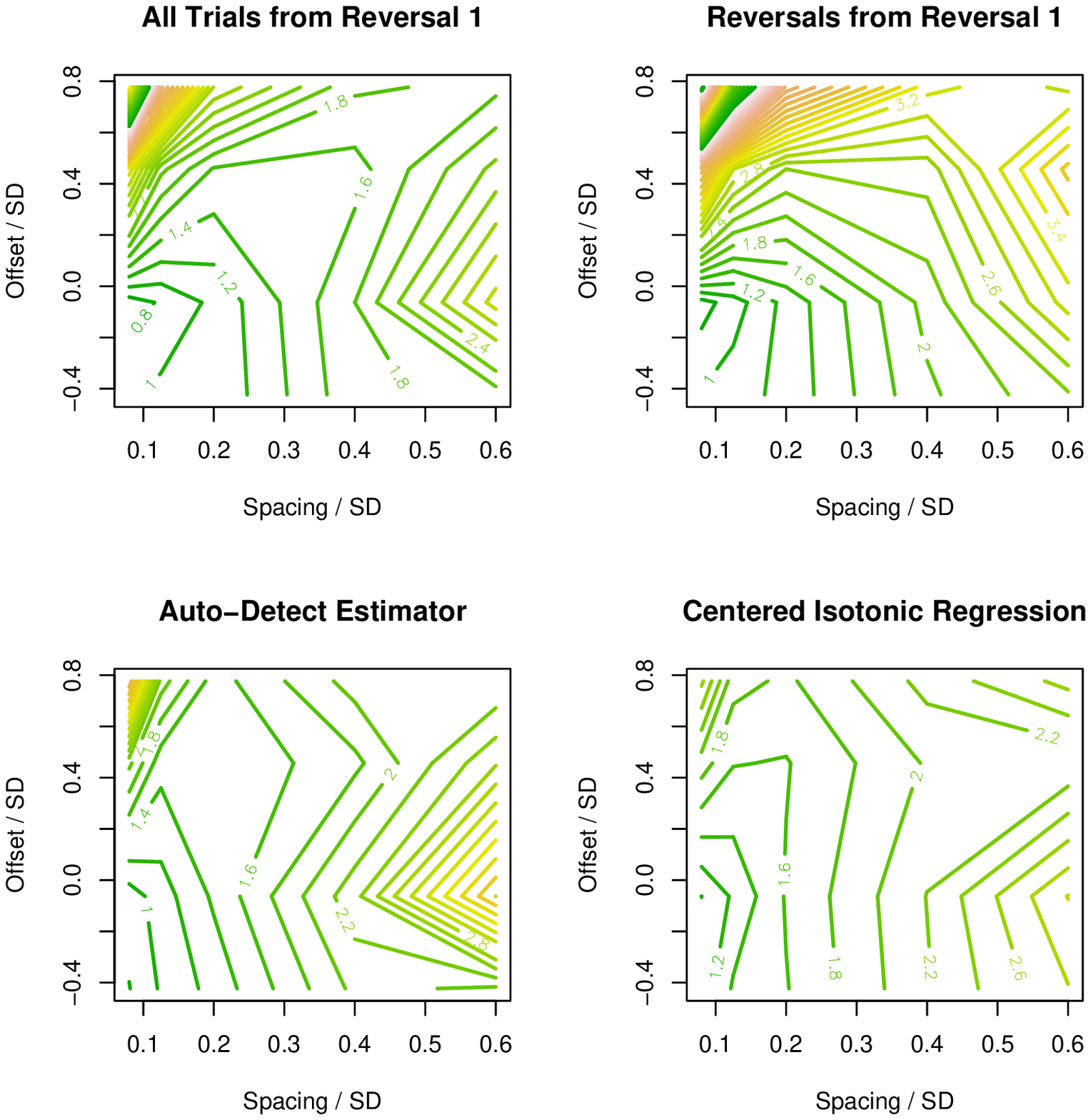}
\caption[Estimator Sensitivity: KR, Exponential Thresholds, MSE]{Sensitivity of selected estimators to normalized spacing $s/\sigma$ (horizontal axis) and to offset $(x_1-Q_p)/\sigma$ (vertical axis). Contours are of KR ($k=3$) ensemble {\bf MSE} w.r.t. $Q_{0.2}$, normalized by the asymptotic percentile-estimation error (\ref{eq:varTq}), for $n=40$. The threshold distribution is exponential. Color-coding is as in Fig.~\ref{fig:sim_sdnorm}. The green bands showing on the top-left corners of the $\hat{v}_1$ and $\hat{w}_1$ maps are artifacts, as the MSE there is so high that the contours need to cycle through the entire spectrum again ($\hat{w}_1$'s normalized MSE reaches as high as $13$). All other details are as in Fig.~\ref{fig:sim_biasnorm}.\label{fig:sim_kmseexp}}
\end{center}
\end{figure}

These advantages are borne out in Fig.~\ref{fig:sim_biasnorm}, mapping estimation bias for $n=30$.\footnote{All scenarios illustrated in Figs. \ref{fig:sim_biasnorm} through \ref{fig:sim_kmseexp} were also mapped for half the sample size, but those maps were omitted here for brevity.} Contours are color-coded to enhance interpretation: in general, \emph{``green is good''} -- i.e., a small bias magnitude. Blue indicates negative bias and yellow positive bias.

The dominant effect is starting-point bias, which gets worse with smaller $s$. The auto-detect estimator (bottom left) mitigates the effect rather well. Note that the bias contours of $\hat{v}_1$ (top left) and $\hat{w}_1$ (top right) are almost identical: as we found in Section~\ref{sec:est2}, for SU\&D the only shift between these two estimators would be due to skew, and here there is no skew.

The starting-point effect even spills over to CIR (bottom right): with small $s$ and $n$ and large offset, most of the information about $F$ is gathered somewhere between $x_1$ and $Q_p$; since away from target $F$ does have a substantial curvature, the resulting linear inverse interpolation is therefore systematically biased towards $x_1$.


If bias was the only ``game in town'' (and normal the only ``distribution in town''), one would be led to believe that coarse spacing and CIR are the way to go. Fig.~\ref{fig:sim_sdnorm} maps the other component of estimation error - standard deviation - in \emph{the same units used for bias}. Note that here the y-axes show absolute offset w.r.t. $x_1$, because of the symmetry of both SU\&D and $F$. Color-coding here still follows \emph{``green is good''}, with the SD becoming yellow, then brown as it increases. The same color-coding is used for MSE contours in subsequent figures.

The first observation is that for most design conditions, estimation SD dominates bias -- for normal thresholds, by a factor of about $10$ (compare SD magnitudes with Fig.~\ref{fig:sim_biasnorm}). Next, we see that as $s$ increases, so does the SD. Finally, note that the estimators have a clear ranking w.r.t. SD - $\hat{v}_1$ tightest, followed by $\hat{v}_{AD}$, $\hat{w}_1$ and CIR, whose SD is larger than $\hat{v}_1$'s by about $10-20\%$.

Next, we look at overall estimation MSE (Fig.~\ref{fig:sim_msenorm}). In the figure, the MSE is normalized by the asymptotic, direct-sampling, threshold percentile estimation error (\ref{eq:varTq}), given in the introduction. Note that when all stars are aligned correctly -- no starting-point offset, fine spacing, tight estimator -- we can sometimes observe what seems like  ``superefficiency'': an error smaller than the asymptotic direct-sampling expression. This happens especially with averaging estimators, through which U\&D converts percentile-finding into a potentially more efficient averaging process.

Since the SD dominates estimation, the less variable $\hat{v}_1$ or $\hat{v}_{AD}$ provide better overall performance. The former holds an edge for coarse spacing ($s>\sigma$), where starting-point bias is small. On the other hand, the auto-detect estimator, since it mitigates starting-point bias while retaining a relatively small SD, offers the best overall combination of performance and robustness around $s\approx\sigma/2$ -- a factor of $2$ smaller than the optimal spacing usually quoted in literature. CIR's large SD bars it from being competitive under this favorable normal-thresholds scenario.

\subsubsection{SU\&D, Exponential Thresholds}
\begin{figure}[!h]
\begin{center}
\Large{\textsf{    SU\&D, Exponential Thresholds, Empirical Bias, $n=30$}}
\\[0.75cm]
\includegraphics[scale=.9]{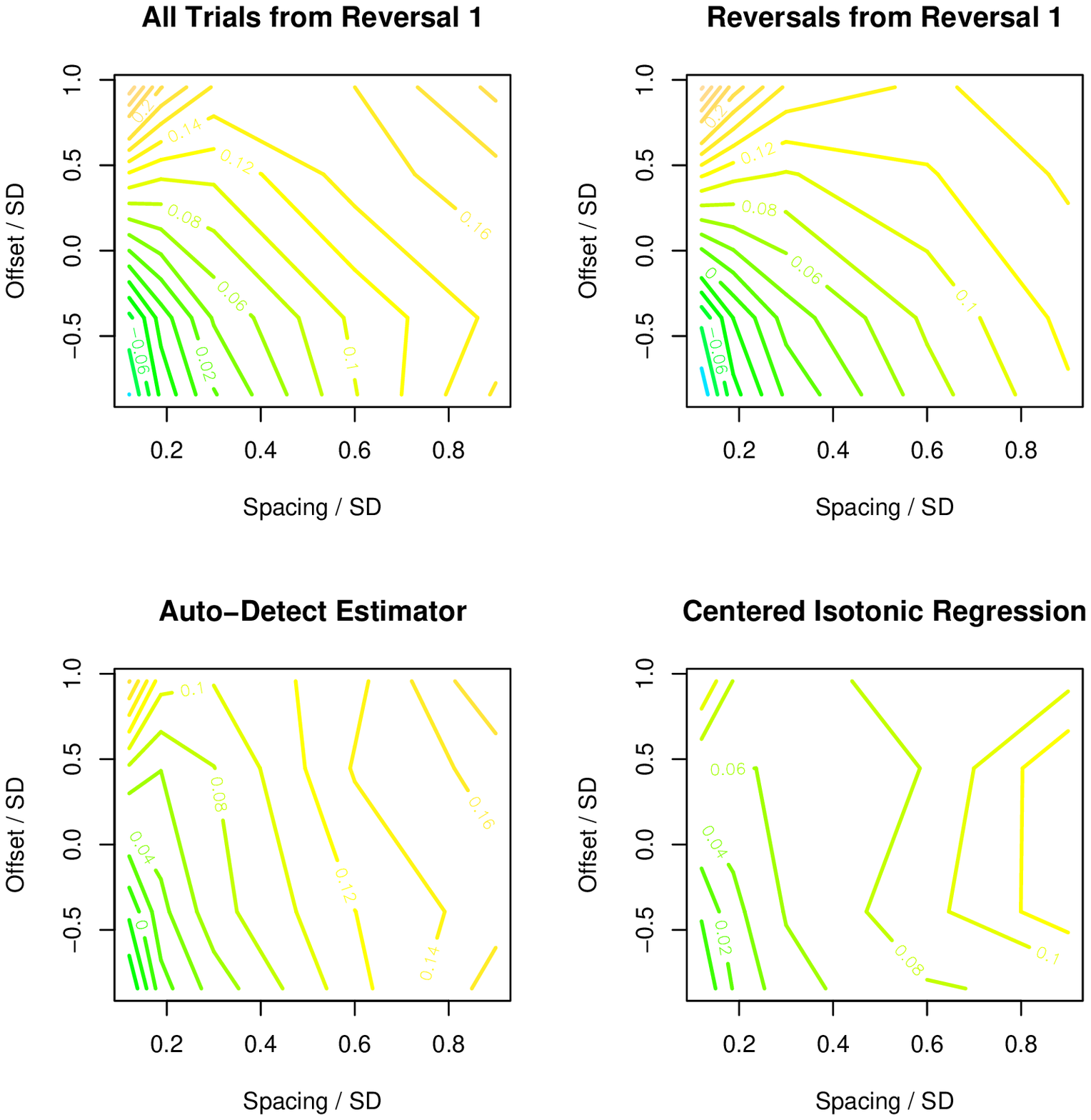}
\caption[Estimator Sensitivity: SU\&D, Exponential Thresholds, Bias]{Sensitivity of selected estimators to normalized spacing $s/\sigma$ (horizontal axis) and to  offset $(x_1-Q_p)/\sigma$ (vertical axis). Contours are of SU\&D ensemble {\bf bias}, normalized by the threshold SD, for $n=30$. The threshold distribution is exponential. All other details are as in Fig.~\ref{fig:sim_biasnorm}, except that here $4$ by $5$ points underly the contours.\label{fig:sim_biasexp}}
\end{center}
\end{figure}

What happens when the distribution is highly asymmetric? I chose the exponential with location parameter $\xi$ and scale parameter $\sigma$ (equal to the threshold SD). The estimation SD plot has been omitted, since the patterns observed in Fig.~\ref{fig:sim_sdnorm} are largely distribution-independent. The bias magnitude is larger now (Fig.~\ref{fig:sim_biasexp}). Averaging-estimator bias is positive almost across the board, since the second-order skew term is positive; it also increases with increasing $s$. Starting-point bias still takes its toll on the finest-spacing designs, but now also -- albeit more mildly -- on coarser designs.\footnote{The latter effect is due to the interaction between starting-point and $f$'s skewness and restricted range. Unlike the simpler starting-point bias, this one can only be positive under exponential thresholds.} The auto-detect scheme can do nothing about the stationary skew-related bias, and therefore $\hat{v}_{AD}$'s bias map shows vertical iso-bias contours.

CIR, too, has a positive spacing-dependent bias -- because $F$ is concave (when $F$'s linear interpolation runs below the true curve, the bias is positive and vice versa). However, CIR's bias is certainly smaller in magnitude.

\begin{figure}[!h]
\begin{center}
\Large{\textsf{    SU\&D, Exponential Thresholds, Empirical MSE, $n=30$}}
\\[0.75cm]
\includegraphics[scale=.9]{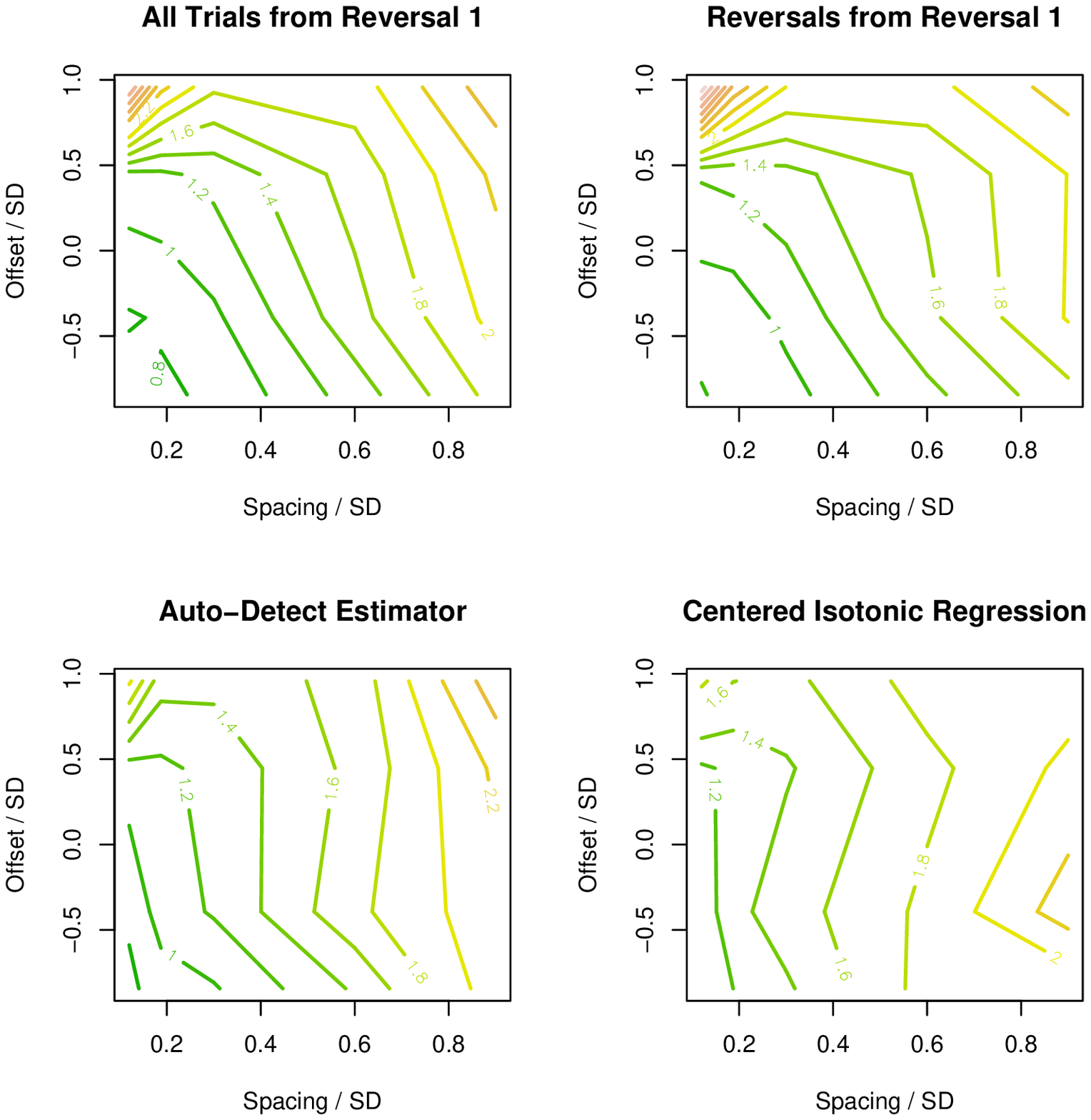}
\caption[Estimator Sensitivity: SU\&D, Exponential Thresholds, MSE]{Sensitivity of selected estimators to normalized spacing $s/\sigma$ (horizontal axis) and to offset $(x_1-Q_p)/\sigma$ (vertical axis). Contours are of SU\&D ensemble {\bf MSE}, normalized by the asymptotic percentile-estimation error (\ref{eq:varTq}), for $n=30$. The threshold distribution is exponential. Color-coding is as in Fig.~\ref{fig:sim_sdnorm}. All other details are as in Fig.~ \ref{fig:sim_biasexp}.\label{fig:sim_mseexp}}
\end{center}
\end{figure}

In spite of larger biases, the magnitude of estimation SD is still typically about twice as large. The best combination of robustness and performance seems to be offered by $\hat{v}_{AD}$ (Fig.~\ref{fig:sim_mseexp}, lower left), in the $s\approx\sigma/4$ to $\sigma/2$ range -- closely followed by CIR. Note that the ``sweet spot'' of quasi-superefficient estimation has moved to the lower left-hand corner, where the negative starting-point bias nearly cancels out the positive stationary biases.

\subsubsection{KR, Normal Thresholds}

We move on to non-median targets. The data shown here are for KR, $k=3$. We look at bias and MSE w.r.t. to $Q_{0.2}$ and not the exact formal target of $Q_{0.2063}$, since there is little applied interest in the latter (the difference is quite small in any case). CIR estimates were, indeed, performed for $Q_{0.2}$ -- giving that estimator a slight \textit{a priori} edge.

\begin{figure}[!h]
\begin{center}
\Large{\textsf{   KR ($k=3$), Normal Thresholds, Empirical Bias, $n=40$}}
\\[0.75cm]
\includegraphics[scale=.9]{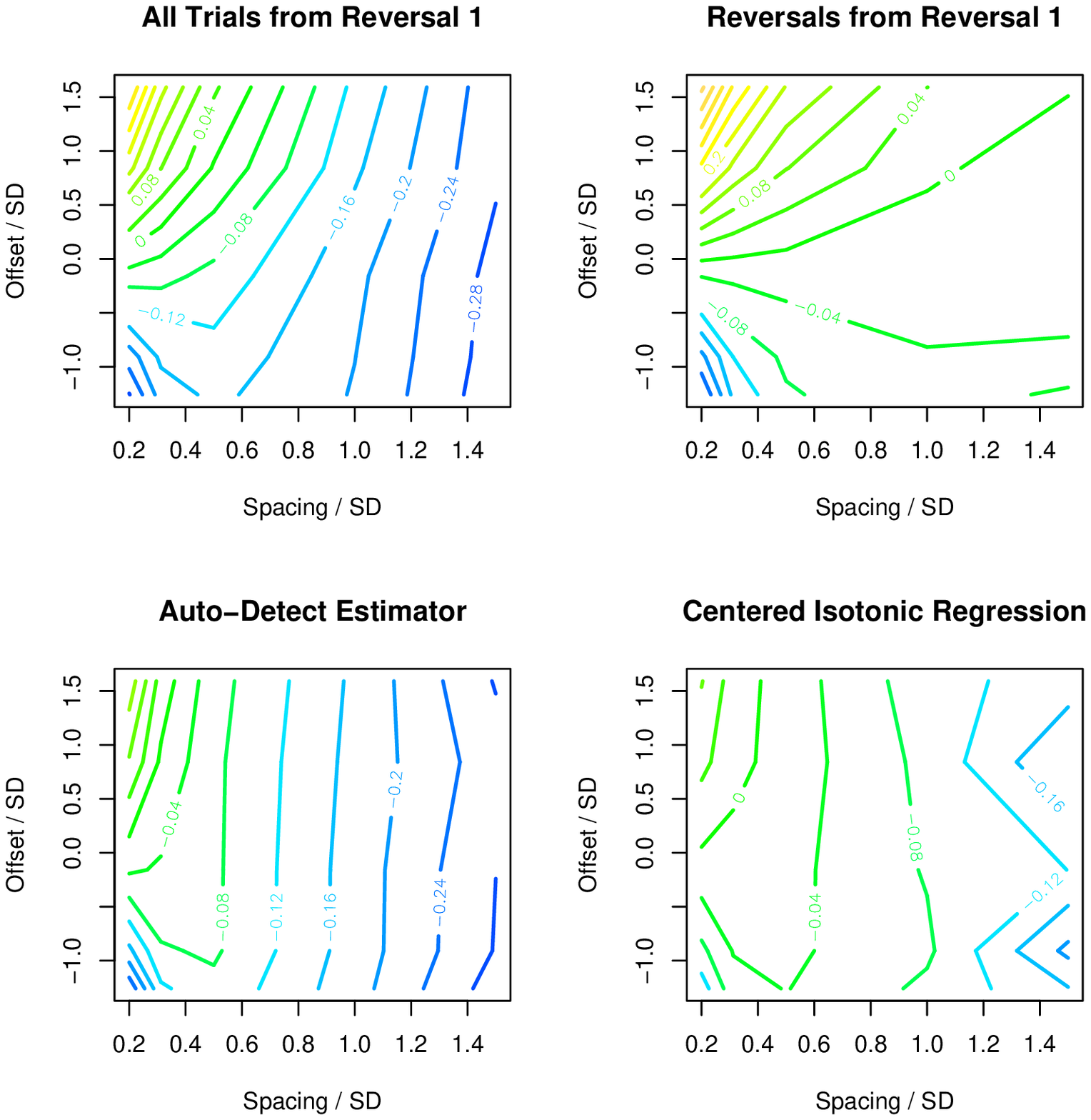}
\caption[Estimator Sensitivity: KR, Normal Thresholds, Bias]{Sensitivity of selected estimators to normalized spacing $s/\sigma$ (horizontal axis) and to offset $(x_1-Q_p)/\sigma$ (vertical axis). Contours are of KR ($k=3$) ensemble {\bf bias} w.r.t. $Q_{0.2}$, normalized by the threshold SD, for $n=40$. The threshold distribution is normal. All other details are as in Fig.~\ref{fig:sim_biasnorm}.\label{fig:sim_kbiasnorm}}
\end{center}
\end{figure}

The starting-point biases make their appearances again, and are greater in magnitude than for SU\&D -- which is expected since KR converges more slowly (Fig.~\ref{fig:sim_kbiasnorm}). The other dominant feature for $\hat{v}_1$ or $\hat{v}_{AD}$ are vertical iso-bias contours. This is the effect of $\mu_\pi$'s first-order stationary bias. As found in Section~\ref{sec:est2}, this bias is negative and linear in $s$. CIR shows a negative spacing-dependent bias as well, because $F$ is locally convex near $Q_{0.2}$. The reversal-average $\hat{w}_1$ shows very little of this effect; it appears that with normal thresholds, $\mu_\pi$'s bias is almost perfectly neutralized by reversal-averaging's positive bias.

\begin{figure}[!h]
\begin{center}
\Large{\textsf{   KR ($k=3$), Normal Thresholds, Empirical MSE, $n=40$}}
\\[0.75cm]
\includegraphics[scale=.9]{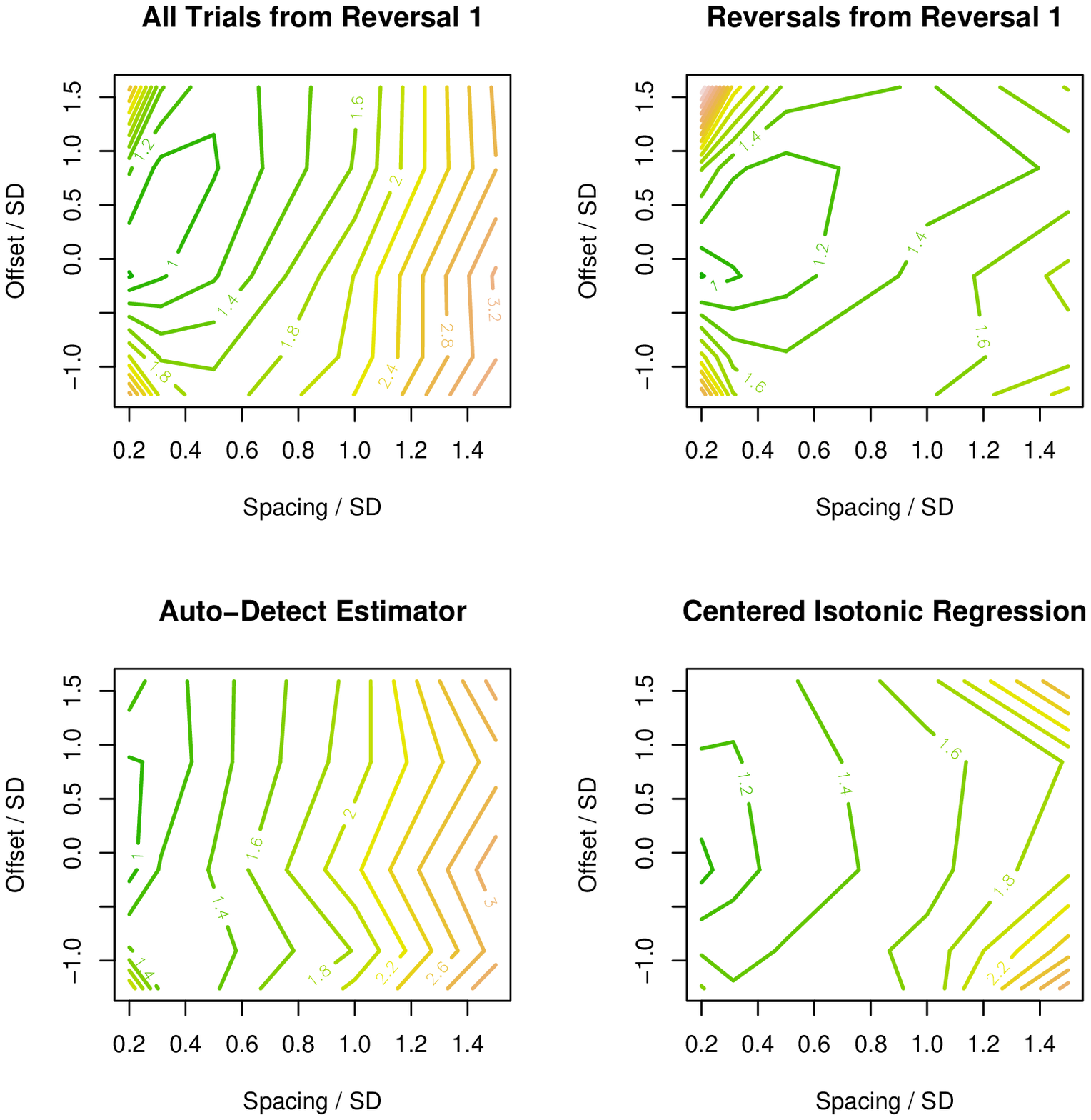}
\caption[Estimator Sensitivity: KR, Normal Thresholds, MSE]{Sensitivity of selected estimators to normalized spacing $s/\sigma$ (horizontal axis) and to offset $(x_1-Q_p)/\sigma$ (vertical axis). Contours are of KR ($k=3$) ensemble {\bf MSE} w.r.t. $Q_{0.2}$,, normalized by the asymptotic percentile-estimation error (\ref{eq:varTq}), for $n=40$. The threshold distribution is normal. Color-coding is as in Fig.~\ref{fig:sim_sdnorm}. All other details are as in Fig.~ \ref{fig:sim_biasnorm}.\label{fig:sim_kmsenorm}}
\end{center}
\end{figure}

Looking at estimation performance, the picture is now more complex (Fig.~\ref{fig:sim_kmsenorm}). Because of the stationary bias, $\hat{v}_{AD}$ is less attractive. Its performance deteriorates quickly as $s$ increases; around $s\approx\sigma/2$ it is matched or beaten by CIR. Moreover, at somewhat larger spacing, as the starting-point bias diminishes and stationary biases kick in, reversal-average $\hat{w}_1$ suddenly appears like a viable option.

However, while $\hat{v}_{AD}$'s and CIR's advantages are grounded in their theoretical robustness, the reversal-averaging surprise is more of a numerical coincidence related to a particular form of $F$, as we shall soon see.

\subsubsection{KR, Exponential Thresholds}

\begin{figure}[!here]
\begin{center}
\Large{\textsf{   KR ($k=3$), Exponential Thresholds, Empirical Bias, $n=40$}}
\\[0.75cm]
\includegraphics[scale=.9]{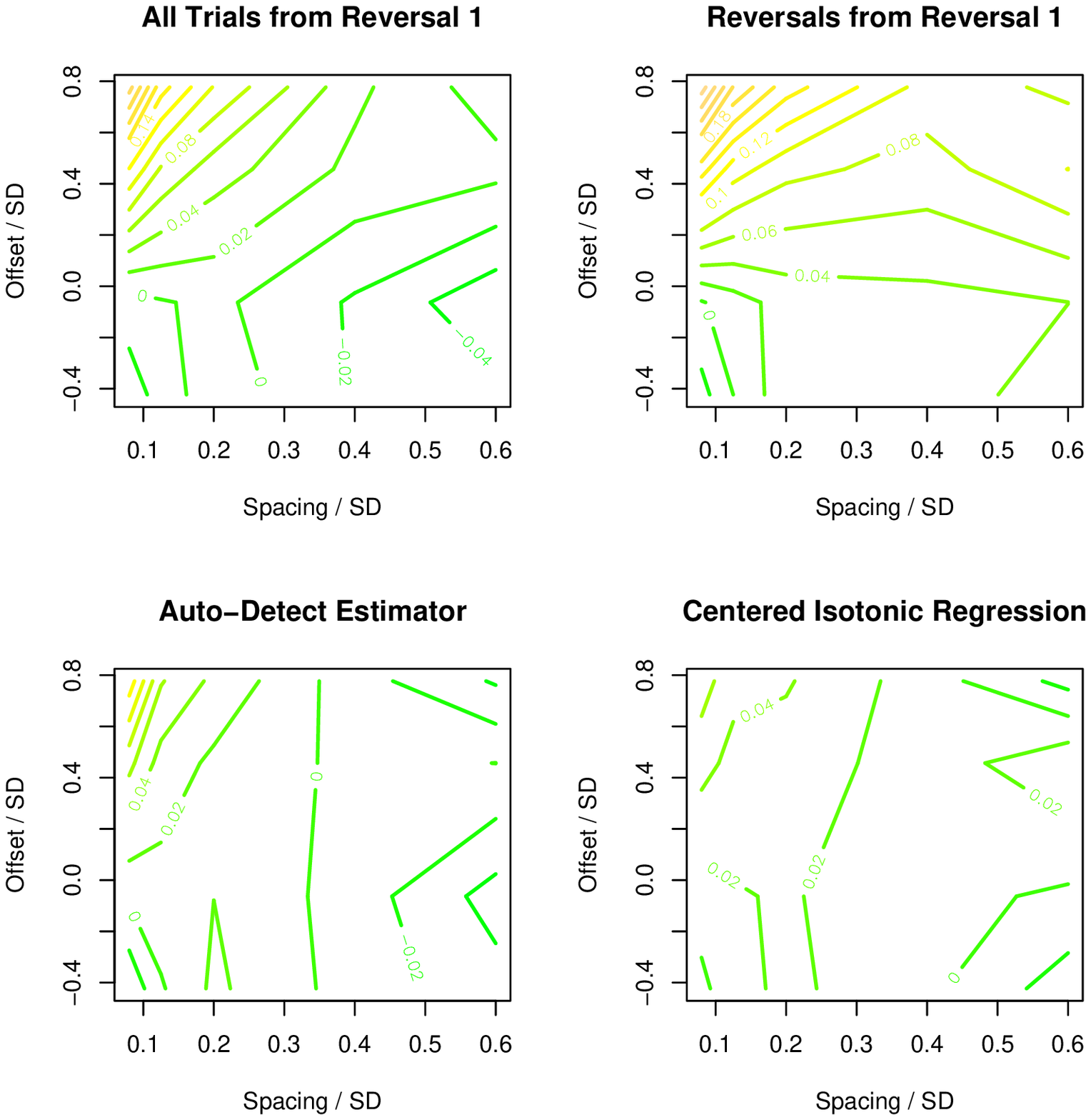}
\caption[Estimator Sensitivity: KR, Exponential Thresholds, Bias]{Sensitivity of selected estimators to normalized spacing $s/\sigma$ (horizontal axis) and to offset $(x_1-Q_p)/\sigma$ (vertical axis). Contours are of KR ($k=3$) ensemble {\bf bias} w.r.t. $Q_{0.2}$, normalized by the threshold SD, for $n=40$. The threshold distribution is exponential. All other details are as in Fig.~\ref{fig:sim_biasnorm}.\label{fig:sim_kbiasexp}}
\end{center}
\end{figure}

Using exponential thresholds, the bias picture changes yet again (Fig.~\ref{fig:sim_kbiasexp}). Now, the first-order spacing-dependent stationary bias has all but disappeared (it is still somewhat visible in the auto-detect map for $n=40$). This bias seems to be neutralized by $f$'s upward skew, which would of course affect the telescopic sum (\ref{eqn:taylortelescope}). This allows the starting-point bias to dominate averaging estimation again.

As far as estimation goes, now $\hat{w}_1$ is by far the worst among the four. $\hat{v}_{AD}$ and CIR are nearly equivalent with fine spacing ($s\approx\sigma/4$ to $\sigma/2$). As $s$ increases, there is a narrow window in which $\hat{v}_1$ appears to overtake them, before the first-order downward bias finally kicks in and hurts all three averaging estimators.

So for KR, normal and exponential thresholds produce almost the opposite outcomes, with different bias sources neutralizing each other for different estimators, depending upon the form of $F$. Note that overall, relative performance with KR is somewhat worse than with SU\&D (when normalized by the asymptotic direct-sampling efficiency) .

The same qualitative patterns were observed for $k=2$ simulation results as well, but have been omitted for brevity.

\subsection{Interval Coverage}
Table~\ref{tbl:sim_cover_1} compares the three interval-estimation options for CIR discussed earlier in the chapter, using data from SU\&D runs. Even though the bootstrap was never used for inverse estimation in the U\&D and IR context (at least not in published literature), it has been recommended for forward estimation by \citet{StylianouEtAl03}. They used the bias-corrected bootstrap; here, after some experimentation with various bootstrap variants, I opted for bootstrap-$t$, which is considered relatively robust and with a rate-$n$ accuracy \citep[Ch. 5]{DavisonHinkley95}.

The bootstrap results are quite disappointing (the same level of coverage was observed with other bootstrap variants, but not shown here). A possible explanation is that the bootstrap process fails to reproduce the hierarchical nature of variability (in sampling, and then in $\hat{F}$ values). On the other hand, the approach developed in Section~\ref{sec:est3} using linearized quantile functions and allowing for the added variability in allocation, seems to be quite conservative with satisfactory overall coverage. As explained earlier, the binomial option fails to distinguish well between quantiles on the tail; the Poisson option does the job better while sacrificing little in coverage. Therefore, subsequently I have used the linearized-Poisson approach for CIR intervals.

Table~\ref{tbl:sim_cover_2} surveys coverage performance across a wider variety of designs, for both CIR and the averaging estimator $\hat{v}_{AD}$. For the latter, the hitting-time approach outlined in Section~\ref{sec:est3} was used (see equation (\ref{eq:varoron})). Overall, coverage is good, tending to be on the conservative side. Even in cases when the average coverage appears to dip below the nominal level, it is usually due to a single difficult scenario out of $6-8$ used for calculations; for all other scenarios coverage was close to, or above, the nominal level.

The interval estimators have their vulnerabilities: CIR coverage suffers as $F$ becomes shallower, since then the $\hat{F}$-s would show many monotonicity violations, resulting in fewer points in CIR's output, points whose location would vary from run to run.\footnote{Using IR would hardly alleviate this: with IR, the output would have wide flat intervals, with inverse estimation highly variable as a result.} If $F$ around target is very shallow indeed, the coverage method would break down; this seems to occur somewhere below $\Delta F\approx 0.05$ between levels.

Averaging-estimator CI's break down most conspicuously when the estimator's bias (whether due to starting-point or to boundary effects) is too large. Cases when target lies on the boundary are not reflected in the averaging-estimator coverage statistics reported here. On the other hand, CIR coverage is quite resistant to boundaries.

\begin{table}
\begin{center}
\caption[SU\&D Interval Coverage Comparison - CIR]{SU\&D interval coverage for CIR estimates under four distribution scenarios, and $n=15,30$. In all scenarios $m=10$, with relatively fine spacing ($s/\sigma<0.5$). Shown are ensemble coverage percentages for $90\%$ (top) and $95\%$ (bottom) CI's, based on ensembles of $2000$ runs per scenario. The bootstrap simulations used $999$ bootstrap repetitions per run.\label{tbl:sim_cover_1}}
\small{
\begin{tabular}{p{4cm}ccccc}
\multicolumn{6}{c} {\Large{\bf Coverage of $90\%$ Confidence Intervals: CIR for SU\&D}}\\
\toprule
& & \multicolumn{4}{c} {{\bf Distribution Scenario Code}}\\
\cmidrule{3-6}
{\bf Interval Method} & & {\bf ``gam5090''} & {\bf ``gam1090''} & {\bf ``log0280''} & {\bf ``norm3330''} \\
  \toprule
  \multirow{2}{4cm}{\bf Bootstrap-$t$}& $n=15$ & 70.6\% & 70.4\% & 71.2\% & 73.6\%\\
\cmidrule{2-6}
& $n=30$ & 79.3\% & 78.7\% & 80.1\% & 78.9\%\\
\midrule[1pt]
  \multirow{2}{4cm}{\bf Linearized Binomial Quantiles}& $n=15$ & 96.7\% & 92.1\% & 95.2\% & 94.8\%\\
\cmidrule{2-6}
& $n=30$ & 97.2\% & 94.4\% & 95.5\% & 95.3\%\\
\midrule[1pt]
  \multirow{2}{4cm}{\bf Linearized Poisson Quantiles} & $n=15$ & 92.1\% & 90.1\% & 92.1\% & 91.6\%\\
\cmidrule{2-6}
& $n=30$ & 96.8\% & 93.9\% & 95.3\% & 95.1\%\\
 \bottomrule
\\
\\
\multicolumn{6}{c} {\Large{\bf Coverage of $95\%$ Confidence Intervals: CIR for SU\&D}}\\
\toprule
& & \multicolumn{4}{c} {{\bf Distribution Scenario Code}}\\
\cmidrule{3-6}
{\bf Interval Method} & & {\bf ``gam5090''} & {\bf ``gam1090''} & {\bf ``log0280''} & {\bf ``norm3330''} \\
  \toprule
  \multirow{2}{4cm}{\bf Bootstrap-$t$}& $n=15$ & 78.9\% & 77.7\% & 77.9\% & 79.4\%\\
\cmidrule{2-6}
& $n=30$ & 85.5\% & 84.7\% & 85.8\% & 86.1\%\\
\midrule[1pt]
  \multirow{2}{4cm}{\bf Linearized Binomial Quantiles} & $n=15$ & 96.9\% & 93.1\% & 96.2\% & 95.9\%\\
\cmidrule{2-6}
& $n=30$ & 97.9\% & 95.4\% & 97.1\% & 96.6\%\\
\midrule[1pt]
  \multirow{2}{4cm}{\bf Linearized Poisson Quantiles} & $n=15$ & 95.8\% & 93.4\% & 95.8\% & 95.3\%\\
\cmidrule{2-6}
& $n=30$ & 98.4\% & 96.4\% & 97.5\% & 97.2\%\\

 \bottomrule
\end{tabular}
}\end{center}
\end{table}

\begin{table}
\begin{center}

\caption[Interval Coverage Summary - Various Settings]{Interval Coverage Summary for various designs and estimators. Shown are empirical coverage statistics of $90\%$ (top) and $95\%$ (bottom) CI's. For non-median designs, the targets of $Q_{0.2}$ and $Q_{0.3}$ were used, rather than each design's actual target. All summaries are of $6$ distribution scenarios each, with an ensemble size $N=2000$ and a relatively fine spacing ($m_{eff}\approx 10$), except for the GU\&D summary (second row in each table), which is from $8$ scenarios, $N=1000$ and a fixed-boundary $m=6$.\label{tbl:sim_cover_2}}
\begin{tabular}{p{5cm}crrrr}
\multicolumn{6}{c} {\Large{\bf $90\%$ Confidence Intervals}}\\
\toprule
{\bf Design and Estimator} & & {\bf Mean} & {\bf S.D.} & {\bf Min.} & {\bf Max.} \\
  \toprule
  \multirow{2}{*}{\bf KR ($k=3$) and CIR}& $n=20$ & 89.2\% & 2.6\% & 86.3\% & 93.1\%\\
\cmidrule{2-6}
& $n=40$ & 92.1\% & 3.2\% & 86.5\% & 95.5\%\\
\midrule
  \multirow{2}{*}{\bf GU\&D$_{(2,0,1)}$ and CIR}& $n=18$ & 94.7\% & 3.3\% & 89.6\% & 99.5\%\\
\cmidrule{2-6}
& $n=32$ & 95.7\% & 1.6\% & 93.8\% & 98.5\%\\
\midrule
  \multirow{2}{*}{\bf KR ($k=3$) and $\hat{v}_{AD}$}& $n=20$ & 94.8\% & 3.0\% & 89.2\% & 97.1\%\\
\cmidrule{2-6} & $n=40$ & 95.8\% & 2.7\% & 92.2\% & 99.6\%\\
\midrule
  \multirow{2}{*}{\bf SU\&D and $\hat{v}_{AD}$}& $n=15$ & 91.8\% & 4.1\% & 84.9\% & 97.1\%\\
\cmidrule{2-6} & $n=30$ & 96.6\% & 3.7\% & 90.0\% & 99.7\%\\
 \bottomrule
\\
\\
\multicolumn{6}{c} {\Large{\bf $95\%$ Confidence Intervals}}\\
\toprule
{\bf Design and Estimator} & & {\bf Mean} & {\bf S.D.} & {\bf Min.} & {\bf Max.} \\
  \toprule
  \multirow{2}{*}{\bf KR ($k=3$) and CIR}& $n=20$ & 93.3\% & 2.2\% & 90.4\% & 95.8\%\\
\cmidrule{2-6} & $n=40$ & 95.4\% & 2.4\% & 91.4\% & 97.8\%\\
\midrule
  \multirow{2}{*}{\bf GU\&D$_{(2,0,1)}$ and CIR}& $n=18$ & 97.1\% & 1.8\% & 94.8\% & 100.0\%\\
\cmidrule{2-6}
& $n=32$ & 97.5\% & 1.3\% & 95.5\% & 99.4\%\\
\midrule
  \multirow{2}{*}{\bf KR ($k=3$) and $\hat{v}_{AD}$}& $n=20$ & 95.5\% & 2.6\% & 90.7\% & 97.9\%\\
\cmidrule{2-6} & $n=40$ & 97.9\% & 1.4\% & 96.4\% & 99.9\%\\
\midrule
  \multirow{2}{*}{\bf SU\&D and $\hat{v}_{AD}$}& $n=15$ & 94.7\% & 2.3\% & 90.6\% & 97.5\%\\
\cmidrule{2-6} & $n=30$ & 98.2\% & 2.0\% & 94.7\% & 99.9\%\\
 \bottomrule
\end{tabular}

\end{center}
\end{table}

\section{The Anesthesiology Experiment}\label{sec:estexp}

As explained in the introduction, this study has been motivated by a consulting collaboration on a real experiment plan \citep{Oron04}. The experiment has eventually been carried out from late 2004 to early 2007, and initial results are being presented \citep{BhanankerEtAl07}. Dr. M.J. Souter has kindly allowed me to present the data here as well. This will help illustrate the theoretical features described so far, and to demonstrate the inevitable gap between theory and practice.

\subsection{Background}
Propofol is the current anesthetic agent of choice for routine surgical procedures in the industrialized world. It has replaced previous agents such as thiopental, because it has no adverse effects on blood pressure like other agents, it enables faster patient recovery from drowsiness after the procedure, and for other reasons. However, propofol has a very unpleasant -- and paradoxical -- side effect: in the minutes before passing out, many patients feel intense pain. Several studies and meta-analyses estimate the overall population frequency of pain responses at $70\%$ \citep{PicardTramer00}. Devising methods to alleviate this side effect is an ongoing practical and academic endeavor.

One interesting approach has been to mix propofol with thiopental -- an older agent with less favorable overall outcomes, but without the pain side-effect. Several studies have shown that the primary therapeutic (anesthesizing) effect for such mixtures is additive -- meaning that efficacy is not adversely affected \citep{VinikEtAl99} -- and that the prevalence of pain responses is reduced \citep{PollardEtAl02}. However, to date no study had yielded a clear, definitive recommendation for mixture proportions to the general patient population. Being familiar with U\&D, which is used extensively in anesthesiology for median-finding \citep{CapognaEtAl01,CamorciaEtAl04,Fisher07}, Dr. Souter approached our Consulting Service to inquire whether U\&D methodology can be used to find a propofol/thiopental mix that reduces pain response frequency to under $20\%$.

\subsection{Design Constraints and Initial Recommendations}
Since propofol is used in routine surgery, and since the proposed treatment mixes it with another FDA-approved agent, this experiment was relatively free of the typical ethical and sample-size constraints encountered in medical research.\footnote{Of course, an appropriate IRB was submitted and approved.} In practice, any adult (within certain health and age boundaries) assigned to surgery, who agreed to participate in the experiment could be recruited. On the other hand, this walk-in recruiting suggested that indeed, a sequential design such as U\&D would work better than a non-sequential one.

Our initial recommendations to Dr. Souter were based upon a literature search and numerical simulations. The search has turned up both KR and BCD. KR exhibited superior performance across the board in simulated runs, in line with the few numerical comparisons appearing in literature. The runs also indicated that for an estimation with CI's of about $\pm 10\%$, sample size should be quite larger than those discussed in the papers presenting BCD \citep{DurhamFlournoy95,DurhamEtAl95} - closer to $100$ than to the samples of $15-35$ simulated there. Additionally, the convergence-stationarity tradeoff with respect to spacing was identified. I recommended $k=4$ (targeting $Q_{0.159}$), narrowly over $k=3$, both because of somewhat better simulation results and to make sure the target is below $Q_{0.2}$. I also recommended a spacing of $10\%$ and a starting point somewhere near the middle of the range between $0$ and $100\%$ propofol.

\subsection{Raw Results}
\begin{figure}
\begin{center}
\includegraphics[scale=.8]{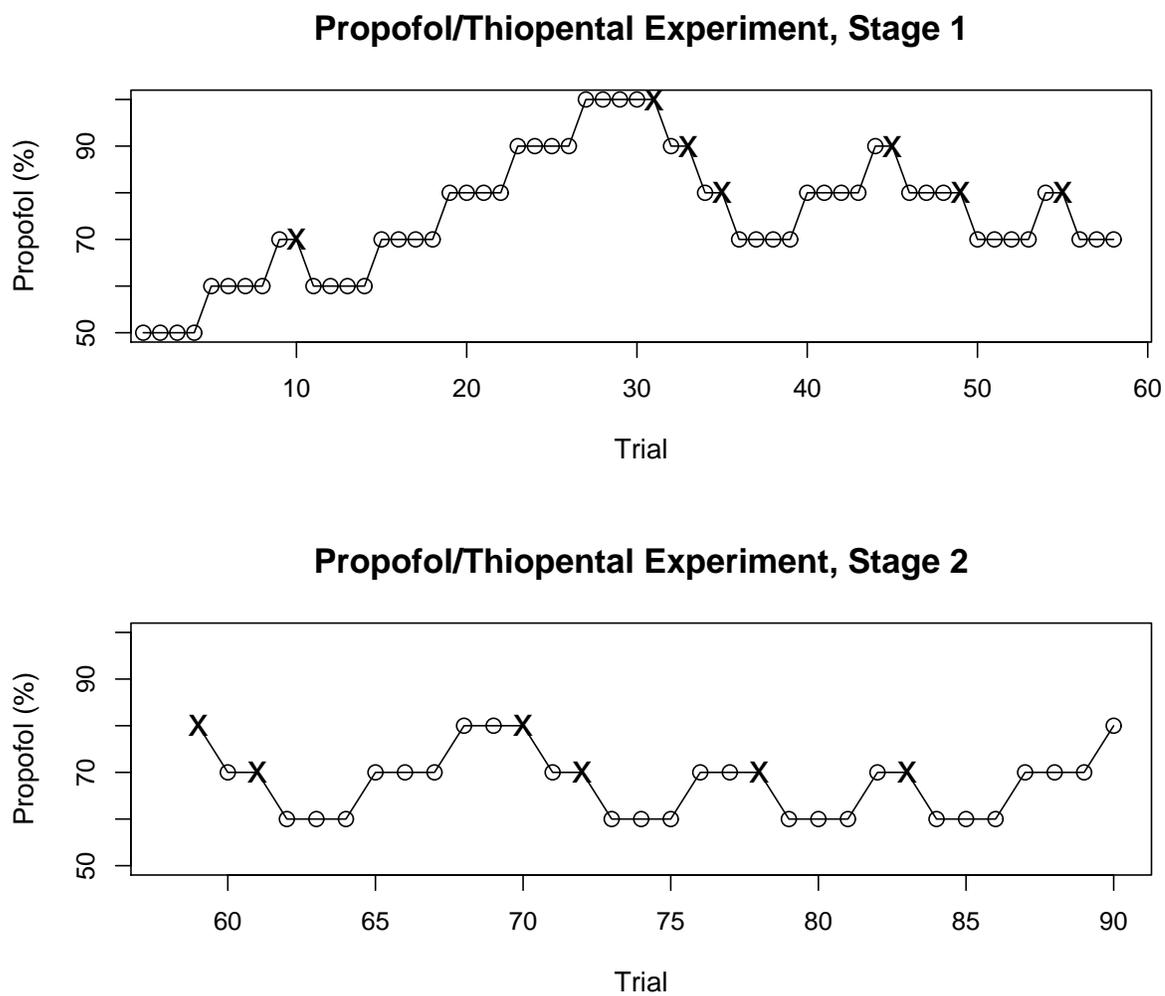}
\caption[Anesthesiology Experiment Runs]{Chains of the propofol/thiopental experiment reported by \citep{BhanankerEtAl07}. Stage $1$ (trials $1-58$) is on top and Stage $2$ (trials $58-90$) on bottom. Responses are marked as in Fig.~\ref{fig:basic}.}\label{fig:souter}
\end{center}
\end{figure}

The experiment went underway in fall 2004 at Harborview Hospital in Seattle, with a KR ($k=4$) design, starting point of $50:50$, and $10\%$ spacing. Treatment assignments were double-blind, and there was a single designated observer to evaluate whether pain response had occurred. The first stage was characterized by very low incidence of pain responses. After $58$ patients, and after several rounds of consultation, the pain evaluator has changed and the design was switched to $k=3$, continuing from the point at which the experiment was at that time ($80\%$ propofol), until trial $90$ when the experiment ended. Trials $59-90$ will be referred to as Stage 2. Some sample demographics are summarized in Table~\ref{tbl:souterD}. The two stages' chains are illustrated in Fig.~\ref{fig:souter}, and the treatment-response summary tables are shown in Table~\ref{tbl:souterF} .

\begin{table}
\begin{center}
\caption[Anesthesiology Experiment Demography Tables]{Age and gender summary table for the propofol/thiopental experiment.\label{tbl:souterD}}
\begin{tabular}{llcc}
   \toprule
&&{\bf Stage 1 (trials 1-58)}& {\bf Stage 2 (trials 59-90)}\\
  \multirow{2}{*}{\bf Gender}&Female &$13$&$10$\\
   &  Male &$45$&$22$\\
  \midrule
  \multirow{6}{*}{\bf Age}& Minimum&$20$&$19$\\
 & First Quartile& $28$& $25$\\
 & Median & $43$& $36$ \\
& Mean  & $41.8$ & $37.3$\\
&Third Quartile& $51$& $47$ \\
 & Maximum & $81$ & $64$\\
  \bottomrule
\end{tabular}
\end{center}
\end{table}

\begin{table}
\begin{center}
\caption[Anesthesiology Experiment Treatment-Response Table]{{Treatment-Response summary table for the propofol/thiopental experiment.}\label{tbl:souterF}}
\begin{tabular}{lcccccccccc}
 \toprule
 \multirow{2}{2.cm}{\bf Propofol} &\multicolumn{3}{c}{{\bf Stage 1 (trials 1-58)}}&\multicolumn{3}{c}{\bf Stage 2 (trials 59-90)}&\multicolumn{3}{c}{\bf Overall}  \\

   & Pain & No Pain & \small{$\hat{F}$} & Pain & No Pain & \small{$\hat{F}$} & Pain & No Pain & \small{$\hat{F}$} \\
 \toprule
$50\%$&$0$&$4$&$0$& & & & $0$&$4$&$0$\\
$60\%$&$0$&$8$&$0$&$0$&$12$&$0$&$0$&$4$&$0$\\
$70\%$&$1$&$16$&$0.06$&$4$&$11$&$0.27$&$5$&$27$&$0.16$\\
$80\%$&$3$&$13$&$0.19$&$2$&$3$&$0.40$&$5$&$16$&$0.24$\\
$90\%$&$2$&$6$&$0.25$& & & & $2$&$6$&$0.25$\\
$100\%$&$1$&$4$&$0.20$& & & & $1$&$4$&$0.20$\\
\midrule
Total&$7$&$51$&$0.12$&$6$&$26$&$0.19$&$13$&$77$&$0.14$\\
\bottomrule
\end{tabular}
\end{center}
\end{table}

\subsection{Conceptual Interlude}
The experiment's rationale can be summarized as follows: we assume there is some general population of potential patients, whose sensitivity to pain given a certain propofol/thiopental mix can be modeled via an overall threshold subdistribution $\mathcal{F}$, monotone increasing with the proportion of propofol, and reaching $\mathcal{F}\approx 0.7$ at a propofol-only treatment. We are looking for the $20$th percentile of $\mathcal{F}$, or more precisely, for an easy-to-administer mix just below that percentile.
Since the experiment was performed at a single convenience-chosen hospital, the sampling population is not the general population. However, $70\%$ pain at propofol-only treatment is a rather universally well-established figure, and Dr. Souter's prior experience at that hospital had been compatible with this frequency as well. Hence, we had no reason to expect sharp deviations from the general-population baseline.

Pain evaluation presents another difficulty. It is not measured via a standard instrument, but evaluated by a subjective observer; thus $\mathcal{F}$ actually describes \textit{reported} pain outcomes from the population of potential \textit{patient-evaluator pairs}. Therefore, from a statistical point of view we should ideally sample (or randomize) several evaluators simultaneously with the sampling of patients. However, the experiment design was constrained to a single, specifically-trained evaluator, with a scheduled change of evaluators at a certain time point. The two stages are in effect samples from two different threshold-evaluator sub-populations, with subdistribution functions $F^{(1)}$ and $F^{(2)}$.

To sum it up, the experiment does not sample the general population of pain thresholds and pain evaluators. Fortunately, the goal is not a purely academic exercise of studying $\mathcal{F}$, but the practical determination of a mix that might reasonably work for general samples out of $\mathcal{F}$. Our hope is that the samples from $F^{(1)}$ and $F^{(2)}$ would show a behavior sufficiently compatible with what we know about $\mathcal{F}$, to make our percentile estimate practically useful for the general population.

\subsection{Analysis and Estimation}
\begin{figure}
\begin{center}
\includegraphics[scale=.8]{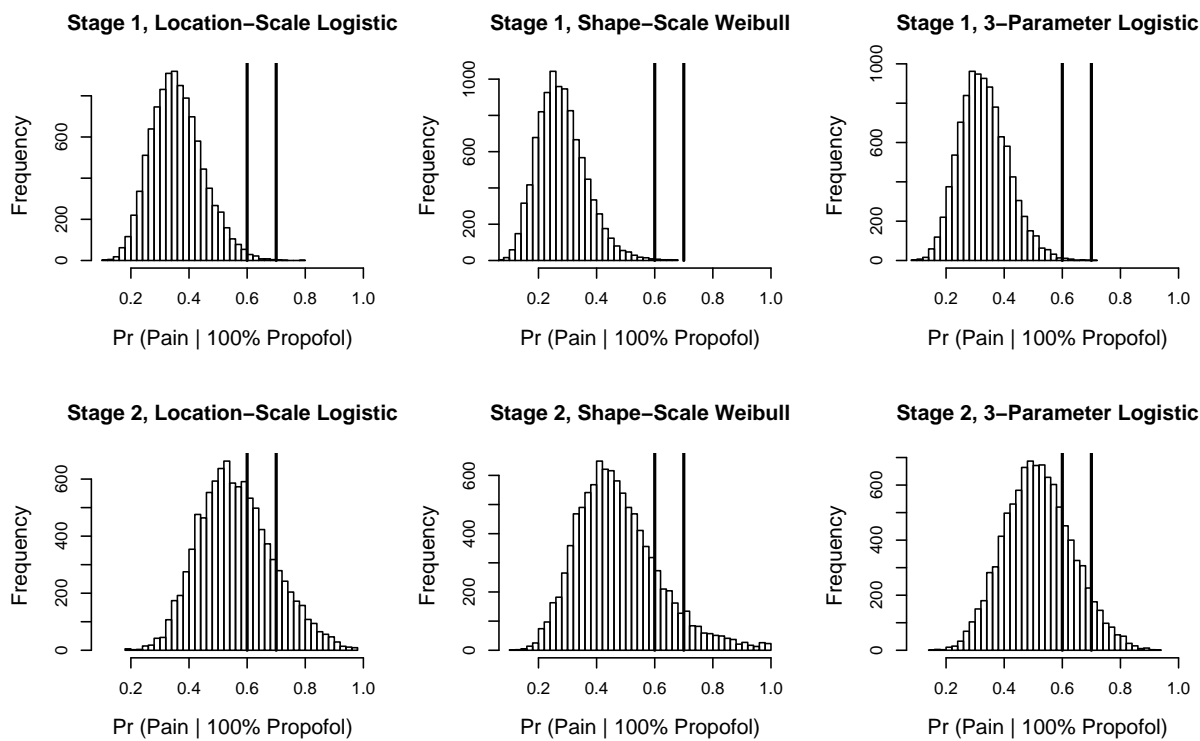}
\caption[Anesthesiology Experiment Sensitivity Analysis]{Histograms of posterior pain frequency at a propofol-only treatment, based on Stage 1 (top) and Stage 2 (bottom) data, and using location-scale logistic (left), shape-scale Weibull (center) and 3-parameter logistic (right) Bayesian models. The thick vertical lines mark pain probabilities of $0.6$ and $0.7$. The posterior probabilities for pain frequency exceeding $60\%$ based on Stage 1 data were $0.008,0.002$ and $0.002$ for the three models, respectively. The analogous probabilities based on Stage 2 data were $0.36,0.17$ and $0.24$. Each posterior distribution was approximated via MCMC, using $400,000$ cycles thinned by a factor of $40$.}\label{fig:souterep}
\end{center}
\end{figure}

The results from Stage $1$ (the first $58$ trials) raise concerns regarding this compatibility. In particular, beginning with subject $11$ there were $20$ consecutive trials with no pain response, culminating in $4$ no-pain trials at propofol-only treatment before finally observing pain on the fifth. Given our knowledge of $\mathcal{F}$ we'd expect a pain probability of $0.7$ for that treatment. For a geometric r.v. with success probability $0.7$, there is less than $0.01$ chance of observing $4$ or more failures before the first success. A more comprehensive sensitivity analysis incorporating all $58$ Stage $1$ trials within a Bayesian framework was carried out, using various functional forms for $F^{(1)}$, and a prior distribution calibrated to predict $F^{(1)}(100\%\textrm{ propofol})\approx 0.7$. The results strongly suggest that Stage $1$ data are not compatible with the population baseline (Fig.~ \ref{fig:souterep}, top). In fact, posterior medians for $F^{(1)}(100\%\textrm{ propofol})$ are around $0.3$.

Later upon closer examination, it was found that Stage 1's patients included subjects aged $76$ and $81$, who should definitely have been excluded from the experiment due to old age, and $3$ more patients $65$ years or older whose inclusion was marginal. None of them exhibited pain.\footnote{The common assumption is that the nervous system loses its sensitivity with age, and hence pain incidence would decrease. The oldest patient out of all $90$ trials to exhibit a pain response was $57$ years old.} Since the design was sequential, this affected allocation to all subsequent patients, hence distorting all averaging estimators. Since Stage 2 had no patients older than $64$, this presents yet another difference between $F^{(1)}$ and $F^{(2)}$.

Stage 2 exhibited a rather different response pattern from Stage 1, a pattern reasonably compatible with the population baseline -- though still a bit on the low side (Fig.~\ref{fig:souterep}, bottom).\footnote{A label-permutation test showed some evidence for different evaluation sensitivities between the two evaluators, even after removing the two oldest patients ($p=0.04$). However, the evidence for this is confounded by the differences in age and gender composition between the two stages (cf. Table~\ref{tbl:souterD}).} Therefore, we decided to estimate $Q_{0.2}$ for the general population using only Stage 2 data.

This estimation is quite straightforward: since $\hat{F}^{(2)}$ is monotone, the IR and CIR estimates are identical - yielding a point estimate of $67.5\%$ propofol for $20\%$ pain response, with a $95\%$ CI of $(55.9\%,79.1\%)$. The auto-detect averaging estimator $\hat{v}_{AD}$ identifies the stage's third trial as the averaging starting point, while the first and third reversals occur on the second and fourth trial, respectively. $\hat{v}_{AD}$ and the reversal-cutoff estimates $\hat{v}_1, \hat{v}_3$ are $67.33\%,67.42\%$ and $67.24\%$ propofol, respectively -- very close to the IR/CIR estimate. The $95\%$ CI for $\hat{v}_{AD}$ is $(51.8\%,82.8\%)$.\footnote{It seems that the IR/CIR interval estimate is a bit too optimistic, because of the steep slope of $\hat{F}$ between $60\%$ and $70\%$ propofol.} Considering convenience of application, my clinical collaborators therefore decided to recommend a $2$ parts ($66.7\%$) propofol, $1$ part thiopental mix for reducing pain response frequency to around $20\%$.

\section{Optimizing Up-and-Down: A Practical Summary}\label{sec:estopt}

Chapters 2-3 have traversed a wide range of U\&D topics. It is time to tie back the loose ends. The optimization recommendations below are split into {\bf ``Clearly Indicated''} -- meaning that solid theory and/or across-the-board numerical results support the conclusion, and {\bf ``Recommended''} -- meaning that my work suggests this direction, but it is not as clear-cut. Within each category, conclusions are arranged by subtopics.

\subsection{Clearly Indicated}
\subsubsection{Method and Sample Size}
\begin{enumerate}
\item Designs with super-small $n<10$ should be avoided. Even though SU\&D is fast-converging, that convergence is probabilistic and does not apply deterministically to each individual run. Under the recommended design decisions (see below), SU\&D convergence would typically take up (on the average) at least half a dozen trials, and at least twice that for other U\&D variants.

From the theoretical and numerical results presented here, a reasonably minimalist sample-size range would be $16-32$ for SU\&D, and around $8-16$ times $(k+1)$ for KR designs.\footnote{The reason for the $(k+1)$ factor is that to move \emph{'up'} and back \emph{'down'} requires at least $k+1$ trials.}

\item For non-median targets, KR is clearly recommended over BCD, unless the target percentile cannot be approximated by a KR design.
\end{enumerate}
\subsubsection{Design: Spacing, Boundaries, Stages}
\begin{enumerate}
\item Boundaries should be avoided. In case there is a real physical boundary (e.g., at zero), a logarithmic (or log-ratio) treatment scale should be used if feasible. If boundaries are unavoidable, then either use a scheme that halves the spacing given some trigger, and/or estimate with the more robust CIR.

\item The recommendation, dominant since \citet{DixonMood48}, to set $s\approx\sigma$, is wrong. The indicated spacing magnitude is about half that, or even a bit smaller. Conceptually, this is because U\&D's stationary advantages do not kick in unless $\pi$ has a clearly defined, sharp peak. For SU\&D with $s\approx\sigma$, $m_{eff.}$ -- the number of levels where $F$ is substantially different from $0$ or $1$ -- is about $4-6$, too little to guarantee such a peak. Technically, the spacing effect manifests itself in estimation variance and bias which both increase with $s$. The main adverse effect of smaller $s$ is larger starting-point biases -- but there are solutions that mitigate much of this effect.

Overall, with knowledge of $\sigma$ one should set a spacing somewhat smaller than $\sigma/2$, and yet a bit smaller for non-median targets. Without such knowledge, aim for $m_{eff.}$ to be somewhere between $8$ and $12$ (for non-median target, one should estimate $m_{eff.}$ as the number of levels where we expect $F$ to be solidly within the interval $(0,2p)$). If one prefers the $\Delta F$ approach, that is, set $s$ w.r.t to the expected increment of $F$ between levels around target, then the recommendations are equivalent to the range $\Delta F\approx 0.1$ to $0.15$ (the currently recommended spacing translates to $\Delta F\approx 0.3$).

\item The optimal starting point minimizes the expected number of trials to target. Whether design boundaries exist or not, visualize a ``realistic range'': the range in which you realistically expect the target might be. The starting point should then be the treatment from which the chain would take the same number of steps to either end. For example, with $k=2$, $x_1$ should be $2/3$ of the way up that range (see Section~\ref{sec:estexp} for my mistaken placement of $x_1$ in the anesthesiology experiment, which cost about a dozen trials).
\end{enumerate}

\subsubsection{Estimation}
\begin{enumerate}
\item Centered Isotonic Regression (CIR) is clearly preferred over IR, unless one knows that $F$ resembles a stair-step function. Even though the bias reduction was proven only for an approximate limit case, the logic underlying the correction is generic, and numerical results point to a $20-30\%$ or even greater improvement in efficiency under most scenarios.

\item BCD and GU\&D should be used with the CIR estimator and not with averaging estimators -- the former because of its slow convergence, and the latter because of the graininess of cohort-based chains.

\item For SU\&D and KR, the reversal-average $\hat{w}$ should be abandoned in favor of $\hat{v}$-type estimators averaging {\bf all} treatments from a cutoff point. They have smaller variance, and one bias source less to worry about.

\item The first reversal is too early for such a cutoff point, as it is far from guaranteeing removal of the starting-point effect. Alternatives will be discussed below under ``Recommended''.

\end{enumerate}
\subsection{Recommended}
\subsubsection{Method and Sample Size}
\begin{itemize}

\item Regarding cohort-based group U\&D (GU\&D) designs: designs with $r_u=0$ ($r_u$ being the ``remain at the same level'' probability) at all levels should inherently converge faster. Conversely, the stationary profile becomes steeper (more ``peaked'') as $r_u$ increases. The GU\&D$_{(k,0,1)}$ design, examined here numerically, seems to be a reasonable candidate for Phase I trials (with $k=2$ or $3$). Note the comment above regarding GU\&D estimation.
\item The question of whether better U\&D methods can be found or whether an optimal one can be formulated -- briefly treated by \citet{BortotGiovagnoli05} for second-order methods using a BCD-type randomization scheme -- remains open. However, it appears that the combination of simplicity, performance and usage track record offered by SU\&D and KR, would be hard to beat.
\end{itemize}
\subsubsection{Design: Spacing, Boundaries, Stages}
\begin{itemize}
\item The ``layover'' boundary conditions suggested by \citet{GarciaPerez98} to mitigate the boundary effect, do not appear to be advantageous on the whole. Whatever bias reduction they offer in some scenarios, is offset by opposite bias in scenarios where the underlying assumption (that $F$ remains constant beyond the boundary) does not hold -- and by increased variance. The milder ``imputation'' fix suggested here suffers from the same problems. Again, the best way to deal with a boundary is to avoid having it, to half the spacing upon brushing up against it, and to estimate using a boundary-resistant estimator such as CIR.
\end{itemize}

\subsubsection{Estimation, Stopping Rules, Multistage Schemes}
\begin{itemize}
\item Instead of the first reversal, one should set the cutoff point for averaging estimators either at the auto-detect point developed here, or possibly at the third reversal. The AD point has the advantage of being adaptive: in case the starting-point effect is small, the AD point will happen early -- while the other alternatives would lose efficiency. However, AD should be used with care: the AD point is not to be allowed to be greater than $n/3$, or than the number of trials required to reach each end of the ``realistic range'' (described above) -- leaving room for one ``switchback'' along the way. For example, for SU\&D if that range is $5$ levels from $x_1$, then the AD point should not be later than $5+2$ steps later, i.e, no later than $x_8$. Clearly, if $n$ is too small to allow for meeting the latter criterion, then it should probably increased (or the spacing made coarser).

\item For KR designs, one may wish to explore incorporating the AD estimator over the zero-state sub-chain only -- an estimator which gets rid of the first-order stationary bias, and introduces a second-order bias in the opposite direction (perhaps some average of $\hat{v}_{AD}$ and $\hat{v}_{AD,0}$ can be used).

\item The AD point could also be used for a stopping rule, by mandating that the experiment continue until AD analysis yields $\tilde{n}$ trials used for averaging.
\item For interval estimation, the bootstrap based on point estimates of $\hat{F}$ does not appear to capture the magnitude of U\&D's inverse-estimation variability. The more direct methods developed here are suggested instead.

\item Regarding multistage schemes: here, too, using the first reversal as a transition point seems too risky. Either the third reversal, or the third or fourth hitting-time at the most frequently visited level are more robust alternatives. Since these events typically occur only a dozen or more trials into the experiment, one should probably refrain from multistage schemes unless $n$ is large enough to a allow for a substantial part of the experiment to take place after the transition point.

\end{itemize}

\newpage
\section*{Additional Glossary for Chapter 3}
\addcontentsline{toc}{section}{\em{Glossary}}

Note: some terms appearing only in the sequence of proofs leading up to Theorem~\ref{thm:updown} are omitted here. They are of no interest beyond that particular result, where they are explained.

\begin{glossary}
\item[CIR] Centered Isotonic Regression.
\item[IR] Isotonic Regression.
\item[NPMLE] The nonparametric MLE.
\item[PAVA] Pooled Adjacent Violators Algorithm. The algorithm used to produce IR estimates.
\\[1cm]
\item[$\overline{B}_{c:d}$] An average of ``$B$'' values, from $B_c$ to $B_d$. ``$B$'' here is a dummy stand-in for any value which is indexed over levels, trials, etc., such as $\hat{F}$, $x$, and so on. Also, the average may be weighted if applicable.
\item[$c(n)$] The location (as index in the treatment chain) of the averaging cutoff point identified by the auto-detect estimation method.
\item[$\hat{F}_u$] The binomial point estimate of $F$ at $l_u$, obtained simply by calculating the proportion of positive responses at $l_u$ accumulated until the point in time at which the estimate is taken.
\item[$\bar{F}$] The linear interpolation of $F$ between design points.
\item[$\mathcal{F}$] The general meta-population (or hyper-population) of potential reported pain outcomes, out of the overall population of patients and pain evaluators. A concept used to interpret the meaning of the anesthesiology experiment in Section~\ref{sec:estexp}
\item[$\mathcal{L}$] The likelihood.
\item[$M$] The length of a monotonicity-violating interval identified by the PAVA or CIR algorithms.
\item[$n_{eff}$] The effective sample size: the number of hypothetical i.i.d. observations needed to achieve the same rate of variance reduction for the sample mean as that observed.
\item[$\hat{v}_{AD}$] Arithmetic average of all treatments, beginning with the ``auto-detect'' point; a suggested estimator of $Q_p$.
\item[$\hat{v}_j$] Arithmetic average of all treatments, beginning with the $j$-th reversal; a possible estimator of $Q_p$.
\item[$\hat{w}_j$] Arithmetic average of treatments at reversal points only, beginning with the $j$-th reversal; a popular estimator of $Q_p$.
\item[$\bar{w}_j$] Similar to $\hat{w}_j$, except that the average is somewhat modified. This was \citet{WetherillEtAl66}'s original reversal estimator (now less commonly used than $\hat{w}_j$).
\item[$z_u$] A value used to denote the running estimate of $F_u$ during the PAVA and CIR algorithms. Its initial value is equal to $\hat{F}_u$.
\\[1cm]
\item[$\mu_\pi$] The mean of the stationary distribution.
\item[$\nu_j$] The length (in trials) of the interval between the $j$-th and $j+1$-th hitting times at a certain Markov-chain state.
\item[$\rho^{j}$] The treatment chain's $j$-th order autocorrelation coefficient. If $j$ is omitted, $\rho$ refers to the first-order coefficient (note the difference from the use of $\rho$ in Chapter~2).
\item[$\sigma$] The standard deviation of thresholds.
\item[$\sigma_{\pi}$] The standard deviation of $\pi$.
\item[$\omega_u$] The weight given to a point estimate at $l_u$ (usually proportional to $n_u$ in the context of PAVA/CIR).
\end{glossary}

\chapter{Bayesian Percentile-Finding Designs}\label{ch:crm}
\section{Overview}\label{sec:crm1}
\subsection{History}

There is evidence for earlier sources, but credit for the first massive and practical introduction of Bayesian designs to percentile-finding goes to Watson and Pelli in psychophysics, circa 1980 \citep{WatsonPelli79,WatsonPelli83}. Their algorithm, nicknamed QUEST, is widely used (though perhaps not as widely as U\&D-descended designs) and even implemented into automated experiment control software. However, the current statistical debate is dominated by QUEST's younger relative CRM (Continual Reassessment Method) -- a design using the same principles, developed independently vis-a-vis the Phase I application \citep{OQuigleyEtAl90}.\footnote{Interestingly, O'Quigley et al. did not cite QUEST, and were probably not aware of its existence.} In that latter paper, the motivation and principles were clearly described, and CRM's advantage over both the `3+3' Phase I design and U\&D was forcefully argued. The main argument is that using only the outcomes of a single cohort to determine allocation is inherently inferior to model-based allocation using information from all trials.

CRM's essential features have not changed much since 1990. However, some modifications have been suggested to cope with what appears to be CRM's greatest obstacle: its perception as risky by the medical community (see, e.g., \citet{Palmer02}). This is not only a perception problem: in statistical literature, as well, researchers showed that CRM may escalate doses too quickly \citep{KornEtAl94}. The simplest fix - limiting transitions to $\pm 1$~level \citep{GoodmanEtAl95} - has been widely accepted, but unconstrained CRM is still practiced by some leading biostatisticians \citep{MathewEtAl04,PistersEtAl04}. Similar concerns have surfaced with respect to QUEST in psychophysics \citep{AlcalaQuintana04,GarciaPerez05}. Recent CRM designs seem to focus on more sophisticated solutions to this problem. One of them, Escalation with Overdose Control \citep[EWOC,][]{BabbEtAl98} will be discussed in the next subsection. Both CRM and EWOC are available online as software packages \citep{XuEtAl07}. An article titled ``CRM tutorial'' for general audiences, including discussion of its modifications, has been recently published \citep{GarrettMayer06}. Two nonparametric or ``model-free'' approaches to CRM have also been suggested -- one of them still using the Bayesian framework with prior and posterior distributions, but with a saturated model \citep{GaspariniEisele00}, and one where allocation decisions are based solely on a nonparametric local estimate of $\hat{F}$ \citep[see below in Section \ref{sec:noncrm}]{LeungWang01,YuanChappell04,IvanovaEtAl07}.

The current popularity of Bayesian approaches in the statistical community notwithstanding, the relative merit of Bayesian percentile-finding designs is hard to assess, especially for Phase I trials. Phase I clinical trials is not an application that easily accommodates repeated or large-scale comparative experiments. To date, the only clear theoretical result regarding CRM properties, is that one-parameter models converge to optimal allocation (to be defined later) under rather restrictive conditions on $F$ \citep{ShenOQuigley96,CheungChappel02}. In most other CRM publications, claims for a given design's advantage are usually supported by logical arguments or by simulation comparisons with competing methods. A recent mini-review co-authored by one of CRM's developers \citep{OQuigleyZohar06} even claims that numerical simulations are the recommended approach to compare percentile-finding designs.

Unfortunately, simulations are prone to `rigging', whether intentional or unintentional. The most common `rigging' in favor of parametric designs is to choose scenarios in which $F$ is correctly specified (or very closely approximated) by the author's parametric model \citep{BabbEtAl98}. Another fallacy is using the wrong estimator for the `competitor' methods. This is almost always the case whenever U\&D appears in CRM publications - not intentionally, but out of ignorance among statisticians regarding common U\&D usage. For example, the paper introducing EWOC \citep{BabbEtAl98} claims its advantage over a host of competitors, including stochastic approximation, CRM, the `3+3' protocol, GU\&D$_{(3,0,2)}$ and KR with $k=2$. However, all simulation scenarios were generated via the location-scale logistic model used by the authors' EWOC, and U\&D `estimation' was defined as the dose allocation following the last trial. By contrast, a recent comparative simulation study in psychophysics, performed by researchers who contributed to both U\&D and Bayesian designs and who are acquainted with `best practices' in both, has yielded the conclusion that QUEST is the worst of the five designs tested, especially as far as robustness (or in their terminology, `usability') is concerned \citep{GarciaPerez05}.

In general, the safe assumption (lacking a clear and verified scientific model for $F$) is that any parametric model used is misspecified. Therefore, when examining the properties of CRM or other parametric designs, our attention will focus on what happens when the model is misspecified.

\subsection{Description of Common Bayesian Designs}
\subsubsection{Definition and Terminology}
\begin{defn}\label{def:crm0} Let a {\bf CRM scheme} be a generic Up-and-Down design (see \ref{def:zero} (i)), also having:
\begin{enumerate}
\item A response model $G(x,\theta)$, which is a (sub)distribution function on $x$ for any $\theta\in\Theta$, and a strictly monotone\footnote{Shorthand for: strictly monotone in each $\theta_j$ holding $x$ and the other parameters constant.} continuous algebraic function of $\theta$, with continuous first and second partial derivatives everywhere in $\Theta$ (hereafter: {\bf the CRM model}). The family of possible curves for a given model will be referred to as $\mathcal{G}$.
\item A prior distribution $\Pi(\theta|\phi)$ for the model's parameters.\footnote{The lower-case $\pi$, usually used to denote prior and posterior in Bayesian literature, has already been taken in this text for denoting stationary distributions.}
\item A two-stage decision rule for allocating the next treatment $x_{i+1}$, using the posterior distribution of $\theta$, $\Pi\left(Q_p|x_1,\ldots x_i,y_1,\ldots y_i,\phi)\right)$.
\item A stopping rule.
\end{enumerate}
\end{defn}

The monotonicity in each parameter is needed for model identifiability. The allocation rule will be further defined and refined in the next section. There has been some work regarding CRM stopping rules \citep[e.g. ][]{ZoharChevret01}; however, here I assume for simplicity that the stopping rule is ``stop after the fixed-size allocated sample is exhausted'' -- which is still the most commonly used rule.

Unlike U\&D, Bayesian designs almost always assume a finite set of treatment levels.

\subsubsection{Allocation rules}

QUEST's original rule is
\begin{equation}\label{eqn:crmbasic}
x_{i+1}=l_{\hat{u}},\hat{u}\equiv\textrm{argmax}\left(\Pi\left(Q_p|x_1,\ldots x_i,y_1,\ldots y_i,\phi)\right)\right),
\end{equation}
with the maximum taken over design points only. That is, allocation goes to the (discrete) posterior mode over design points.\footnote{Doubtlessly, this rule was chosen for computation reasons. It allowed calculation to proceed as a simple updating of the previous posterior, instead of tedious integrals (MCMC in its current form did not exist at the time). So this was probably the only option enabling individual researchers to use QUEST for their experiments.} The original CRM rule is
\begin{equation}\label{eqn:crmbasic}
x_{i+1}=l_{\hat{u}},\hat{u}\equiv\textrm{argmin}\left|l_v-E\left[Q_p|x_1,\ldots x_i,y_1,\ldots y_i,\Pi(\theta)\right]\right|,
\end{equation}
or in words: allocate the next treatment to the level `closest' to $Q_p$'s posterior mean \citep{OQuigleyEtAl90}. `Closest' may be measured either on the treatment scale or on the response scale. Technically, the estimate is calculated either by viewing $Q_p$ as an algebraic function of $\theta$ and estimating $\theta$, by calculating at $G$'s posterior estimates at design points, or directly by looking at $Q_p$'s posterior distribution.

Following concerns regarding potential jumps by CRM to high-toxicity territory \citep{KornEtAl94}, \citet{GoodmanEtAl95} suggested modifying this rule to: move one level in direction of $Q_p$'s posterior mean (or no move, if the current level is the closest). EWOC's rule is even more conservative: after setting an acceptable risk level $\alpha$, allocation is
\begin{equation}\label{eqn:ewocbasic}
x_{i+1}=\max\left\{l_v:\Pr\left(G(x=l_v,\theta)\geq Q_p|x_1,\ldots x_i,y_1,\ldots y_i\right)\leq\alpha\right\},
\end{equation}
or in words: allocate the next treatment to the largest level whose posterior probability of exceeding $Q_p$ is not greater than the predetermined risk level \citep{BabbEtAl98}.\footnote{EWOC's originators actually assume a continuous treatment range, in which case the next dose can be allocated so that the risk is exactly $\alpha$.}. This allocation rule can be traced to the asymmetric loss function
\begin{equation}\label{eq:ewocloss}
l_\alpha(x,Q_p)=\left\{\begin{array}{rr}\alpha(Q_p-x) & x\leq Q_p \\
(1-\alpha)(x-Q_p) & x>Q_p \end{array}\right.
\end{equation}
EWOC's originators used a risk level of $\alpha=0.25$. Note that since $G$ is monotone in $x$, the EWOC estimator allocates to the level just below the $100\alpha\%$ percentile of $Q_p$'s posterior distribution. Therefore, using median instead of mean with the `plain' CRM approach is basically equivalent to using EWOC with $\alpha=0.5$, showing that these two CRM variants are very closely related.

\subsubsection{Models and estimation}

QUEST was introduced with a $4$-parameter Weibull model.\footnote{Two of these parameters are the `false positive' and `false negative' rates used in psychophysics, and two are the familiar shape and scale Weibull parameters.} By contrast, the original model used by O'Quigley et al. was a one-parameter ``hyperbolic tangent'' or ``power'' model, which is essentially a $3$-parameter logistic model with location and scale fixed and a free shape parameter. Since then, O'Quigley and others have repeatedly argued in favor of a one-parameter CRM and against using more parameters. The reasoning is that since $Q_p$ is a single parameter, using more than $1$ parameter in the model that tracks it down is dangerously redundant and less robust. A currently popular one-parameter CRM model is $G(l_u)=\left(G^{(0)}(l_u)\right)^\theta$, where the $G^{(0)}$ values at design points are determined according to prior knowledge, and $\theta$ is a single parameter \citep{MathewEtAl04}. Clearly, the original ``power'' model is a special case of this family. Another variation is one-parameter scale logistic, with fixed location \citep{GarrettMayer06}. However, many other CRM researchers use the 2-parameter, location-scale logistic model, perhaps simply because it is the most commonly used model in medical dose-response applications, and can be generalized to logistic regression. EWOC developers recommend this model as well.

Following the CRM experiment, estimation is a straightforward: most often, {\bf CRM estimation is simply the next allocation}. This is in stark contrast to U\&D, where allocation and estimation are quite separate. Sometime minor variations appear, such as using the MLE for estimation (which is equivalent to a uniform prior), but the principle remains the same. The practice of using allocation for estimation may seem self-evident, but it may also raise questions, especially with regards to CRM. CRM's developers still claim that a one-parameter model should be used, regardless of researchers' beliefs about the properties of the true $F$. This is understandable for allocation, but once the experiment is over one would think that the final estimate should be generated using the most realistic model, not the one deemed most effective for mid-experiment allocation. However, that point is not often debated nowadays.

\subsection{Nonparametric CRM}\label{sec:noncrm}
A radically different modeling approach was recently suggested: here the main argument is that misspecified models impose too many constraints, preventing the CRM scheme from adapting to the data \citep{GaspariniEisele00}. The proposed solution is to use a {\bf saturated model:} a model with $m$ parameters, that can exactly match $F$ on $\{l_m\}$. This model has not been generally accepted in CRM circles, and its performance is not quite spectacular, even in the simulations presented by its originators. One counter-argument to saturated CRM was that saturated models, too, imposed constraints via the model's prior - as claimed by \citet{OQuigley02} in a direct rejoinder to \citet{GaspariniEisele00} -- but that unlike parametric models, these constraints are less easily tractable.

A truly nonparametric approach was suggested by \citet{LeungWang01}. Here, instead of a Bayesian posterior estimate, the point estimates to be used at each stage are derived from isotonic regression. At each step, the next level is chosen from the current level and its two immediate neighbors, according to whose IR estimate is closest to $p$. This design can also be seen as a constrained CRM with a saturated model and no prior (i.e., a constrained MLE-based allocation).

Two modifications to this design have since appeared, both of them relaxing the conditions somewhat to include an interval of $\hat{F}$ values that allows remaining at the same level. \citet{YuanChappell04} suggest the interval $[p,2p]$, tailored to very low targets. \citet{IvanovaEtAl07} suggest a more generic window: $\left[p-\Delta_1(p),p+\Delta_2(p)\right]$. For convenience, the authors preferred to use $\Delta_1(p)=\Delta_2(p)=\Delta (p)$. Recommended values for $\Delta (p)$ were determined by intense numerical scenario searches run on the tpm's of GU\&D designs with increasing cohort size (the authors view a decision at $l_u$ as a GU\&D decision with a cohort size equal to $n_u$). For $0.1\leq p \leq 0.25$, $\Delta (p)=0.09$; $\Delta (0.3)=0.10$ and $\Delta (0.5)=0.13$. This design has been dubbed {\bf Cumulative Cohort Design (CCD)}, and will be discussed later on.

\section{Theoretical Study: Convergence}

As CRM researchers admit, since the application is a small-sample one good asymptotic properties are no guarantee for actual performance. On the other hand, convergence properties may help us understand important aspects of small-sample behavior. The main existing result regarding CRM convergence, from \citet{ShenOQuigley96}, deals with misspecified one-parameter models. These models can guarantee to match the true $F$ at one point;\footnote{It is possible that more points would match, but this would be a lucky coincidence, not a matching guaranteed by the model.} the hope is that the CRM scheme ensures this point will be the design point closest to $(Q_p,p)$. \citet{ShenOQuigley96} were able to prove convergence to this point almost surely as $n\rightarrow\infty$, only by assuming that the family of $G$ curves be ``close'' enough to $F$, so that no matter which value of $F$ is matched by $G$, the optimal level is still allocated. \citet{CheungChappel02} illustrate this restriction, noting that it is quite unrealistic for one-parameter models; essentially, it means that the model is only ``mildly misspecified.'' They conjecture that the condition may be relaxed, but do not present a proof. As far as I know, no convergence result exists for multi-parameter models.

During my work I have attempted to find more general CRM convergence proofs, as a tool to better understand the properties of this design approach. Following is a summary of my main results, beginning with one-parameter models. For the remainder of this section, I assume that in all cases, regardless of the allocation method, there is only a single optimal level (i.e., we rule out ties). Additionally, I assume that the design is constrained CRM (i.e., CRM limited to transitions of $\pm 1$~level), rather than the original unconstrained design.\footnote{It can be proven that the two become equivalent as $n$ increases, and therefore the proofs hold for both; but this thesis has enough text in it already, and unconstrained CRM is not recommended in any case.}

\subsection{Preamble}

First, I find that CRM's input data are better described by running proportions and point estimates than by the treatment and response chains.
\begin{lem}\label{lem:xy_trans_pF}
The statistics $\left(\{\hat{p}_m\},\{\hat{F}_m\}\right)$, with $\{\hat{p}_m\}$ defined via
\begin{equation}\label{eq:phat}
\hat{p}_u=\frac{n_u}{n},
\end{equation}
and $\{\hat{F}_m\}$ defined in eq. (\ref{eq:Fhat}) in Chapter \ref{ch:est}, are sufficient for $\theta$ in the CRM likelihood.
\end{lem}
The proof follows easily from the factorization theorem.

Next, we look more closely at the CRM allocation rule.
\begin{defn}\label{def:alloc}
The CRM allocation rule is composed of two stages:
\begin{enumerate}
\item A point estimate $\hat{\theta}$, generically defined as
\begin{equation}\label{eq:Qpdef}
\hat{\theta}=A_1\left(\{\hat{p}_m\},\{\hat{F}_m\}|G,\phi\right).
\end{equation}
{\bf $\hat{\theta}$ \underline{must} be a continuous function of $\{\hat{p}_m\},\{\hat{F}_m\}$, with continuous first-order partial derivatives.}

This estimate is then used to generate $\hat{G}=G(\hat{\theta})$ and $\hat{Q_p}=\hat{G}^{-1}(p)$. Or, these statistics can be estimated directly from their posterior distributions, in such a way that they obey the same smoothness requirements w.r.t. $\{\hat{p}_m\},\{\hat{F}_m\}$.
\item A distance-based allocation $x_{n+1}=A_2(\hat{G})$. Common options include
\begin{enumerate}
\item {\bf ``Closest treatment''}: $\textrm{argmin}_u\left|l_u-\hat{Q_p}\right|$;
\item {\bf ``Closest response''}: $\textrm{argmin}_u\left|\hat{G}_u-p\right|$;
\item {\bf ``Just under''}: $\max\left\{l_u:l_u\leq\hat{Q_p}\right\}$.
\end{enumerate}
\end{enumerate}
\end{defn}

Part 1 of this definition ensures that the CRM estimate is a smooth function of the sufficient statistics. If one takes the posterior mean, median or (continuous) mode of $\Pi\left(\theta |\{\hat{p}_m\},\{\hat{F}_m\},\phi\right)$ to estimate $\theta$, this smoothness condition should hold.\footnote{If for some reason it does not, then probably that specific CRM scheme is inadequate for use.}

Finally, let us define a key property of parametric models in general and of CRM in particular.
\begin{defn}\label{def:df}
\begin{itemize}
\item A CRM model $G(x,\theta)$ will be said to have {\bf $d$ degrees of freedom (d.f.)}, if for any $\epsilon>0$, any set of $d$ indices $\left(u_1<\ldots <u_d\right)\in\{1\ldots m\}$ and any set of corresponding real numbers $0<a_1<\ldots <a_d<1$, there exists a domain $\mathcal{D}\in\Theta$, containing a $d$-dimensional open rectangle, such that $\left|G_{u_j}-a_j\right|<\epsilon\ \ \forall \theta\in\mathcal{D},j\in 1\ldots d$. Also, $\Pi(\theta)>0$ $\forall\theta\in\Theta$.
\item A CRM model will be called {\bf economical} if $d$ is equal to the number of parameters.
\item A CRM model will be called {\bf misspecified} if there exists $\epsilon>0$, such that there is no $\theta\in\Theta$ for which $\left|G_{u}-F_{u}\right|<\epsilon$ $\forall u\in 1\ldots m$, where $F$ is the true response-threshold CDF. Otherwise, the model will be called correctly-specified.
\end{itemize}
\end{defn}

In words, a model with $d$ d.f. can always fit up to $d$ values of $F$ arbitrarily closely at any given time. All CRM models discussed here are assumed to be economical, i.e. that $d$ equals the number of parameters. Typical standard parametric models usually answer this requirement. For example, the $1$-parameter shape logistic model known as ``power'' \citep{OQuigleyEtAl90} has $d=1$, a standard location-scale logistic or normal model has $d=2$, and a shape-scale Weibull model has $d=2$ if responses are known to be positive.\footnote{Note that in literature one may encounter non-economical model choices - for example, the $1$-parameter scale logistic \citep{OQuigleyChevret91}, for which a careless choice of $\mu_0$, the fixed location parameter, can prevent the model from fitting all design levels as $Q_p$. Such a model will have $d=0$ d.f.} Clearly, a CRM model with $d=m$ d.f. cannot be misspecified. The case $d=m$ is of course a {\bf saturated model}.

\subsection{Convergence Proofs for Misspecified One-Parameter Models}\label{sec:crmconv1}

Note that the posterior distribution
\begin{equation}\label{eqn:crmlike1}
\Pi\left(\theta|n,\{\hat{p}_m\},\{\hat{F}_m\},\phi\right)\propto f\left(x_1,\ldots x_n,y_1,\ldots y_n\mid \theta\right)\Pi(\theta\mid\phi)
\end{equation}
can be written as
\begin{equation}\label{eqn:crmlike2}
\Pi\left(\theta|n,\{\hat{p}_m\},\{\hat{F}_m\},\phi\right)=\frac{1}{C}\mathcal{L}\left(x_1,\ldots x_n,y_1,\ldots y_n; \theta\right)\Pi(\theta\mid\phi),
\end{equation}
where $C$, a shorthand for the marginal distribution of the data, is not a function of $\theta$. As $n\rightarrow\infty$, the prior's relative weight diminishes, and the posterior's form converges to that of the likelihood for all parameters about which our information continues to increase. Therefore, Bayesian estimators -- whether the mean, mode or a given posterior percentile -- will converge to the MLE. Moreover, this means that asymptotically it does not matter whether the statistics $\hat{G},\hat{Q_p}$ are estimated directly or via $\hat{\theta}$. With one parameter this assumption always holds, and it has been used by \citet{ShenOQuigley96} in their proof.\footnote{I will discuss the multi-parameter case more carefully later on.} Similarly, my inspection of CRM asymptotics will be likelihood-based.

\citet{ShenOQuigley96} have shown that correctly specified and ``mildly misspecified'' CRM schemes with $d=1$ converge almost surely to the optimal allocation. Instead of constraining $F$, I now constrain the treatment chain.

\begin{lem}\label{lem:pi1} Suppose that a CRM experiment has a limiting distribution $\mathbf{\pi}=\left(\pi_1\ldots\pi_m\right)'$ (in the sense that as $n\rightarrow\infty$, $\hat{p}_u\rightarrow \pi_u$ $\forall u\in\{1\ldots m\}$ in probability; of course, $\sum_u \pi_u=1$). Then at most $2$ elements of $\pi$ are nonzero, and the corresponding levels must be adjacent to each other.
\end{lem}
\begin{proof} Let us first assume allocation according to the ``Closest treatment'' or ``Just under'' rules. We use the properties of $\hat{G},\hat{Q}_p$ outlined in Definitions \ref{def:crm0},\ref{def:alloc}. Since both are continuously (partially) differentiable in $\{\hat{p}_m\},\{\hat{F}_m\}$, as the former tend to a limit so will they: $\hat{G}\rightarrow\tilde{G}$, and in particular $\hat{Q}_p\rightarrow\tilde{Q}_p$, in probability. Therefore, for every $\epsilon>0$ there is an $n_0(\epsilon)$ such that for all $n>n_0$, $\Pr\left(\hat{Q}_p\in[\tilde{Q}_p-s/3,\tilde{Q}_p+s/3]\right)>1-\epsilon$.  But all points in this interval imply an assignment by stage $A_2$ in Definition \ref{def:alloc} either a single level or to one of two adjacent levels. Thus, at most $2$ levels have a nonvanishing probability of being assigned. As to the ``Closest response'' rule, the conditions on $\hat{G}$ imply that $\hat{G}\rightarrow\tilde{G}$ as well, and therefore allocation would behave in a similar manner.\qed\end{proof}

\begin{figure}[!h]
\begin{center}
\includegraphics[scale=.6,angle=-90]{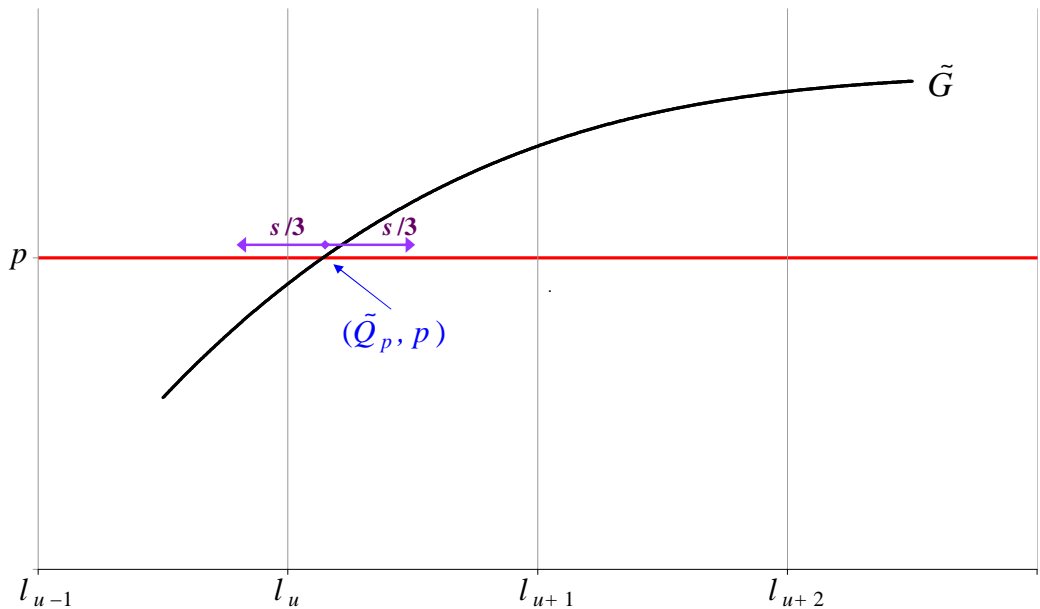}
\caption[Illustration of First CRM Lemma]{Illustration of the proof in Lemma \ref{lem:pi1}. From the lemma's conditions and from generic CRM model smoothness properties, the model curve $\hat{G}$ converges to a limit curve $\tilde{G}$, with a limit target-estimate $\tilde{Q}_p$ (which is generally not the true target). This estimate must lie between two design points, depicted here as $l_u,l_{u+1}$. Due to the curve's convergence, the probability of $\hat{Q}_p$ straying far enough from $\tilde{Q}_p$ to allocate other levels beside these two (the distance $s/3$  for ``Closest treatment'' allocation is used in the proof and shown here), is diminishing.\label{fig:pi1}}
\end{center}
\end{figure}

The proof's rationale is illustrated in Fig.~\ref{fig:pi1}. Note that the lemma says nothing about whether the method converges correctly.

Lemma~\ref{lem:pi1}'s proof emphasizes the importance of checking the model $G$'s properties. If we are not assured that outputs (estimates) are smooth in the inputs (raw data), as specified in Definition \ref{def:alloc}, we cannot guarantee convergence. This is because the maximal-likelihood set may include a positive-measure region in $\Theta$. If the allocation's first step hops arbitrarily within that region, the allocation itself may end up spread out across several levels. Fortunately, a posterior-based decision process (posterior mean, percentile or mode) should usually fulfill the necessary requirements.

Another consequence of Lemma~\ref{lem:pi1} is that for every $\epsilon>0$ there is a sufficiently large $n$, such that the score equations can be rewritten as
\begin{equation}\label{eq:score2}
\left|\frac{\partial l}{n\partial\theta_j}-\left[\frac{\partial G(l_{v},\theta)}{\partial\theta_j}\frac{\hat{p}_{v}\left(\hat{F}_{v}-G_{v}\right)}{G_{v}(1-G_{v})}+\frac{\partial G(l_{v+1},\theta)}{\partial\theta_j}\frac{\hat{p}_{v+1}\left(\hat{F}_{v+1}-G_{v+1}\right)}{G_{v+1}(1-G_{v+1})}\right]\right|<\epsilon,\\
j=1\ldots d,
\end{equation}
assuming of course that the two nonzero elements are $\pi_v,\pi_{v+1}$.

Now, recall that for a given $d=1$ curve family $\mathcal{G}$, only a single curve can (via the likelihood equation) be guaranteed to pass through a specified point $\left(x,F(x)\right)$. Let this curve be called the CRM curve that {\bf matches $F$ at $x$}.

\begin{thm}\label{thm:d1} For a CRM scheme with $d=1$, assume a limiting distribution $\mathbf{\pi}$ exists. Then:

(i) If $\mathbf{\pi}$ has exactly   two nonzero elements indexed $u,u+1$, and if all CRM curves $G$ matching $F$ in $[l_u,l_{u+1}]$ obey $G_{u+1}-G_u\leq F_{u+1}-F_u$ (i.e., the model is {\bf ``shallower''} than the true curve in the vicinity of $[l_u,l_{u+1}]$), then $l_u,l_{u+1}$ must be the two levels around target.

(ii) Under the ``Closest response'' or ``Just under'' allocation rules, if $\mathbf{\pi}$ has a single nonzero element indexed $u$, and if the model is ``shallower'' than the true curve in the vicinities of $[l_{u-1},l_u]$ and $[l_u,l_{u+1}]$, (assuming w.l.o.g. $1<u<m$), then {\bf the scheme converges in probability to optimal allocation}, i.e. allocation only to the closest level to target.
\end{thm}
\begin{proof} (i) Since there are two elements, the solution to the score equations (\ref{eq:score2}) tends to a limit curve $\tilde{G}$ obeying
$$
\tilde{G}_{u+1}-F_{u+1}=-K\left(\tilde{G}_u-F_u\right)
$$
for some positive constant $K$. Clearly, $\tilde{G}$ must match $F$ somewhere in $(l_u,l_{u+1})$. Since all $G$ curves answering this criterion are ``shallower'' than $F$ in the vicinity of this segment, we have $\tilde{G}_u\geq F_u$ and $\tilde{G}_{u+1}\leq F_{u+1}$. Now assume by contradiction and w.l.o.g. that $F_u>p$ and $u>1$, so that $l_u,l_{u+1}$ are not the design points around target. Then, obviously, $p<F_u\leq\tilde{G}_u<\tilde{G}_{u+1}$. The probability of assigning $l_{u+1}$ diminishes, reaching a contradiction.\qed

(ii) The proof here is for the ``Closest response'' rule, but it can be easily tailored for the ``Just under'' rule as well. Since $\pi_u$ is the dominant element, $\tilde{G}$ will match $F$ exactly at $l_u$. Now, assume by contradiction that this is not the optimal level, and further assume w.l.o.g. that $l_{u-1}$ is closer to target. This means that $p<\tilde{G}_u$. Since $\tilde{G}_{u-1}\in \left[F_{u-1},\tilde{G}_u\right)$, $\tilde{G}_{u-1}$ is closer to target than $\tilde{G}_u$.  $l_{u-1}$ becomes a superior assignment under $\tilde{G}$, reaching a contradiction.\qed
\end{proof}

\begin{figure}[!hb]
\begin{center}
\includegraphics[scale=.6,angle=-90]{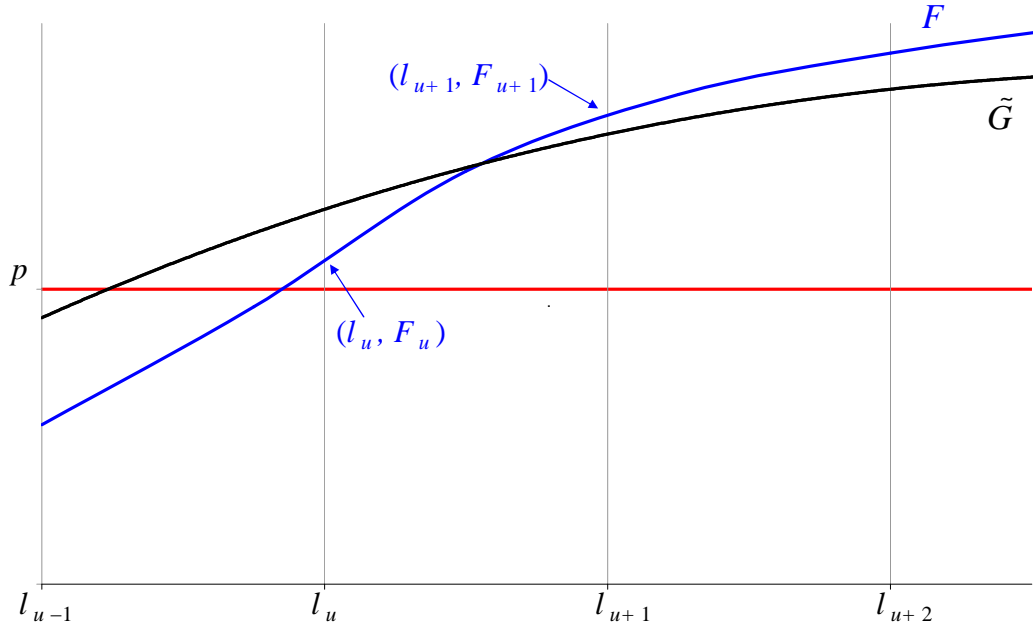}
\caption[Illustration of CRM $d=1$ Theorem]{Illustration of the proof in Theorem \ref{thm:d1}, part (i). We assume by contradiction that levels $l_u,l_{u+1}$ each receive an non-diminishing fraction of the allocations, and that both are on the same side of target. Since $d=1$, the model's limit curve $\tilde{G}$ cannot match the true curve $F$ at both design points, and instead matches it somewhere between them. Since $\tilde{G}$ is ``shallower'' than $F$ in this region, both $\tilde{G}$ values are on the same side of target as well, ensuring the gradual elimination of allocations to $l_{u+1}$ - yielding a contradiction.\label{fig:crmd1}}
\end{center}
\end{figure}

The proof's rationale for part (i) is illustrated on Fig.~\ref{fig:crmd1}.

\subsection{Convergence Proofs for Multi-Parameter Models}\label{sec:crmconv1}

First, we must note that the likelihood-based approach to analyze CRM convergence is not automatically correct with $d>1$. We also need that $G(x)$ will be a nontrivial function of all parameters at all design points. I will assume this is the case (though it may not be true for some saturated model forms), and continue using the likelihood-based approach. Then Lemma \ref{lem:pi1}'s result can be further strengthened with $2$ or more d.f.

\begin{lem}\label{lem:pi2} Suppose that a CRM experiment has a limiting distribution $\mathbf{\pi}=\left(\pi_1\ldots\pi_m\right)'$. Then for $d\geq 2$ d.f., only a single element of $\pi$ is positive (i.e., it is equal to $1$ and the rest are zero).
\end{lem}

\begin{proof}
Following Lemma \ref{lem:pi1}, let us assume that the only nonzero elements are $\pi_v,\pi_{v+1}$.
For every $\epsilon>0$ and sufficiently large $n$, the score equations can be approximated by (\ref{eq:score2}).
The MLE is found by setting the score to zero. Since there are $d\geq 2$ such equations, the MLE converges to the degenerate solution $\hat{G}_{v}=\hat{F}_{v}, \hat{G}_{v+1}=\hat{F}_{v+1}$. This means that as $n\rightarrow\infty$, $\hat{G}_{v}\rightarrow F_{v}, \hat{G}_{v+1}\rightarrow F_{v+1}$. Since we have ruled out ties in the introduction to this section, one of these level is closer to the true target $Q_p$. Therefore, for sufficiently large $n$, this level would dominate.\qed
\end{proof}

\begin{figure}[!h]
\begin{center}
\includegraphics[scale=.6,angle=-90]{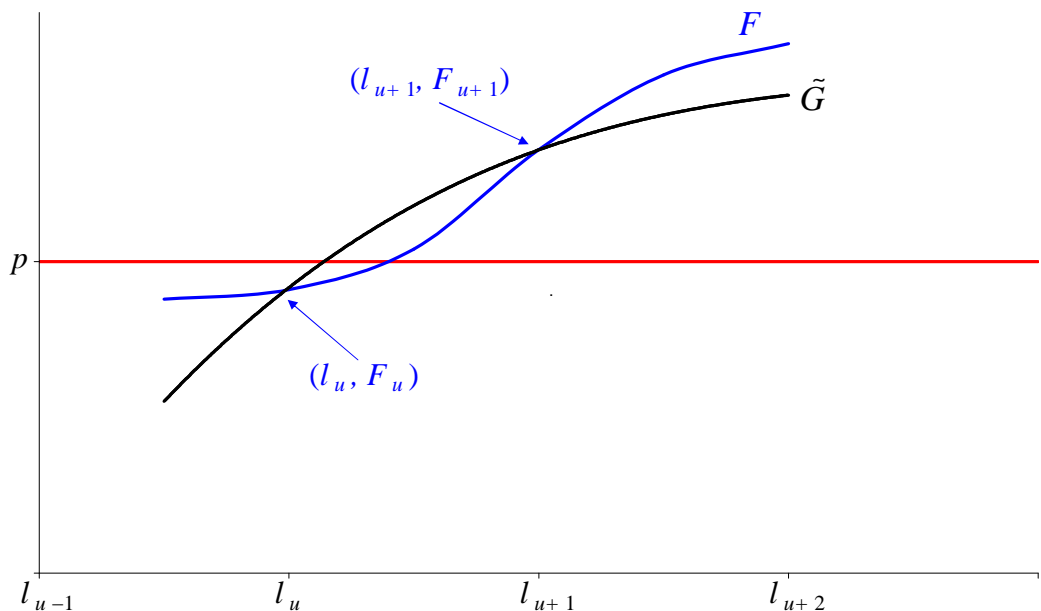}
\caption[Illustration of Second CRM Lemma]{Illustration of the proof in Lemma \ref{lem:pi2}. Assuming by contradiction that levels $l_u,l_{u+1}$ each receive an non-diminishing fraction of the allocations, and since $d\geq 2$, the model's limit curve $\tilde{G}$ will match the true curve $F$ at both design points. But then, since we have ruled out ties, one level will clearly emerge as closer to target (according to the allocation definition of ``closer''), and from a certain point onward would dominate the allocations. Note that this would happen whether $l_u,l_{u+1}$ are the two levels around target (as in the illustration) or not.\label{fig:pi2}

This figure can also be used to understand Theorem \ref{thm:littlebig}'s proof. Say that $l_{u+1}$ is CRM's limit level, but $l_u$ is closer to target. If $n_u,n_{u+2}$ are also unbounded, then $n_u\gg n_{u+2}$. $\hat{G}$ will tend to match $F$ at $l_{u+1}$ and $l_u$, and ultimately $l_u$ (being closer to target) will supplant $l_{u+1}$ as the most frequently allocated level.}
\end{center}
\end{figure}

The proof's rationale is illustrated in Fig.~\ref{fig:pi2}. Note that here, too, the lemma says nothing about whether the method converges correctly.

We are now ready to prove convergence for misspecified-model CRM using the limiting-distribution approach, under certain conditions -- regardless of how misspecified the model is.

\begin{thm}\label{thm:littlebig} Suppose that a CRM experiment has
\begin{enumerate}
\item A limiting distribution $\mathbf{\pi}=\left(\pi_1\ldots\pi_m\right)'$;
\item $d\geq 2$ degrees of freedom;
\item The ``Closest response'' or ``Just under'' allocation rule; and
\item If $\pi_u >0$, then $l_{u-1}$ and $l_{u+1}$ (when applicable) also belong to $S$, defined as
\begin{equation}\label{eq:Sdef}
S\equiv\left\{l_u:n_u\to\infty\ \ \textrm{ as }\ n\to\infty\right\}.
\end{equation}
\end{enumerate}
Then $\pi_{u^*}=1$, with $u^*$ being the index of the level closest to target. In other words, {\bf the CRM scheme converges to optimal allocation.}
\end{thm}
\begin{proof}
Again by contradiction.

First, note that a level belonging to $S$ does not automatically have a positive stationary frequency; all we know is that it is allocated infinitely often. In fact, due to Lemma \ref{lem:pi2}, we know that only a single level, say $l_v$, has $\pi_v>0$ (in fact, $\pi_v=1$).

Now assume w.l.o.g. that $v>u^*$, meaning that $F_v>p$. Since
$$n_v\gg\max\left(n_{v-1},n_{v+1}\right)
$$
(the relation $a\gg b$ is introduced here as a shorthand for $b/a\rightarrow 0$ as $a\rightarrow\infty$), we can show using similar reasonings to that used above in Lemma \ref{lem:pi2}, that $\hat{G}_{v}\rightarrow F_{v}$. Therefore, $\Pr(\hat{G}_{v}<p)\rightarrow 0$ and $\Pr(\hat{G}_{v}>p)\rightarrow 1$. Now, $l_{v+1}$ can only be allocated if $\hat{G}_{v}<p$, and $l_{v-1}$ can only be allocated if $\hat{G}_{v}>p$. Hence, not only $n_v\gg n_{v-1}$ (given), but also $n_{v-1}\gg n_{v+1}$. So if any contradiction can arise, it can do so only from allocations to $l_{v-1}$. We focus our attention there.

Since $d\geq 2$, $\hat{G}$ can match $\hat{F}$ in at least two points -- the first being of course $l_v$. Since $n_{v-1}\gg n_{v+1}$, matching at $l_{v-1}$ (rather than at $l_{v+1}$) will yield higher likelihood. At the same time, since $n_{v-1}\rightarrow\infty$ (given), $\hat{F}_{v-1}\rightarrow F_{v-1}$. Since it is given that $l_{v-1}$ is closer to target, then (again similarly to Lemma \ref{lem:pi2}'s proof) $\Pr\left(\left|\hat{G}_{v-1}-p\right|<\left|\hat{G}_v-p\right|\right)\rightarrow 1$, meaning that CRM allocations to $l_{v-1}$ will eventually dominate allocations to $l_v$ -- producing a contradiction.\qed
\end{proof}

The rationale for this proof, as well, can be understood via Fig.~\ref{fig:pi2}.

Why does Theorem \ref{thm:littlebig} not apply to the ``Closest treatment'' allocation rule? Because then the allocation is based on modeling $F$ between design points, where we have no direct information. A misspecified model would in general model the curve incorrectly. Even when $\hat{G}=F$ at the two design points around $Q_p$, the misspecification error between them can be large enough to perceive the second-closest level as closest. However, it can be shown that under the ``Closest treatment'' rule and assuming all other conditions hold, we can do no worse than second-closest.

Lemma \ref{lem:pi2} and Theorem \ref{thm:littlebig} hold regardless of whether the model is correctly specified. Can we do better assuming the model is correctly specified - as is the case with $d=1$?

In this problem, the difference between misspecified and correctly-specified models boils down to the number of $F$ values that can be simultaneously fitted -- up to $d$ for misspecified models, and up to $m$ for correctly-specified models. For the latter, one would expect that the MLE's consistency -- a standard textbook result -- would guarantee CRM's optimal allocation convergence. However, the textbook result refers to i.i.d. sampling. Consistency under U\&D sampling was proven by \citet{Tsutakawa67block}; the key to the proof was the fact that allocation proportions converge to $\mathbf{\pi}$, similarly to the condition used in the misspecified-model proofs above.

Our case is yet a bit more different. For U\&D, every treatment level for which $0<F_u<1$ has a positive $\pi_u$. For CRM, we have just proven that if $\mathbf{\pi}$ exists then it has only $1$ nonzero element if $d>1$. This means that for $d>1$ we cannot rely on the likelihood equations to provide a consistent MLE for $\theta$, because there is no identifiability. With over one d.f. you can run an infinite number of curves through a single point. So at present, Theorem \ref{thm:littlebig} is the most I can offer using the limiting-distribution approach, regardless of whether or not the model is correctly specified.

\subsection{Nonparametric CRM-type Models}

Finally, we turn to the nonparametric CRM-style designs mentioned earlier. \citet{LeungWang01}'s ``isotonic regression'' design can be seen as a special (prior-less) case of a $d=m$ CRM scheme. Since there is no prior, the likelihood-based anlysis used here is applicable. However, as discussed above, the addition of more degrees of freedom beyond the second one does not appear to help guarantee convergence; in fact, it may hinder it due to premature over-fitting (see further below). Regarding the more sophisticated CCD scheme and similar designs \citep{IvanovaEtAl07,YuanChappell04}, I have the following result.

\begin{thm}\label{thm:ccd} Schemes that repeatedly allocate level $l_u$ as long as $\hat{F}_u\in\left[p-\Delta_1(p),p+\Delta_2(p)\right]$, and move ``up'' or ``down'' if $\hat{F}_u$ is below or above this interval, respectively, will converge optimally if $l_{u^*}$ is the only level whose true $F$ value is in the said interval.\end{thm}
\begin{proof} Assume the condition specified in the theorem holds. Then, any level $l_u$ that is allocated infinitely often will ultimately see $\hat{F}_u\rightarrow F_u$. If $F_u$ lies outside the interval, allocation will eventually move away from $l_u$ -- upward if $F_u$ is below the interval, and vice versa. As $n\rightarrow\infty$, this will happen for all levels except $l_{u^*}$.\qed
\end{proof}

\subsection{Conceptual Interpretation of the Results}\label{sec:crminterp}
The various proofs may leave worrying about conditions like ``having an asymptotically dominant level, whose neighbors still receive allocations infinitely often'' being too vague, not directly observable from the design or perhaps even circular.

First, let us frame this in context. As far as I know, to date, a full generation since Bayesian percentile-finding schemes were first implemented, there has been one (1) convergence proof published \citep{ShenOQuigley96}, with much narrower applicability and under conditions more restrictive than those postulated here \citep{CheungChappel02}. Moreover, this proof -- and the subject of CRM proofs in general -- has not played a central role in the CRM debate.\footnote{According to an August 14, 2007, ISI Web of Science search, 151 articles have cited \citet{OQuigleyEtAl90}, CRM's ``founding manifesto'', since 2000. Over the same time period, only 23 articles cited \citet{ShenOQuigley96}, the paper presenting the above-mentioned proof -- and 9 of these were co-authored by O'Quigley. This compares, e.g., with 85 articles citing \citet{GoodmanEtAl95} and 49 citing \citet{BabbEtAl98} -- two other key CRM articles published around the same time frame.} The absence of theoretical results may indicate that they are not easy to come by, or that they require inconvenient conditions such as those I used.

For example, the existence of $\pi$ was specified as a condition, though it perhaps can be proven in some cases. At face value, it may be hard to envision a treatment chain for this application that does not converge to some $\pi$; however, quasi-periodical chains that gravitate between levels without stabilizing may be possible. In general the existence of $\pi$ is a {\bf random condition}. The same goes for specifying that certain levels are allocated infinitely often. Whether or not a random condition holds, may depend upon an individual run's trajectory.

In spite of using various approaches, I could not bring about a hermetic convergence proof for general CRM designs, regardless of the form of $F$, without resorting to random conditions. Without these conditions, a CRM experiment may find itself dug in a ``hole'' with very misguided point estimates surrounding the optimal level -- point estimates which cannot be corrected over time with more information, since they themselves prevent the allocation of any trials to these levels. This means, that for many designs and distribution scenarios, some runs may converge and some not, depending upon their particular history. This is an important point, whose significance will be more fully understood in the next section.

Meanwhile, one thing that can be learned from the proofs is CRM's operating principle. CRM takes advantage of two salient features of percentile-finding experiments:
\begin{enumerate}
\item The pointwise convergence of $\{\hat{F}_m\}$, as mandated by the laws of large numbers.
\item The problem's basic geometry in 2 dimensions, i.e. a continuous increasing $F$, a horizontal line corresponding to $F=p$ which it must cross at some point, and the discrete grid -- on which both decisions and observations must take place (\emph{cf.} Fig.~\ref{fig:pi1}).
\end{enumerate}

As seen above, CRM designs take advantage of the second (geometric) feature only if they have $2$ or more d.f. This point has not been noticed so far by CRM theorists.

Inverting the conditions spelled out in the definitions and the theorems, can help provide more details about what might stop CRM from converging:
\begin{enumerate}
\item An allocation scheme whose posterior $\hat{Q}_p$ is not smooth in the sufficient-statistic inputs (see Definition \ref{def:alloc}), or is not monotone in all parameters (see Definition \ref{def:crm0}) cannot be guaranteed to converge.
\item Non-economical models (see Definition \ref{def:df}) should be avoided.
\item One-parameter models seem better off with a ``shallow'' curve family. ``Shallowness'' improves the chain's mobility: the experiment is less likely to lock into a single level based on flimsy evidence. On the other hand, ``shallow'' models are more likely to end up oscillating between $2$ levels rather than converging to a single level. The model's slope around target can be easily checked via sensitivity analysis before the experiment.
\item Schemes are more likely to converge correctly when using the ``Closest response'' rule than when using the ``Closest treatment'' rule. This is because the former does not require correct specification between design points. To my knowledge, this has not been explicitly mentioned in CRM literature, though ``Closest response'' does appear to be more widely used.
\item If the experiment locks into a single level early on and appears to never leave it, there is no guarantee that this is indeed the optimal level. This point can be inferred from the conditions to most theorems and lemmas in the previous section, and will be discussed in more detail in the next section.
\end{enumerate}

It seems that for CRM convergence, misspecification and over-fitting are two sides of the same coin. With misguided point estimates, both a ``steep'' one-parameter model and a saturated model are prone to lock onto the wrong level, perhaps indefinitely.

At bottom line, from a convergence point of view and with no knowledge of $F$'s true form, a flexible-shape $2-3$ parameter model appears to be most suitable. However, there is no guarantee that this model type has an advantage for small $n$ -- where quick response may be more crucial than long-term fit, and therefore a one-parameter model may be more suitable.

The nonparametric ``isotonic regression'' design is essentially a saturated CRM scheme, so it is prone to over-fitting and therefore not recommended. The modified CCD method of \citet{IvanovaEtAl07}, which allows a tolerance window, avoids some of the convergence pitfalls of saturated models because it forces transitions once the local point estimate is outside the window -- regardless of the estimates at neighboring levels. However, it too is not guaranteed to converge optimally, unless level spacing is very coarse.

\section{Theoretical Study: Small-Sample Behavior}

\subsection{Conceptual Interlude}\label{sec:crmconcept}

Looking back at the two core features driving CRM's convergence (mentioned at the previous section's close), they seem simple and reliable enough. However, while feature $2$ (geometry) always holds, feature $1$ (point estimate precision) kicks in only when $n$ becomes quite large. Before this happens, things can go wrong -- most often via the ``locking'' phenomenon.

More broadly, CRM is an {\bf estimation-based allocation scheme}. At each point in the experiment, we choose to observe thresholds at the level most likely to be closest to target according to our model. This is in sharp contrast with U\&D, where allocation is sampling-oriented: there, we quickly converge to random-walk sampling from $\mathbf{\pi}$, without insisting that each single trial be performed at the best level. Sampling proceeds regardless of where we believe the exact target lies. Through this sampling, U\&D provides us with gradually better information about the location of $\mathbf{\pi}$'s mean, and also about $F$ values in the $3-4$ levels closest to it -- enabling response-based estimation as well.

CRM does not consider any given observation's value for estimation further down the road, but focuses on the here and now. This should be seen as a vulnerability, and is of course related to the ``locking'' phenomenon.
From a common-sense perspective, we certainly should not presume that our model is good enough for meaningful estimation at, say, $n=5$ -- especially if we acknowledge that the model is misspecified. {\bf However, CRM is based precisely upon this presumption.} Nonparametric CRM approaches, including CCD, share this basic presumption, and therefore also display CRM's core vulnerability.

\subsection{''Locking'', ``Gambling'' and ``Unlucky Batches''}

I introduce the notion of {\bf ``lucky'' and ``unlucky batches'' of thresholds}. We can quantify these terms: a ``lucky batch'' is a sequentially-sampled group of, say, $5-10$ thresholds, whose sample moments (most importantly the first two: mean and variance) are ``close'' to the population moments. Again, the meaning of ``close'' can be quantified: close enough so that, with a given spacing, the level closest to the batch's $100p$-th percentile is indeed the optimal level.\footnote{Keep in mind, though, that thresholds are not directly observed.} An ``unlucky batch'' is the opposite - a similarly sized sample whose moments are far enough from population moments, to throw $\{\hat{F}_m\}$ off in a manner that causes CRM to allocate to the wrong level. ``Unlucky batches'' are to binary-response sampling what outliers are to ordinary direct sampling.

\begin{figure}
\begin{center}
\includegraphics[scale=.6]{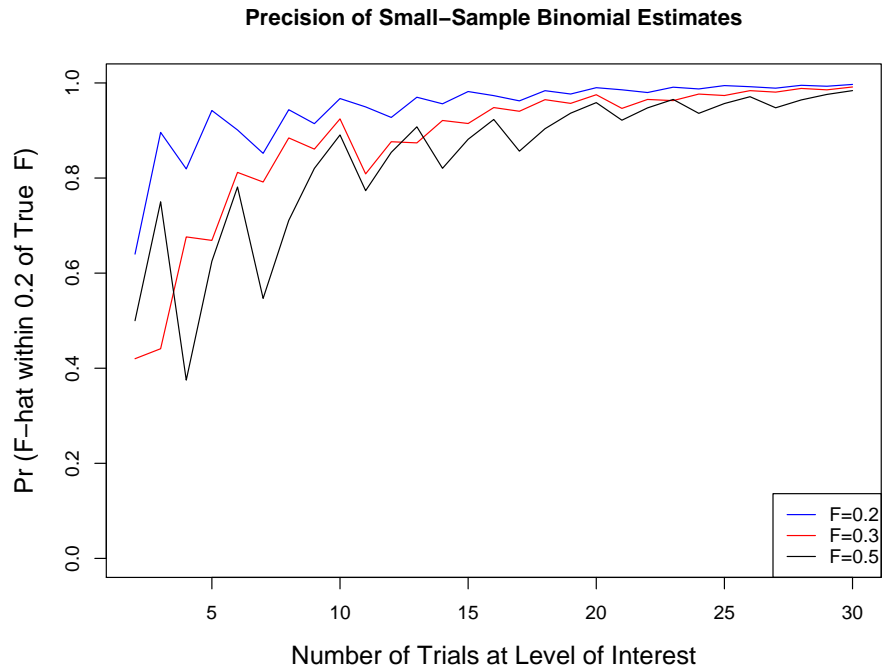}
\\[1cm]
\includegraphics[scale=.6]{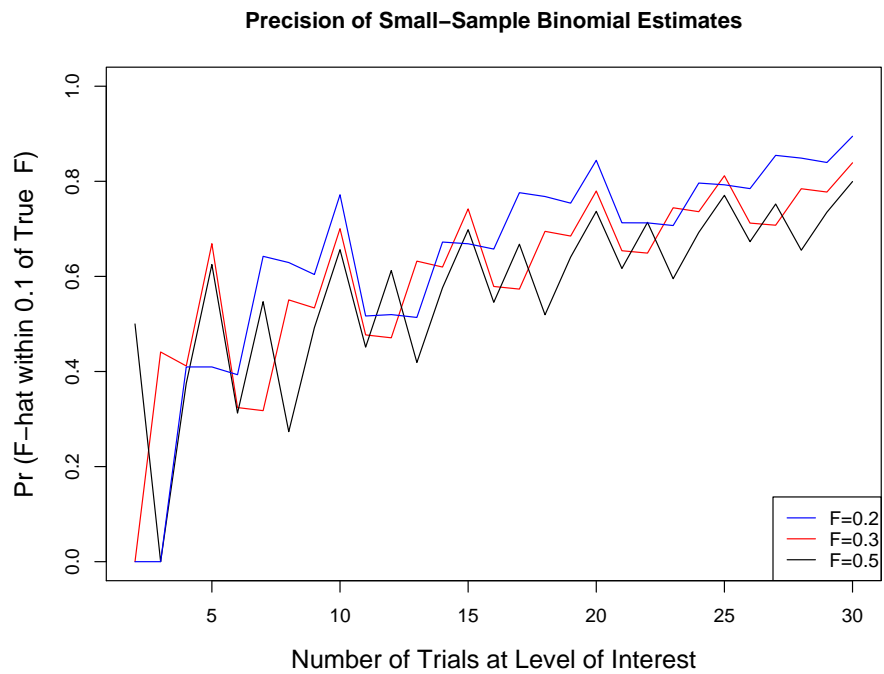}
\caption[Precision of Binomial Point Estimates]{Graphs of Probabilities of $\Pr\left(|\hat{F}_u-p|\leq\Delta F\right)$ as a function of $n_u$, for $\Delta F=0.2$ (top) and $0.1$ (bottom), and true $F$ values of $0.2$ (blue), $0.3$ (red) and $0.5$ (black).}\label{fig:binomials}
\end{center}
\end{figure}

Whether a batch is deemed ``unlucky'' in a particular experiment depends on the interplay between the threshold variance $Var(T)$, the level spacing $s$, the CRM model (including its prior specification) and also the ``luck of the draw'' -- the particular sequence in which the batch is observed, and the treatment levels at which it is observed. The first two factors can be summarized via $\Delta F$, the difference in $F$ values between adjacent levels around target. Once $\Delta F$ is fixed, some notion of how likely ``unlucky batches'' are can be gained by looking at $\Pr\left(|\hat{F}_u-F|\leq\Delta F\right)$ at a single level $l_u$, for various values of $p$ and $\Delta F$, as a function of $n_u$: The higher the probability, the smaller the chance of an ``unlucky batch''. This is a simple, straightforward calculation of binomial probabilities. Fig.~\ref{fig:binomials} shows this for $\Delta F=0.2$ (top) and $0.1$ (bottom), and true values $p=0.2,0.3$ and $0.5$.

If $\Delta F$ is large -- $0.2$ or greater -- the probability that a given batch would be ``unlucky'' decreases to about $20\%$ by the time $n_u\approx 5$ for $p=0.2$ (Fig.~\ref{fig:binomials}, top). If the target is closer to the median, it takes more trials -- around $n_u\approx 10$ -- to reach the same risk level. What if the design spacing is finer, so that $\Delta F=0.1$? The risk of ``unlucky batches'' dramatically increases: it hovers around $50\%$ for $n_u\approx 5-15$, regardless of target. A risk of less than $20\%$ is not attained until well after the typical sample size for CRM experiments is exhausted (Fig.~ \ref{fig:binomials}, bottom).
All in all, it is reasonable to expect that in an experiment of size $20-40$, at least one ``unlucky batch'' of $5-10$ observations will occur.

''Unlucky batches'' can throw any design off. An U\&D chain, too, will trend up when encountering a sequence of uncharacteristically high thresholds, and vice versa. However, its Markov-chain short memory becomes an asset: by the next hitting time at a level close to target, the probabilistic effect of previous excursions upon future allocations becomes zero \citep{Tsutakawa67}.\footnote{The effect on {\bf estimation} is still retained, whether averaging or IR/CIR estimation is used.}

On the other hand, CRM is a long-memory method -- a property that its developers are all too happy to point out as an advantage \citep{OQuigleyZohar06}. The long memory itself is not necessarily detrimental, but combined with CRM's estimation-oriented allocation scheme, CRM turns into a something of a gambling endeavor. The effect of outliers on estimation can be devastating for small $n$, but diminishes as $n$ increases. Similarly, the effect of an ``unlucky batch'' on CRM's estimation-allocation {\bf depends mostly upon when it occurs in the experiment}. If the ``unlucky batch'' happens late enough, after point estimates have stabilized reasonably close to their true values, especially around target -- the CRM chain will be quite resistant to subsequent ``unlucky batches''. On the other hand, an early ``unlucky batch'' can create a ``perfect storm''. CRM would trend away from target and lock onto the wrong levels, collecting subsequent information at the wrong place -- thereby slowing down the self-correction process. Moreover, since the estimation-allocation uses all previous data, it will resist the correcting influence of more well-behaved batches following the ``unlucky'' one.

To sum it up: CRM, with its long memory and estimation-based allocation, gambles that ``unlucky batches'' would occur late in the experiment -- or not at all. If the gamble succeeds, we are quite likely to have a successful experiment. If it fails, we are almost assured of the opposite.

\begin{figure}[!h]
\begin{center}
\includegraphics[scale=.5,angle=-90]{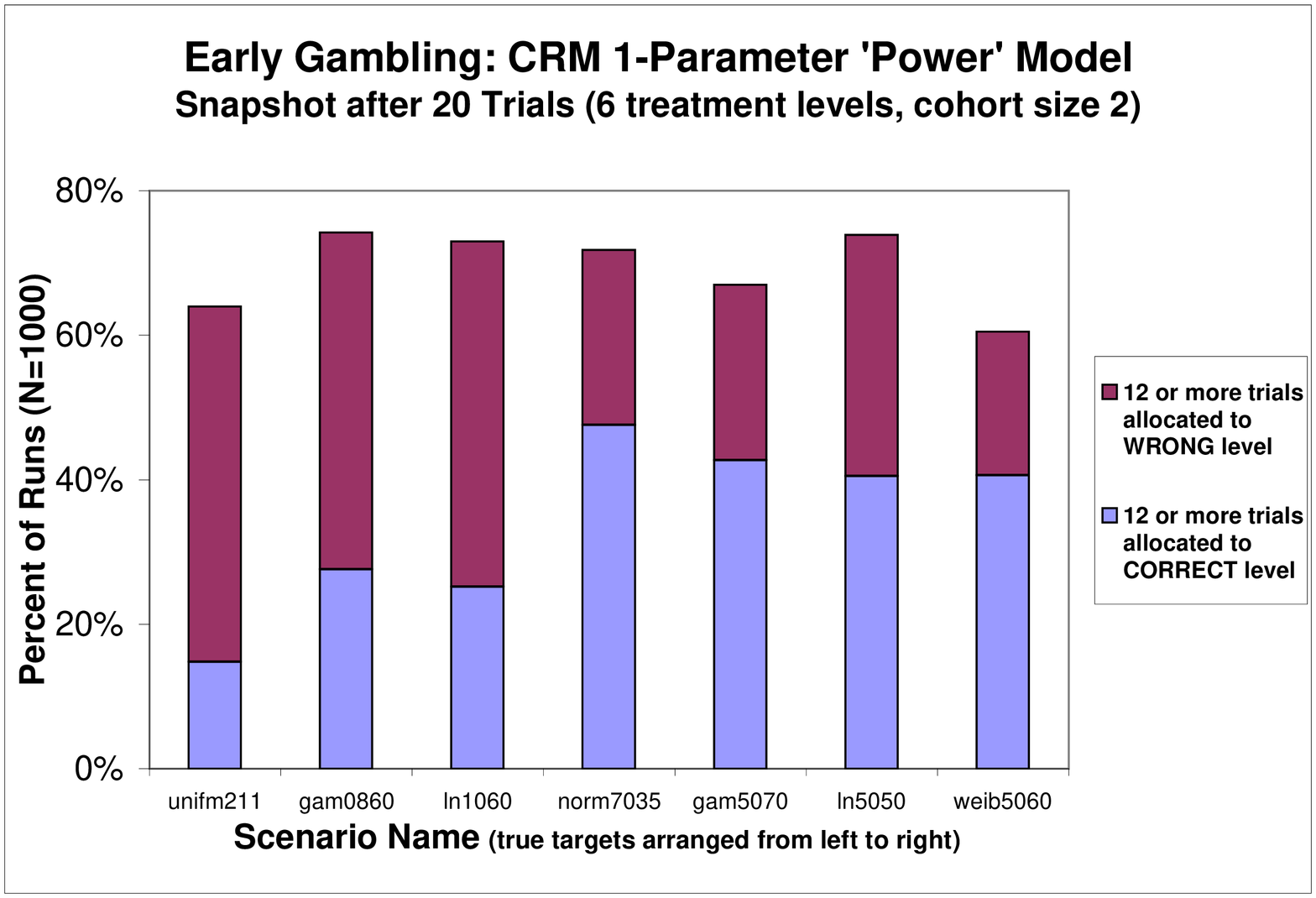}
\\[1cm]
\includegraphics[scale=.5,angle=-90]{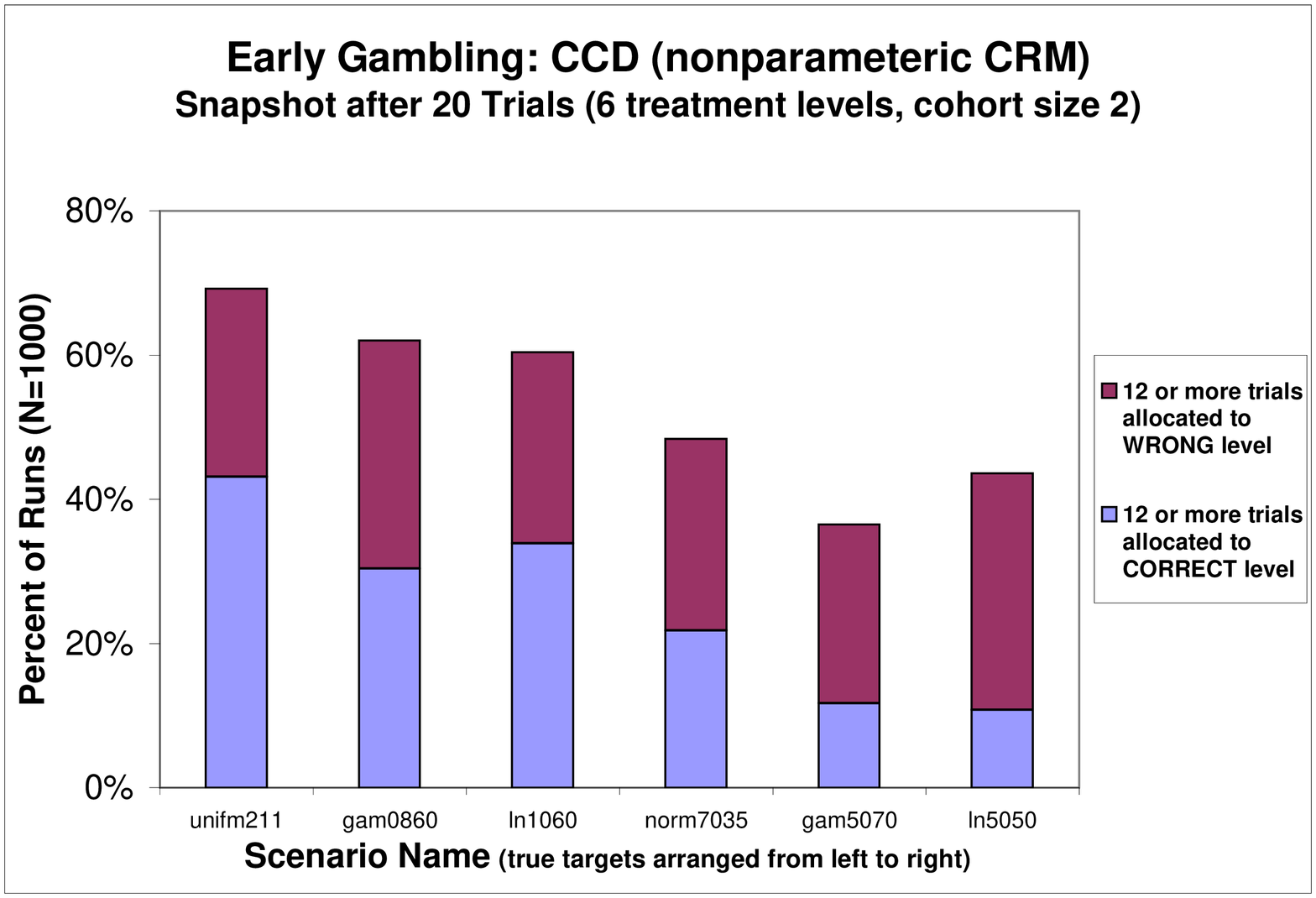}
\caption[The ``Gambling'' Property of CRM-type Designs]{Charts depicting the proportion of CRM-type runs, in which at least $12$ of the first $20$ trials were allocated to the same level. This proportion is then split cases when this level was correct (optimal, light blue) and wrong (maroon). The experiment simulated was a cohort design with cohort size $2$, $p=0.3$, $m=6$ and seven different distributions arranged according to target location. Shown are the one-parameter ``power'' (logistic with free shape) model of \citet{OQuigleyEtAl90}, optimized with ``shallow'' slope (top); and the CCD method of \citet{IvanovaEtAl07}, with their recommended window width $\Delta=0.1$ (bottom) -- over the same simulated thresholds.\label{fig:gambling}}
\end{center}
\end{figure}

\begin{figure}[!h]
\begin{center}
\includegraphics[scale=.5]{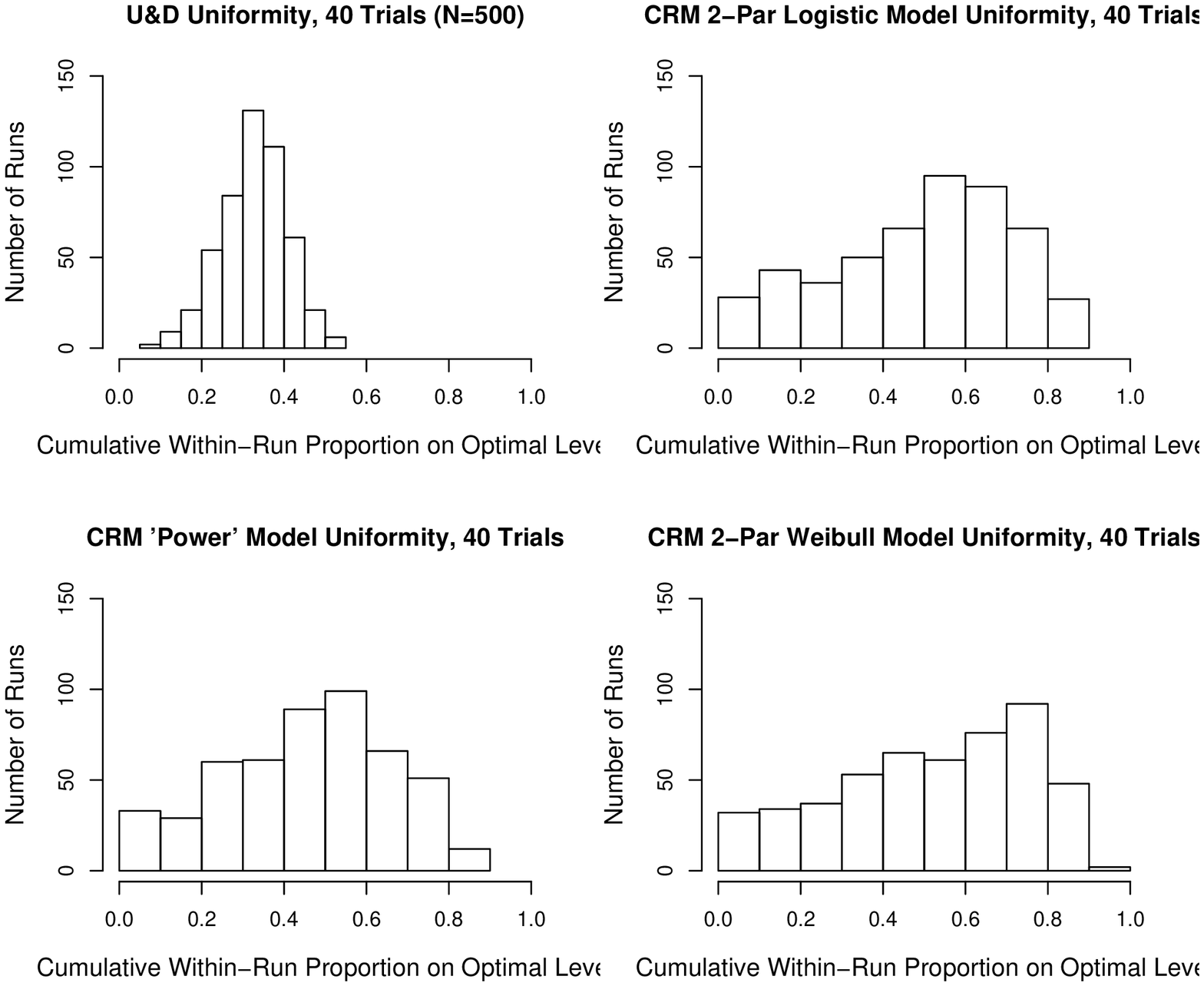}
\\[1cm]
\includegraphics[scale=.5]{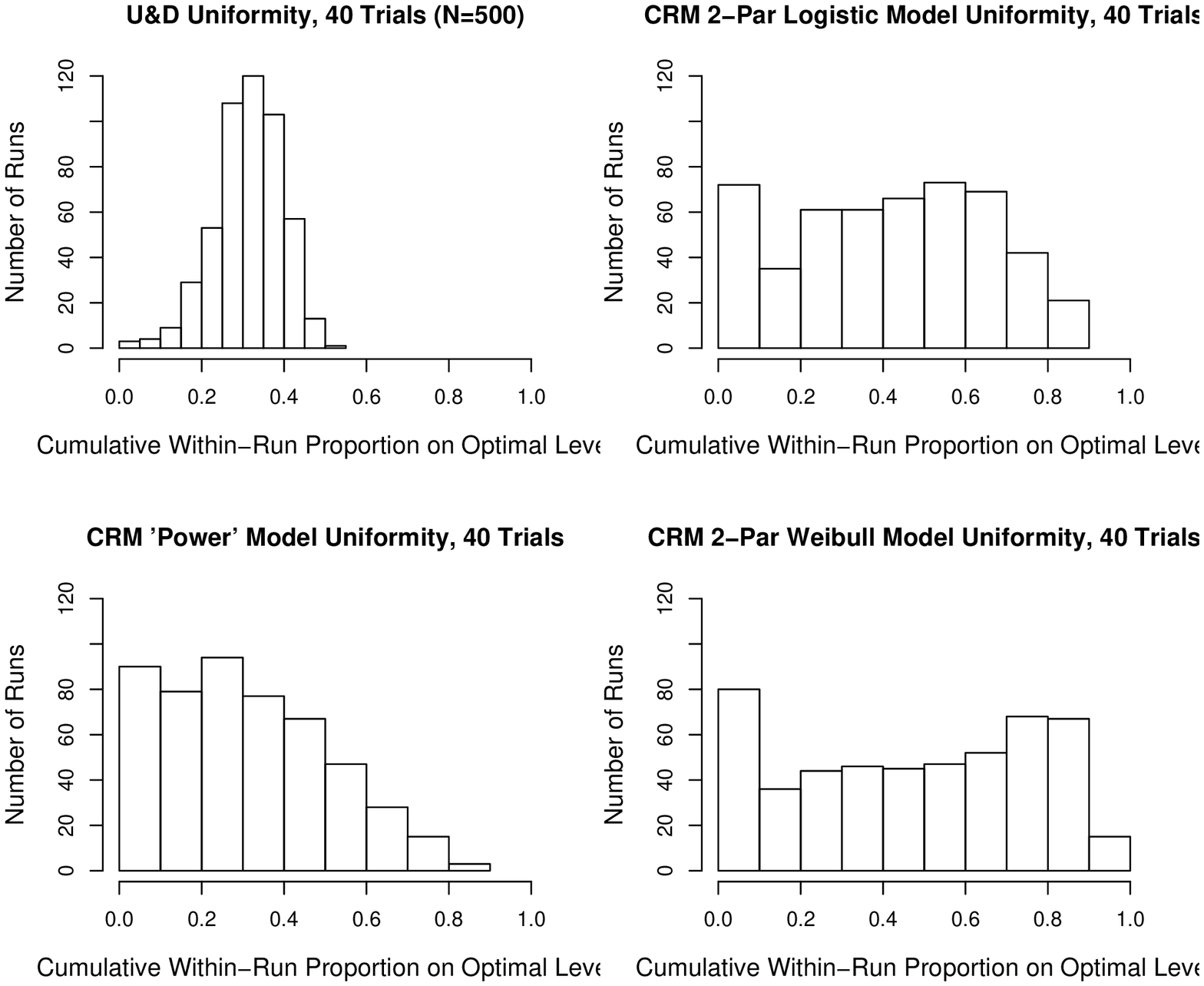}
{\small
\caption[Uniformity of CRM Allocations]{Uniformity of U\&D and CRM allocations, shown via histograms of $\hat{p}_{u^*}$ from $500$ runs, after $40$ trials. Simulations were non-cohort (or cohort size 1), with normal (top) and gamma, shape $1.2$ (bottom) thresholds. In both cases $m=8$. Each frame shows histograms of U\&D runs (top left-hand corner), vs. CRM with the one-parameter ``power'' model (bottom left-hand corner), two-parameter logistic (top right-hand corner) and shape-scale Weibull (bottom right-hand corner).\label{fig:CRMunif}}
}
\end{center}
\end{figure}

\begin{figure}[!h]
\begin{center}
\includegraphics[scale=.5]{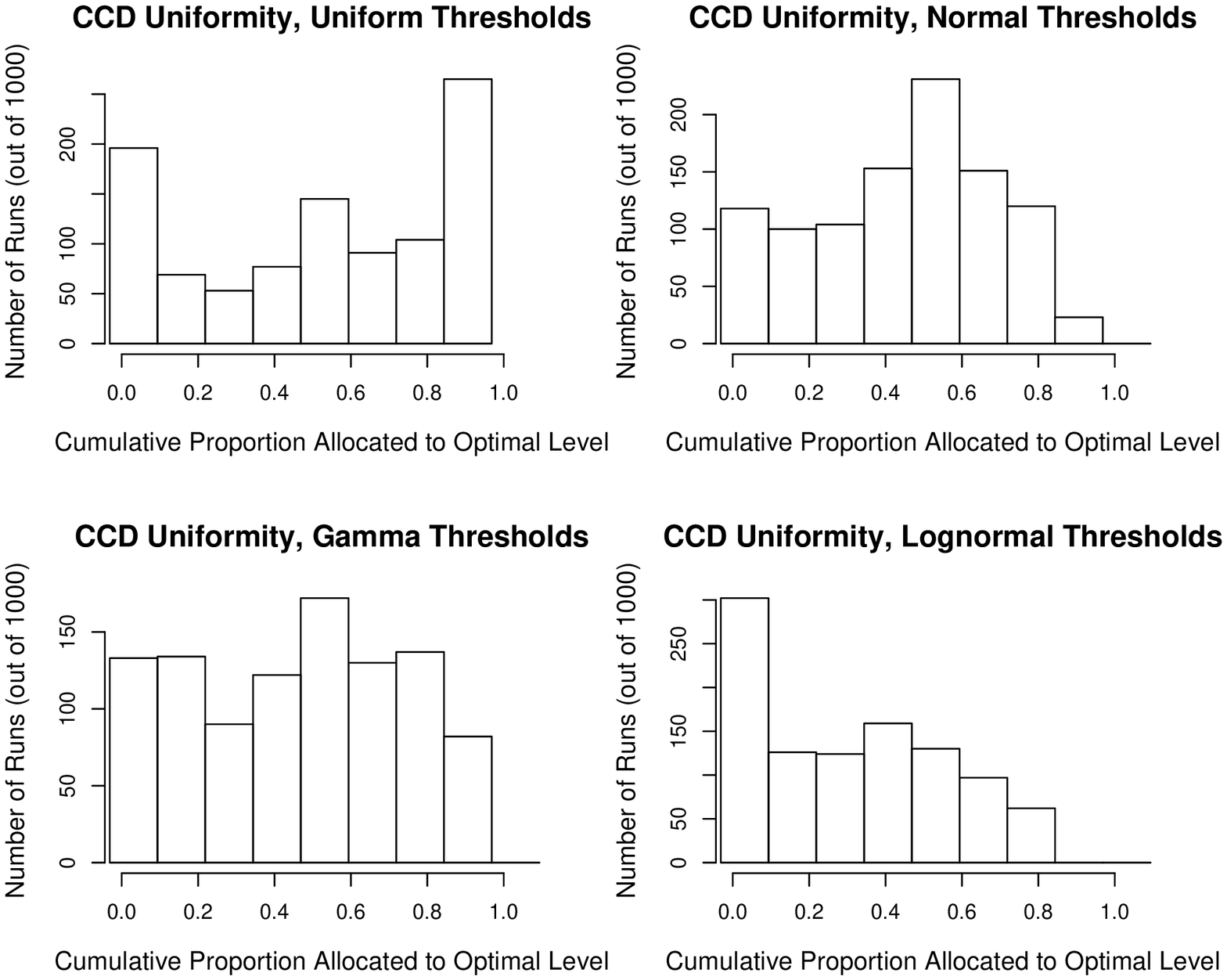}
\\[1cm]
\includegraphics[scale=.5]{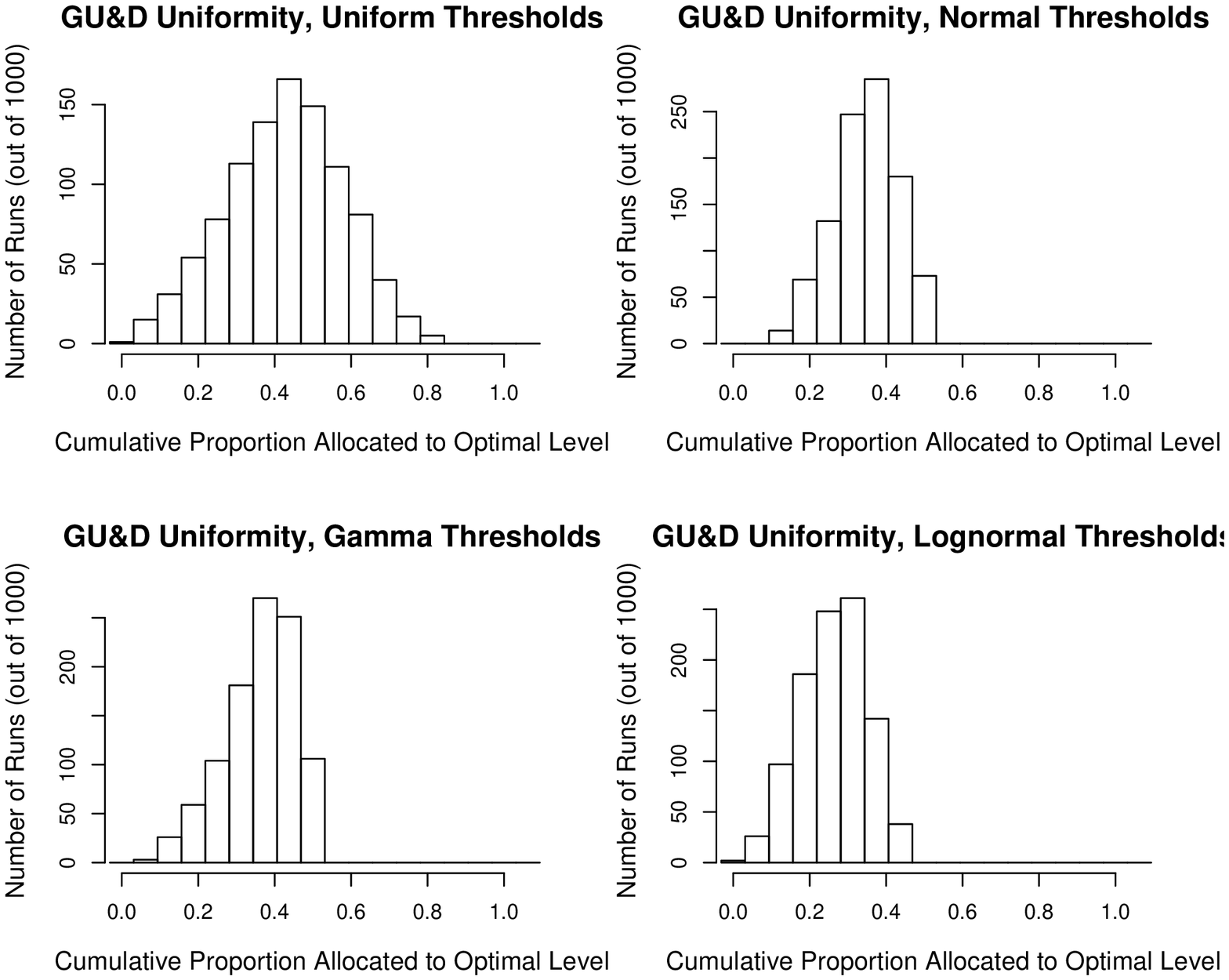}
{\small
\caption[Uniformity of CCD Allocations]{Uniformity of CCD (top) and GU\&D (bottom) allocations, shown via histograms of $\hat{p}_{u^*}$ from $1000$ runs, after $32$ trials. Cohort size is 2 in designs targeting $Q_{0.3}$, with uniform (top-left frames), and gamma, shape $0.8$ (bottom-left), normal (top-right) and lognormal (bottom-left) thresholds. In all cases $m=6$ and $x_1=l_2$. The location of $l_{u^*}$ in the frames was $l_1,l_2,l_3$ and $l_4$, respectively.\label{fig:CCDunif}}
}\end{center}
\end{figure}

This gambling property of CRM is demonstrated in Figs. \ref{fig:gambling} through \ref{fig:CCDunif}. In in Fig. \ref{fig:gambling} we see the proportion of runs for which, out of the first $20$ trials, at least $12$ were allocated to the same single level (in symbols: the proportion of runs for which, at $n=20$, $\hat{p}_u\geq 0.6$ for some $u$). This is the proportion of runs for which CRM was quite ``sure'', during the early phase, that it ``knew'' where the target is. This proportion varies with target location (relative to the prior's weighting), and with distribution, but for one-parameter CRM (top) it remains at or above $60\%$ of the runs. However, in about half of the cases, the gamble fails: the bulk of the first $20$ trials were allocated to the wrong level (in all scenarios depicted, there is only a single optimal level). Also shown are data from a CCD experiment \citep{IvanovaEtAl07}, simulated over the same thresholds (Fig.~\ref{fig:gambling}, bottom). The ``gambling'' tendency is somewhat subdued, but around $50\%$ of runs, on average, are still dominated early on by a single level. The success rate is, if anything, even worse than CRM's. Note that for U\&D designs, such an exclusive allocation pattern is all but impossible.

In Fig.~\ref{fig:CRMunif}, we can see that the effect of early ``gambling'' lingers on. Shown are histograms of $\hat{p}_{u^*}$ after $40$ trials. That is: after $40$ trials, the cumulative proportion allocated to the correct optimal level is calculated separately for each run; the proportion from all runs are then tallied into a histogram. These histograms can be interpreted as descriptive measures of {\bf convergence uniformity}. This is done for U\&D and for 3 different CRM models. If the design converges uniformly, we expect the distribution of $\hat{p}_{u^*}$ to be tight. This indeed occurs for U\&D designs (top left-hand corner of each group of $4$ histograms). The peak is rather modest in its location:  $\hat{p}_{u^*}<0.4$ in most runs -- but on the other hand, $\hat{p}_{u^*}$ rarely falls below $0.2$. CRM designs over the same thresholds (regardless of specific model) show a much large variability. With an ``easy'' distribution (normal, top), the ensemble mean of $\hat{p}_{u^*}$ for all 3 models is larger than U\&D's. However, a sizable proportion of runs have little or no allocation at $l_{u^*}$ after $40$ trials. With a less convenient distribution (gamma with shape parameter $1.2$, bottom), the ensemble mean is approximately the same for U\&D and CRM, but the proportion of poorly-allocated CRM runs shoots up, especially with the one-parameter model.

It should be noted, that while the ensemble mean of $\hat{p}_{u^*}$ is often reported in CRM simulation studies as a measure of the design's success, to my knowledge no study has looked at $\hat{p}_{u^*}$'s variability between runs.

Figure \ref{fig:CCDunif} shows a similar analysis comparing GU\&D$_{2,0,1}$ and the nonparametric CCD with cohort the same cohort size and target. The patterns are similar, with CCD having much larger variability. I used the same cohort size to make comparisons easier. Similar analysis with CCD and cohort size 3 (the size recommended by its developers) showed the same pattern. This indicates, that CRM's allocation robustness problem is not primarily caused by the use of a model, but rather by the reliance upon blunt point estimates to drive allocations early in the experiment.

\subsection{Effect of Prior}
In the previous section discussing convergence, the prior was almost always neglected. However, for small samples it plays a decisive role. The first few allocations are not only driven by point estimates, but also by the prior. From numerical studies, it is clear that CRM performs much better when the true target corresponds to a high-density region of the prior, and vice versa. Naturally, the more parameters to the model, the ``heavier'' the prior becomes and the more time it takes for its effect to wear off. Since ultimately we'd like the data to play the leading role, this is one place where one-parameter models have a distinct advantage. In this work I have not attempted to tinker with priors too much, but usually optimized them so that $\hat{Q}_p$ has a broad unimodal distribution centered around the middle of the design range, and so that the boundary levels have a smallest prior probability, but not a negligible one. The existence of a peak in $\hat{Q}_p$'s prior, even a broad one, was sufficient to show a clear performance gap between scenarios in which in coincided with the true targets, and scenarios in which it did not.

\newpage
\section*{Additional Glossary for Chapters 4-5}
\addcontentsline{toc}{section}{\em{Glossary}}

\begin{glossary}

\item[BUD] Bayesian Up-and-Down: a family of methods combining U\&D and CRM allocation principles.
\item[CCD] Cumulative Cohort Design, a method recently developed by \citet{IvanovaEtAl07}. I characterize it here as a nonparametric Bayesian design.
\item[CRM] Continual Reassessment Method, the Bayesian design developed by \citet{OQuigleyEtAl90}. Due to its prominence in statistical circles, it is also used here as shorthand for any Bayesian percentile-finding design.
\item[EWOC] Escalation With Overdose Control, the Bayesian design developed by \citet{BabbEtAl98}. Essentially it is a CRM variant that aims to reduce the incidence of visits to treatments above target.
\item[QUEST] The Bayesian design developed by \citet{WatsonPelli79,WatsonPelli83}. Very similar to CRM, but predates it by a decade and is almost unknown outside of psychophysics.
\\[1cm]
\item[$d$] The number of degrees of freedom in the model (see Definition~\ref{def:df}).
\item[$G$] The Bayesian model curve for $F$.
\item[$\tilde{G}$] The limit curve of $G$ as $n\to\infty$ (in case it has a limit curve).
\item[$\mathcal{G}$] The space of possible model curves under a given specification.
\item[$\hat{p}_u$] The empirical cumulative proportion of allocations to $l_u$ for a given sample size.
\item[$q$] The cardinality of the set $S$ (see below).
\item[$S$] The set of all levels receiving an unbounded number of allocation in a given experiment. In general this set is random.
\item[$u^*$] The index of the level closest to target according to the metric used (a.k.a. the optimal level).
\\[1cm]
\item[$\alpha$] A key parameter in EWOC: allocation is to the $\alpha$ posterior quantile of $Q_p$.
\item[$\beta$] A key parameter in BUD: if the U\&D allocation falls outside the (quantile-based) $100(1-2\beta)\%$ posterior credible interval for $Q_p$, it can be overridden by the Bayesian allocation.
\item[$\Delta_1(p),\Delta_2(p)$] Parameters used in nonparametric Bayesian designs
\item[$\theta$] The parameter vector used to construct the model $G$, to create a tolerance window of $\hat{F}$ values, for which the current allocation is retained.
\item[$\Theta$] The parameter space of $\theta$.
\item[$\Pi$] Prior and posterior distributions of $\theta$.
\item[$\phi$] The parameter vector in $\theta$'s prior distribution.

\end{glossary}


\chapter{Combined Design Approaches}\label{ch:bud}
\section{Existing Designs and Current Knowledge}

Designs combining U\&D with Bayesian or other long-memory methods are a recent and increasingly popular area of research. These designs focus mostly on the Phase I application. They are fueled by the dissatisfaction with traditional '3+3' type designs among statisticians, and by the failure of purely Bayesian designs to win over many practitioners due to their perceived risk. Since the Phase I field is rife with creative statisticians, I am sure that newer designs will continue to be published as this thesis goes to print.

A few researchers \citep{Storer01,Potter02} have suggested the simple ``two-stage'' template familiar from U\&D designs: start the experiment as U\&D, then after the first reversal change to a CRM or other model-based design. The two cited studies suggest starting with a median-targeting SU\&D, then switch to a design that targets a lower percentile.

An interesting combination has originated from one of the few statistical groups studying U\&D \citep{IvanovaEtAl03}. They suggested a novel U\&D variant that uses information from all previous trials in transition decisions. Attributed to an idea from an unpublished 1950's thesis by Narayana, this method (dubbed here NM for ``Narayana Method'') requires the experiment to fulfill two conditions in order to move 'up' or 'down' at trial $i$: 1. Some U\&D transition rule (e.g., BCD, KR, etc.), and 2. $\hat{F}(x_i)\leq p$ for 'up' transitions, and vice versa for 'down' transitions. If only one condition is fulfilled, the experiment remains at the same level, i.e. $x_{i+1}=x_i$. NM's condition 2 is essentially nonparametric CRM, as discussed in Chapter \ref{ch:crm}. Asymptotically, NM was proven to converge to a $2-3$ level Markov chain around target ($3$ levels corresponds to the case when $Q_p$ is very close to a design level).

NM appears to have been abandoned, but may have helped inspire the ideas of \citet{IvanovaEtAl07}. Even though the latters' recent CCD method is presented as a fusion of U\&D and CRM principles, it is in fact a nonparametric CRM, with U\&D affecting it only indirectly in the determination of the window width $\Delta(p)$ (see Section \ref{sec:noncrm}).

\section{Theoretical Discussion}

At this point in the dissertation, the problems with the two-stage approach cited above are apparent. The first reversal happens too early in the experiment to guarantee reliable model-based estimation-allocation. Moreover, both authors chose to use median-targeting SU\&D for the first stage, even though the experiment's ultimate target is lower. Even though this may seem like a clever way to offset the first reversal's typical downward bias (assuming the experiment starts low as is customary in Phase I), it means that a sizable proportion of these two-stage experiments would end up with a transition point close to the median. This is highly undesirable, especially for the toxicity-averse Phase I application. A safer two-stage hybrid scheme would begin with an U\&D whose target is closer to $Q_p$, and then transition at reversal $3$ or $5$, or at a hitting time (see Section~3.6).

Similarly, by now we may suspect what could go wrong with NM. First, being a CRM rule it suffers from the ``locking'' and ``gambling'' tendencies described in Section \ref{sec:crmconcept}. Moreover, even among CRM variants the saturated nonparametric model seems to have inferior convergence properties (see Section \ref{sec:crminterp}), since it shares little information between point estimates. Finally, the use of two criteria, both of which must be satisfied for a transition to occur, further amplifies the CRM-related ``locking'' tendency. Therefore, even though the asymptotic proof in \citet{IvanovaEtAl03} is valid once the chain actually reaches the vicinity of target, it does not preclude the experiment being held up away from target for an indefinite amount of time. These problems have emerged in some numerical trials I performed (data not shown; see more discussion in \citet{Oron05}).
\section{New Methodologies and Optimization}
\subsection{Conceptual Overview}\label{sec:budconcept}
Before proceeding to new combined Markovian-Bayesian designs, it may be helpful to list each approach's pros and cons. Using the insight gained by prior researchers and the added results reported in this thesis, the picture is clearer now. The table below compares the properties of "best in class" implementations of each approach.

\begin{table}
\begin{center}
\caption[Comparison of U\&D and CRM Properties]{Side-by-side comparison of Markovian and Bayesian design properties.\label{tbl:proscons}}
\small{
\begin{tabular}{p{3cm}p{2.5cm}p{3cm}p{3cm}c}
\toprule
\multicolumn{2}{l}{\bf Category} & {\bf U\&D} & {\bf Bayesian} & {\bf Advantage}\\
\toprule[1pt]

\multirow{5}{3cm}{\bf Allocation} & Convergence Rate & Geometric & $\sqrt{n}$ & U\&D \\
\cmidrule{2-5}
& Guaranteed? & By Markov Properties & Generally, No & U\&D \\
\cmidrule{2-5}
& Sharpness & Only as good as $\pi$ & Asymptotically maximal sharpness & Bayesian \\
\cmidrule{2-5}
& Robustness to Misspecification & Sensitive to slope of $F$ around target & Sensitive to slope, to shape deviations from model family and to bad prior & U\&D \\
\cmidrule{2-5}
& Robustness to ``unlucky batches'' & Short memory; reset at next hitting time & Depends on when they occur & U\&D \\
\toprule[1pt]
\multirow{4}{3cm}{\bf Estimation} & Method & Method choice a 'free-for-all' & 'Natural' Estimation as next allocation, but misspecified & None\\
\cmidrule{2-5}
& Efficiency & Approximately $\sqrt{n}$ & Approximately $\sqrt{n}$ & None \\
\cmidrule{2-5}
& Robustness & Averaging sensitive to boundaries, CIR robust but less efficient & Sensitive (see allocation above) & U\&D \\
\cmidrule{2-5}
& Interval Estimation & Complicated but feasible & 'Naturally' available & Bayesian \\
\toprule[1pt]
\multirow{3}{3cm}{\bf Implementation Aspects} & Intuitive Simplicity & Simple; no counter-intuitive transitions & Sophisticated; Counter-intuitive transitions possible at start & U\&D \\
\cmidrule{2-5}
& Step-by-Step & Practitioners can be self-sufficient & Requires constant statistician/software support & U\&D \\
\cmidrule{2-5}
& Flexibility & Cohort designs and best-in-class single-trial designs limited;  & Flexible & Bayesian \\
\cmidrule{2-5}
& Sensitivity to Spacing and Boundaries & Optimal around $\Delta F\approx 0.1-0.15$, sensitive to boundaries & Needs $\Delta F\approx 0.2$ or more, no problem with boundaries & None \\
\bottomrule[1pt]
\end{tabular}
}\end{center}
\end{table}

The main conclusion from Table \ref{tbl:proscons} is that there is probably no magic solution to the small-sample percentile-finding problem. In particular, the Achilles heel of both approaches is allocation. Even though U\&D is listed as superior to CRM on most line items under ``allocation'', it is only a relative advantage. Run-to-run variability cannot be eliminated given the problem's nature. On the other hand, post-experiment estimation is not as problematic as it may appear. For U\&D, the pros and cons of averaging estimators vs. CIR were clearly outlined in Chapter \ref{ch:est}, and both are satisfactory if used ``as directed''. Moreover, there is nothing preventing us from choosing model-based estimation after an U\&D experiment, if we trust the model. Similarly, while CRM's ``natural'' estimator is the next allocation, we may supersede it with a more sophisticated model if warranted -- or with CIR if we trust no model.

Returning to allocation, both approaches are vulnerable to outlying ``unlucky batches'' and to variations in the form of $F$ (though U\&D is in general less sensitive). Fortunately, to a large degree the two approaches are complementary. U\&D is relatively more robust early on, and converges faster to its stationary behavior. As to CRM, its allocation performance improves with time. As $n$ increases, it actually becomes more robust than U\&D.

As discussed in the previous section, current proposals to combine U\&D and Bayesian designs fail to alleviate the worst small-sample risks. Instead of these ideas, I propose a gradual transition scheme, which is applicable regardless of sample size. Two options are described below.

\subsection{Randomized Transition}
Since CRM relies upon point estimates, its model's performance is indexed to $n$. The following randomization scheme utilizes this property:

{\bf Randomized Bayesian Up-and-Down (R-BUD):}

\begin{itemize}
\item Start with a $5-10$ trial U\&D 'burn-in';
\item For each subsequent trial, apply CRM allocation with probability $n/(n+n_0)$; otherwise apply U\&D allocation.
\end{itemize}

The factor $n$ can be replaced by $\sqrt{n}$, or any other increasing function (preferably one that has theoretical justification). Another optional safety precaution is to mandate that the first several allocations at each level will be U\&D.

Even though individual allocations are random, the R-BUD design actually guarantees transition to CRM-dominated allocations at a predetermined pace. This pace is controlled by the tuning parameter $n_0$. In view of results presented earlier in the thesis, good values for $n_0$ are in the range $10-30$.

\subsection{Inference-Based Transition}
\subsubsection{Description}
R-BUD provides a reasonable combination template. However, it does not adapt to the data. In some cases, a ``lucky'' early run or a successful CRM model may provide opportunity for earlier transition; in other cases the opposite happens.

At least to some extent, CRM itself can provide testimony about how well it is doing. This is sketched in Algorithm~3 below.

\begin{algorithm}[h]
\begin{algorithmic}\label{alg:CBUD}
\caption{Credible Bayesian Up-and-Down (C-BUD)}
\Procedure{C-BUD}{$x_1,\{l_m\}$,cutoff,$n$}
\Statex
\State $i\leftarrow 1$
\While {$i\leq$ cutoff}
    \State $x_{i+1}\leftarrow x^{(U\&D)}_{i+1}$
    \State $i\leftarrow i+1$
\EndWhile
\While {$i\leq n$}
    \If{$x^{(U\&D)}_{i+1}\neq x^{(CRM)}_{i+1}$}
        \State Calculate $\hat{Q}_{p,\beta},\hat{Q}_{p,1-\beta}$ : the CRM $\beta$ and $1-\beta$ posterior quantiles of $Q_p$
        \If{$x^{(U\&D)}_{i+1}\in\left[\hat{Q}_{p,\beta}-s/2,\hat{Q}_{p,1-\beta}+s/2\right]$}
            \State $x_{i+1}\leftarrow x^{(U\&D)}_{i+1}$
        \Else $\textrm{ }x_{i+1}\leftarrow x^{(CRM)}_{i+1}$
        \EndIf
    \EndIf
    \State $i\leftarrow i+1$
\EndWhile
\EndProcedure
\end{algorithmic}
\end{algorithm}

Notes:
\begin{enumerate}
\item The C-BUD scheme allows CRM to override U\&D, only when a quantile-based $100(1-2\beta)\%$ credible interval around the former is narrow enough to exclude a CRM allocation identical to the U\&D one.\footnote{This exclusion is ensured by adding the $s/2$ buffers on each side of the Cr.I.; the method could be implemented without them, though this is not advisable; see note 2.} As $\beta$ increases, a CRM override becomes easier.
\item This allows the following easy interpretation: for $\beta\geq 0.5$, C-BUD becomes CRM. For $\beta=0$, C-BUD becomes U\&D. Hence, it would seem that $\beta=0.25$ is a roughly equal mix of the two designs. However, if the two allocations differ, then due to the discrete design CRM automatically begins with some positive baseline credibility level (on the order of $1/m$). This is conceptually balanced by demanding the exclusion buffer around the $100(1-2\beta)\%$ credible interval. In practice, the range $0.15-0.25$ for $\beta$ seems to reflect a relatively balanced combination. With these $\beta$ values we may expect around $~10-15$ trials to pass before observing the first Bayesian allocation override.
\item An equivalent definition of the C-BUD rule is: take the tail closest to the U\&D allocation (including the U\&D-allocated level itself), and calculate the posterior probability of $Q_p$ falling within that tail. If it is less than $100\beta\%$, override the U\&D allocation.
\item Posterior-median CRM allocation fits naturally within this scheme and is strongly recommended over posterior mean or mode - though, of course, the method could be implemented with any CRM allocation rule.
\item Usually, if the two allocations differ then $x^{(U\&D)}_{i+1}=x^{(CRM)}_{i+1}\pm s$. However, occasionally the U\&D and CRM allocations point in different directions (i.e., $x^{(U\&D)}_{i+1}=x_i+s,x^{(CRM)}_{i+1}=x_i-s$ or vice versa). In that case we have two options:
\begin{enumerate}
\item Proceed as in the usual case, i.e. use the CRM allocation whenever the U\&D one is outside the calculated credible interval;
\item If the U\&D allocation is ``rejected'', use the CRM allocation only if $$x_i\not\in\left[\hat{Q}_{p,\beta}-s/2,\hat{Q}_{p,1-\beta}+s/2\right]
$$ as well. Otherwise $x_{i+1}=x_i$.
\end{enumerate}
\item For toxicity-averse applications such as Phase I, the design can be modified to have two different $\beta$ values on the two tails. If we desire to minimize upward excursions, then the right-hand $\beta$ should be larger.
\end{enumerate}

Lifelong (or converted) Bayesians would now ask, ``what is the loss function?'' First we should note that in spite of the use of Bayesian calculations, our framework is hybrid: one which evaluates a non-Bayesian procedure using Bayesian methods, but which gives the non-Bayesian decision a ``home-court advantage''. Therefore, it is not clear that formulating a loss function provides added value, being mostly an exercise in reverse engineering. At this moment, I have no satisfactory formulation for the loss function.

Because of C-BUD's greater theoretical appeal and flexibility, from this point on we limit discussion of BUD to this variant and leave the randomized R-BUD behind. From now on, referring to BUD will imply C-BUD, unless otherwise noted.

\subsection{Estimation and Stopping Rules}
In view of CRM's coherence advantage, one is tempted to use the $n+1$-th CRM allocation as the BUD estimate. However, this is somewhat self-contradictory. After all, the BUD framework is about not accepting the CRM decision at face value; why, then use it as the final estimate? Moreover, the CRM allocation model is knowingly misspecified. If we use a model-based estimation, it better be the most realistic model we can suggest for the problem, rather than the CRM model. An alternative option is to use U\&D estimators -- either $\hat{v}_{AD}$, which can be applied to any chain (not necessarily a purely U\&D one), or CIR.

The BUD framework suggests some interesting stopping rule options. The most straightforward one would be to stop after a specified number of successful CRM overrides (defined as allocation that exceeds the required credibility threshold, whether or not it coincides with the U\&D allocation), indicating that the model is fitting reasonably well around target. This subject is left for subsequent research.

\subsection{CCD-BUD Combination}

The frequentist interpretation suggests a way to combine CCD \citep{IvanovaEtAl07} with U\&D. We recall that CCD forces a transition whenever $\hat{F}(x_i)$ is outside a pre-specified tolerance window around $p$. This is still CRM-style estimation-driven allocation, but it is somewhat less prone to ``locking'' than CRM (see Fig.~\ref{fig:gambling}). Additionally, since it has no prior and no curve assumptions, it is more robust to misspecification and to boundaries. In order to combine it with U\&D, we use frequentist binomial confidence intervals for $Q_p$, instead of the credible intervals.

\subsection{Numerical Examples}
\begin{figure}
\begin{center}
\includegraphics[scale=.5,angle=-90]{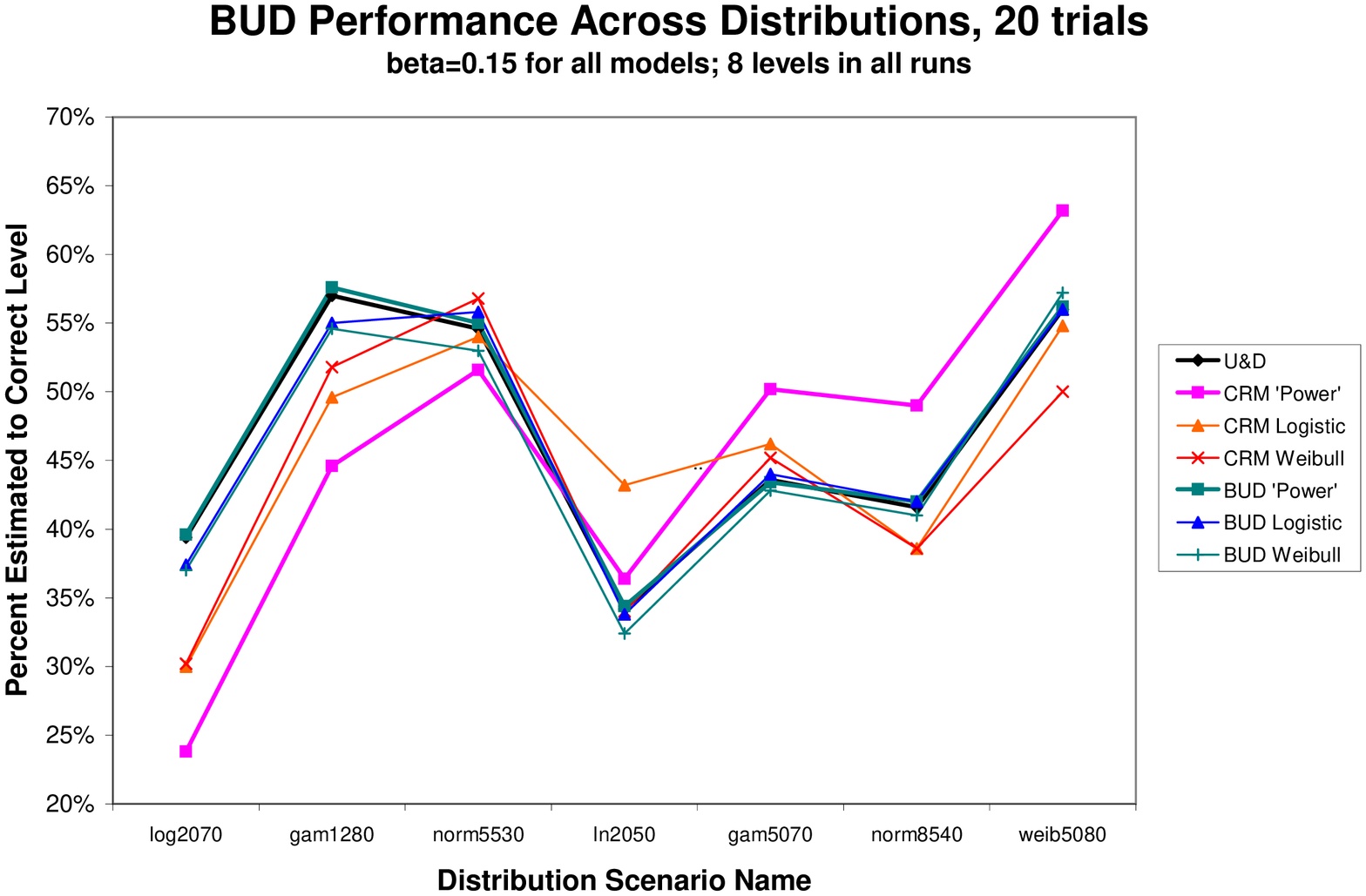}
\\[1cm]
\includegraphics[scale=.5,angle=-90]{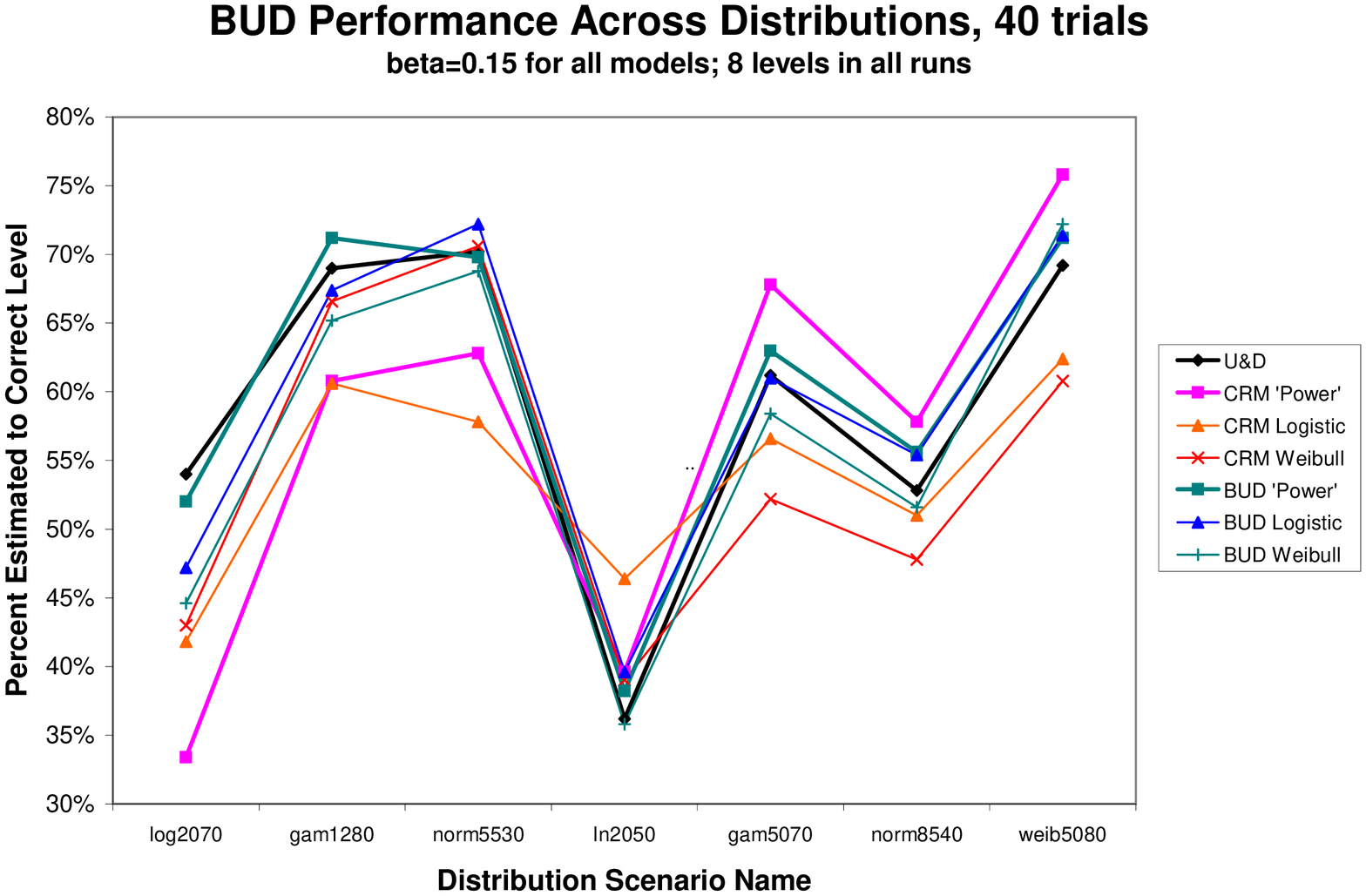}
\caption[KR, CRM, BUD Comparison]{Comparison of U\&D (KR, $k=2$), CRM and BUD ($\beta=0.15$) estimation on a variety of simulated distributions, and $n=20$ (top) and $40$ (bottom). ``Correct level'' was defined as the level whose $F$ value is closest to $0.3$. The design had $8$ levels, with all runs starting at level $4$. The 'power' model is logistic with shape parameter and fixed location and scale ($d=1$), introduced by \cite{OQuigleyEtAl90}. The other two models have $d=2$: 'Logistic' is location-scale, and 'Weibull' is shape-scale. The auto-detect averaging estimator $\hat{v}_{AD}$ was used for U\&D and BUD estimation.\label{fig:bud2040}}
\end{center}
\end{figure}

\begin{figure}
\begin{center}
\includegraphics[scale=.5]{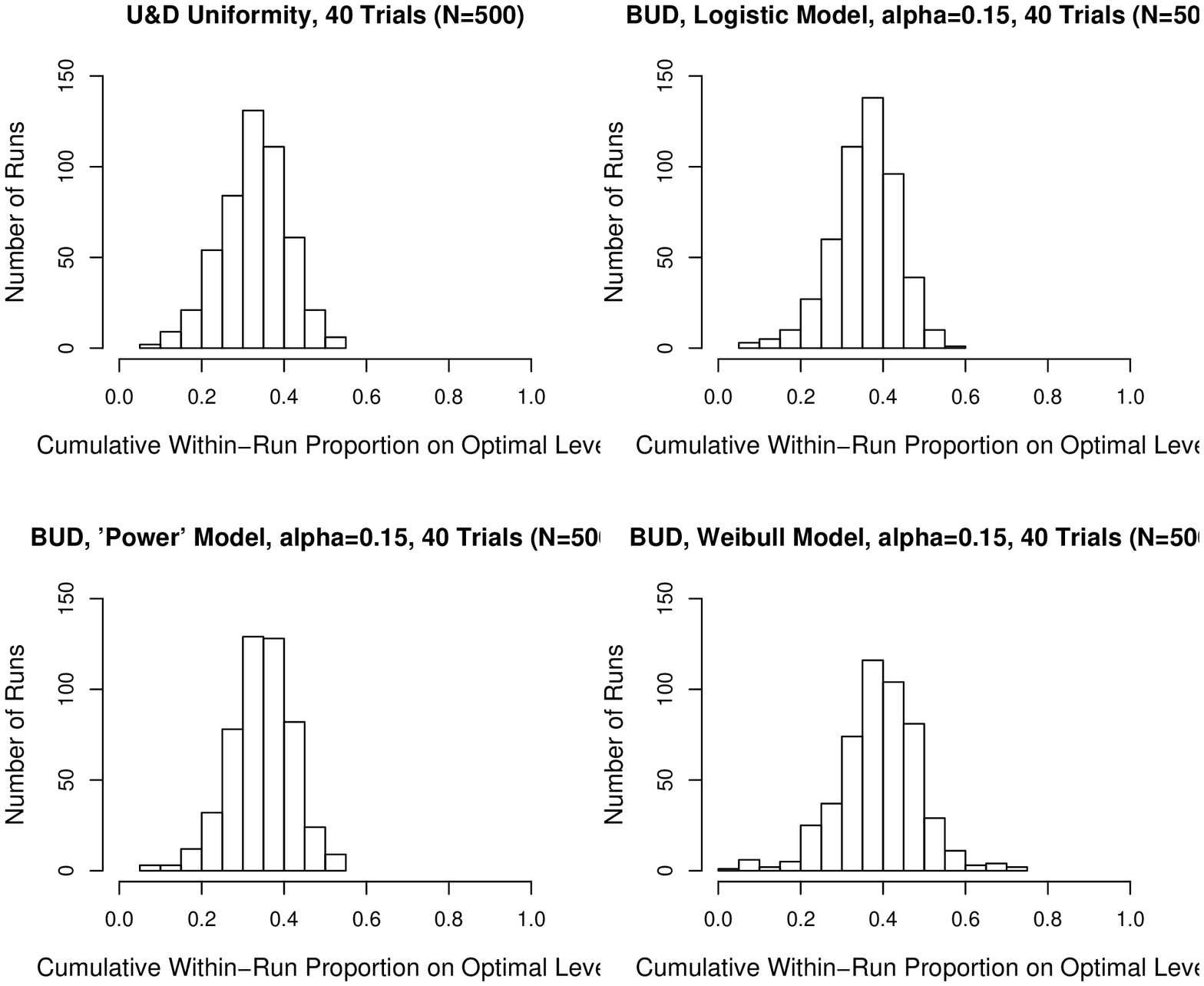}
\\[1cm]
\includegraphics[scale=.5]{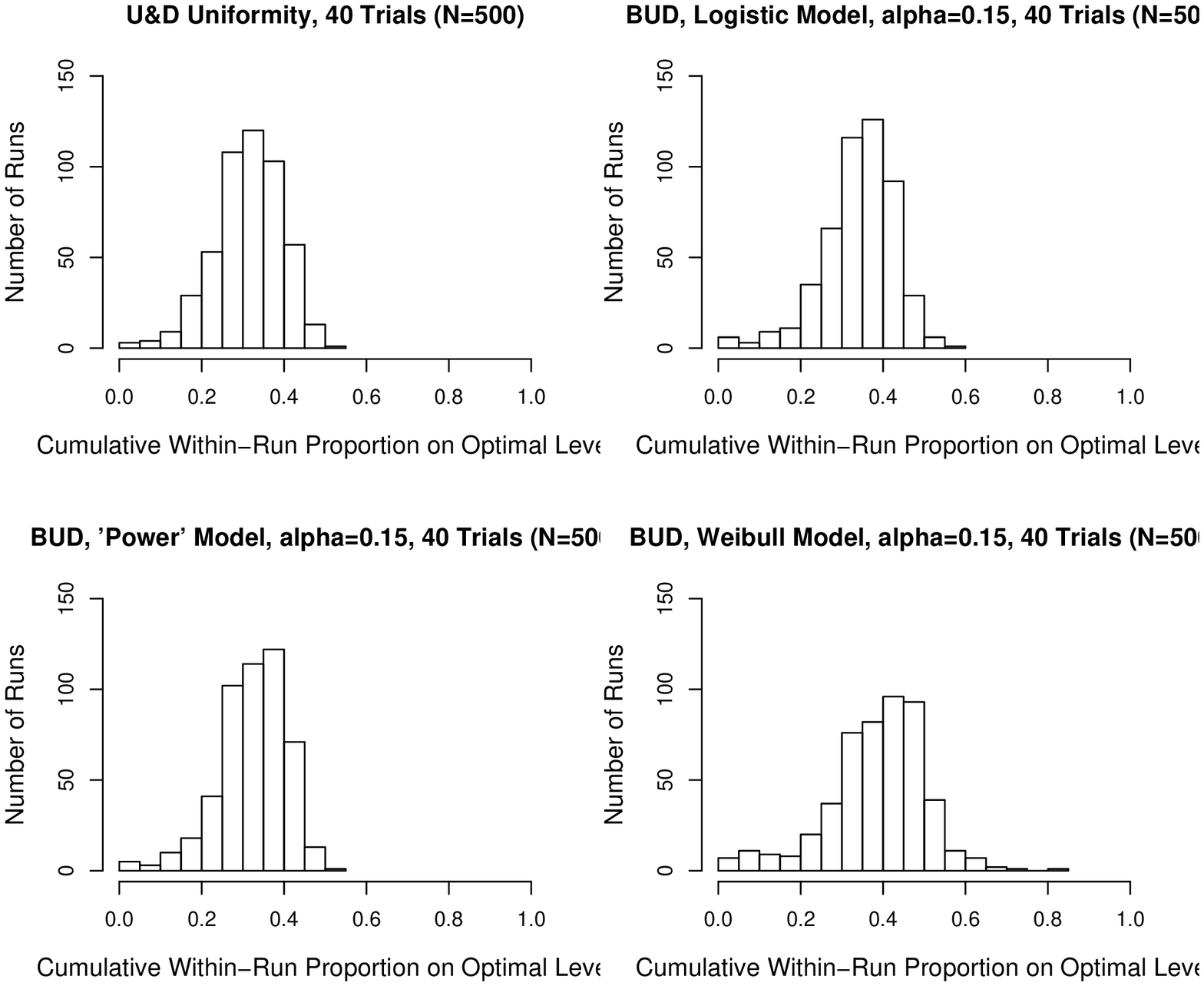}
\caption[Uniformity of BUD Allocations]{Uniformity of U\&D and BUD ($\beta=0.15$) allocations, shown via histograms of $\hat{p}_{u^*}$ from $500$ runs, after $40$ trials. Simulations were non-cohort (or cohort size 1), with normal (top) and gamma, shape $1.2$ (bottom) thresholds. In both cases $m=8$. These are the same scenarios depicted in Fig.~\ref{fig:CRMunif}. Each frame shows histograms of U\&D runs (top left-hand corner), vs. BUD combined with the one-parameter ``power'' model (bottom left-hand corner), two-parameter logistic (top right-hand corner) and shape-scale Weibull (bottom right-hand corner).\label{fig:BUDunif}}
\end{center}
\end{figure}

\begin{figure}
\begin{center}
\includegraphics[scale=.5,angle=-90]{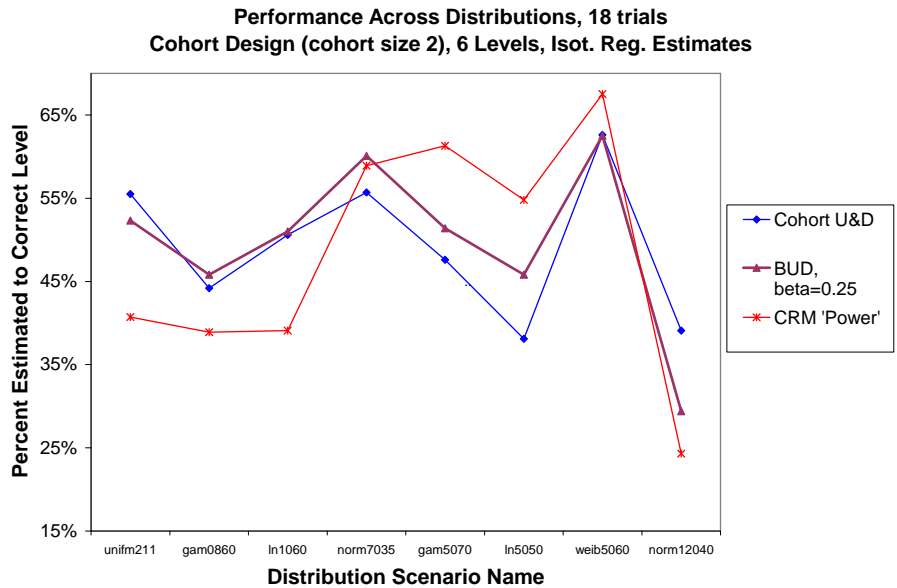}
\\[1cm]
\includegraphics[scale=.5,angle=-90]{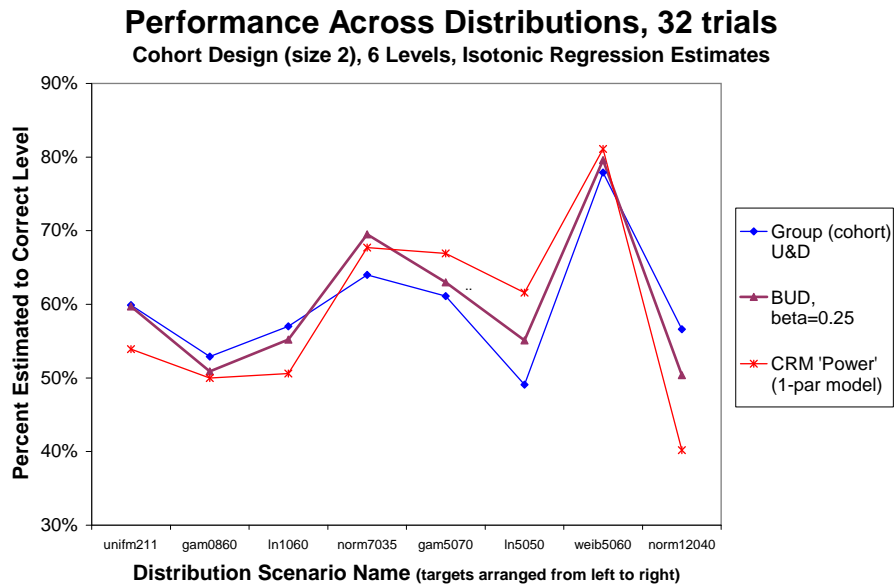}
\caption[GU\&D, CRM, BUD Comparison]{Comparison of U\&D (GU\&D$_{(2,0,1)}$), CRM (with cohorts of size $2$) and BUD ($\beta=0.15,0.25,0.35$) designs on a variety of simulated distributions, and $n=18$ (top) and $32$ (bottom). 'Correct level' was defined as the level whose $F$ value is closest to $0.3$. The design had $6$ levels, with all runs starting at level $2$. Only the 'power' model ($d=1$) was used for CRM and BUD. The CIR estimator was used for U\&D and BUD. Distributions are arranged by target location, so that the leftmost distribution has level $1$ as the optimal level, and the rightmost distribution has level $6$.\label{fig:bud1832}}
\end{center}
\end{figure}

\begin{figure}
\begin{center}
\includegraphics[scale=.5,angle=-90]{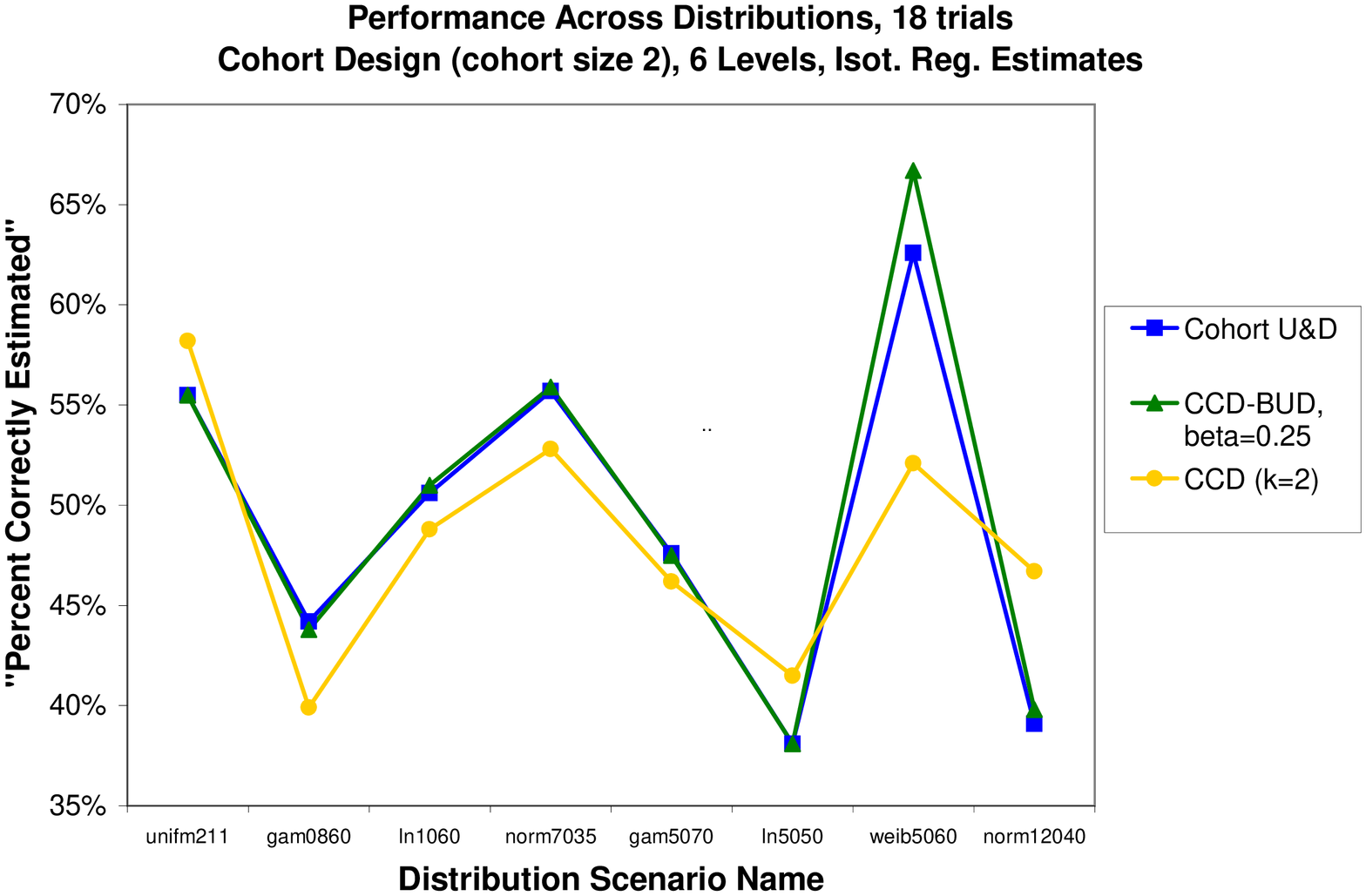}
\\[1cm]
\includegraphics[scale=.5,angle=-90]{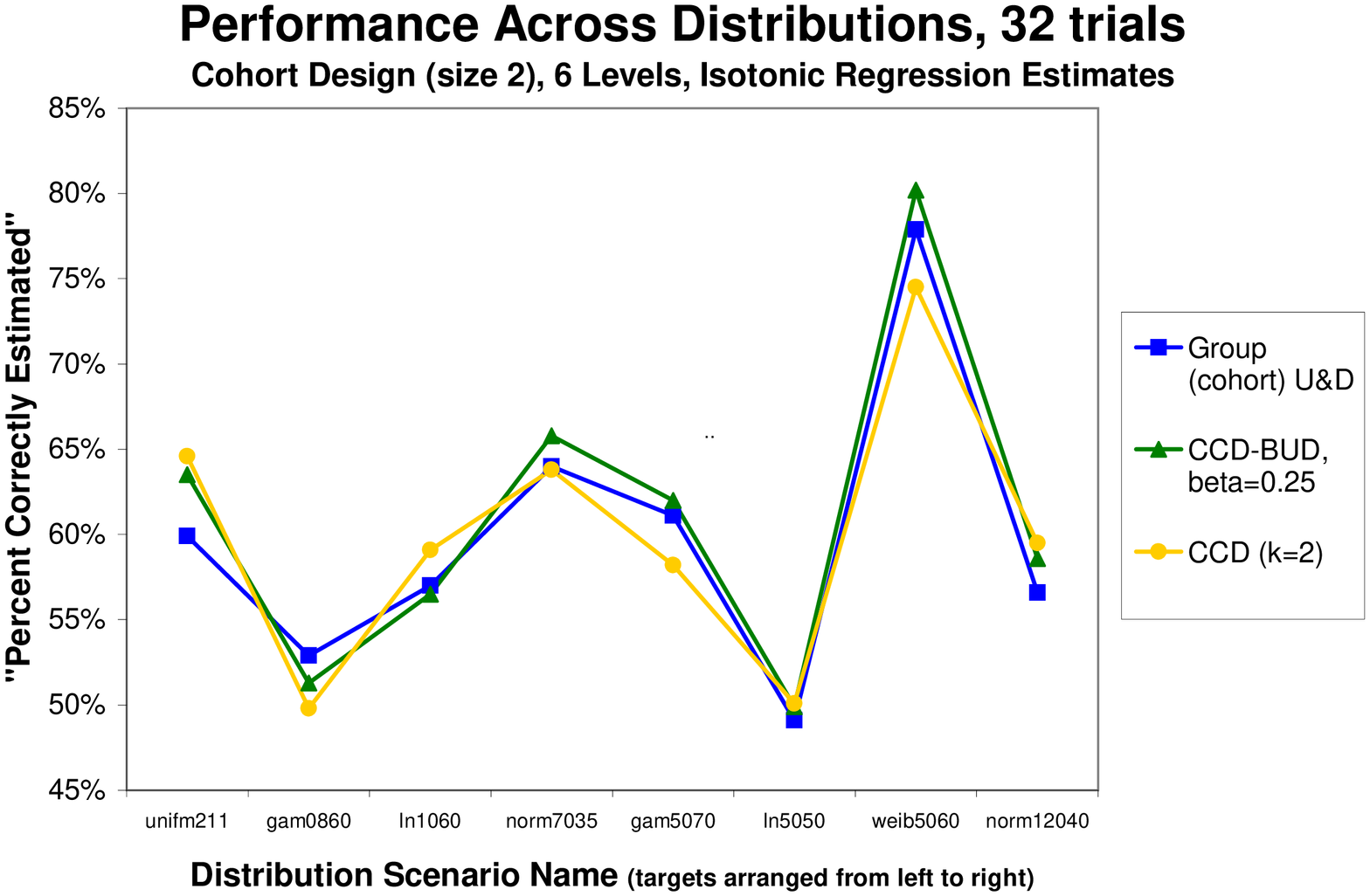}
\caption[GU\&D, CCD, CCD-BUD Comparison]{Comparison of U\&D (GU\&D$_{(2,0,1)}$), CCD (with cohorts of size $2$) and CCD-BUD ($\beta=0.25$) designs on the same thresholds and under the same conditions as in Fig.~\ref{fig:bud1832}.\label{fig:ccd1832}}
\end{center}
\end{figure}

As a demonstration of the pros and cons of CRM and BUD, I present graphically some summaries from numerical simulations. The first (Figs.~\ref{fig:bud2040}, \ref{fig:BUDunif}) is of a one-trial-at-a-time experiment with $m=8$; the second (Figs.~\ref{fig:bud1832}, \ref{fig:ccd1832}) is of a cohort ($k=2$) experiment with $m=6$ and a low starting point, \emph{a l\`{a}} Phase I. Shown are estimation ``success'' probabilities, with ``success'' declared whenever the level closest to $\hat{Q}_p$ is also the level closest to $Q_p$ (a rather standard measure of success in Phase I literature). The most salient observation is that success is very far from being assured; indeed, even after $30-40$ trials, expecting $60\%$ success is a bit optimistic under both frameworks.

In the first example, the focus was on inspecting both BUD (color-coded to blue hues, $\beta=0.15$ in all cases) and various CRM models (shades of red and pink). Here I refrained from simulating conditions with target on the boundary, which is equivalent to assuming that the design did not really have a physical boundary (i.e., treatments could in principle be increased or decreased beyond the nominal boundaries). The three CRM models differ wildly from each other in their performance pattern: one-parameter ``power'' tends to strongly over- or under-perform, while the two-parameter logistic seems most robust, but overall its success rate is relatively low, and seems to improve more slowly than others over time (compare the top figure for $n=20$ with the bottom, $n=40$). The BUD models track pretty closely to U\&D at $n=20$, but post visible gains over it at $n=40$. Interestingly, the one-parameter BUD combination appears to be most robust.

From the same simulations, Fig.~\ref{fig:BUDunif} demonstrates convergence uniformity or robustness to ``unlucky batches'', as in Fig.~\ref{fig:CRMunif}. For BUD runs, $\hat{p}_{u^*}$ is almost as tightly distributed around its mean as for U\&D runs -- but the peak shifts (sometimes almost imperceptibly) to the right. The better the model fit, the more substantial the rightward shift. The risk of very bad runs (very low $\hat{p}_{u^*}$) only slightly increases compared to U\&D.

In the second numerical example I retained only the ``power'' model, since the focus was on Phase I style constraints, with smaller $m$ ($6$) and a low starting level ($l_2$). Here (Fig.~\ref{fig:bud1832}) GU\&D was color-coded blue, CRM red, and BUD in shades of purple and pink, according to the value of $\beta$ ($0.15$ to $0.35$). Additionally, the $x-$ axis was arranged so that the true targets are in increasing order. 'Hard', reflecting physical boundaries were assumed, as is the case in these applications. This immediately suggests using the boundary-resistant CIR for U\&D and BUD estimation.

A clear pattern emerges, with CRM holding the edge when targets are at or above the center of the treatment range, and U\&D better for targets lower than center or on the boundaries. The explanation is quite mundane: CRM's prior has a (broad, but distinct) peak on levels $3-4$. Hence, it has a forfeit advantage if these are indeed the targets, and also to level $5$ (which U\&D has to climb $3$ levels in order to reach). On the boundaries, CRM suffers from having a low prior weight placed there (which is intuitively natural; otherwise, the boundaries would be wider).

How does BUD fare in these simulations? It seems that the design most exposed to CRM override ($\beta=0.35$) is not as robust as smaller exposures, without gaining the advantages of full CRM. Smaller values of $\beta$ seem to remain well within the performance envelope created by U\&D and CRM, regardless of who's on top. Overall, they track closer to U\&D which is understandable given the small $n$, and perform better on the average than either U\&D or CRM.

Fig.~\ref{fig:ccd1832} summarizes simulation results under the same conditions, but comparing U\&D to CCD and CCD-BUD with $\beta=0.25$. Note that the vertical scale is smaller here, because all designs are substantially more robust than CRM in terms of estimation success. CCD ($k=2$), in particular, is the most robust design I've observed from that respect.\footnote{This does not indicate all-around robustness; in terms of run-to-run variability discussed on Section \ref{sec:crmconcept}, CCD is almost as variable as CRM and much less robust than U\&D or BUD. See Fig.~\ref{fig:CCDunif}.} At $n=18$, there are very few overrides with $\beta=0.25$, and as a result CCD-BUD tracks very closely to U\&D. Later on some differences emerge, and again CCD-BUD seems to perform, on the average, better than either U\&D or CCD.

\chapter{Conclusions and Future Directions}\label{ch:conc}

\subsubsection{Percentile-Finding: Limitations and Potential}

Revisiting the framework outlined in the introduction, we now have a better grasp of the problem's limitations and of the way they play out in practice. As the anesthesiology experiment demonstrated (Section~3.4), the variability encountered with live subjects is usually much greater and less tractable than when using a simplistic functional form for $F$, and experimental runs are never as smooth as computer simulation output.

Considering the two main solution paths currently available (U\&D and Bayesian), the choice is between quick convergence, albeit to a random walk that is by definition prone to excursions; and a design that locks onto a single level, without guarantee that it is in fact the best one. In the longer run, as far as estimation precision is concerned the two approaches would usually agree; however, the sequence of treatment allocations during the early phase is almost as different as it gets.

Since the settings are those of an experiment, even if toxicity to live subjects is involved the decision whether to take a risk over the allotted sample has already been taken. Hence, placing ``allocation purity'' above all other considerations appears to be detrimental to the goal of controlling toxicity over a larger population (thanks to Barry Storer for emphasizing this point). Moreover, as Section~4.3 demonstrated, CRM's attempt to identify the best level early on may actually result in increased risk compared with U\&D.

On the other hand, the list of ``best in class'' U\&D designs is short and imposes practical constraints. For example, the fastest-converging cohort design targeting $Q_{0.3}$ has a cohort size of $2$; the next fast-converging option is with cohort size $5$ (GU\&D$_{(5,1,2)}$; \citet{GezmuFlournoy06}). Designs with $k=3$, the most popular cohort size for Phase I studies, are much slower. similarly, for single-trial experiments the best design (KR) can approximate $Q_{0.3},Q_{0.2}$ or smaller percentiles (somewhat underestimating the former and overestimating the latter) -- but cannot work for other below-median targets such as $Q_{0.4}$. Another constraint has to do with boundaries, whose effect must be avoided or mitigated in one of the ways outlined on Chapter 3.

CRM and its nonparametric offshoots (CCD, etc.) are much less constrained in this respect. They perform best (besides the more detailed conditions outlined in Chapter~4) when the spacing is coarse so that $\Delta F$ between levels is around $0.2$ or more. Still, given their large small-sample variability, I would not recommend using an exclusive CRM-type design.

If one plans a highly constrained experiment (i.e., with constraints that hamper U\&D performance), a BUD-type hybrid design offers less risk than CRM/CCD alone. Without these constraints, ``best in class'' U\&D should suffice, possibly with a down-shift scheme to smaller spacing midway through the experiment. The finer points of further optimizing U\&D remain to be explored.

\subsubsection{Centered Isotonic Regression}
Rather early in my work I accidentally discovered the fix described here as centered isotonic regression (CIR). It immediately showed a clear performance improvement; however, its theoretical justification and properties were not as clear. The theoretical treatment presented in Sections 3.2 and 3.3 has been developed gradually since then (aided in part by Marloes Maathuis). Many other ideas and directions related to CIR were not discussed here. This estimator may have applicability beyond U\&D, perhaps as yet another modified-MLE estimator, to be almost automatically preferred over the MLE in ordinary settings due to its practical advantage (such as using $S^2$ rather than $\hat{\sigma}^2$ for normal sample variance estimates).

Tailoring CIR to other applications may mandate different solutions for finding the optimal $x$-coordinate of pooled estimates. This is probably true with binary current-status data. Another interesting perspective is viewing CIR as a form of monotone linear splines, for which the node location is determined automatically. The latter idea may be extended to higher-order splines.

\printendnotes

%
\bibliographystyle{elsart-harv}
\addcontentsline{toc}{chapter}{Bibliography}
\bibliography{phd}
%
%
\appendix
\raggedbottom\sloppy

\addcontentsline{toc}{chapter}{Appendices}
\chapter*{Appendices}
\section*{Appendix A: Proof of KR Unimodality}
In principle this appendix is redundant, but for some reason the question of KR's mode has become controversial and resulted in at least two article rejections - one by Gezmu and Flournoy and one by yours truly. Therefore, a detailed derivation proving that KR has a single stationary mode follows.

As indicated in Theorem \ref{thm:kr1}, the proof looks at the stationary profile $\gamma$ as a continuous function of $F$. Since $F$ is a continuous strictly increasing function of $x$, it suffices to show that $\gamma(F)$ is monotone decreasing. We simplify equation (\ref{eqn:krgamma}) to the form

\begin{equation}\label{eqn:krgammaF}
\gamma(F)=\frac{F(1-F)^k}{F_+\left\{1-(1-F)]^k\right\}},
\end{equation}
where $F_+=F(x+s)$. The derivative $\gamma^{'}(F)$ has the same sign as
\begin{equation}\label{eqn:gamprime1}
\begin{array}{l}
F_+\left\{(1-F)^k+kF(1-F)^{2k-1}\right\}+F_+^{'}F(1-F)^{2k}\\
-F_+\left\{(1-F)^{2k}+kF\left[(1-F)^{k-1}+(1-F)^{2k-1}\right]\right\}-F_+^{'}F(1-F)^k.
\end{array}
\end{equation}
Canceling terms and dividing through by $F_+(1-F)^{k-1}$ we get
\begin{equation}\label{eqn:gamprime2}
\begin{array}{l}
1-F+\frac{F_+^{'}}{F_+}F(1-F)^{k+1}
-(1-F)^{k+1}-kF-\frac{F_+^{'}}{F_+}F(1-F).\\
<1-F-(1-F)^{k+1}-kF.
\end{array}
\end{equation}
This expression is zero at $F=0$, and has derivative $(k+1)\left[(1-F)^k-1\right]<0$ for $F>0$. Therefore it is negative for $F>0$, and by extension $\gamma(x)$ is decreasing. Therefore, by Lemma \ref{lem:mode}, KR has a single stationary mode as claimed in Theorem \ref{thm:kr1}. 
\section*{Appendix B: Code for The New Estimators}
\subsection*{Code for Auto-Detect Estimator}

Note that the AD estimator can be used not only for U\&D chains, but for any vector that has a starting-point effect and eventually converges to some stationary behavior. In that latter case, however, care must be taken in interval estimation. One should use the autocorrelation-based estimate of $n_{eff}$ (\verb"se1" in the code) and not the ``hitting time'' estimate which is based on U\&D theory.

\singlespacing
{\small
\begin{verbatim}
ADmean<-function (x,before=T,safe=length(x)/4,full=T) {

#### REMOVING 'FRONT TAIL' AND RETURNING MEAN OF REMAINING VALUES
# Assaf Oron, 10/2007
# ARGUMENT LIST
# x: The vector to be averaged
# before: if true (default), averaging begins from one position before the
# identified transition point
# safe: a position beyond which the algorithm stops looking. This means that
# at least length(x)-safe+1 elements will be included in averaging. Defaults
# to length(x)/4.
# full: if true, output includes information needed for for interval estimation.
# Otherwise, only point estimate is returned.

cutoff=round(safe)+1
n=length(x)
base=sign(x[1]-mean(x[2:n]))
a=2

while (a<cutoff & sign(x[a]-mean(x[(a+1):n]))!= -(base)) a=a+1

if (before) a=a-1
if (!full) return(outmean) ### POINT ESTIMATE ONLY

#### NOW THE TOUGH PART - VARIANCE ESTIMATION
#### WE BASE IT ON THE SD AND TWO ESTIMATES OF EFFECTIVE SAMPLE SIZE
#### IN ANY CASE, WE USE ONLY THE AVERAGED PART OF x

### THE FIRST IS AUTOCORRELATION BASED; WE LOOK AT 1ST AND 2ND DEGREE

xx=x[a:n]
nn=length(xx)

cor1=max(cor(xx[1:(nn-1)],xx[2:nn]),0)
cor2=max(cor(xx[1:(nn-2)],xx[3:nn]),0)
corfac=max(cor1,sqrt(cor2))

### THE FOLLOWING SAMPLE SIZE ESTIMATE IS BASED ON DERIVATION IN THESIS
eff_n1=max(nn*(1-corfac)/(1+corfac),1)

sig2=var(xx)
se1=sqrt(sig2/eff_n1)

############
### NOW, CALCULATING S.E. VIA INTERVALS BETWEEN HITTING TIMES
### AT A GIVEN LEVEL (MARKOV CHAIN THEORY)

firsts=c(ifelse(a==1,1,ifelse(x[a]==x[a-1],0,1)),diff(xx))

tmp=table(xx[firsts!=0])
unq=sort(unique(xx[firsts!=0]))
dists=sort(abs(unq-mean(xx)),index.return=T)

### The effective i.i.d. sample size is the mean number of
### hitting intervals for the two levels closest to our estimate
eff_n2=tmp[dists$ix[1]]-1
closest=unq[dists$ix[1]]

piececut=ifelse(firsts!=0 & xx==closest,1,0)
#cat(xx,'\n')
#cat(piececut,'\n')
pieceid=cumsum(piececut)
piecelens=sapply(split(xx,pieceid),length)

#### THIS, TOO, IS BASED ON A DERIVATION FOUND IN THE THESIS
se2=sqrt(eff_n2*(mean(piecelens^2)*(sig2+(outmean-closest)^2)
+max(outmean,closest)^2*var(piecelens))/nn^2)

####### NOW WE TAKE THE MORE CONSERVATIVE S.E. ESTIMATE; BOTH WILL BE RETURNED ANYWAY
se=max(se1,se2)

#### THE D.F. ESTIMATE FOR T-STAT IS TAKEN FROM THE HITTING TIME APPROACH
df=max(eff_n2-1,1)
return(c(outmean,se,df,a,se1,se2,eff_n1,eff_n2))
}
\end{verbatim} }
\onehalfspacing
\subsection*{Code for CIR Estimator}

First, a generic code for forward estimation (point estimate only). This code is completely analogous to the well-known PAVA code; in fact, parts of the R \verb"pava()" code are copied and pasted here.

\singlespacing
{\small
\begin{verbatim}
cir.pava <- function (y,x, wt=rep(1,length(x)),boundary=2,full=FALSE) {
# Returns centered-isotonic-regression y values at original x points #
# Assaf Oron, 10/2007
#
### ARGUMENTS:
# y: y values (raw point estimates). Must match the x values.
# They are given as first argument, both for compatibility with 'pava'
# and to enable parallel running via 'apply' type routines
# x: treatments. Need to be pre-sorted in increasing order
# wt: weights.
# boundary: action on boundaries. Defaults to 2,
# which is analogous to 'rule=2' on function 'approx', i.e.
# returned y values are constant outside the boundaries.
# boundary=1 does linear extrapolation.
# In addition, one can impose boundaries as inputs or
# augment the output with boundaries, as discussed in
# the dissertation text.
# full: if FALSE, only point estimates at x values are returned

### Validation stuff
n <- length(x)
if (n <= 1) {
if (!full) return (y)
else return(list(x=x,y=y,z=x))
}
if (any(is.na(x)) || any(is.na(y))) {
    stop ("Missing values in 'x' or 'y' not allowed")    }
if (any(diff(x)<=0)) {stop ("x must be strictly increasing")}

z<-x  # Keeping a 'clean' copy of x for final stage

lvlsets <- (1:n)
repeat {
    viol <- (as.vector(diff(y)) <= 0) # Find adjacent violators
    if (!(any(viol))) break
    i <- min( (1:(n-1))[viol]) # Pool first pair of violators
    y[i] <- (y[i]*wt[i]+y[i+1]*wt[i+1]) / (wt[i]+wt[i+1])
    x[i] <- (x[i]*wt[i]+x[i+1]*wt[i+1]) / (wt[i]+wt[i+1])  # new x is calculated
    wt[i]<-wt[i]+wt[i+1]  # weights are combined

# Deleting the i-1-th element
    y<-y[-(i+1)]
    x<-x[-(i+1)]
    wt<-wt[-(i+1)]
    n <- length(y)
    if (n <= 1) break
  }

if (boundary==1) {

### Utilize this option if you wish to use linear extrapolation
### outside the boundary
### (the 'approx' function does not have this option)
### In general, this is *not* recommended;
### rather, impose boundary conditions whenever possible
### (as inputs or after output of this function)
### or use the default, constant boundary conditions

    if (x[n]<max(z)) {
        x<-c(x,max(z))
        y<-c(y,y[n]+(y[n]-y[n-1])*(x[n+1]-x[n])/(x[n]-x[n-1])) }
    if (x[1]>min(z)) {
        x<-c(min(z),x)
        y<-c(y[1]-(y[2]-y[1])*(x[2]-x[1])/(x[3]-x[2]),y) }
}

# Now we re-interpolate to original x values, stored in z
# If we didn't set boundary=1, then this will give constant
# y values for x values falling outside new range of x

if (!full) return(approx(x,y,z,rule=2)$y)

else return(list(x=x,y=y,z=z,wt=wt))
}
\end{verbatim} }
\onehalfspacing

Now, inverse estimation specifically tailored for the U\&D application. The input here is as a yes-no summary table of the binary responses. The output is a vector, with the first element an inverse point estimate of $Q_p$, and the other elements percentile estimates of $Q_p$ for percentiles given by the user (for interval estimation).

This code includes interval estimation as discussed in the text - again, tailored for U\&D in which treatment allocation is random. The same principles can be applied to generate interval estimates for non-random allocations.

The code uses some small utilities that follow below.
\singlespacing
{\small
\begin{verbatim}
cir.upndown<-function(yesno,xseq,target,xbounds=c(0,1),ybounds=c(0,1),
    full=F,cioption="poisson",plist=c(.025,.975)) {
# Centered-isotonic-regression for Up-and-Down
# Assaf Oron, 10/2007
# Returns Point & Interval Estimates of Target percentile

### ARGUMENTS:
# yesno: Yes-no table of binary responses. Can contain rows of zeros
# xseq: Treatment values matched to the responses
# target: The target response rate (between 0 and 1)
# xbounds,ybounds: used for interpolation in case (estimated)
# target falls outside CIR output boundaries.
# full: complete output or estimates only
# cioption: which method to use for interval estimation
# can choose from 'poisson', 't' or binomial (any other string)
# plist: percentile list for interval estimation. Is a vector
# of any length

n_m<-yesno[,1]+yesno[,2]
x<-xseq[n_m>0]
y<-yesno[n_m>0,1]/n_m[n_m>0]

xbounds[1]<-min(xbounds[1],min(x))
xbounds[2]<-max(xbounds[2],max(x))

### Point Estimate  ###########################

# We start via forward estimation, just calling 'cir.pava'

pavout<-cir.pava(y,x,wt=n_m[n_m>0],full=TRUE)

newx<-pavout$x
newy<-pavout$y
newn=pavout$wt

### Error control if the interpolation target is outside boundary
if (min(newy)>target) {
    newx<-c(xbounds[1],pavout$x)
    newy=c(ybounds[1],pavout$y)
    newy=c(ybounds[1],pavout$y)
    newn<-c(1,pavout$wt)
}
mm=length(newy)
if (max(newy)<target) {
    newx<-c(newx,xbounds[2])
    newy=c(newy,ybounds[2])
    newn=c(newn,1)
}
### The estimate is generated by using 'approx' with x and y interchanged

out=ifelse(length(newy)==1,newx,approx(x=newy,y=newx,
    xout=target,ties="ordered",rule=2)$y)

### Now to Interval Estimate ###########################

# We find the point estimate's location w.r.t. our grid

yplace=min(max(rank(c(target,newy),ties.method="max")[1]-1,1),length(newy)-1)
minn=min(newn[yplace],newn[yplace+1],na.rm=T)

if (length(newy)==1 || newy[yplace+1]==newy[yplace])  {
### degenerate cases (error control)

    cigaps=ifelse(plist<0.5,xbounds[1]-xbounds[2],xbounds[2]-xbounds[1])
#   }
} else {

    yplratio=(target-newy[yplace])/(newy[yplace+1]-newy[yplace])

    if(cioption=='t') {
        width=smoothqt(plist,minn,target)/minn

    } else if (cioption=='poisson') {
        width=smoothqpois(plist,minn,target)/minn-target

    } else width=smoothqbinom(plist,minn,target)/minn-target

### Inverting the forward interval estimates

    cigaps=width*(newx[yplace+1]-newx[yplace])/(newy[yplace+1]-newy[yplace])
}

#### Output

if (!full) { return(c(out,out+cigaps))
} else return (list(raw=yesno,paved=pavout,out=out,ci=out+cigaps))
}
\end{verbatim} }
\onehalfspacing

Finally, the utilities used for interval estimation.

\singlespacing
{\small
\begin{verbatim}
smoothqbinom=function(p,size,prob,add=T,half=F) {
q1=qbinom(p,size,prob)
p1=pbinom(q1,size,prob)

### We smooth out the binomial quantile function, because n is random

p2=pbinom(q1-1,size,prob)
out=q1-(p1-p)/(p1-p2)
if(add==T) {
        out[p>0.5]=out[p>0.5]+2
        out[p<0.5]=out[p<0.5]-1

}
if (half==T) out=out-0.5

#### This part to ensure we are still conservative compared with
#### the traditional quantile function

out[p<0.5]=ifelse(out[p<0.5]<q1[p<0.5],out[p<0.5],q1[p<0.5])
out[p>0.5]=ifelse(out[p>0.5]>q1[p>0.5],out[p>0.5],q1[p>0.5])

out
}
####
smoothqt=function(p,size,prob) { qt(p,df=size-1)*sqrt(size*prob*(1-prob)) }
####

smoothqpois=function(p,size,prob,add=F) {

refp=c(0.5,ifelse(p<0.5,1-p,p))
q1=qpois(refp,lambda=size*prob)
p1=ppois(q1,lambda=size*prob)
p2=ppois(q1-1,lambda=size*prob)

### We smooth out the Poisson quantile function, because n is random

out=q1-(p1-refp)/(p1-p2)

extra=ifelse(add==T,1,0)

ifelse(p>0.5,out[2:length(out)]+extra,2*out[1]-out[2:length(out)]-extra)+
    size*prob-out[1]

}
\end{verbatim} }
\onehalfspacing

\section*{Appendix C: Some Technical Details of CRM and BUD simulations}
A wider range of CRM models than that which appears in Chapter \ref{ch:crm} figures was simulated and contemplated during my work. In the end, I presented results from only three models -- \citet{OQuigleyEtAl90}'s original one-parameter model, location-scale logistic and shape-scale Weibull -- for the following reasons:
\begin{itemize}
\item Only these three have been used in practice a reasonable number of times (the Weibull model is popular in sensory studies' QUEST);\footnote{Here I implicitly count the ''power'' model used nowadays as a generalization of \citet{OQuigleyEtAl90}'s one-parameter model.}
\item Chapter \ref{ch:crm}'s theoretical results indicate little benefit from models with more than 2-3 parameters;
\item Simulation running times increase exponentially with the number of parameters, as will be detailed below.
\end{itemize}

The implementation involves calculation of posterior statistics. In terms of both computing time and precision, the worst-case default is MCMC. This is reasonably feasible option for a single actual experiment. However, in simulation (or sensitivity analysis) the calculation time could be prohibitive: for each single run, one has to simulate quite a few MCMC cycles (it seemed that no fewer than $20,000$ cycles are usually needed), for each single \emph{trial}. Moreover, as stated we are only interested in a handful of statistics at most -- so having the complete posterior on our hands (as MCMC provides) is quite an overkill. Therefore, researchers in this field naturally look for shortcuts. Computationally the easiest solution is to find a closed-form expression. For example, \citet{GaspariniEisele00} simulated a saturated model using a ''product of betas'' prior that enabled them to find a closed-form expression for the posterior mean.

The next-fastest solution is numerical integration. The posterior mean, either of $\theta$, of $\{G_m\}$ or of $Q_p$, can be easily calculated using a ratio of two integrals. Posterior percentiles of $Q_p$ can also be found this way -- by converting the likelihood equation so that $Q_p$ is one of the parameters, and integrating it over intervals (or parameter-space slices) of the form $\left[l_u-s/2,l_u+s/2\right]$. Numerical integration is typically more accurate than MCMC (tested over known prior distributions). The computing time increases exponentially with $d$, the number of parameters. For $d=1$, a calculation that would take days using MCMC takes only minutes via R's \verb"integrate()" function. For $d=2$, it takes hours using \verb"adapt()" -- which is still a factor of $5-10$ shorter than MCMC. For $d\geq 3$ the integration approach becomes more cumbersome symbolically, less stable numerically, and may in fact take more time than MCMC.

In implementing BUD, I cut a few corners -- while the CRM allocation was usually calculated using the ``closest response'' criterion mandated by theory, I calculated posterior percentiles (to determine override) using the ``closest treatment'' criterion, because this could be done via numerical integration rather than MCMC. Sensitivity analysis showed that the difference from doing a ``closest response'' calculation are quite small, and since the focus was on a proof of concept the extra computing expense was not worthwhile.

\vita{
\singlespacing
\begin{footnotesize}
Assaf Peretz Oron was born in Boston on May 23, 1966, to Rachel and Mosh\'{e} Oron -- a young Israeli couple in America for studies. Assaf's first name is from the Bible: the Biblical Assaf was the author of many Psalms. Assaf's middle name is in memory of his paternal great-grandfather Peretz Hochman, who was murdered by the Nazis in March 1942 at the Be{\l}\.{z}ec death camp, together with most inhabitants of his hometown of Piaski in Poland, where Mosh\'{e} was born four years earlier (Assaf's grandparents had fortunately escaped Poland with their children in fall 1939).\footnote{\scriptsize{About half a million Polish and Ukrainian Jews were murdered in Be{\l}\.{z}ec during its single year of operation, with only two known survivors -- a blood-curdling death-factory efficiency of over $99.9995\%$. Be{\l}\.{z}ec is virtually unknown to the general public, despite being the place where the industrial death-camp template was launched.}}

Assaf's last name ``Oron'' literally means ``small light'', and was chosen by his parents shortly after their marriage, sensing -- like many young Israelis at the time -- that a modern Hebrew name reflects their self-identity better than their parents' European names. Assaf is the second of four, having an older brother and younger sister and brother.
He grew up mostly in Jerusalem, but was exposed to American culture as well thanks to a sabbatical year spent in California in 1979-1980.\footnote{\scriptsize{At the time, the ``Silicon Valley'' was not so multicultural, and teaching Americans how to pronounce his name was a mission beyond Assaf's capabilities. He settled for whatever they came up with, which included ``Ozzie''.}}

From 1985 to 1988, Assaf served 3 mandatory years in the Israel Defense Forces (IDF), volunteering to infantry. During the time this dissertation was written in summer 2007, a debate raged in the Israeli media around a reported increase in draft evasions. Assaf weighed in with a piece published on the Israeli conscience-objector website YeshGvul.org, recalling his IDF experience which he summarized in these words:
\begin{quote}
... a variation on ``Catch 22'':  a hollow experience, stupid to the point of absurdity and beyond, callous to human life, brutal, mentally destabilizing, humiliating and corrupting.
\end{quote}
Despite this, Assaf continued to serve in the IDF's combat reserve. In 2001 Assaf pledged to refuse serving in what he saw (and still sees) as the oppression of Palestinians under Israeli Occupation. However, to the best of his knowledge, Assaf is still formally on the IDF combat reserve roster.

In 1989, Assaf undertook academic studies at the Hebrew University in Jerusalem. He majored in Physics and the ``Amirim'' interdisciplinary excellence program, graduating \emph{cum laude} in 1992. Also during these years, Assaf took elective courses in archaeology, where he was fortunate to meet the love of his life and future partner, Orna Eliyahu -- an Archaeology and Art History major. They were married in 1994, and have since produced Daniel (1997), Guy (1998) and Ben (2006).

After a stint as an outdoor-ed instructor in the Israeli public school system, Assaf entered graduate studies in the Department for Environmental Sciences and Energy Research (ESER) at the Weizmann Institute of Science in Rehovot, Israel. His thesis, completed in 1997, was about theoretical (or, as he calls it, ``dry'') hydrology, under the guidance of Brian Berkowitz.
Assaf's next 5 years were spent in ``proper jobs'' -- first as a process engineer in Intel factories, then as the algorithm project manager in a bioinformatic startup currently known as Procognia (Israel).

In September 2002, Assaf returned to studies, taking up the statistics Ph.D. track at the University of Washington in Seattle.\footnote{\scriptsize{His family arrived with him; Orna is currently making pottery and selling it at Pike Place Market.}} After a grueling first year, and a couple of years beating around the bush with potential dissertation subjects such as game theory and clustering, Assaf realized that the one topic with which he's become obsessed is that pesky consulting project about 'Up-and-Down'. Fortunately for him, his advisor Peter Hoff and the graduate coordinator Jon Wellner were quite accommodating, and that is how you ended up with this book.

\end{footnotesize}
}

\end{document}